\documentclass[12pt]{article}

\usepackage{verbatim,color,amssymb}
\usepackage{amsmath}					
\usepackage{amsthm}					
\usepackage{natbib}
\usepackage{multirow}
\usepackage{setspace}
\usepackage[mathscr]{euscript}
\usepackage{fancyhdr}
\usepackage{enumitem}
\usepackage{graphicx}
\usepackage{geometry}
    \usepackage[bookmarks=false]{hyperref}
\usepackage{soul}
\usepackage{sidecap}

\usepackage{lscape}

\usepackage{subfig}
\usepackage{tikz}
\usetikzlibrary{arrows,chains,backgrounds,fit}

\newtheorem{Thm}{\underline{\bf Theorem}}

\newtheorem{Cond}{\underline{\bf Conditions}}

\newtheorem*{Proof*}{Proof}

\newtheorem{Lem}{\underline{\bf Lemma}}
\newtheorem{Cor}{\underline{\bf Corollary}}

\def\nN{\mathbb{N}}
\def\rR{\mathbb{R}}

\def\C{{\cal C}}

\def\F{{\cal F}}

\def\N{{\cal N}}
\def\S{{\cal S}}
\def\T{{\cal T}}

\def\X{{\cal X}}

\def\diag{\hbox{diag}}
\def\Ind{\hbox{I}}
\def\wh{\widehat}
\def\wt{\widetilde}

\def\diag{\hbox{diag}}
\def\log{\hbox{log}}

\def\var{\hbox{var}}
\def\cov{\hbox{cov}}
\def\corr{\hbox{corr}}
\def\sign{\hbox{sign}}
\def\trace{\hbox{trace}}

\def\Beta{\hbox{Beta}}

\def\Dir{\hbox{Dir}}

\def\Ga{\hbox{Ga}}

\def\IG{\hbox{Inv-Ga}}
\def\IW{\hbox{IW}}
\def\MVN{\hbox{MVN}}
\def\MVL{\hbox{MVL}}
\def\MVT{\hbox{MVT}}
\def\Normal{\hbox{Normal}}

\def\Mult{\hbox{Mult}}
\def\Wish{\hbox{W}}

\def\DP{\hbox{DP}}

\def\ANNALS{{\it Annals of Statistics}}

\def\ANNALSAS{{\it Annals of Applied Statistics}}
\def\ANNALSISM{{\it Annals of the Institute of Statistical Mathematics}}
\def\AJE{{\it American Journal of Epidemiology}}

\def\BA{{\it Bayesian Analysis}}

\def\BIOK{{\it Biometrika}}

\def\CANADAJS{{\it Canadian Journal of Statistics}}

\def\CDA{{\it Computational Statistics \& Data Analysis}}

\def\ECTH{{\it Econometric Theory}}

\def\JASA{{\it Journal of the American Statistical Association}}
\def\JRSSB{{\it Journal of the Royal Statistical Society, Series B}}

\def\JCGS{{\it Journal of Computational and Graphical Statistics}}

\def\JMA{{\it Journal of Multivariate Analysis}}
\def\JNS{{\it Journal of Nonparametric Statistics}}
\def\JSS{{\it Journal of Statistical Software}}
\def\JECM{{\it Journal of Econometrics}}

\def\IEEESPL{{\it IEEE Signal Processing Letters}}
\def\IEEETIT{{\it IEEE Transactions on Information Theory}}

\def\P_25_ICML{{\it Proceedings of the 25th international conference on Machine learning}}

\def\SCAN{{\it Scandinavian Journal of Statistics}}

\def\STATSCI{{\it Statistical Science}}
\def\SSNC{{\it Statistica Sinica}}
\def\SaC{{\it Statistics and Computing}}
\def\STATSCI{{\it Statistical Science}}

\def\refhg{\hangindent=20pt\hangafter=1}
\def\refmark{\par\vskip 2mm\noindent\refhg}

\def\refhg{\hangindent=20pt\hangafter=1}
\def\refmark{\par\vskip 2mm\noindent\refhg}

\def\bse{\begin{eqnarray*}}
\def\ese{\end{eqnarray*}}
\def\be{\begin{eqnarray}}
\def\ee{\end{eqnarray}}
\def\bq{\begin{equation}}
\def\eq{\end{equation}}

\def\wh{\widehat}

\def\trans{^{\rm T}}

\def\th{^{th}}
\def\bone{{\mathbf 1}}

\def\b1e{{\mathbf e}}
\def\bA{{\mathbf A}}
\def\bB{{\mathbf B}}
\def\bc{{\mathbf c}}
\def\bC{{\mathbf C}}
\def\bd{{\mathbf d}}
\def\bD{{\mathbf D}}
\def\bI{{\mathbf I}}

\def\bP{{\mathbf P}}

\def\bs{{\mathbf s}}
\def\bS{{\mathbf S}}
\def\bt{{\mathbf t}}

\def\bU{{\mathbf U}}
\def\bV{{\mathbf V}}

\def\bW{{\mathbf W}}

\def\bX{{\mathbf X}}

\def\bY{{\mathbf Y}}
\def\bz{{\mathbf z}}
\def\bZ{{\mathbf Z}}
\def\bS{{\mathbf S}}
\def\bzero{{\mathbf 0}}

\def\tmu{\widetilde{\mu}}

\newcommand{\etam}{\mbox{\boldmath $\eta$}}
\newcommand{\bmu}{\mbox{\boldmath $\mu$}}
\newcommand{\bDelta}{\mbox{\boldmath $\Delta$}}

\newcommand{\bpi}{\mbox{\boldmath $\pi$}}

\newcommand{\bxi}{\mbox{\boldmath $\xi$}}
\newcommand{\bepsilon}{\mbox{\boldmath $\epsilon$}}
\newcommand{\btheta}{\mbox{\boldmath $\theta$}}
\newcommand{\bbeta}{\mbox{\boldmath $\beta$}}

\newcommand{\bzeta}{\mbox{\boldmath $\zeta$}}

\newcommand{\bSigma}{\mbox{\boldmath $\Sigma$}}

\newcommand{\blambda}{\mbox{\boldmath $\lambda$}}
\newcommand{\bLambda}{\mbox{\boldmath $\Lambda$}}
\newcommand{\bOmega}{\mbox{\boldmath $\Omega$}}
\newcommand{\bPsi}{\mbox{\boldmath $\Psi$}}

\newcommand{\bGamma}{\mbox{\boldmath $\Gamma$}}
\newcommand{\btau}{\mbox{\boldmath $\tau$}}

\newcommand{\abs}[1]{\left\vert#1\right\vert}
\newcommand{\norm}[1]{\left\Vert#1\right\Vert}

\renewcommand\footnoterule{\kern-3pt \hrule \textwidth 2in \kern 2.6pt}

\def\boxit#1{\vbox{\hrule\hbox{\vrule\kern6pt \vbox{\kern6pt \textcolor{blue}{#1}\kern6pt}\kern6pt\vrule}\hrule}}

\def\authorfootnote#1{{\let\thefootnote\relax\footnotetext{#1}}}

\allowdisplaybreaks

\begin{document}
\thispagestyle{empty}
\baselineskip=28pt

\begin{center}
{\LARGE{\bf Bayesian Semiparametric Multivariate Density Deconvolution}}
\end{center}
\baselineskip=12pt

\vskip 2mm
\begin{center}
Abhra Sarkar\\
Department of Statistical Science, Duke University, Durham,\\ NC 27708-0251, USA\\
abhra.sarkar@duke.edu \\
\hskip 3mm \\
Debdeep Pati\\
Department of Statistics, Florida State University, Tallahassee,\\ FL 32306-4330, USA\\
debdeep@stat.fsu.edu\\
\hskip 3mm \\
Bani K. Mallick\\
Department of Statistics, Texas A\&M University, 3143 TAMU, College Station,\\ TX 77843-3143, USA\\
bmallick@stat.tamu.edu\\
\hskip 3mm \\
Raymond J. Carroll\\
Department of Statistics, Texas A\&M University, 3143 TAMU, College Station,\\ TX 77843-3143, USA\\
and School of Mathematical and Physical Sciences, University of Technology Sydney, Broadway NSW 2007, Australia\\
carroll@stat.tamu.edu\\
\end{center}

\vskip 4mm
\begin{center}
{\Large{\bf Abstract}}
\end{center}
\baselineskip=12pt
We consider the problem of multivariate density deconvolution when interest lies in estimating the distribution of a vector valued random variable $\bX$
but precise measurements on $\bX$ are not available, observations being contaminated by measurement errors $\bU$.
The existing sparse literature on the problem assumes the density of the measurement errors to be completely known.
We propose robust Bayesian semiparametric multivariate deconvolution approaches 
when the measurement error density of $\bU$ is not known but replicated proxies are available for at least some individuals. 
Additionally, we allow the variability of $\bU$ to depend on the associated unobserved values of $\bX$ through unknown relationships, 
which also automatically includes the case of multivariate multiplicative measurement errors. 
Basic properties of finite mixture models, multivariate normal kernels and exchangeable priors are exploited in novel ways 
to meet modeling and computational challenges. 
Theoretical results showing the flexibility of the proposed methods in capturing a wide variety of data generating processes are provided. 
We illustrate the efficiency of the proposed methods in recovering the density of $\bX$ through simulation experiments.
The methodology is applied to estimate the joint consumption pattern of different dietary components from contaminated 24 hour recalls.
%
%
Supplementary materials present substantive additional details.

\baselineskip=12pt
\par\vfill\noindent
\underline{\bf Some Key Words}: B-splines, Conditional heteroscedasticity, Latent factor analyzers, Measurement errors, Mixture models, Multivariate density deconvolution,
Regularization, Shrinkage.
\par\medskip\noindent
\underline{\bf Short Title}: Multivariate Density Deconvolution
\par\medskip\noindent

\clearpage\pagebreak\newpage
\pagenumbering{arabic}
\newlength{\gnat}
\setlength{\gnat}{17pt}
\baselineskip=\gnat

\section{Introduction}

Many problems of practical importance require estimation of the density $f_{\bX}$ of a vector valued random variable $\bX$.
Precise measurements on $\bX$ may not, however, be available, observations being contaminated by measurement errors $\bU$.
Under the assumption of additive measurement errors, 
the observations are generated from a convolution of the density $f_{\bX}$ of $\bX$ and the density $f_{\bU}$ of the measurement errors $\bU$.
The problem of estimating the density $f_{\bX}$ from available contaminated measurements then becomes a problem of multivariate density deconvolution.

This article proposes novel Bayesian semiparametric density deconvolution approaches 
based on finite mixtures of latent factor analyzers 
for robust estimation of the density $f_{\bX}$
when the measurement error density $f_{\bU}$ is not known, 
but replicated proxies contaminated with measurement errors $\bU$ are available for at least some individuals. 
The proposed deconvolution approaches are highly robust, not having to impose restrictive parametric assumptions on $f_{\bX}$ or $f_{\bU}$. 
Additionally, the variability of $\bU$ is allowed to depend on the associated unobserved values of $\bX$ through unknown relationships.

While the focus of the article will primarily be on additive measurement errors,
importantly, the methodology for additive conditionally heteroscedastic measurement errors developed here also automatically encompasses the case of multivariate multiplicative measurement errors. 

To the best of our knowledge, all existing multivariate deconvolution approaches assume that $\bU$ is independent of $\bX$ and that the error density $f_{\bU}$ is completely known. 
Ours is thus the first paper that allows the density of the measurement errors to be unknown and free from parametric laws
and additionally also accommodates conditional heteroscedasticity in the measurement errors.

The literature on the problem of univariate density deconvolution, in which context we denote the variable of interest by $X$ and the measurement errors by $U$, is vast. 
Most of the early literature considered scenarios when the measurement error density $f_{U}$ is completely known.
Fourier inversion based deconvoluting kernel density estimators have been studied by 
Carroll and Hall (1988), Liu and Taylor (1989), Devroye (1989), Fan (1991a, 1991b, 1992) and Hesse (1999) among many others.
For a review of these methods, the reader may be referred to Section 12.1 in Carroll, et al. (2006) and Section 10.2.3 in Buonaccorsi (2010).
In reality $f_{U}$ is rarely known.
The problem of deconvolution when the errors are homoscedastic with an unknown density and replicated proxies are available for each subject has been addressed by Li and Vuong (1998). 
See also Diggle and Hall (1993), Neumann (1997), Carroll and Hall (2004) and the references therein.
The assumptions of homoscedasticity of $U$ and their independence from $X$ are also often unrealistic.
Flexible Bayesian density deconvolution approaches that allow $U$ to be conditionally heteroscedastic 
have recently been developed in Staudenmayer, et al. (2008) and Sarkar, et al. (2014).
Staudenmayer, et al. (2008) assumed the measurement errors to be normally distributed 
and used finite mixtures of B-splines to estimate $f_{X}$ and a variance function that captured the conditional heteroscedasticity.
Sarkar, et al. (2014) further relaxed the assumption of normality of $U$
employing flexible infinite mixtures of normal kernels induced by Dirichlet processes to estimate both $f_{X}$ and $f_{U}$.
Sieve based methods developed in Schennach (2004) and Hu and Schennach (2008) can also handle conditional heteroscedasticity.  


In sharp contrast to the univariate case, the literature on multivariate density deconvolution is quite sparse.
We can only mention Masry (1991), Youndj\'e and Wells (2008), Comte and Lacour (2013), Hazelton and Turlach (2009, 2010) and Bovy, et al. (2011). 
The first three considered deconvoluting kernel based approaches assuming the measurement errors $\bU$ to be distributed independently from $\bX$  
according to a known probability law.
Hazelton and Turlach (2009, 2011), working with the same assumptions on $\bU$, proposed weighted kernel based methods. 
Bovy, et al. (2011) modeled the density $f_{\bX}$ using flexible mixtures of multivariate normal kernels, 
but they assumed $f_{\bU}$ to be multivariate normal with known covariance matrices, 
independent from $\bX$.
As in the case of univariate problems, the assumptions of a fully specified $f_{\bU}$, 
known covariance matrices, and independence from $\bX$ are highly restrictive for most practical applications.

The focus of this article is on multivariate density deconvolution when $f_{\bU}$ is not known but replicated proxies are available for at least some individuals. 
The proposed deconvolution approaches can additionally accommodate conditional heteroscedasticity in $\bU$.
The problem is important, for instance, in nutritional epidemiology,   
where nutritionists are typically interested not just in the consumption behaviors of individual dietary components but also in their joint consumption patterns.  
The data are often available in the form of dietary recalls and are contaminated by measurement errors that show strong patterns of conditional heteroscedasticity.

As in Sarkar, et al. (2014), we use mixture models to estimate both $f_{\bX}$ and $f_{\bU}$   
but the multivariate nature of the problem brings in new modeling challenges and computational obstacles
that preclude straightforward extension of their univariate deconvolution approaches. 
Instead of using infinite mixtures induced by Dirichlet processes, we use finite mixtures of multivariate normal kernels with exchangeable Dirichlet priors on the mixture probabilities.  
The use of finite mixtures and exchangeable priors greatly reduces computational complexity while retaining essentially the same flexibility. 
Carefully constructed priors also allow automatic model selection and model averaging. 
To save space, detailed discussions on these important issues are moved to Section \ref{sec: mvt finite vs infinite mixture models} in the Supplementary Materials. 

We also exploit symmetric Dirichlet priors and properties of multivariate normal distributions and finite mixture models 
to develop a novel strategy that enables us to enforce a required zero mean restriction on the measurement errors.  
Our proposed technique, as opposed to the one adopted by Sarkar, et al. (2014), 
is particularly suitable for high dimensional applications and can be easily generalized to enforce moment restrictions on other types of finite mixture models. 

It is well known that inverse Wishart priors, due to their dense parametrization, are not suitable for modeling covariance matrices in high dimensional applications.  
In deconvolution problems the issue is further complicated since $\bX$ and $\bU$ are both latent. 
This results in numerically unstable estimates even for small and moderate dimensions, 
particularly when the true covariance matrices are sparse and the likelihood function is of complicated form.
To reduce the effective number of parameters required to be estimated, 
we consider factor-analytic representation of the component specific covariance matrices with sparsity inducing shrinkage priors on the factor loading matrices.

Models for multivariate regression errors that assume normality but allow the covariance matrix to vary flexibly 
with associated precisely measured and possibly multivariate predictors have recently been developed in the literature (Hoff and Niu, 2012; Fox and Dunson, 2016, etc.). 
Unlike regression settings, exclusive relationships exist between different components of multivariate measurement errors $\bU$ 
and different components of the associated multivariate latent `predictor' $\bX$ 
- the $\ell\th$ component $U_{\ell}$ of $\bU$ 
contaminates only the $\ell\th$ component $X_{\ell}$ of $\bX$ but not others. 
We thus deem covariance regression models that allow $\cov(\bU\vert\bX)$ to vary arbitrarily with all components of $\bX$ 
to be inappropriate in multivariate measurement error settings. 
As discussed above, the assumption of multivariate normality is also particularly restrictive in measurement error problems. 
In this article, we develop a semiparametric approach that appropriately highlights the exclusive associations between $U_{\ell}$ and $X_{\ell}$
while allowing the distribution of $(\bU \vert \bX)$ to depart from normality.
Importantly, the model also arises naturally from multivariate multiplicative measurement error settings, automatically encompassing such cases. 
Diagnostic tools for checking model adequacy are also discussed.


The likelihood function for the conditional heteroscedastic model poses significant computational challenges.
We overcome these obstacles by designing a novel two-stage procedure that exploits the unique properties of conditionally heteroscedastic multivariate measurement errors to our advantage. 
The procedure first estimates the variance functions characterizing $\var(U_{\ell} \vert X_{\ell})$
using reparametrized versions of the corresponding univariate submodels.
The estimates obtained in the first stage are then plugged-in to estimate the remaining parameters in the second stage.
Having two estimation stages, our deconvolution method for conditionally heteroscedastic measurement errors is not purely Bayesian.  
But they show good empirical performance and, with no other solution available in the existing literature, 
they provide at least workable starting points towards more sophisticated methodology.

The article is organized as follows. 
Section \ref{sec: mvt density deconvolution models} details the models.  
Model identifiability issues and implementation details, including the choice of hyper-parameters and Markov chain Monte Carlo (MCMC) algorithms to sample from the posterior, are discussed in the Supplementary Materials. 
Section \ref{sec: mvt model identifiability} discusses model identifiability issues.
Section \ref{sec: mvt model flexibility} presents theoretical results showing flexibility of the proposed models.
Simulation studies comparing the proposed deconvolution methods to a naive method that ignores measurement errors are presented in Section \ref{sec: mvt simulation studies}. 
Section \ref{sec: mvt data analysis} presents an application of the proposed methodology in estimation of the joint consumption pattern of dietary intakes 
from contaminated 24 hour recalls in a nutritional epidemiologic study. 
Section \ref{sec: mvt discussion} includes a discussion.
An unnumbered section concludes the article with a description of the Supplementary Materials. 
%

\vspace{-0.10cm}
\section{Deconvolution Models} \label{sec: mvt density deconvolution models}
The goal is to estimate the unknown joint density of a $p$-dimensional multivariate random variable $\bX$.
There are $i=1,\dots,n$ subjects.
Precise measurements of $\bX$ are not available.
Instead, for $j=1,\dots,m_{i}$, replicated proxies $\bW_{ij}$
contaminated with measurement errors $\bU_{ij}$  are available for each subject $i$.
The replicates are assumed to be generated by the model
\vspace{-5ex}
\be
\bW_{ij} &=& \bX_{i} + \bU_{ij}. 
\ee
\vspace{-5ex}\\
Given $\bX_{i}$, $\bU_{ij}$ are independently distributed with $E(\bU_{ij}\vert\bX_{i}) = \bzero$.    \label{eq: additive error}
The marginal density of $\bW_{ij}$ is denoted by $f_{\bW}$.
The implied conditional distributions of $\bW_{ij}$ and $\bU_{ij}$, given $\bX_{i}$,
are denoted by $f_{\bW\vert \bX}$ and $f_{\bU\vert \bX}$, respectively.

%

\subsection{Modeling the Density $f_{\bX}$} \label{sec: mvt density of interest}
In this article $f_{\bX}$ is specified as a mixture of multivariate normal kernels
\vspace{-5ex}\\
\be
f_{\bX}(\bX) = \textstyle\sum_{k=1}^{K_{\bX}} \pi_{\bX,k} ~ \MVN_{p}(\bX \vert \bmu_{\bX,k},\bSigma_{\bX,k}), \label{eq: mixture model for f_X}
\ee
\vspace{-5ex}\\
where $\MVN_{p}(\cdot \vert \bmu,\bSigma)$ denotes a $p$-dimensional multivariate normal density with mean $\bmu$ and covariance matrix $\bSigma$. 
For the rest of this subsection, the subscript $\bX$ is kept implicit to keep the notation clean. 

We assign a finite Dirichlet prior to the mixture probability vector $\bpi = (\pi_{1},\dots,\pi_{K})\trans$ as 
\vspace{-5ex}\\
\be
\bpi &\sim&  \Dir(\alpha/K,\dots,\alpha/K). \label{eq: symmetric Dirichlet prior}
\ee
\vspace{-5ex}\\
Here $\Dir(\alpha_{1},\dots,\alpha_{K})$ denotes a finite dimensional Dirichlet distribution on the $K$-dimensional unit simplex with concentration parameter $(\alpha_{1},\dots,\alpha_{K})$. 
Given $K$ and the latent cluster membership indices, the prior is conjugate. 
The symmetry of the assumed Dirichlet prior helps in additional reduction of computational complexity by simplifying MCMC mixing issues.
Provided $K$ is sufficiently large, a carefully chosen $\alpha$ can impart the posterior with certain properties  
that simplify model selection and model averaging issues by influencing the posterior to concentrate in regions that favor empty redundant components, 
see Section \ref{sec: mvt choice of hyper-parameters} and Section \ref{sec: mvt finite vs infinite mixture models} of the Supplementary Materials.
We assign conjugate multivariate normal priors to the component specific mean vectors $\bmu_{k}$, so that  
\vspace{-5ex}\\
\be
\bmu_{k} &\sim& \MVN_{p}(\bmu_{0},\bSigma_{0}). 
\ee
\vspace{-5ex}\\
The conjugacy again helps in simplifying posterior calculations. 
Later on, we will employ similar mixture models for the density of the measurement errors, 
and this conjugacy, along with some basic properties of multivariate normal kernels, 
will also help us enforce the mean zero restriction on the measurement errors. 
For the component specific covariance matrices $\bSigma_{k}$, we first consider conjugate inverse Wishart priors
\vspace{-5ex}\\
\be\
\bSigma_{k} \sim \IW_{p}(\nu_{0},\bPsi_{0}).
\ee
\vspace{-5ex}\\
Here $\IW_{p}(\nu,\bPsi)$ denotes an inverse Wishart density on the space of $p\times p$ positive definite matrices with mean $\bPsi/(\nu-p-1)$.
While the conjugacy of the inverse Wishart priors helps in simplifying posterior calculations, 
in complex high dimensional problems its dense parameterization may result in numerically unstable estimates, 
particularly when the covariance matrices are sparse.
In a deconvolution problem the issue is compounded further by the nonavailability of the true $\bX_{i}$'s.
To reduce the effective number of parameters to be estimated, we consider a parsimonious factor-analytic representation of the component specific covariance matrices:
\vspace{-7ex}\\
\be
\bSigma_{k} = \bLambda_{k}\bLambda_{k}\trans +\bOmega,    \label{eq: density of X latent factor characterization}
\ee 
\vspace{-5ex}\\
where $\bLambda_{k}$ are $p\times q_{k}$ factor loading matrices 
and $\bOmega$ is a diagonal matrix with non-negative entries. 
In practical applications $q_{k}$ will typically be much smaller than $p$, inducing parsimonious characterizations of the unknown covariance matrices $\bSigma_{k}$. 
Model (\ref{eq: mixture model for f_X}) can be equivalently represented as 
\vspace{-7ex}\\
\be
&&\Pr(C_{i} = k) = \pi_{k}, \\
&&(\bX_{i}\vert C_{i} = k) = \bmu_{k} + \bLambda_{k}\etam_{i} + \bDelta_{i},     \\
&& \etam_{i} \sim \MVN_{p}(\bzero,\bI_{p}), 
~~~~~\bDelta_{i} \sim \MVN_{p}(\bzero,\bOmega),
\ee
\vspace{-5ex}\\
where $C_{i}$ are the mixture labels associated with $\bX_{i}$, 
$\etam_{i}$ are latent factors, and $\bDelta_{i}$ are errors with covariance $\bOmega = \diag(\sigma_{1}^{2},\dots,\sigma_{p}^{2})$.

The above characterization of $\bSigma_{k}$ is not unique, 
since for any semi-orthogonal matrix $\bP$ the loading matrix $\bLambda_{k}^{1} = \bLambda_{k}\bP$ also satisfies (\ref{eq: density of X latent factor characterization}).
Since interest lies primarily in estimating the density $f_{\bX}$, identifiability of the latent factors is, however, not required.
This also allows the loading matrices to have a-priori a potentially infinite number of columns. 
Sparsity inducing priors, that favor more shrinkage as the column index increases, can then be used to shrink the redundant columns towards zero. 
In this article, we  do this by adapting the shrinkage priors proposed in Bhattacharya and Dunson (2011) that allow easy posterior computation. 
Let $\bLambda_{k} = ((\lambda_{k,jh}))_{j=1,h=1}^{p,\infty}$, where $j$ and $h$ denote the row and the column indices, respectively.
For $h=1,\dots,\infty$, we assign priors as follows
\vspace{-5ex}\\
\be
\lambda_{k,jh}  &\sim& \Normal(0,\phi_{k,jh}^{-1}\tau_{k,h}^{-1}),   
~~~~~\phi_{k,jh} \sim \Ga(\nu/2,\nu/2), \\
\tau_{k,h} &\sim& \textstyle\prod_{\ell=1}^{h} \delta_{k,\ell},   
~~~~~\delta_{k,\ell} \sim \Ga(a_{\ell},1),
~~~~~\sigma_{j}^{2} \sim \IG(a_{\sigma},b_{\sigma}).
\ee
\vspace{-5ex}\\
Here $\Ga(\alpha,\beta)$ denotes a Gamma distribution with shape parameter $\alpha$ and rate parameter $\beta$ and $\hbox{IG}(a,b)$ denotes an inverse-Gamma distribution with shape parameter $a$ and scale parameter $b$.
In the $k\th$ component factor loading matrix $\bLambda_{k}$, the parameters $\{\phi_{k,jh}\}_{j=1}^{p}$ control the local shrinkage of the elements in the $h\th$ column, 
whereas $\tau_{k,h}$ controls the global shrinkage. 
When $a_{h} > 1$ for $h=2,\dots,\infty$, the sequence $\{\tau_{k,h}\}_{h=1}^{\infty}$ becomes stochastically increasing 
and thus favors more shrinkage as the column index $h$ increases.

In addition to inducing adaptive sparsity and hence numerical stability, 
by favoring more shrinkage as the column index increases, the shrinkage priors  
play another important role in making the proposed factor analytic model highly robust to misspecification of the number of latent factors, 
allowing us to adopt simple strategies to determine the number of latent factors to be included in the model in practice. 
Details are deferred to Section \ref{sec: mvt choice of hyper-parameters}.

Throughout the rest of the paper, mixtures with inverse Wishart prior on the covariance matrices will be referred to as MIW models 
and mixtures of latent factor analyzers will be referred to as MLFA models.

%
%
For a review of finite mixture models and mixtures of latent factor analyzers, without moment restrictions or sparsity inducing priors and with applications in measurement error free scenarios, 
see Fokou\'{e} and Titterington (2003), Fr\"{u}hwirth-Schnatter (2006), Mengersen, et al. (2011) and the references therein. 
For other types of shrinkage priors, see Brown and Griffin (2010), Carvalho, et al. (2010), Bhattacharya, et al. (2014) etc. 

\subsection{Modeling the Density of the Measurement Errors}  \label{sec: mvt density of errors}

\subsubsection{Independently Distributed Measurement Errors}  \label{sec: mvt density of homoscedastic errors}
In this section, we develop models for the measurement errors $\bU$ assuming them to be independent from $\bX$. 
That is, we assume $f_{\bU\vert \bX} = f_{\bU}$ for all $\bX$. 
This remains the most extensively researched deconvolution problem for both univariate and multivariate cases.  
The techniques developed in this section will also provide crucial building blocks for 
more realistic models in Section \ref{sec: mvt density of heteroscedastic errors}.
The measurement errors and their density are now denoted by $\bepsilon_{ij}$ and $f_{\bepsilon}$, respectively, for reasons to become obvious shortly in Section \ref{sec: mvt density of heteroscedastic errors}. 

As in Section \ref{sec: mvt density of interest}, a mixture of multivariate normals can be used to model the density $f_{\bepsilon}$ 
but the model now has to satisfy a mean zero constraint.
That is 
\vspace{-5ex}\\
\be
f_{\bepsilon}(\bepsilon) = \textstyle\sum_{k=1}^{K_{\bepsilon}}\pi_{\bepsilon,k} ~ \MVN_{p}(\bepsilon \vert \bmu_{\bepsilon,k},\bSigma_{\bepsilon,k}),  \\
\text{subject to}~ \textstyle\sum_{k=1}^{K_{\bepsilon}}\pi_{\bepsilon,k} \bmu_{\bepsilon,k} = \bzero.
\ee
\vspace{-5ex}\\
To get numerically stable estimates of the density of the errors, latent factor characterization of the covariance matrices with sparsity inducing shrinkage priors as in Section \ref{sec: mvt density of interest} may again be used.
Details are curtailed to avoid unnecessary repetition and we only present the mechanism to enforce the zero mean restriction on the model. 
The subscript $\bepsilon$ is again dropped in favor of cleaner notation. 
In later sections, the subscripts $\bX$ and $\bepsilon$ reappear to distinguish between the parameters associated with $f_{\bX}$ and $f_{\bepsilon}$, when necessary. 

Without the mean restriction and under conjugate multivariate normal priors $\bmu_{k}\sim \MVN_{p}(\bmu_{0},\bSigma_{0})$, 
the posterior full conditional of $\bmu^{Kp \times 1} = (\bmu_{1}\trans,\dots,\bmu_{K}\trans)\trans$ is given by
\vspace{-4ex}\\
\be
\MVN_{K p} \left\{\left(\begin{array}{c}
\bmu_{1}^{0} \\
\bmu_{2}^{0}\\
\vdots\\
\bmu_{K}^{0}
\end{array} \right),
\left(\begin{array}{c c c c}
\bSigma_{1}^{0} & \bzero & \dots & \bzero\\
\bzero & \bSigma_{2}^{0} & \dots & \bzero\\
\vdots & \vdots & & \vdots\\
\bzero & \bzero & \dots & \bSigma_{K}^{0}
\end{array} \right)\right\} \equiv \MVN_{K p} (\bmu^{0},\bSigma^{0}),   \label{eq: joint mvt normal posterior of the mean vector}
\ee
\vspace{-3ex}\\
where $\bepsilon_{ij}$ and other conditioning variables are implicitly understood. 
Explicit expressions of $\bmu^{0}$ and $\bSigma^{0}$ in terms of the conditioning variables can be found in Section \ref{sec: mvt choice of hyper-parameters}. 
The posterior full conditional of $\bmu$ under the mean restriction can then be obtained easily by further conditioning the distribution in (\ref{eq: joint mvt normal posterior of the mean vector}) by $\bmu_{R} = \sum_{k=1}^{K}\pi_{k} \bmu_{k} = 0$ 
and is given by 
\vspace{-5ex}\\
\be
(\bmu \vert \bmu_R = \bzero)  \sim \MVN_{Kp}\{\bmu^{0} - \bSigma_{1,R}^{0}(\bSigma_{R,R}^{0})^{-1}\bmu_R^{0},\bSigma^{0} - \bSigma_{1,R}^{0}(\bSigma_{R,R}^{0})^{-1}\bSigma_{R,1}^{0}\}, \label{eq: conditional mvt normal posterior of the mean vector}
\ee
\vspace{-5ex}\\
where $\bmu_{R}^{0} = \sum_{k=1}^{K}\pi_{k} \bmu_{k}^{0} = E(\bmu_{R})$, 
$\bSigma_{k,K} = \pi_{k}\bSigma_{k}^{0} = \cov(\bmu_{k},\bmu_{R})$, 
$\bSigma_{R,R}^{0} = \bSigma_{K+1,K+1} = \sum_{k=1}^{K}\pi_{k}^{2}\bSigma_{k}^{0} = \cov(\bmu_{R})$, 
and $\bSigma_{R,1}^{0} = ( \bSigma_{1,K+1}, \bSigma_{2,K+1}, \dots, \bSigma_{K,K+1} )$.
To sample from this singular density,
we can first sample from the non-singular distribution of $\{(\bmu_{1}\trans,\bmu_{2}\trans,\dots,\bmu_{K-1}\trans)\trans \vert \bmu_{R} = \bzero\}$, 
which can also be trivially obtained from (\ref{eq: conditional mvt normal posterior of the mean vector}), and then set $\bmu_{K} = - \sum_{k=1}^{K-1}\pi_{k}\bmu_{k}/\pi_{K}$.

\vskip 0pt
\begin{figure}[!ht]
\centering
\includegraphics[height=5.5cm, width=16cm, trim=2cm 1cm 1cm 1cm, clip=true]{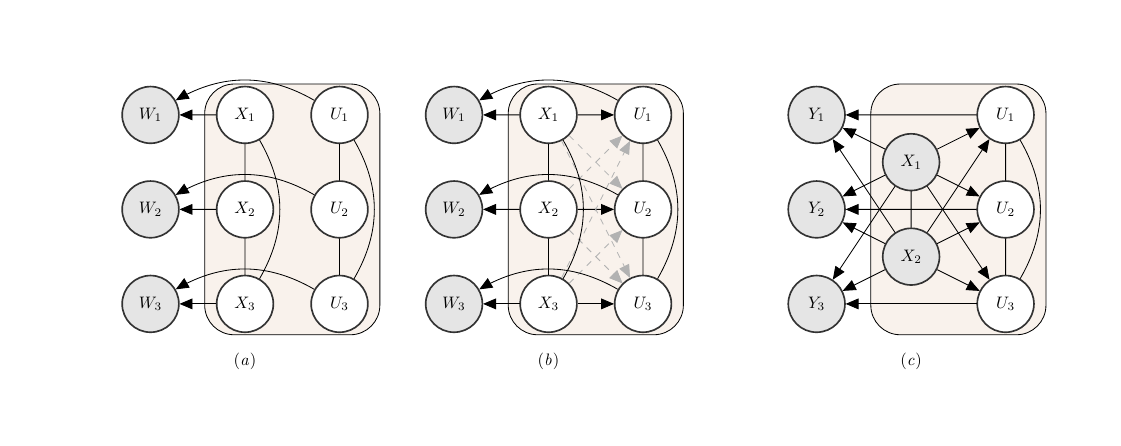}
\vskip -10pt
\caption{
Dependency structures in trivariate deconvolution problems with (a) independently distributed and (b) conditionally varying measurement errors.
(c) Dependency structure in a trivariate regression problem with response $\bY$, regression errors $\bU$ and bivariate predictor $\bX$. 
The filled rectangular regions focus on the relationship between the (potentially conditionally varying) errors $\bU$ 
and the (corresponding conditioning) variable $\bX$. 
The unfilled and the shaded nodes signify latent and observable variables, respectively. 
The directed and the undirected edges represent one and two-way relationships, respectively. 
The solid black and the dashed gray edges in panel (b) 
signify strong and weak dependencies, respectively. 
}
\label{fig: graphical models}
\end{figure}

\subsubsection{Conditionally Heteroscedastic Measurement Errors} \label{sec: mvt density of heteroscedastic errors}
We now consider the case when the variances of the measurement errors depend on the associated unknown values of $\bX$ through unknown relationships.

Interpreting the conditioning variables $\bX$ broadly as predictors, one can loosely connect our problem of modeling conditionally heteroscedastic $\bU$ to the problem of covariance regression (Hoff and Niu, 2012; Fox and Dunson, 2016, etc.),
where the covariance of the multivariate regression errors are allowed to vary flexibly with precisely measured and possibly multivariate predictors.
In such problems, the dimension of the regression errors is unrelated to the dimension of the predictors 
and different components of the regression errors are assumed to be equally influenced by different components of the predictors. 
In multivariate deconvolution problems, in contrast, the dimension of $\bU_{ij}$ is exactly the same as the dimension of $\bX_{i}$, 
the $\ell\th$ component $U_{ij\ell}$ being the measurement error associated exclusively with $X_{i\ell}$.
See Figure \ref{fig: graphical models}. 
While different components of $\bU_{ij}$ may be correlated, 
this exclusive association between $U_{ij\ell}$ and $X_{i\ell}$ implies that 
the dependence of $U_{ij\ell}$ on $\bX_{i}$ should be explained primarily through $X_{i\ell}$.
Figure \ref{fig: mvt EATS data results VFn}, for instance, suggests strong conditional heteroscedasticity patterns 
and it is plausible to assume that this conditional variability in $U_{ij\ell}$ can be explained mostly through $X_{i\ell}$ only. 
It is interesting to note these contrasts between conditionally varying regression and measurement errors  
become particularly prominent in the multivariate set up. 
Additionally, the aforementioned covariance regression approaches all assume multivariate normality of the regression errors. 
As discussed in the introduction, such strong parametric assumptions on the error distribution are particularly restrictive in measurement error problems. 
Additional detailed discussions of these important issues and resulting modeling implications 
can be found in Section \ref{sec: mvt comments on the model for U given X} of the Supplementary Materials.
They preclude direct application of existing covariance regression approaches to multivariate deconvolution problems   
but warrant models that can highlight the aforementioned unique dependence relationships, 
accommodate distributional flexibility while enforcing the mean zero restriction, and 
produce computationally stable estimates even in the absence of precise information on the conditioning variable $\bX$. 

The semiparametric approach that we adopt in this article achieves distributional flexibility, enforces the mean zero restriction, 
accommodates the exclusive relationships between $U_{ij\ell}$ and $X_{i\ell}$ 
but ignores the weak dependencies of $U_{ij\ell}$ on $\{X_{im}\}_{m\neq \ell}$ depicted in Figure \ref{fig: graphical models}(b). 
Specifically, we let 
\vspace{-7ex}\\
\be
(\bU_{ij}\vert \bX_{i}) =  \bS(\bX_{i})\bepsilon_{ij},  	\label{eq: mvt multiplicative structure}
\ee
\vspace{-5ex}\\
where $\bS(\bX_{i}) = \diag\{s_{1}(X_{i1}),s_{2}(X_{i2}),\dots,s_{p}(X_{ip})\}$ and $\bepsilon_{ij}$, henceforth referred to as the `scaled errors', are distributed independently of $\bX_{i}$.
Model (\ref{eq: mvt multiplicative structure}) implies that $\cov(\bU_{ij}\vert \bX_{i}) = \bS(\bX_{i})~ \cov(\bepsilon_{ij}) ~ \bS(\bX_{i})$ and marginally $\var(U_{ij\ell}\vert \bX_{i}) = s_{\ell}^{2}(X_{i\ell})\var(\epsilon_{ij\ell})$, a function of $X_{i\ell}$ only. 
The techniques developed in Section \ref{sec: mvt density of homoscedastic errors} can now be employed to model the density of $\bepsilon_{ij}$, allowing different components of $\bU_{ij}$ to be correlated and their joint density to deviate from multivariate normality.

We model the variance functions $s_{\ell}^{2}$, denoted also by $v_{\ell}$, using positive mixtures of B-spline basis functions
with smoothness inducing priors on the coefficients as in Staudenmayer, et al. (2008).
For the $\ell\th$ component, partition an interval $[A_{\ell},B_{\ell}]$ of interest into $L_{\ell}$ subintervals using knot points
$A_{\ell} = t_{\ell,1}=\dots=t_{\ell,q+1} < t_{\ell,q+2} < t_{\ell,q+3} < \dots < t_{\ell,q+L_{k}} < t_{\ell,q+L_{\ell}+1} = \dots = t_{\ell,2q+L_{\ell}+1}=B_{\ell}$.
A flexible model for the variance functions is given by
\vspace{-5ex}\\
\be
&& v_{\ell}(X_{i\ell}) = s_{\ell}^{2}(X_{i\ell}) = \textstyle\sum_{j=1}^{J_{\ell}} b_{q,j,\ell}(X_{i\ell}) \exp(\xi_{j\ell})= \bB_{q,J_{\ell},\ell}(X_{i\ell}) \exp(\bxi_{\ell}), \label{eq: mvt models for variance functions} \\
&& (\bxi_{\ell}\vert J_{\ell}, \sigma_{\xi,\ell}^{2}) \propto (2\pi\sigma_{\xi,\ell}^{2})^{-J_{\ell}/2} \exp\{-\bxi_{\ell}\trans P_{\ell}\bxi_{\ell}/(2\sigma_{\xi,\ell}^{2})\},~~~~ \sigma_{\xi,\ell}^{2} \sim \IG(a_{\xi},b_{\xi}).
\ee
\vspace{-5ex}\\
Here $\{b_{q,j,\ell}\}_{j=1}^{J_{\ell}}$ denote $J_{\ell}= (q+L_{\ell})$ B-spline bases of degree $q$ as defined in de Boor (2000), $\bxi_{\ell} = \{\xi_{1\ell},\xi_{2\ell},\dots,\xi_{J_{\ell}\ell}\}\trans$; 
$\exp(\bxi_{\ell}) = \{\exp(\xi_{1\ell}), \exp(\xi_{2\ell}),\dots,\exp(\xi_{J_{\ell}\ell})\}\trans$;
and $P_{\ell}=D_{\ell}\trans D_{\ell}$, where $D_{\ell}$ is a $J_{\ell}\times(J_{\ell}+2)$ matrix such that $D_{\ell}\bxi_{\ell}$ computes the second differences in $\bxi_{\ell}$.
The prior $P_{0}(\bxi_{\ell}\vert\sigma_{\xi,\ell}^{2})$ induces smoothness in the coefficients because it penalizes $\sum_{j=1}^{J_{k}}(\Delta^2 \xi_{j\ell})^2 = \bxi_{\ell}\trans P_{\ell}\bxi_{\ell}$, the sum of
squares of the second order differences in $\bxi_{\ell}$ (Eilers and Marx, 1996).
The parameters $\sigma_{\xi,\ell}^{2}$ play the role of smoothing parameter -
the smaller the value of $\sigma_{\xi,\ell}^{2}$, the stronger the penalty and the smoother the variance function.
The inverse-Gamma hyper-priors on $\sigma_{\xi,\ell}^{2}$ allow the data to have influence on the posterior smoothness and make the approach data adaptive.

Since $s_{\ell}^{2}(X_{i\ell}) \var(\epsilon_{ij\ell})=\{s_{\ell}^{2}(X_{i\ell})c\} \{\var(\epsilon_{ij\ell})/c\}$ for any $c>0$, 
the variance functions $s_{\ell}^{2}$'s can not be uniquely determined  
without additional restrictions on $\var(\epsilon_{ij\ell})$. 
Separate identifiability of $\bS$ and $f_{\bepsilon}$ is, however, not required for inference on $f_{\bX}$ 
or to assess the conditional variability in $U_{ij\ell}$.
The latter, for instance, may simply be obtained as $\var(U_{ij\ell} \vert X_{i}) = s_{\ell}^{2}(X_{i\ell}) \var(\epsilon_{ij\ell})$. 
We thus avoid additional identifiability restrictions that would further compound modeling challenges.  
Adjustments made to the estimates of $s_{\ell}^{2}$ and $f_{\bepsilon}$ 
to enable comparisons with the corresponding true values in simulation experiments are discussed in 
Section \ref{sec: mvt estimation of variance functions} in the Supplementary Materials.

\subsubsection{Multiplicative Measurement Errors} \label{sec: mvt multiplicative errors}

In this section we consider the case of multivariate multiplicative measurement errors. 
The replicates are now assumed to be generated by the model
\vspace{-2ex}
\be
\bW_{ij} &=& \bX_{i} \circ \wt\bU_{ij},     \label{eq: multiplicative error}
\ee
\vspace{-5ex}\\
where $\circ$ denotes element wise product 
and the errors $\wt\bU_{ij}$ are distributed independently of $\bX_{i}$ with $E(\wt\bU_{ij}) = \bone$. 
Importantly, model (\ref{eq: multiplicative error}) can be reformulated to arrive at model (\ref{eq: mvt multiplicative structure}) as 
\vspace{-5ex}
\be
\bW_{ij} &=& \bX_{i} \circ \wt\bU_{ij} = \bX_{i} + \bU_{ij}, ~~~\text{with}~~~\bU_{ij}=\bX_{i}\circ(\wt\bU_{ij}-\bone) = \bS(\bX_{i})\bepsilon_{ij}.    \label{eq: multiplicative error2}
\ee
\vspace{-5ex}\\
with $E(\bU_{ij} \vert \bX_{i})=\bX_{i} \circ E(\wt\bU_{ij}-\bone)=\bzero$, 
$\bS(\bX_{i})=\diag\{s_{1}(X_{i1}),\dots,s_{p}(X_{ip})\}$ with $s_{\ell}(X_{i\ell})=X_{i\ell}$ 
and $\bepsilon_{ij}=(\wt\bU_{ij}-1)$ are independent of $\bX_{i}$ with $E(\bepsilon_{ij})=\bzero$. 
This observation precludes the need for separate methodology to be developed 
for the problem of multivariate density deconvolution in the presence of multiplicative measurement errors 
and further emphasizes the importance of the additive conditionally heteroscedastic measurement error model (\ref{eq: mvt multiplicative structure}) developed in Section \ref{sec: mvt density of heteroscedastic errors}.  

\section{Posterior Inference} \label{sec: mvt posterior inference}
Inference is based on samples drawn from the posterior using MCMC algorithms. 
A Gibbs sampler for the independent error case discussed in Section \ref{sec: mvt density of homoscedastic errors} is presented in 
Section \ref{sec: mvt posterior computation} of the Supplementary Materials. 
For the conditionally heteroscedastic case discussed in Section \ref{sec: mvt density of heteroscedastic errors}, 
the full conditionals of the parameters characterizing the variance functions do not have closed form expressions. 
MCMC algorithms where we tried to integrate Metropolis-Hastings (MH) steps within the Gibbs sampler to generate samples from the full posterior 
were numerically unstable and failed to converge sufficiently quickly.
To address this challenge, we designed a novel two-stage procedure. 
For each $k$, we first estimate the functions $s_{\ell}(X_{i\ell})$ by fitting the univariate deconvolution models $W_{ij\ell} = X_{i\ell} + s_{\ell}(X_{i\ell})\epsilon_{ij\ell}$. 
High precision estimates of the variance functions $s_{\ell}^{2}(X_{i\ell})$ can be obtained using the univariate deconvolution models.  
See Figure \ref{fig: mvt simulation results VFn d4 n1000 m3 MLFA X1 E1 Ind} in the main article 
and Figure \ref{fig: mvt simulation results VFn d4 n1000 m3 MIW X1 E1 AR} in the Supplementary Materials for illustrations.
Parameters characterizing other components of the full model are then sampled using a Gibbs sampler keeping the estimates of the variance functions fixed. 
Additional details are deferred to Sections \ref{sec: mvt estimation of variance functions} and \ref{sec: mvt two-stage sampler} of the Supplementary Materials.

\section{Model Identifiability} \label{sec: mvt model identifiability}
This section presents a discussion of model identifiability issues. 
The density of interest $f_{\bX}$ is identifiable under mild technical assumptions. 
In the case of independently distributed measurement errors considered in Section \ref{sec: mvt density of homoscedastic errors} of the main paper, 
appealing to Li and Vuong (1998), the densities $f_{\bX}$ and $f_{\bepsilon}$ are identifiable provided $m_{i}\geq 2$ replicates are available for some individuals, 
and the characteristics functions $\phi_{\bX}(\bt)=E \{\exp(\iota \bt\trans\bX)\}$ and $\phi_{\bepsilon}(\bt)=E \{\exp(\iota \bt\trans\bepsilon)\}$ are non-vanishing everywhere.  

In the case of conditionally heteroscedastic measurement errors considered in Section \ref{sec: mvt density of heteroscedastic errors} of the main paper, 
appealing to Hu and Schennach (2004), the densities $f_{\bX}$ and $f_{\bU \vert \bX}$ are identifiable provided $m_{i}\geq 3$ replicates are available for some individuals, 
the joint, conditional and marginal densities of $\bW_{1},\bW_{2},\bW_{3},\bX$ are all bounded, and the density $f_{\bX\vert\bW}$ is bounded complete
in the sense that the unique solution to $\int f_{\bX\vert\bW}(\bX)g(\bX)d\bX=0$ for all $\bW$ and for all bounded $g(\bX)$ is $g(\bX)=0$ for all $\bX$. 
The following lemma provides a sufficient condition for the density $f_{\bX\vert\bW}$ to be bounded complete.  

\begin{Lem} \label{lem: identifiability}
$f_{\bX\vert\bW}$ is bounded complete if $E\{\exp(\iota \bt\trans\bX\vert \bW)\}$ is non-vanishing everywhere for all $\bW$.
\end{Lem}
\begin{proof}
By Theorem 10C of Goldberg (1961), since $E\{\exp(\iota \bt\trans\bX\vert \bW)\}$ is non-vanishing everywhere for all $\bW$, the closed linear span of $f_{\bX\vert\bW}(\cdot)$ is $L_1(\rR)$.  By Hahn-Banach Theorem,  the dual space of $L_1(\rR)$ is $L_{\infty}(\rR)$ and there is an  isometric isomorphism from $L_\infty(\rR)$ to $L_{1}(\rR)$ given by $g \mapsto \Phi_g$ where 
 $\Phi_g(f_{\bX\vert\bW}) = \int f_{\bX\vert\bW}(\bX)g(\bX)d\bX $ for all $\bW$.   
Since the closed linear span of $f_{\bX\vert\bW}(\cdot)$  for all $\bW$  is $L_1(\rR)$,   
 $\int f_{\bX\vert\bW}(\bX)g(\bX)d\bX = 0$ for all $\bW$ implies that the mapping $\Phi_g$ is identically $0$.  By the isometric isomorphism above, it follows that $g$ should be identically $0$. 
\end{proof}

Different types of completeness of densities are often used as key identifying conditions in measurement error problems. 
See, for example, d'Haultfoeuille (2011) and Carroll, et al. (2010).
Here, we have provided a general sufficient condition for bounded completeness to hold true and a novel proof using functional analysis techniques. 
Loosely speaking, if the density $f_{\bX\vert\bW}(\bX)$ varies with $\bX$, its characteristic function does not vanish. 
Without sufficient variability of the density of $\bX\vert\bW$, observations on $\bW$ do not have enough information to recover the density of $\bX$.  

Model parameters specifying the components $f_{\bX}$, $f_{\bepsilon}$, $s_{\ell}$ etc. are not separately identifiable. 
For inference on identifiable functional model components, identifiability of individual parameters is, however, not required. 
Indeed, the mixture models and the associated priors were so chosen that the mixture components remain unidentifiable.
This helps simplify MCMC mixing issues. See Section \ref{sec: mvt finite vs infinite mixture models} of the Supplementary Materials.

\section{Model Flexibility}\label{sec: mvt model flexibility}
This section presents a theoretical study of the flexibility of the proposed models.
Proofs of the results are presented in the Supplementary Materials. 
We focus on the deconvolution models for conditionally heteroscedastic measurement errors, 
the case of independently distributed errors following as a special case. 
First we show that componentwise our models for the density $f_{\bX}$ of $\bX$, the density $f_{\bepsilon}$ of the scaled errors $\bepsilon$, and the variance functions $v_{\ell}$ are all highly flexible.  
Building on these results, we then show that our proposed deconvolution models can accommodate a large class of data generating processes.

Let the generic notation $\Pi$ denote a prior on some class of random functions.
Also let  $\T$ denote the target class of functions to be modeled by $\Pi$.
The support of $\Pi$ throws light on the flexibility of $\Pi$.
For $\Pi$ to be a flexible prior, one would expect that $\T$ or a large subset of $\T$ would be contained in the support of $\Pi$.

For investigating the flexibility of priors for density functions, a relevant concept is that of Kullback-Leibler (KL) support.
The  KL divergence between two densities $f_{0}$ and $f$, denoted by $d_{KL}(f_0,f)$, is defined as
$d_{KL}(f_0,f) = \int f_{0}(Z) ~ \log ~\{f_{0}(Z)/f(Z)\} dZ$.
Let $\Pi_{f}$ denote a prior assigned to a random density $f$.
A density $f_0$ is said to belong to the KL support of $\Pi_{f}$ if
$\Pi_{f}\{f:d_{KL}(f_0,f)<\delta\}>0~\forall \delta>0$.
The class of densities in the KL support of $\Pi_{f}$ is denoted by $KL(\Pi_{f})$.

Let $\F$ be the class of target densities to be modeled by the prior $\Pi_{f}$.
Let $\S$ denote the support of $\F$ and $\wt{\F}\subseteq\F$ denote the class of densities that satisfy the following fairly minimal set of regularity conditions.
Since $\wt\F$ is a large subclass of $\F$, its inclusion in the KL support of $\Pi_{f}$ would establish the flexibility of $\Pi_{f}$.

\begin{Cond}\label{cond: mvt regularity conditions on the density}
1. $f_{0}$ is continuous on $\S$ except on a set of measure zero.\\
2. The second order moments of $f_{0}$ are finite.\\
3. For some $r>0$ and for all $\bz\in\S$, there exist hypercubes $C_{r}(\bz)$ with side length $r$ and $\bz \in C_{r}(\bz)$ such that
\vspace{-8ex}\\
\bse
\int f_{0}(\bz)  ~ \log \left\{\frac{f_{0}(\bz)}{\inf_{\bt\in C_{r}(\bz)} f_{0}(\bt)}\right\} d\bz < \infty.
\ese 
\end{Cond}

Let $\Pi_{\bX}$ be a generic notation for both the MIW and the MLFA prior on $f_{\bX}$ 
defined in Section \ref{sec: mvt density of interest}.
Similarly, let $\Pi_{\bepsilon}$ be a generic notation for both the MIW and the MLFA prior on $f_{\bepsilon}$  
defined in Section \ref{sec: mvt density of errors}.
When the measurement errors are distributed independently of $\bX$, the support of $f_{\bX}$, say $\X$, may be taken to be any subset of $\rR^{p}$. 
For conditionally heteroscedastic measurement errors, 
the variance functions $s_{\ell}^{2}(\cdot)$ that capture the conditional variability are modeled by mixtures of B-splines defined on closed intervals $[A_{k},B_{k}]$.
In this case, the support of $f_{\bX}$ is assumed to be the closed hypercube $\X = [A_{1},B_{1}]\times \dots \times [A_{p},B_{p}]$.
Let $\F_{\bX}$ denote the set of all densities on $\X$, the target class of densities to be modeled by $\Pi_{\bX}$
and $\wt\F_{\bX} \subseteq \F_{\bX}$ denote the class of densities $f_{0\bX}$ that satisfy Conditions \ref{cond: mvt regularity conditions on the density}.
Similarly, let $\F_{\bepsilon}$ denote the set of all densities on $\rR^{p}$ that have mean zero 
and $\wt\F_{\bepsilon} \subseteq \F_{\bepsilon}$ denote the class of densities $f_{0\bepsilon}$ that satisfy Conditions \ref{cond: mvt regularity conditions on the density}.
The following Lemma establishes the flexibility of the models for $f_{\bX}$ and $f_{\bepsilon}$.

\begin{Lem} \label{Lem: mvt KL support of the priors}
1. $\wt{\F}_{\bX}\subseteq KL(\Pi_{\bX})$
2. $\wt{\F}_{\bepsilon} \subseteq KL(\Pi_{\bepsilon})$. 
\end{Lem}

For investigating the flexibility of models for general classes of functions, a relevant concept is that of sup norm support.
The sup norm distance between two functions $g_{0}$ and $g$, denoted by $||g_0-g||_{\infty}$, is defined as
$||g_0-g||_{\infty} = \sup_{Z} |g_{0}(Z)-g(Z)|$.
Let $\Pi_{g}$ denote a prior assigned to a random function $g$.
A function $g_0$ is said to belong to the sup norm support of $\Pi_{g}$ if
$\Pi_{g}(g: ||g_0-g||_{\infty}<\delta)>0~\forall \delta>0$.
The class of functions in the sup norm support of $\Pi_{g}$ is denoted by $SN(\Pi_{g})$.

Let $\Pi_{\bV}$ denote the prior on the variance functions based on mixtures of B-spline basis functions defined in Section \ref{sec: mvt density of heteroscedastic errors}.
For notational convenience we consider the case of a univariate variance function supported on $[A,B]$.
Extension to the multivariate case with variance functions supported on $\X$ is technically trivial. 
Let $\C_{+}[A,B]$ denote the set of continuous functions from $[A,B]$ to $\rR^{+}$.
Also, for $\alpha \leq (q+1)$, let $\C_{+}^{\alpha}[A,B] \subseteq \C_{+}[A,B]$ denote the set of functions
that are $\alpha_{0}$ times continuously differentiable, and for all $v_{0} \in \C_{+}^{\alpha}[A,B]$, $\norm{v_{0}}_{\alpha}<\infty$,
where $\alpha_{0}$ is largest integer less than or equals to $\alpha$ and the seminorm is defined by
$\norm{v_{0}}_{\alpha} = \sup_{X,X'\in[A,B], X\neq X'}  \{ |v_{0}^{(\alpha_0)}(X)-v_{0}^{(\alpha_0)}(X')|  /  |X-X'|^{\alpha-\alpha_0}\}$.
The local support properties of B-splines make the models for the variance functions very flexible as is indicated by the following lemma.

\begin{Lem} \label{Lem: mvt sup norm support of priors on variance functions}
$ \C_{+}^{\alpha}[A,B] \subseteq \C_{+}[A,B] \subseteq SN(\Pi_{\bV})$. 
\end{Lem}

Although technically the sup norm distance between linear combinations of B-splines and any continuous function can be made arbitrarily small by increasing the number of knots,
for obvious reasons the actual bounds for the sup norm distance may not be very sharp if the function to be modeled is wiggly.
However, for most applications of practical importance, the true variance function may be assumed to be smooth,
that is, to belong to some $\C_{+}^{\alpha}[A,B]$ with $\alpha\geq 1$.
Therefore, for practical reasons, it is only important that the smaller H\"{o}lder class of functions $\C_{+}^{\alpha}[A,B]$ belongs to the sup norm support of $\Pi_{\bV}$.
As shown in Section \ref{sec: mvt proof of sup norm support of priors on variance functions} of the Supplementary Materials, 
the bounds for sup norm distance in this case will also be much sharper.

Since the models for the variance functions $v_{\ell}$ and the models for the density of the scaled errors $f_{\bepsilon}$ are separately very flexible,
under model (\ref{eq: mvt multiplicative structure}) on the measurement errors,
the implied conditional and joint densities are also expected to be very flexible.
This is investigated in the next lemma.
For a given $\bX$, let $\Pi_{\bU\vert \bX}$ denote the prior for $f_{\bU\vert \bX}$ induced by $\Pi_{\bepsilon}$ and $\Pi_{\bV}$ under model (\ref{eq: mvt multiplicative structure}).
Define $\wt\F_{\bU\vert \bX} = \{f_{0\bU\vert \bX}: f_{0\bU\vert \bX}(\bU) = \prod_{k=1}^{p}s_{0k}^{-1}(X_{k}) f_{0\bepsilon}\{\bS_{0}^{-1}(\bX)\bU\}, s_{0k}^{2}\in \C_{+}[A_{k},B_{k}]~\hbox{for}~ k=1,\dots,p, f_{0\bepsilon}\in \wt{\F}_{\bepsilon}\}$.
Also let $\Pi_{\bU\vert \bV}$ denote the prior for the unknown conditional density of $\bU$ induced by $\Pi_{\bepsilon}$ and $\Pi_{\bV}$ under model (\ref{eq: mvt multiplicative structure}).
Define $\wt\F_{\bU\vert \bullet} = \{f_{0\bU\vert \bullet}: ~\hbox{for any given}~\bX\in\X, ~ f_{0\bU\vert\bullet} = f_{0\bU\vert \bX}\in \wt\F_{\bU\vert \bX}\}$.
Finally, let $\Pi_{\bX,\bU}$  denote the prior for the joint density of $(\bX,\bU)$ induced by $\Pi_{\bX}$, 
$\Pi_{\bepsilon}$ and $\Pi_{\bV}$ under model (\ref{eq: mvt multiplicative structure}).
Define $\wt\F_{\bX,\bU} = \{f_{0,\bX,\bU}: f_{0,\bX,\bU}(\bX,\bU) = f_{0,\bX}(\bX)f_{0,\bU \vert \bX}(\bU\vert \bX),~\hbox{where}~ f_{0\bX}\in \wt\F_{\bX} ~\hbox{and}~ f_{0\bU\vert \bX} \in \wt\F_{\bU\vert \bX} ~\hbox{for all}~ \bX\in \X\}$.

\begin{Lem} \label{Lem: mvt KL support of the prior on the density of U|X}
1. $\wt\F_{\bU\vert \bX} \subseteq KL(\Pi_{\bU\vert \bX})$ for any given $\bX\in \X$.\\
2. For any $f_{0\bU\vert \bullet}\in\wt\F_{\bU\vert \bV}$, $\Pi_{\bU\vert \bV}\{\sup_{\bX\in\X} d_{KL}(f_{0\bU\vert \bX},f_{\bU\vert \bX})<\delta\}>0$ for all $\delta>0$.\\
3. $\wt\F_{\bX,\bU} \subseteq KL(\Pi_{\bX,\bU})$.
\end{Lem}

The flexibility of the implied model for the marginal density $f_{\bW}$ is the subject of our final result.
Since the only observed quantities are $\bW_{ij}$, the support of the induced prior on $f_{\bW}$ tells us about the types of likelihood functions the model can approximate.

Let $\Pi_{\bW}$ denote the prior for the density of $\bW$ induced by $\Pi_{\bX}$, $\Pi_{\bepsilon}$ and $\Pi_{\bV}$ under model (\ref{eq: mvt multiplicative structure}).
Also let $\wt\F_{\bW} = \{f_{0\bW}: f_{0\bW}(\bW) = \int f_{0\bX}(\bX)f_{0\bU\vert \bX}(\bW-\bX)d\bX, f_{0\bX} \in \wt\F_{\bX}, f_{0\bU\vert \bullet} \in \wt\F_{\bU\vert \bullet}\}$,
the class of densities $f_{0\bW}$ that can be obtained as the convolution of two densities $f_{0\bX}$ and $f_{0\bU\vert \bullet}$, where $f_{0\bX}\in \wt\F_{\bX}$ and $f_{0\bU\vert \bullet} \in \wt\F_{\bU\vert \bullet}$.

Since the supports of $\Pi_{\bX}$ and $\Pi_{\bU\vert \bX}$ are large, 
it is expected that the support of $\Pi_{\bW}$ will also be large.
However, because convolution is involved, investigation of KL support of $\Pi_{\bW}$ is a difficult problem.
A weaker but relevant concept is that of $L_1$ support.
The $L_1$ distance between two densities $f_{0}$ and $f$, denoted by $||f_{0}-f||_{1}$, is defined as
$||f_0-f||_{1} = \int |f_{0}(Z) - f(Z)| dZ$.
A density $f_0$ is said to belong to the $L_1$ support of $\Pi_{f}$ if
$\Pi_{f}(f: ||f_0-f||_{1}<\delta)>0~\forall \delta>0$.
The class of densities in the $L_1$ support of $\Pi_{f}$ is denoted by $L_{1}(\Pi_{f})$.
The following theorem shows that the $L_1$ support of $\Pi_{\bW}$ is large.

\begin{Thm} \label{Thm: mvt L1 support of induced prior on density of W}
$\wt\F_{\bW} \subseteq L_{1}(\Pi_{\bW})$.
\end{Thm}

The proofs of these results are deferred to Section \ref{sec: mvt proofs of theoretical results} of the Supplementary Materials.
The proofs require that the number of mixture components $K$ be allowed to vary over $\nN$, the set of all positive integers, 
through priors, denoted by the generic notation $P_{0}(K)$, that assign positive probability to all $K\in \nN$.
Posterior computation for such methods will be computationally intensive, specially in a complicated multivariate set up like ours. 
In our implementation, we thus keep the number of mixture components fixed at finite values.
%

\newpage
\section{Simulation Experiments} \label{sec: mvt simulation studies}

The mean integrated squared error (MISE) of estimation of $f_{\bX}$ by $\wh{f}_{\bX}$ is defined as $MISE = E_{f_{\bX}} \int \{f_{\bX}(\bX)-\widehat{f}_{\bX}(\bX)\}^{2}d\bX$.
Based on $B$ simulated data sets, a Monte Carlo estimate of MISE is given by
$MISE_{est} = ~ B^{-1}~\sum_{b=1}^{B}\sum_{m=1}^{M}\{f_{\bX}(\bX_{b,m})-\widehat{f}_{\bX}^{(b)}(\bX_{b,m})\}^{2}/p_{0}(\bX_{b,m})$,
where $\{\bX_{b,m}\}_{b=1,m=1}^{B,M}$ are random samples from the density $p_{0}$.
We designed simulation experiments to evaluate the MISE performance of the proposed models for a wide range of possibilities. 
The MISEs we report here are all based on $100$ simulated data sets 
and $M=10^6$ samples generated from each of the two densities (a) $p_{0} = f_{\bX}$, the true density of $\bX$, 
and (b) $p_{0}$ that is uniform on the hypercube with edges $\min_{k}\{\bmu_{\bX,k}-3\bone_{p}\}$ and $\max_{k}\{\bmu_{\bX,k}+3\bone_{p}\}$.
With carefully chosen initial values and proposal densities for the MH steps, we were able to achieve quick convergence for the MCMC samplers.
The use of exchangeable Dirichlet priors helped simplify mixing issues (Geweke, 2007). 
See Section \ref{sec: mvt computational complexity} in the Supplementary Materials for additional discussions.
We programmed our methods in {R}.
In each case, we ran $3000$ MCMC iterations and discarded the initial $1000$ iterations as burn-in.
The post burn-in samples were thinned by a thinning interval of length $5$. 
For the univariate samplers, $1000$ MCMC iterations with a burn-in of $500$ sufficed to produce stable estimates of the variance functions.
In our experiments with much larger iteration numbers and burn-ins, the MISE performances remained practically the same.
This being the first article that tries to solve the problem of multivariate density deconvolution when the measurement error density is unknown,
the proposed MIW and MLFA models have no competitors. 
We thus compared our models with a naive Bayesian method that ignores measurement errors and treats the subject specific means 
as precisely measured observations instead, modeling $f_{\bX}$ by a finite mixture of multivariate normals as in (\ref{eq: mixture model for f_X}) 
with inverse Wishart priors on the component specific covariance matrices.

\begin{figure}[!ht]
\begin{center}
\includegraphics[width=16cm, trim=0cm 0cm 0cm 0cm, clip=true]{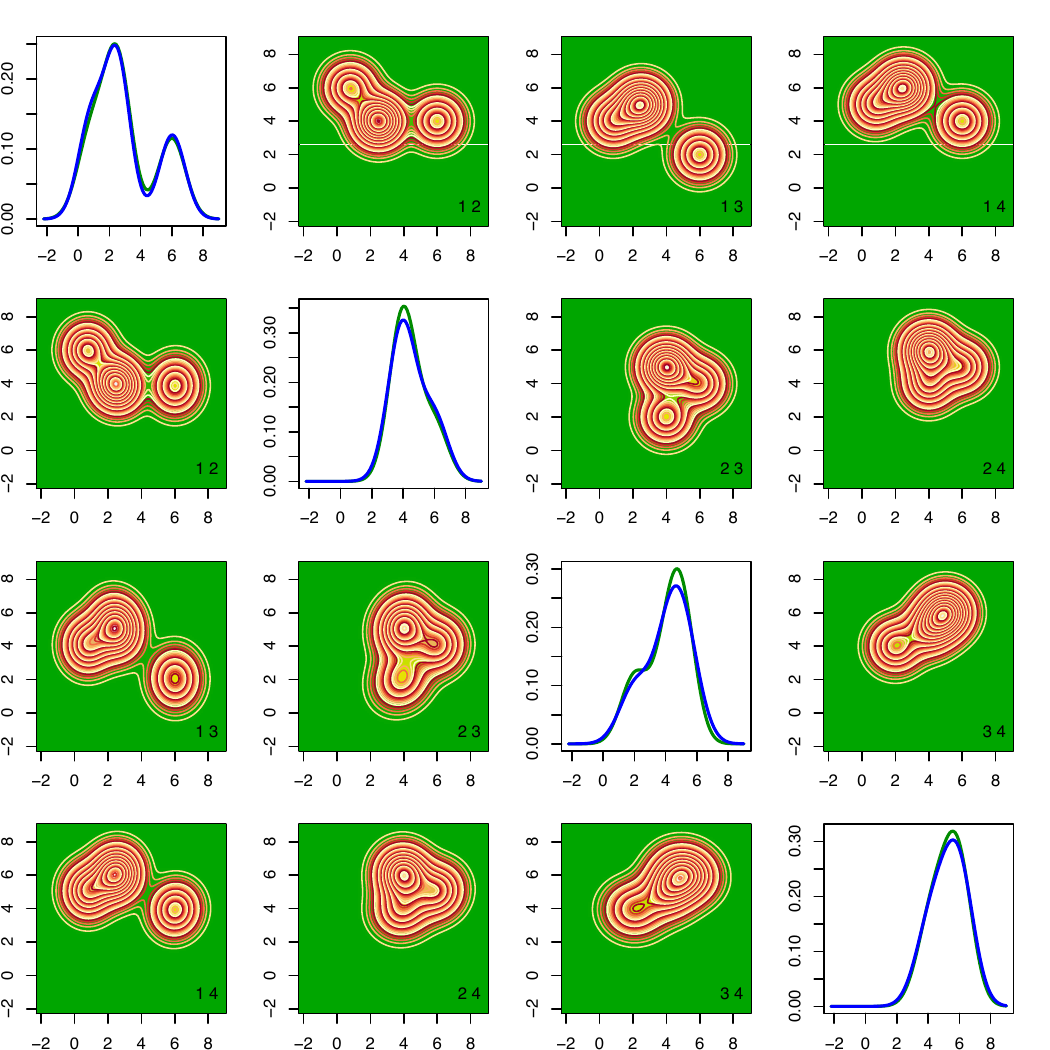}
\end{center}
\caption{\baselineskip=10pt Results for the  $f_{\bX}$ produced by the MLFA (mixtures of latent factor analyzers) method for the conditionally heteroscedastic error distribution $f_{\bepsilon}^{(2)}$ with sample size $n=1000$, $m_{i}=3$ replicates for each subject and identity matrix (I) for the component specific covariance matrices. The results correspond to the data set that produced the median of the estimated integrated squared errors (ISE) out of a total of 100 simulated data sets.
See Section \ref{sec: mvt simulation studies} for additional details.
The upper triangular panels show the contour plots of the true two dimensional marginal densities.
The lower triangular diagonally opposite panels show the corresponding estimates.
The numbers $i,j$ at the bottom right corners of the off-diagonal panels show that the marginal densities $f_{X_{i},X_{j}}$ are plotted in those panels. 
The diagonal panels show the true (lighter shaded green lines) and the estimated (darker shaded blue lines) one dimensional marginals.
The figure is in color in the electronic version of this article.
}
\label{fig: mvt simulation results XS d4 n1000 m3 MLFA X1 E1 Ind}
\end{figure}

\begin{figure}[!ht]
\begin{center}
\includegraphics[height=16cm, width=16cm, trim=0cm 0cm 0cm 0cm, clip=true]{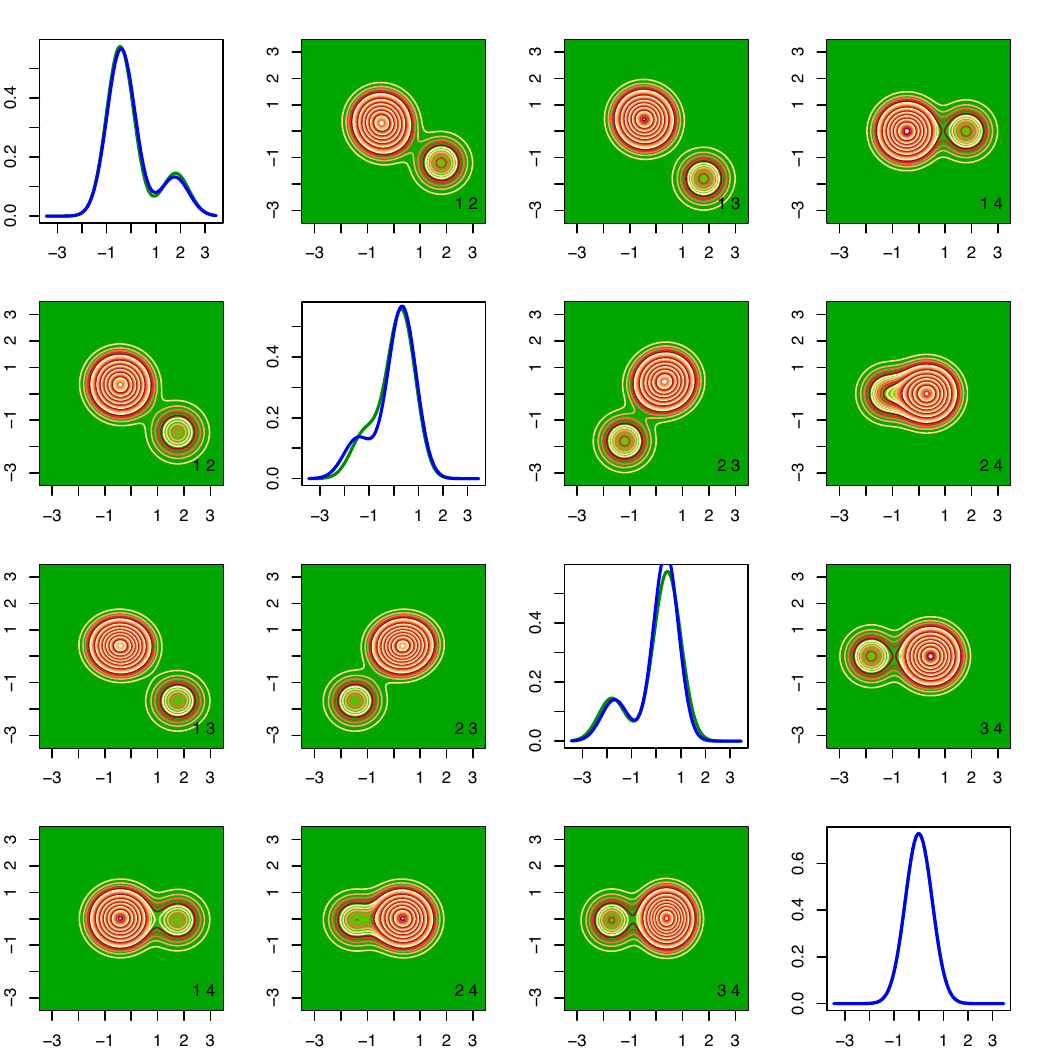}
\end{center}
\caption{\baselineskip=10pt Results for the density of the scaled measurement errors $f_{\bepsilon}$ produced by the MLFA (mixtures of latent factor analyzers) method for the conditionally heteroscedastic error distribution $f_{\bepsilon}^{(2)}$ with sample size $n=1000$, $m_{i}=3$ replicates for each subject and identity matrix (I) for the component specific covariance matrices. The results correspond to the data set that produced the median of the estimated integrated squared errors (ISE) out of a total of 100 simulated data sets.
See Section \ref{sec: mvt simulation studies} for additional details.
The upper triangular panels show the contour plots of the true two dimensional marginal densities.
The lower triangular diagonally opposite panels show the corresponding estimates.
The numbers $i,j$ at the bottom right corners of the off-diagonal panels show that the marginal densities $f_{\epsilon_{i},\epsilon_{j}}$ are plotted in those panels. 
The diagonal panels show the true (lighter shaded green lines) and the estimated (darker shaded blue lines) one dimensional marginals.
The figure is in color in the electronic version of this article.
}
\label{fig: mvt simulation results ES d4 n1000 m3 MLFA X1 E1 Ind}
\end{figure}

We considered two choices for the sample size $n = 500, 1000$.
For each subject, we simulated $m_{i}=3$ replicates.
The true density of $\bX$ was chosen to be
 $f_{\bX}(\bX) = \sum_{k=1}^{K_{\bX}} \pi_{\bX,k}~ \MVN_{p}(\bX \vert \bmu_{\bX,k},\bSigma_{\bX,k})$ with $p=4$, $K_{\bX}=3$, $\bpi_{\bX} = (0.25,0.50,0.25)\trans$, $\bmu_{\bX,1} = (0.8,6,4,5)\trans$, $\bmu_{\bX,2} = (2.5,4,5,6)\trans$ and $\bmu_{\bX,3} = (6,4,2,4)\trans$.
For the density of the measurement errors $f_{\bepsilon}$ we considered two choices, namely
\begin{enumerate}
\item $f_{\bepsilon}^{(1)}(\bepsilon) = \MVN_{p}(\bepsilon \vert \bzero,\bSigma_{\bepsilon})$, and
\item $f_{\bepsilon}^{(2)}(\bepsilon) = \sum_{k=1}^{K_{\bepsilon}} \pi_{\bepsilon,k}~ \MVN_{p}(\bepsilon \vert \bmu_{\bepsilon,k},\bSigma_{\bepsilon,k})$ with $K_{\bepsilon}=3$, $\bpi_{\bepsilon} = (0.2,0.6,0.2)\trans$, $\bmu_{\bepsilon,1} = (-0.3,0,0.3,0)\trans$, $\bmu_{\bepsilon,2} = (-0.5,0.4,0.5,0)\trans$ and 
$\bmu_{\bepsilon,3} = -(\pi_{\bepsilon,1}\bmu_{\bepsilon,1}+\pi_{\bepsilon,2}\bmu_{\bepsilon,2})/\pi_{\bepsilon,3}$.
\end{enumerate}
For the component specific covariance matrices, we set $\bSigma_{\bX,k} = \bD_{\bX}\bSigma_{\bX,0}\bD_{\bX}$ for each $k$, where $\bD_{\bX} = \diag(0.75^{1/2},\dots,0.75^{1/2})$. 
Similarly, $\bSigma_{\bepsilon, k} = \bD_{\bepsilon}\bSigma_{\bepsilon,0}\bD_{\bepsilon}$ for each $k$, where $\bD_{\bepsilon} = \diag(0.3^{1/2},\dots,0.3^{1/2})$.
For each pair of $f_{\bX}$ and $f_{\bepsilon}$, we considered four types of covariance structures for $\bSigma_{\bX,0} = \{(\sigma_{ij}^{\bX,0})\}$ and $\bSigma_{\bepsilon,0} = \{(\sigma_{ij}^{\bepsilon,0})\}$, namely
\begin{enumerate}
\item Identity (I): $\bSigma_{\bX,0} = \bSigma_{\bepsilon,0} = \Ind_{p}$, 
\item Latent Factor (LF): $\bSigma_{\bX,0} = \bLambda_{\bX}\bLambda_{\bX} + \bOmega_{\bX}$, with $\bLambda_{\bX} = (0.7,\dots,0.7)\trans$ and $\bOmega_{\bX} = \diag(0.51,\dots,0.51)$, and $\bSigma_{\bepsilon,0} = \bLambda_{\bepsilon}\bLambda_{\bepsilon} + \bOmega_{\bepsilon}$, with $\bLambda_{\bepsilon} = (0.5,\dots,0.5)\trans$ and $\bOmega_{\bepsilon}=\diag(0.75,\dots,0.75)$,
\item Autoregressive (AR): $\sigma_{ij}^{\bX,0} = 0.7^{\abs{i-j}}$ and $\sigma_{ij}^{\bepsilon,0} = 0.5^{\abs{i-j}}$ for each $(i,j)$, and
\item Exponential (EXP): $\sigma_{ij}^{\bX,0} = \exp(-0.5\abs{i-j})$ and $\sigma_{ij}^{\bepsilon,0} = \exp(-0.9\abs{i-j})$ for each $(i,j)$.
\end{enumerate}
The parameters were chosen to produce a wide variety of one and two dimensional marginal densities, 
see Figure \ref{fig: mvt simulation results XS d4 n1000 m3 MLFA X1 E1 Ind} and also Figure \ref{fig: mvt simulation results ES d4 n1000 m3 MLFA X1 E1 Ind}.
Scale adjustments by multiplication with $\bD_{\bX}$ and $\bD_{\bepsilon}$ were done so that the simulated values of each component of $\bX$ fall essentially in the range $(-2,6)$
and the simulated values of all components of $\bepsilon$ fall essentially in the range $(-3,3)$.
For conditionally heteroscedastic measurement errors, we set the true variance functions at $s_{\ell}^{2}(X)=(1+X/4)^{2}$ for each component $\ell$.
A total of $16~ (2\times 1\times 2\times 4)$ cases were thus considered for both independent and conditionally heteroscedastic measurement errors.

We first discuss the results of the simulation experiments when the measurement errors $\bU$ were independent of $\bX$.
The estimated MISEs are presented in Table \ref{tab: mvt MISEs homoscedastic}.
When the true $f_{\bepsilon}$ was a single component multivariate normal, 
the MLFA model produced the lowest MISE when the true covariance matrices were diagonal. 
In all other cases the MIW model produced the best results.
When the true $f_{\bepsilon}$ was a mixture of multivariate normals, the model complexity increases and the performance of the MIW model started to deteriorate.
In this case, the MLFA model dominated the MIW model when the true covariance matrices were either diagonal or had a latent factor characterization.

\begin{figure}[!ht]
\begin{center}
\includegraphics[width=16cm, trim=0cm 0cm 0cm 0cm, clip=true]{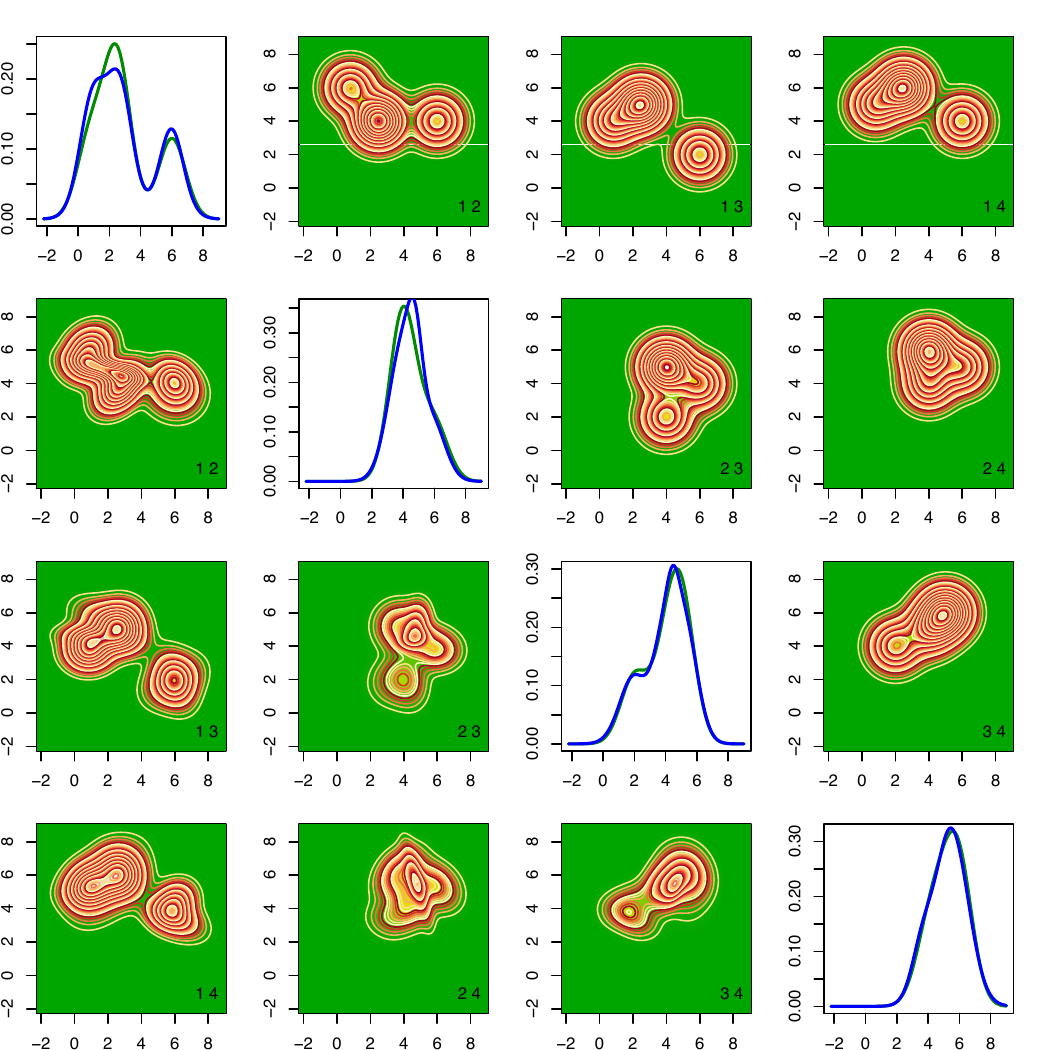}
\end{center}
\caption{\baselineskip=10pt Results for $f_{\bX}$ produced by the MIW (mixtures with inverse Wishart priors) method for the conditionally heteroscedastic error distribution $f_{\bepsilon}^{(2)}$ with sample size $n=1000$, $m_{i}=3$ replicates for each subject and identity matrix (I) for the component specific covariance matrices. The results correspond to the data set that produced the median of the estimated integrated squared errors (ISE) out of a total of 100 simulated data sets.
See Section \ref{sec: mvt simulation studies} for additional details.
The upper triangular panels show the contour plots of the true two dimensional marginal densities.
The lower triangular diagonally opposite panels show the corresponding estimates.
The numbers $i,j$ at the bottom right corners of the off-diagonal panels show that the marginal densities $f_{X_{i},X_{j}}$ are plotted in those panels. 
The diagonal panels show the true (lighter shaded green lines) and the estimated (darker shaded blue lines) one dimensional marginals.
The figure is in color in the electronic version of this article.
}
\label{fig: mvt simulation results XS d4 n1000 m3 MIW X1 E1 Ind}
\end{figure}

\begin{figure}[!ht]
\begin{center}
\includegraphics[height=16cm, width=16cm, trim=0cm 0cm 0cm 0cm, clip=true]{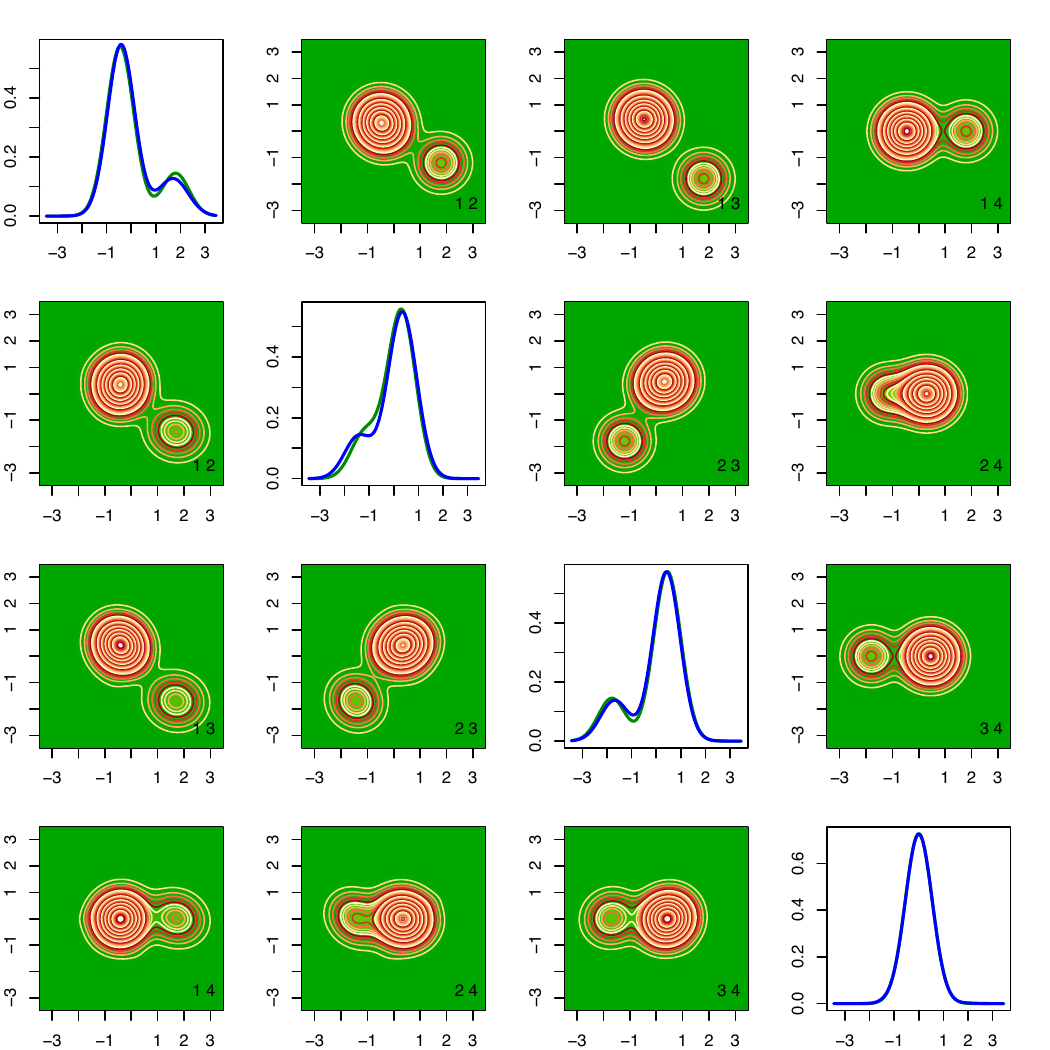}
\end{center}
\caption{\baselineskip=10pt Results for the density of the scaled measurement errors $f_{\bepsilon}$ produced by the MIW (mixtures with inverse Wishart priors) method for the conditionally heteroscedastic error distribution $f_{\bepsilon}^{(2)}$ with sample size $n=1000$, $m_{i}=3$ replicates for each subject and identity matrix (I) for the component specific covariance matrices. The results correspond to the data set that produced the median of the estimated integrated squared errors (ISE) out of a total of 100 simulated data sets.
See Section \ref{sec: mvt simulation studies} for additional details.
The upper triangular panels show the contour plots of the true two dimensional marginal densities.
The lower triangular diagonally opposite panels show the corresponding estimates.
The numbers $i,j$ at the bottom right corners of the off-diagonal panels show that the marginal densities $f_{\epsilon_{i},\epsilon_{j}}$ are plotted in those panels. 
The diagonal panels show the true (lighter shaded green lines) and the estimated (darker shaded blue lines) one dimensional marginals.
The figure is in color in the electronic version of this article.
}
\label{fig: mvt simulation results ES d4 n1000 m3 MIW X1 E1 Ind}
\end{figure}

The estimated MISEs for the cases when $\bU$ were conditionally heteroscedastic
are presented in Table \ref{tab: mvt MISEs heteroscedastic}.
Models that accommodate conditional heteroscedasticity are significantly more complex 
compared to models that assume independence of the measurement errors from $\bX$.  
The numerically more stable MLFA model thus out-performed the MIW model in all 32 cases. 
The improvements were particularly significant when the true covariance matrices were sparse and the number of subjects was small ($n=500$).
The true and estimated univariate and bivariate marginals of $f_{\bX}$ 
produced by the MIW and the MLFA methods 
when the true density of the scaled errors was a mixture of multivariate normals ($f_{\bepsilon}^{(2)}$) and the component specific covariance matrices were diagonal ($\Ind$)
are summarized in Figure \ref{fig: mvt simulation results XS d4 n1000 m3 MIW X1 E1 Ind} and Figure \ref{fig: mvt simulation results XS d4 n1000 m3 MLFA X1 E1 Ind}, respectively. 
The true and estimated univariate and bivariate marginals for the density of the scaled errors $f_{\bepsilon}$ for this case  
produced by the two methods 
are summarized in Figure \ref{fig: mvt simulation results ES d4 n1000 m3 MIW X1 E1 Ind} and Figure \ref{fig: mvt simulation results ES d4 n1000 m3 MLFA X1 E1 Ind}, respectively.  
The true and the estimated variance functions produced by the univariate submodels are summarized in Figure \ref{fig: mvt simulation results VFn d4 n1000 m3 MLFA X1 E1 Ind}.
Comparisons between Figure \ref{fig: mvt simulation results XS d4 n1000 m3 MIW X1 E1 Ind} and Figure \ref{fig: mvt simulation results XS d4 n1000 m3 MLFA X1 E1 Ind}
illustrate the limitations of the MIW models in capturing high dimensional sparse covariance matrices and the improvements that can be achieved by the MLFA models. 
The estimates of $f_{\bepsilon}$ produced by the two methods are in better agreement. 
This may be attributed to the fact that many more residuals are available for estimating $f_{\bepsilon}$ than there are $\bX_{i}$'s to estimate $f_{\bX}$. 
Figure \ref{fig: mvt simulation results VFn d4 n1000 m3 MLFA X1 E1 Ind} in the main paper and Figures \ref{fig: mvt simulation results VFn d4 n1000 m3 MIW X1 E1 AR} and \ref{fig: mvt simulation results VFn d4 n1000 m3 MIW X1 HT_E0 Ind} in the Supplementary Materials 
show that the univariate submodels can recover the true variance functions well. 
Additional figures when the true covariance matrices had auto-regressive structure (AR) are presented in the Supplementary Materials.
In this case the true covariance matrices were not sparse. 
The MLFA method still vastly dominated the MIW method when the sample size was small ($n=500$). 
When the sample size was large ($n=1000$)  the two methods produced comparable results.

The proposed deconvolution methods, in particular the MLFA method, are highly scalable. 
In small scale simulations, not reported here, we tried $p=6,8$ and $10$ and observed good empirical performance. 
We have focused here on $p=4$ dimensional problems since with $p=4$ the numbers of univariate and bivariate marginals, $p=4$ and ${p\choose 2}=6$, 
remain manageable and the results are conveniently graphically summarized. 

Additional small scale simulations for a variety of other distributions with similar MISE patterns are presented in the Supplementary Materials.

\section{Example}  \label{sec: mvt data analysis}

Dietary habits are known to be leading causes of many chronic diseases.
Accurate estimation of the distributions of dietary intakes is thus important
in nutritional epidemiologic surveillance and epidemiology.
Nutritionists are typically interested not just in the consumption patterns of individual dietary components but also in their joint consumption patterns. 
By the very nature of the problem, $\bX$, the average long term daily intakes of the dietary components, can never be directly observed.
Data are thus typically collected from a representative sample of the population in the form of dietary recalls, 
the subjects participating in the study remembering and reporting the type and amount of food they had consumed in the past 24 hours.
The problem of estimating the joint consumption pattern of the dietary components from the contaminated 24-hour recalls then becomes a problem of multivariate density deconvolution.  

A large scale epidemiologic study conducted by the National Cancer Institute,
the Eating at America's Table (EATS) study (Subar, et al. 2001),
serves as the motivation for this paper.
In this study $n=965$ participants were interviewed $m_i=4$ times over the course of a year 
and their 24 hour dietary recalls ($\bW_{ij}$'s) were recorded.
The goal is to estimate the joint consumption patterns of the true daily intakes ($\bX_{i}$'s).

\begin{figure}[!ht]
\begin{center}
\includegraphics[height=16cm, width=16cm, trim=0cm 0cm 0cm 0cm, clip=true]{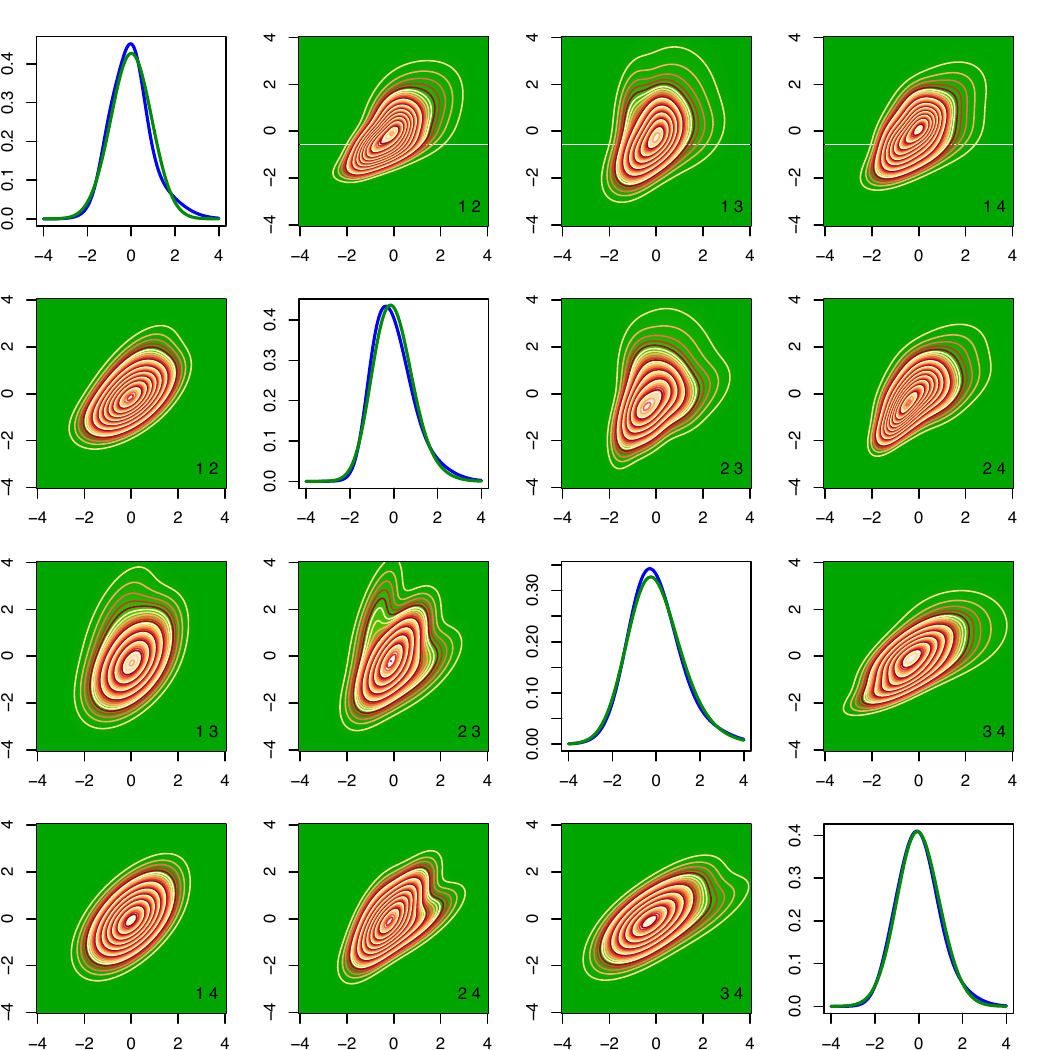}
\end{center}
\caption{\baselineskip=10pt Results for the EATS data set for the density of the scaled errors $f_{\bepsilon}$. 
The off-diagonal panels show the contour plots of two-dimensional marginals estimated by the MIW method (upper triangular panels) and the MLFA method (lower triangular panels).
The numbers $i,j$ at the bottom right corners of the off-diagonal panels show that the marginal densities $f_{\epsilon_{i},\epsilon_{j}}$ are plotted in those panels. 
The diagonal panels show the one dimensional marginal densities estimated by the MIW method (darker shaded blue lines) 
and the MLFA method (lighter shaded green lines).
The figure is in color in the electronic version of this article.
}
\label{fig: mvt EATS data results ES}
\end{figure}

To illustrate our methodology, we consider the problem of estimating the joint consumption pattern of four dietary components, 
namely (a) carbohydrate, (b) fiber, (c) protein and (d) a mineral potassium. 
%
%
Figure \ref{fig: mvt EATS data results VFn} shows the plots of subject-specific means versus subject-specific variances for daily intakes of the dietary components 
with the estimates of the variance functions produced by univariate submodels superimposed over them.
As is clearly identifiable from this plot, conditional heteroscedasticity is a very prominent feature of the measurements errors contaminating the 24 hour recalls. 
%
%
The estimated univariate and bivariate marginal densities of average long term daily intakes of the dietary components produced 
by the MIW method and the MLFA method are summarized in Figure \ref{fig: mvt EATS data results XS}. 
The estimated univariate and bivariate marginal densities for the scaled errors are summarized in Figure \ref{fig: mvt EATS data results ES}.
The estimated marginals of $\bX$ produced by the two methods look quite different, 
while the estimated marginals of $\bepsilon$ are in close agreement. 
The estimated univariate and bivariate marginal densities of the long term intakes of the dietary components produced by the MIW model look irregular and unstable, 
whereas the estimates produced by the MLFA model look relatively more regular and stable.
In experiments not reported here, we observed that the estimates produced by the MIW method were sensitive to the choice of the number of mixture components, 
but the estimates produced by the MLFA model were quite robust. 
The trace plots and the frequency distributions of the of the numbers of nonempty mixture components are summarized 
in Figures \ref{fig: mvt EATS  data results Trace Plots MIW} and \ref{fig: mvt EATS  data results Trace Plots MLFA} in the Supplementary Materials  
and provide some idea about the relative stability of the two methods.
These observations are similar to that made in Section \ref{sec: mvt simulation studies} for conditionally heteroscedastic measurement errors and sparse covariance matrices.

We next comment only on the estimates produced by the MLFA method assuming them to be closer to the truth. 
The estimates show that the long term daily intakes of the four dietary components are strongly correlated. 
The shapes of the bivariate consumption patterns suggest deviations from normality. 
Similarly, the shapes of the bivariate marginals for the scaled errors suggest that the measurement errors in the reported 24 hour recalls are positively correlated and deviate from normality.  
People who consume more are expected to do so for most dietary components. 
Strong correlations between the intakes of the dietary components are thus somewhat expected. 
The correlations among different components of the measurement errors suggest that people usually have a tendency to either over-report or under-report the daily intakes.  
These findings illustrate the importance of robust but numerically stable multivariate deconvolution methods in nutritional epidemiologic studies. 

Additional discussions on potentially far-reaching impact of our work on nutritional epidemiology studies are deferred to 
Section \ref{sec: mvt potential impact} in the Supplementary Materials.

\section{Discussion} \label{sec: mvt discussion}
We considered the problem of multivariate density deconvolution when the measurement error density is not known 
but replicated proxies are available for some individuals.
We used flexible finite mixtures of multivariate normal kernels with symmetric Dirichlet priors on the mixture probabilities to model both the density of interest and the density of the measurement errors.  
We proposed a novel technique to make the model for the density of the errors satisfy a zero mean restriction. 
We showed that the dense parametrization of inverse Wishart priors are not suitable for modeling covariance matrices in the presence of measurement errors.  
We proposed a numerically more stable approach based on latent factor characterization of the covariance matrices with sparsity inducing priors on the factor loading matrices.
We built models for conditionally heteroscedastic additive measurement errors that also automatically accommodate multivariate multiplicative measurement errors.



The methodological contributions of this article are not limited to deconvolution problems. 
Mixtures of latent factor analyzers with sparsity inducing priors on the factor loading matrices can be used in other high dimensional applications including ordinary density estimation. 
The techniques proposed in Section \ref{sec: mvt density of homoscedastic errors} to enforce the mean zero moment restriction on the measurement errors can be readily used to model multivariate regression errors that are distributed independently of the predictors. 
The technique can also be adapted to relax the strong assumption of multivariate normality made by Hoff and Niu (2012) and Fox and Dunson (2016) in covariance regression problems. 


As explained in Sections \ref{sec: mvt density of heteroscedastic errors} and \ref{sec: mvt multiplicative errors} in the main paper and also in Section \ref{sec: mvt comments on the model for U given X} in the Supplementary Materials, 
the structural separability assumption (\ref{eq: mvt multiplicative structure})
arises naturally in both additive and multiplicative multivariate measurement error settings. 
%
It would still be interesting, in future work, to consider more general covariance models that allow $\var(U_{ij\ell}\vert \bX)$ 
to be explained primarily by $X_{i\ell}$, as in the current approach,  
but would allow the residual variability to be explained by the remaining components $\{X_{im}\}_{m\neq \ell}$ of $\bX$. 
The current MCMC based implementation of the proposed methodology is computationally intensive. 
We are pursuing the development of faster algorithms for approximate posterior inference as the subject of a separate manuscript.

The question of consistency of Bayesian procedures is intimately related to the flexibility of the priors.
For instance, in ordinary density estimation problems inclusion of the true density in the KL support of the prior is a  sufficient condition to ensure weak consistency 
via the Schwartz theorem. 
In density deconvolution problems such a condition is not sufficient but is still required. 
The results from Section \ref{sec: mvt model flexibility} thus provide crucial first steps in that direction.
We have not pursued the question of consistency of the proposed deconvolution methods any further in this article. 
It remains an important direction for future research.

\baselineskip=14pt
\section*{Supplementary Materials}
The Supplementary Materials discuss 
the choice of hyper-parameters and MCMC algorithms to sample from the posterior, 
including the two-stage estimation procedure for conditionally heteroscedastic measurement errors.
The Supplementary Materials also present our arguments in favor of finite mixture models, 
pointing out how their close connections and their subtle differences with possible infinite dimensional alternatives  
are exploited to achieve significant reduction in computational complexity 
while retaining the major advantages of infinite dimensional mixture models including model flexibility and automated model selection and model averaging.
The Supplementary Materials additionally present 
discussions on the contrasts between regression and measurement errors 
that preclude the use of covariance regression techniques to model conditionally heteroscedastic measurement errors,   
the proofs of the theoretical results presented in Section \ref{sec: mvt model flexibility}, 
some additional figures, 
and results of additional simulation experiments. 
R programs implementing the deconvolution methods for conditionally heteroscedastic errors are included as part of the Supplementary Materials. 
The EATS data analyzed in Section \ref{sec: mvt data analysis} can be accessed from National Cancer Institute by arranging a Material Transfer Agreement.
A simulated data set, simulated according to one of the designs described in Section \ref{sec: mvt simulation studies}, 
and a `readme' file providing additional details are also included in the Supplementary Materials.

\baselineskip=14pt
\section*{Acknowledgments}
Pati's research was supported by Award No. N00014-14-1-0186 from the Office of Naval Research. 
Carroll's research was supported in part by a grant U01-CA057030 from the National Cancer Institute.
Mallick's research was supported in part by National Cancer Institute of the National Institutes of Health under award number R01CA194391. 
We acknowledge the Texas A\&M University Brazos HPC cluster that contributed to the research reported here.


\section*{References}
\refmark
Bhattacharya, A. and Dunson, D. B. (2011). Sparse Bayesian infinite factor models.
\BIOK, 98, 291-306.
\refmark
Bhattacharya, A., Pati, D., Pillai, N. and Dunson, D. B. (2014). Bayesian shrinkage. 
\emph{Unpublished manuscript}.
\refmark
Brown, P. J. and Griffin, J. E. (2010). Inference with normal-gamma prior distributions in regression problems.
\BA, 5, 171-188.
\refmark
Bovy, J., Hogg, D. W. and Rowies, S. T. (2011). Extreme deconvolution: inferring complete distribution functions from noisy, heterogeneous and incomplete observations.
\ANNALSAS, 5, 1657-1677.
\refmark
Buonaccorsi, J. P. (2010). \emph{Measurement Error: Models, Methods and Applications}.
New York: \emph{Chapman and Hall/CRC}.
\refmark
Carroll, R. J. and Hall, P. (1988). Optimal rates of convergence for deconvolving a density.
\JASA, 83, 1184-1186.
\refmark
Carroll, R. J. and Hall, P. (2004). Low order approximations in deconvolution and regression with errors in variables.
\JRSSB, 66, 31-46.
\refmark
Carroll, R. J., Ruppert, D., Stefanski, L. A. and Crainiceanu, C. M. (2006).
\emph{Measurement Error in Nonlinear Models} (2nd ed.). Boca Raton: \emph{Chapman and Hall/CRC Press}.
\refmark
Carvalho, M. C., Polson, N. G. and  Scott, J. G. (2010). The horseshoe estimator for sparse signals.
\BIOK, 97, 465-480.
\refmark
Comte, F. and Lacour, C. (2013). Anisotropic adaptive density deconvolution. 
\emph{Annales de l'Institut Henri Poincar\'e - Probabilit\'{e}s et Statistiques}, 49, 569-609.
\refmark
Devroye, L. (1989). Consistent deconvolution in density estimation.
\CANADAJS, 17, 235-239.
\refmark
Diggle, P. J. and Hall, P. (1993). A Fourier approach to nonparametric deconvolution of a density estimate.
\JRSSB, 55, 523-531.
\refmark
Eilers, P. H. C. and Marx, B. D. (1996). Flexible smoothing with B-splines and penalties.
\STATSCI, 11, 89-121.
%
%
\refmark
Fan, J. (1991a). On the optimal rates of convergence for nonparametric deconvolution problems.
\ANNALS, 19, 1257-1272.
\refmark
Fan, J. (1991b). Global behavior of deconvolution kernel estimators.
\SSNC, 1, 541-551.
\refmark
Fan, J. (1992). Deconvolution with supersmooth distributions.
\CANADAJS, 20, 155-169.
\refmark
Fokou\'{e}, E. and Titterington, D. M. (2003). Mixtures of factor analyzers. Bayesian estimation and inference by stochastic simulation. 
\emph{Machine Learning}, 50, 73-94.
\refmark
Fox, E. B. and Dunson, D. (2016). Bayesian nonparametric covariance regression.
To appear in \emph{Journal of Machine Learning Research}.
%
%
\refmark
Fr\"{u}hwirth-Schnatter, S. (2006). \emph{Finite Mixture and Markov Switching Models}.
New York: \emph{Springer}.
\refmark
Geweke, J. (2007). Interpretation and inference in mixture models: Simple MCMC works.
\CDA, 51, 3529-3550.
\refmark
Hazelton, M.L. and Turlach, B.A. (2009). Nonparametric density deconvolution by weighted kernel estimators. 
\SaC, 19, 217-228.
\refmark
Hazelton, M.L. and Turlach, B.A. (2010). Semiparametric density deconvolution. 
\SCAN, 37, 91-108.
\refmark
Hesse, C. H. (1999). Data driven deconvolution.
\JNS, 10, 343-373.
\refmark
Hoff, P. D. and Niu, X. (2012). A covariance regression model.
\SSNC, 22, 729-753.
\refmark
Hu, Y and Schennach, S. (2008). Instrumental Variable Treatment of Nonclassical Measurement Error Models. \emph{Econometrica}, 76, 195-216.
\refmark
Li, T. and Vuong, Q. (1998). Nonparametric estimation of the measurement error model using multiple indicators.
\JMA, 65, 139-165.
\refmark
Liu, M. C. and Taylor, R. L. (1989). A consistent nonparametric density estimator for the deconvolution problem.
\CANADAJS, 17, 427-438.
%
%
\refmark
Masry, E. (1991). Multivariate probability density deconvolution for stationary random processes.
\IEEETIT, 37, 1105-1115.
\refmark
Mengersen, K. L., Robert, C. P. and Titterington, D. M. (eds) (2011). \emph{Mixtures - Estimation and Applications}.
Chichester: \emph{John Wiley}.
\refmark
Neumann, M. H. (1997). On the effect of estimating the error density in nonparametric deconvolution.
\JNS, 7, 307-330.
%
%
%
%
%
\refmark
Sarkar, A., Mallick, B. K., Staudenmayer, J., Pati, D. and Carroll, R. J. (2014).  
Bayesian semiparametric density deconvolution in the presence of conditionally heteroscedastic measurement errors. 
\JCGS, 23, 1101-1125. 
\refmark
Schennach, S. (2004). Nonparametric regression in the presence of measurement error. 
\ECTH, 20, 1046-1093.
\refmark
Staudenmayer, J., Ruppert, D. and Buonaccorsi, J. P. (2008). Density estimation in the presence of heteroscedastic measurement error.
\JASA, 103, 726-736.
\refmark
Subar, A. F., Thompson, F. E., Kipnis, V., Midthune, D., Hurwitz, P. McNutt, S., McIntosh, A. and Rosenfeld, S. (2001).
Comparative validation of the block, Willet, and National Cancer Institute food frequency questionnaires.
\AJE, 154, 1089-1099.
\refmark
Youndj\'e, E. and Wells, M. T. (2008). Optimal bandwidth selection for multivariate kernel deconvolution.
\emph{TEST}, 17, 138-162.

\newgeometry{left=2cm,right=2.5cm,top=2.5cm,bottom=0.1cm}
\newpage
\thispagestyle{empty}

\begin{table}[!ht]\footnotesize
\begin{center}
\begin{tabular}{|c|c|c|c c c|}
\hline
\multirow{2}{75pt}{True Error Distribution} & \multirow{2}{50pt}{Covariance Structure} & \multirow{2}{*}{Sample Size}	& \multicolumn{3}{|c|}{MISE $\times 10^4$} \\ \cline{4-6}
				&												&		& MLFA		& MIW		& Naive			  \\  \hline\hline
\multirow{8}{75pt}{(a) Multivariate Normal}	& \multirow{2}{50pt}{\centering I}	
																& 500	& \bf{1.24}		& 3.05		& 8.01		\\
				&												& 1000	& \bf{0.59}		& 1.33		& 6.58		\\ \cline{2-6}
				& \multirow{2}{50pt}{\centering LF}						& 500	& 6.88		& \bf{6.33}		& 33.41		\\
				&												& 1000	& 5.15		&\bf{3.10}		& 32.42		\\  \cline{2-6} 
				& \multirow{2}{50pt}{\centering AR}  						& 500	& 11.91		& \bf{5.51}		& 27.17		\\
				&												& 1000	& 9.82		& \bf{2.78}		& 26.01		\\  \cline{2-6}
				& \multirow{2}{50pt}{\centering EXP}  					& 500	& 7.15		& \bf{4.40}		& 17.82		\\
				&												& 1000	& 5.46		& \bf{2.19}		& 17.40		\\ \hline
\multirow{8}{75pt}{(b) Mixture of Multivariate Normal}	& \multirow{2}{50pt}{\centering I}	
																& 500	& \bf{1.28}		& 3.24		& 5.97	\\
				&												& 1000	& \bf{0.64}		& 1.37		& 4.99	\\ \cline{2-6}
				& \multirow{2}{50pt}{\centering LF}  						& 500	& \bf{7.28}		& 7.51		& 31.62	\\
				&												& 1000	& \bf{4.17}		& 4.34		& 31.48	\\  \cline{2-6}
				& \multirow{2}{50pt}{\centering AR}  						& 500	& 10.43		& \bf{6.66}		& 30.74	\\
				&												& 1000	& 7.75		& \bf{4.35}		& 28.90	\\  \cline{2-6}
				& \multirow{2}{50pt}{\centering EXP}  					& 500	& 7.16		& \bf{5.18}		& 17.85	\\
				&												& 1000	& 4.87		& \bf{2.66}		& 17.26	\\ \hline

\end{tabular}
\caption{\baselineskip=10pt Mean integrated squared error (MISE) performance
of MLFA (mixtures of latent factor analyzers) and MIW (mixtures with inverse Wishart priors) density deconvolution models described in Section \ref{sec: mvt density deconvolution models} of this article
for {\bf homoscedastic} errors
compared with a naive method that ignores measurement errors 
for different measurement error distributions.
The minimum value in each row is highlighted.
}
\label{tab: mvt MISEs homoscedastic}
\end{center}
\end{table}

\thispagestyle{empty}

\begin{table}[!ht]\footnotesize
\begin{center}
\begin{tabular}{|c|c|c|c c c|}
\hline
\multirow{2}{75pt}{True Error Distribution} & \multirow{2}{50pt}{Covariance Structure} & \multirow{2}{*}{Sample Size}	& \multicolumn{3}{|c|}{MISE $\times 10^4$} \\ \cline{4-6}
				&												&		& MLFA		& MIW		& Naive	  \\  \hline\hline
\multirow{8}{75pt}{(a) Multivariate Normal}	& \multirow{2}{50pt}{\centering I}	
																& 500	& \bf{2.53}		& 19.08	& 10.64			\\
				&												& 1000	& \bf{1.15}		& 9.43	& 9.14			\\ \cline{2-6}
				& \multirow{2}{50pt}{\centering LF}  						& 500	& \bf{11.46}	& 34.21	& 21.33			\\
				&												& 1000	& \bf{5.78}		& 15.98	& 20.75			\\  \cline{2-6} 
				& \multirow{2}{50pt}{\centering AR}  						& 500	& \bf{17.11}	& 30.83	& 36.44			\\
				&												& 1000	& \bf{10.77}	& 12.46	& 36.37			\\ \cline{2-6}
				& \multirow{2}{50pt}{\centering EXP}  					& 500	& \bf{11.63}	& 26.99	& 24.28			\\
				&												& 1000	& \bf{6.67}		& 10.56	& 23.36			\\ \hline
\multirow{8}{75pt}{(b) Mixture of Multivariate Normal}	& \multirow{2}{50pt}{\centering I}	
																& 500	& \bf{2.79}		& 22.17	& 20.16		\\
				&												& 1000	& \bf{1.38} 	& 10.55	& 19.39		\\ \cline{2-6}
				& \multirow{2}{50pt}{\centering LF}  						& 500	& \bf{13.39} 	& 35.67	& 43.43		\\
				&												& 1000	& \bf{7.50} 	& 20.86 	& 43.28		\\  \cline{2-6}
				& \multirow{2}{50pt}{\centering AR}  						& 500	& \bf{18.27}	& 35.70	& 75.26		\\
				&												& 1000	& \bf{12.06}	& 16.64	& 77.55		\\ \cline{2-6}
				& \multirow{2}{50pt}{\centering EXP}  					& 500	& \bf{12.11}	& 34.50	& 48.76		\\
				&												& 1000	& \bf{7.59}		& 13.74	& 50.02		\\ \hline

\end{tabular}
\caption{\baselineskip=10pt Mean integrated squared error (MISE) performance
of MLFA (mixtures of latent factor analyzers) and MIW (mixtures with inverse Wishart priors) density deconvolution models described in Section \ref{sec: mvt density deconvolution models} of this article
for {\bf conditionally heteroscedastic} errors
compared with a naive method that ignores measurement errors 
for different measurement error distributions.
The minimum value in each row is highlighted.
}
\label{tab: mvt MISEs heteroscedastic}
\end{center}
\end{table}
\restoregeometry



\newpage
\thispagestyle{empty}

\begin{figure}[!ht]
\begin{center}
\includegraphics[height=12cm, width=16cm, trim=0cm 0cm 0cm 0cm, clip=true]{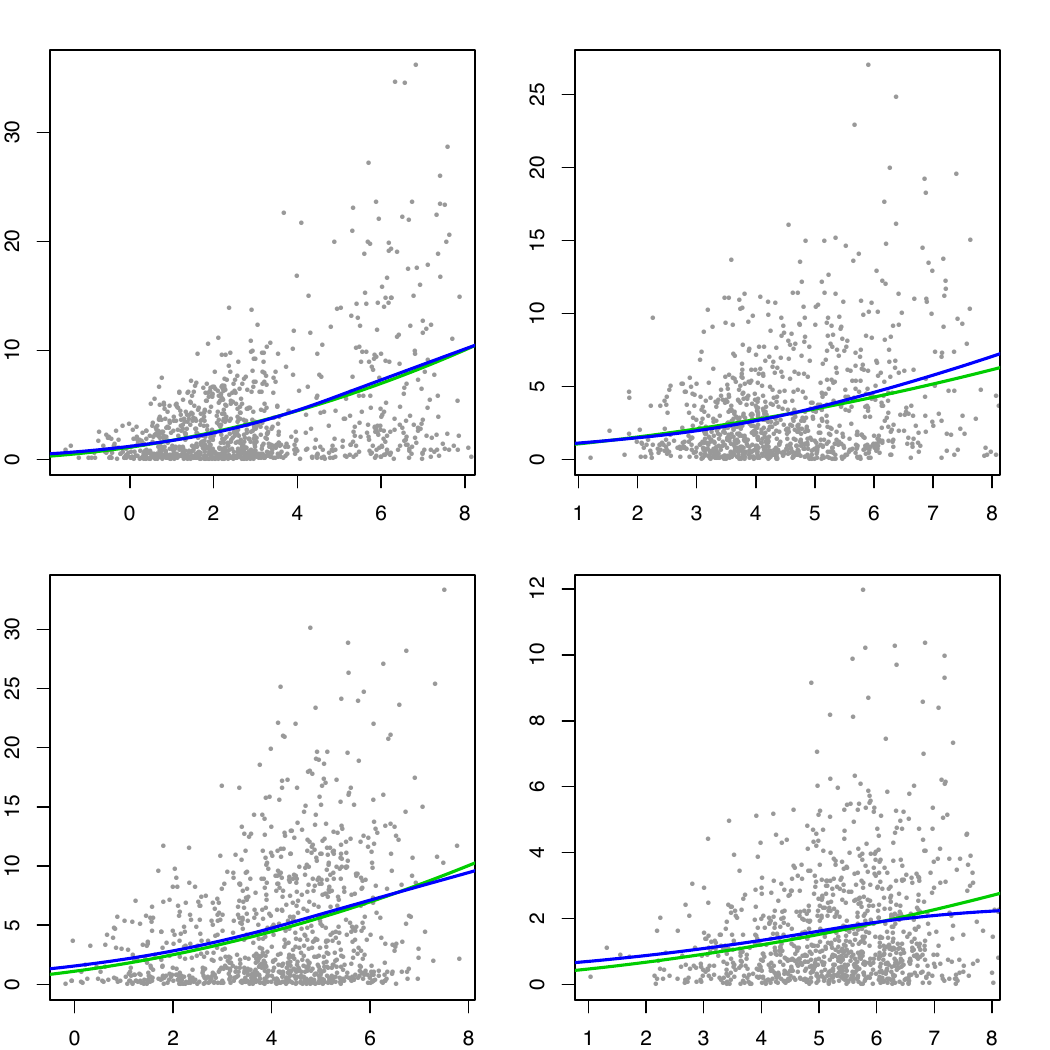}
\end{center}
\caption{\baselineskip=10pt Results for conditional variability $\var(U \vert X)=s^{2}(X)\var(\epsilon)$ produced by the univariate density deconvolution method for each component of $\bX$ for the conditionally heteroscedastic error distribution $f_{\bepsilon}^{(2)}$ with sample size $n=1000$, $m_{i}=3$ replicates for each subject and identity matrix (I) for the component specific covariance matrices. The results correspond to the data set that produced the median of the estimated integrated squared errors (ISE) out of a total of 100 simulated data sets for the MLFA (mixtures of latent factor analyzers) method.
For each component of $\bX$, the true variance function is $s^{2}(X) = (1+X/4)^{2}$.
See Section \ref{sec: mvt density of heteroscedastic errors} and Section \ref{sec: mvt estimation of variance functions} for additional details.
In each panel, the true (lighter shaded green lines) and the estimated (darker shaded blue lines) variance functions 
are superimposed over a plot of subject specific sample means vs subject specific sample variances.
The figure is in color in the electronic version of this article.
}
\label{fig: mvt simulation results VFn d4 n1000 m3 MLFA X1 E1 Ind}
\end{figure}

\newpage
\thispagestyle{empty}

\newpage
\thispagestyle{empty}

\newpage
\thispagestyle{empty}

\vspace{-0.25cm}

\newpage
\thispagestyle{empty}

\newpage
\thispagestyle{empty}

\begin{figure}[!ht]
\begin{center}
\includegraphics[height=16cm, width=16cm, trim=0cm 0cm 0cm 0cm, clip=true]{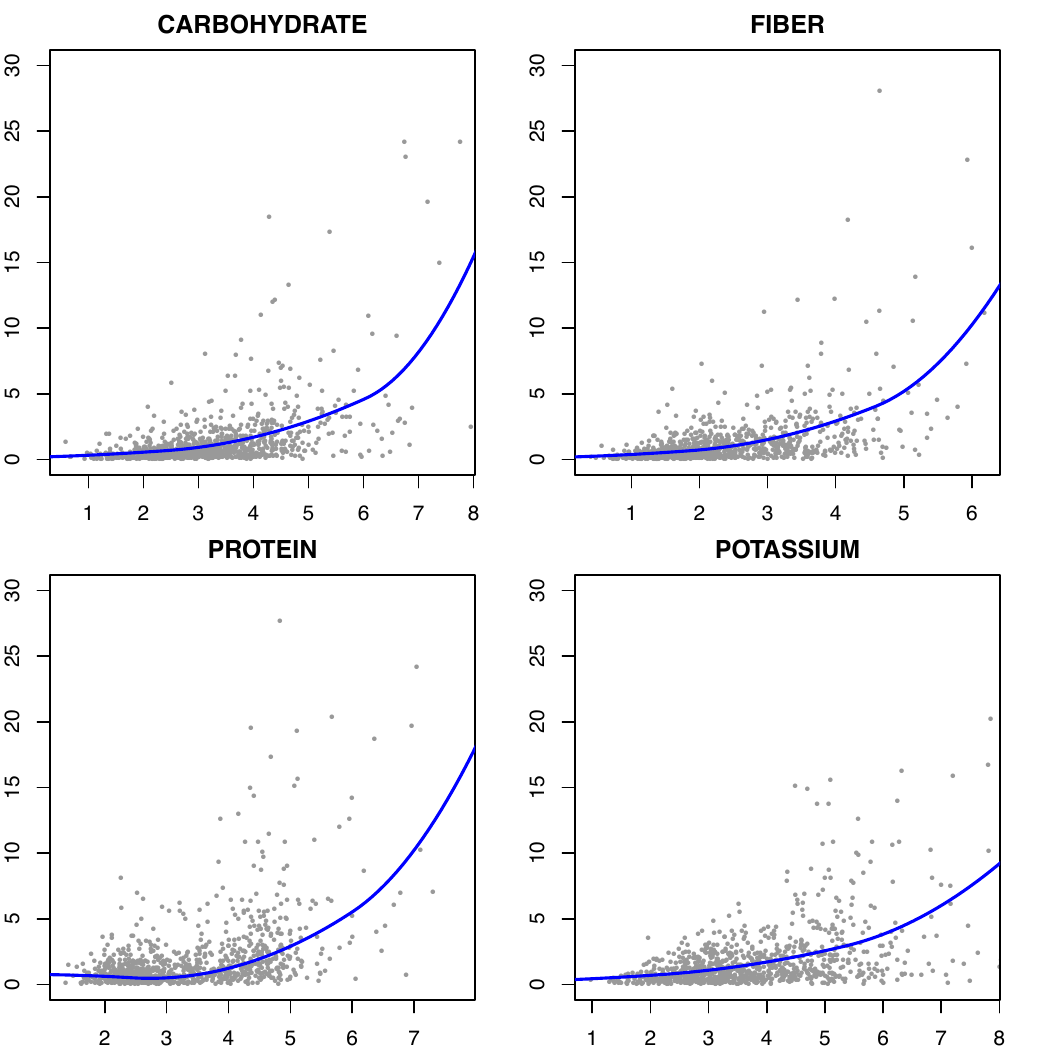}
\end{center}
\caption{\baselineskip=10pt Estimated variance functions $\var(U\vert X) = s^{2}(X)\var(\epsilon)$ produced by the univariate density deconvolution method for each component of $\bX$ for the EATS data set with sample size $n=965$, $m_{i}=4$ replicates for each subject.
See Section \ref{sec: mvt data analysis} for additional details.
The figure is in color in the electronic version of this article.
}
\label{fig: mvt EATS data results VFn}
\end{figure}

\newpage
\thispagestyle{empty}

\begin{figure}[!ht]
\begin{center}
\includegraphics[height=16cm, width=16cm, trim=0cm 0cm 0cm 0cm, clip=true]{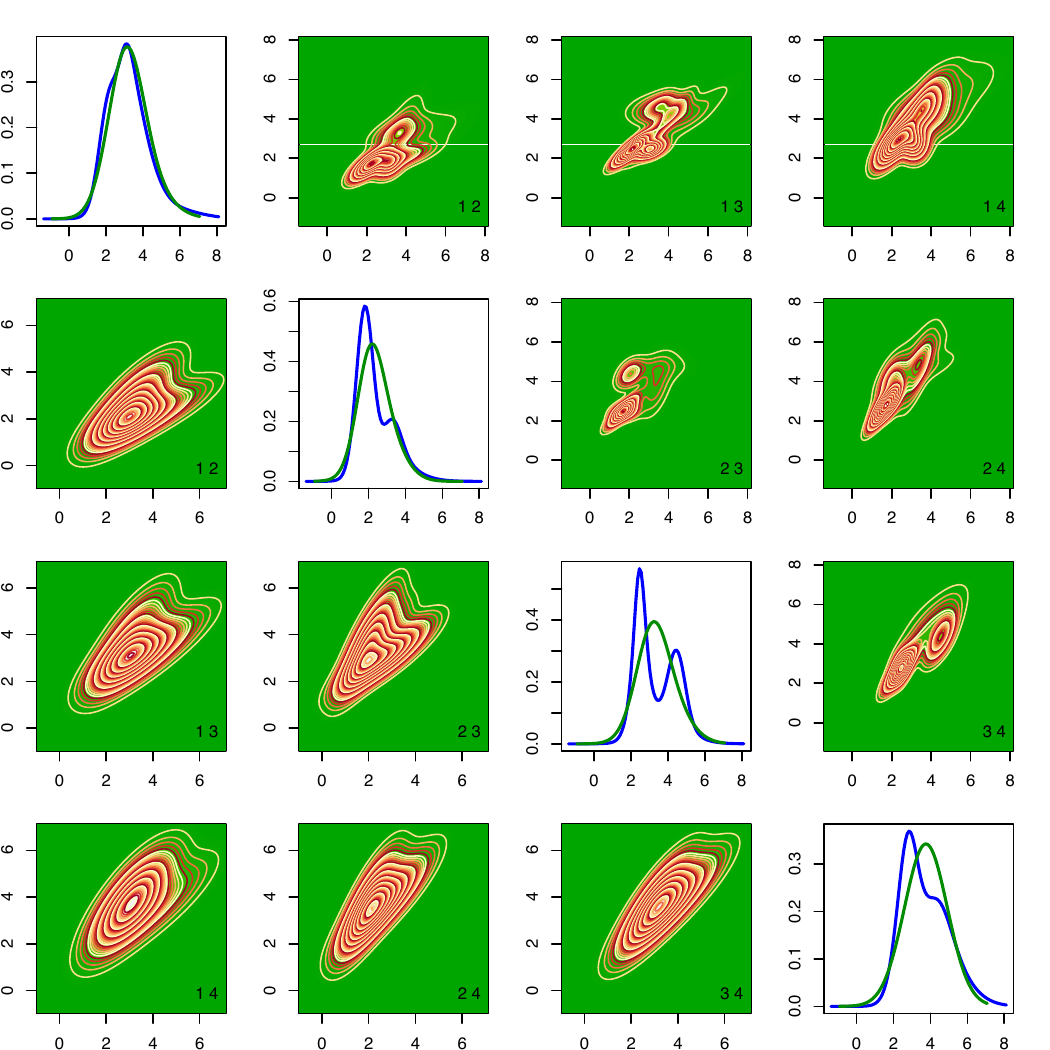}
\end{center}
\caption{\baselineskip=10pt Results for the EATS data set for the  $f_{\bX}$. 
The off-diagonal panels show the contour plots of two-dimensional marginals estimated by the MIW method (upper triangular panels) and the MLFA method (lower triangular panels).
The numbers $i,j$ at the bottom right corners of the off-diagonal panels show that the marginal densities $f_{X_{i},X_{j}}$ are plotted in those panels. 
The diagonal panels show the one dimensional marginal densities estimated by the MIW method (darker shaded blue lines) 
and the MLFA method (lighter shaded green lines).
The figure is in color in the electronic version of this article.
}
\label{fig: mvt EATS data results XS}
\end{figure}

\newpage
\thispagestyle{empty}


\clearpage\pagebreak\newpage
\pagestyle{fancy}
\fancyhf{}
\rhead{\bfseries\thepage}
\lhead{\bfseries SUPPLEMENTARY MATERIALS}

\baselineskip 15pt
\vspace{-0.5cm}
\begin{center}
{\LARGE{Supplementary Materials} 
for\\ {\bf Bayesian Semiparametric Multivariate Density Deconvolution}}
\end{center}

\setcounter{equation}{0}
\setcounter{page}{1}
\setcounter{table}{1}
\setcounter{figure}{0}
\setcounter{section}{0}
\numberwithin{table}{section}
\renewcommand{\theequation}{S.\arabic{equation}}
\renewcommand{\thesubsection}{S.\arabic{section}.\arabic{subsection}}
\renewcommand{\thesection}{S.\arabic{section}}
\renewcommand{\thepage}{S.\arabic{page}}
\renewcommand{\thetable}{S.\arabic{table}}
\renewcommand{\thefigure}{S.\arabic{figure}}
\baselineskip=15pt

\begin{center}
Abhra Sarkar\\
Department of Statistical Science, Duke University, Durham, NC 27708-0251, USA\\
abhra.sarkar@duke.edu \\
\vspace{2mm}
Debdeep Pati\\
Department of Statistics, Florida State University, Tallahassee, FL 32306-4330, USA\\
debdeep@stat.fsu.edu\\
\vspace{2mm}
Bani K. Mallick\\
Department of Statistics, Texas A\&M University, 3143 TAMU, College Station,\\ TX 77843-3143, USA\\
bmallick@stat.tamu.edu\\
\vspace{2mm}
Raymond J. Carroll\\
Department of Statistics, Texas A\&M University, 3143 TAMU, College Station,\\ TX 77843-3143, USA\\
and School of Mathematical and Physical Sciences, University of Technology Sydney, Broadway NSW 2007, Australia\\
carroll@stat.tamu.edu\\
\end{center}

\baselineskip=14pt

The Supplementary Materials are organized as follows. 
Section \ref{sec: mvt choice of hyper-parameters} discusses the choice of hyper-parameters. 
In Section \ref{sec: mvt posterior computation}, we describe a Gibbs sampler for drawing samples from the posterior 
of the deconvolution model for multivariate independently distributed homoscedastic errors, described in Section \ref{sec: mvt density of homoscedastic errors} of the main paper.
In Section \ref{sec: mvt estimation of variance functions}, we detail a two stage estimation procedure for drawing samples from the posterior 
of the deconvolution model for multivariate conditionally heteroscedastic measurement errors described in Section \ref{sec: mvt density of heteroscedastic errors} of the main paper. 
Section \ref{sec: mvt two-stage sampler} provides heuristic justification for the two-stage sampler. 
In Section \ref{sec: mvt comments on the model for U given X}, we provide additional detailed discussion 
of the model for multivariate conditionally heteroscedastic measurement errors described in Section \ref{sec: mvt density of heteroscedastic errors} of the main paper, 
contrasting it with models for multivariate conditionally varying regression errors (Section \ref{sec: mvt regression vs measurement errors}), 
its connections with latent factor models (Section \ref{sec: latent factor models for different covariance classes}), 
its flexibility, limitations, and plausible generalizations (Section \ref{sec: mvt cov mat model}), 
and tools for model adequacy checks (Section \ref{sec: mvt model adequacy checks}).
Section \ref{sec: mvt finite vs infinite mixture models} presents our arguments in favor of finite mixture models, 
pointing out how their close connections and their subtle differences with possible infinite dimensional alternatives  
are exploited to achieve significant reduction in computational complexity (Section \ref{sec: mvt computational complexity})
while retaining the major advantages of infinite dimensional mixture models including model flexibility (Section \ref{sec: mvt supp mat model flexibility}) 
and automated model selection and model averaging (Section \ref{sec: mvt model selection and model averaging}).
Section \ref{sec: mvt proofs of theoretical results} details proofs of the theoretical results presented in Section \ref{sec: mvt model flexibility} of the main paper. 
Section \ref{sec: mvt additional figures} presents additional figures related to the simulation experiments discussed in Section \ref{sec: mvt simulation studies} of the main paper. 
Section \ref{sec: mvt additional simulation studies} presents results of additional simulation experiments. 
Section \ref{sec: mvt potential impact} discusses potentially far-reaching impact of our work in nutritional epidemiology.

\baselineskip=17pt
\newpage
\section{Choice of Hyper-Parameters} \label{sec: mvt choice of hyper-parameters}

We discuss the choice of hyper-parameters in this section. 
To avoid unnecessary repetition, in this section and onwards, 
symbols sans the subscripts $\bX$ and $\bepsilon$ are sometimes used as generics for similar components and parameters of the models.
For example, $K$ is a generic for $K_{\bX}$ and $K_{\bepsilon}$; $\bmu_{k}$ is a generic for $\bmu_{\bX,k}$ and $\bmu_{\bepsilon,k}$; and so on.

\begin{enumerate}[topsep=0pt,itemsep=-1ex,partopsep=2ex,parsep=2ex, leftmargin=0cm, rightmargin=0cm, wide=3ex]
\item {\bf Number of mixture components:} 
Practical application of our method requires that a decision be made on the number of mixture components $K_{\bX}$ and $K_{\bepsilon}$ in the models for the densities $f_{\bX}$ and $f_{\bepsilon}$, respectively.

Our simulation experiments suggest that when the true densities are finite mixtures of multivariate normals and $K_{\bX}$ and $K_{\bepsilon}$ are assigned values greater than the corresponding true numbers, 
the MCMC chain often quickly reaches a steady state where the redundant components become empty.
See Figures \ref{fig: mvt simulation results Trace Plots d4 n1000 m3 MLFA X1 E1 Ind}, 
\ref{fig: mvt simulation results Trace Plots d4 n1000 m3 MIW X1 E1 AR} and 
\ref{fig: mvt simulation results Trace Plots d4 n1000 m3 MLFA X1 E1 AR} in the Supplementary Materials for illustrations.
These observations are similar to that made in the context of ordinary density estimation by Rousseau and Mengersen (2011)
who studied the asymptotic behavior of the posterior for overfitted mixture models and showed that when $\alpha/K< L/2$, 
where $L$ denotes the number of parameters specifying the component kernels, 
the posterior is stable and concentrates in regions with empty redundant components. 
We set $\alpha_{\bX}=\alpha_{\bepsilon} =1$ so that the condition $\alpha/K < L/2$ is satisfied.

Educated guesses about $K_{\bX}$ and $K_{\bepsilon}$ may nevertheless be useful in safeguarding against gross overfitting that would result in a wastage of computation time and resources.
The following simple strategies may be employed. 
Model based cluster analysis techniques as implemented by the mclust package in R (Fraley and Raftery, 2007) may be applied to
the starting values of $\bX_{i}$ and the corresponding residuals, obtained by fitting univariate submodels for each component of $\bX$, 
to get some idea about $K_{\bX}$ and $K_{\bepsilon}$.
The chain may be started with larger values of $K_{\bX}$ and $K_{\bepsilon}$ and after a few hundred iterations the redundant empty components may be deleted on the fly. 

As shown in Section \ref{sec: mvt model flexibility}, 
our methods can approximate a large class of data generating densities, 
and we found the strategy described above to be very effective in all cases we experimented with. 
The parameter $\alpha$ now plays the role of a smoothing parameter, smaller values favoring a smaller number of mixture components and thus smoother densities. 
In simulation experiments involving multivariate t and multivariate Laplace distributions reported in the Supplementary Materials, 
and in some other cases not reported here, the values $\alpha_{\bX}=\alpha_{\bepsilon} =1$ worked well. 

As we discuss in Section \ref{sec: mvt simulation studies}, the MIW method becomes highly numerically unstable 
when the measurement errors are conditionally heteroscedastic and the true covariance matrices are highly sparse. 
In these cases in particular, the MIW method usually requires much larger sample sizes for the asymptotic results to hold 
and in finite samples the above mentioned strategy usually overestimates the required number of mixture components. 
See Figure \ref{fig: mvt simulation results Trace Plots d4 n1000 m3 MIW X1 E1 Ind} in the Supplementary Materials for an illustration.
Since mixtures based on $(K+1)$ components are at least as flexible as mixtures based on $K$ components, as far as model flexibility is concerned, such overestimation is not an issue.  
But since this also results in clusters of smaller sizes, 
the estimates of the component specific covariance matrices become numerically even more unstable, 
further compounding the stability issues of the MIW model. 
In contrast, for the numerically more stable MLFA model, for the exact opposite reasons, 
the asymptotic results are valid for moderate sample sizes and such models are also more robust to overestimation of the number of nonempty clusters.

\item {\bf Number of latent factors:} 
For the MLFA method, the MCMC algorithm summarized in Section \ref{sec: mvt posterior computation} also requires that the component specific infinite factor models be truncated at some appropriate truncation level. 
The shrinkage prior again makes the model highly robust to overfitting allowing us to adopt a simple strategy.
Since a latent factor characterization leads to a reduction in the number or parameters only when $q_{k} \leq \lceil (p+1)/2 \rceil$, 
where $\lceil s \rceil$ denotes the largest integer smaller than or equals to $s$, 
we simply set the truncation level at $q_{k} = q = \max\{2,\lceil (p+1)/2 \rceil\}$ for all the components. 
We also experimented by setting the truncation level at $q_{k} = q = p$ for all $k$ with the results remaining practically the same. 
The shrinkage prior, being continuous in nature, does not set the redundant columns to exact zeroes, 
but it adaptively shrinks the redundant parameters sufficiently towards zero, thus producing stable and efficient estimates of the densities being modeled.

\item {\bf Other hyper-parameters:} 
We take an empirical Bayes type approach to assign values to other hyper-parameters.
We set $\bmu_{\bX,0} = \overline{\bX}^{(0)}$, the overall mean of ${\bX}_{1:n}^{(0)}$, 
where $\bX_{1:n}^{(0)}$ denote the starting values of $\bX_{1:n}$ for the MCMC sampler discussed in Section \ref{sec: mvt posterior computation}. 
For the scaled errors we set $\bmu_{\bepsilon,0} = \bzero$.
For the MIW model we take $\nu_{0} = (p+2)$, the smallest possible integral value of $\nu_{0}$ for which the prior mean of $\bSigma_{k}$ exists. 
We then take $\bSigma_{\bX,0}/2 = \bPsi_{\bX,0}=\cov(\overline{\bX}_{1:n}^{(0)})$.
These choices imply $E(\bSigma_{\bX,k}) = \bPsi_{\bX,0} = \cov(\overline{\bX}^{(0)})$
and, since the variability of each component is expected to be significantly less than the overall variability, ensure noninformativeness. 
Similarly, for the scaled errors we take $\bSigma_{\bepsilon,0}/2 = \bPsi_{\bepsilon,0} = \cov(\bepsilon_{1:N}^{(0)})$.
For the MLFA model, the hyper-parameters specifying the prior for $\bLambda$  are set at $a_{1} = 1, a_{h} = 2$ for all $h\geq2$, and $\nu = 1$.
Inverse gamma priors with parameters $a_{\sigma} = 1.1, b_{\sigma} = 1$ are placed on the elements of $\bOmega$.
For each $k$, the variance functions were modeled using quadratic (q=2) B-splines based on $(2\times2+5+1)=10$ equidistant knot points 
on $[A_{k},B_{k}] = [\hbox{min}(\overline{\bW}_{k,1:n})-0.1~\hbox{range}(\overline{\bW}_{k,1:n}),\hbox{max}(\overline{\bW}_{k,1:n})+0.1~\hbox{range}(\overline{\bW}_{k,1:n})]$,
where $\overline{\bW}_{\ell,1:n}$ denotes the subject specific means corresponding to $\ell\th$ component.
\end{enumerate}

\section{Posterior Computation}  \label{sec: mvt posterior computation}

Samples from the posterior can be drawn using Gibbs sampling techniques. 
In what follows $\bzeta$ denotes a generic variable that collects the observed proxies $\bW_{1:N}$ 
and all the parameters of a model, including the imputed values of $\bX_{1:n}$ and $\bepsilon_{1:N}$, that are not explicitly mentioned.

Carefully chosen starting values can facilitate convergence of the sampler.
The posterior means of the $X_{i\ell}$'s, obtained by fitting univariate submodels, 
are used as the starting values for the multivariate sampler.  
The number of mixture components are initialized at $K_{\bX}=(m_{\bX}+2)$, where $m_{\bX}$ denotes the optimal number of clusters 
returned by model based clustering algorithm implemented by the mclust package in R applied to the corresponding initial values $\bX_{1:n}^{(0)}$.
The component specific mean vectors of the nonempty clusters are set at the mean of $\bX_{i}^{(0)}$ values that belong to that cluster. 
The component specific mean vectors of the two empty clusters are set at $\overline{\bX}^{(0)}$, the overall mean of ${\bX}_{1:n}^{(0)}$. 
For the MIW model, the initial values of the cluster specific covariance matrices are chosen in a similar fashion.
The mixture probabilities for the $k\th$ nonempty cluster is set at $\bpi_{\bX,k} = n_{k}/n$, where $n_{k}$ denotes the number of $\bX_{i}^{(0)}$ belonging to the $k\th$ cluster.
The mixture probabilities of the empty clusters are initialized at zero.
For the MLFA method, the starting values of all elements of $\bLambda$ and $\etam$ are set at zero. 
The starting values for the elements of $\bOmega$ are chosen to equal the variances of the corresponding starting values.
The parameters specifying the density of the scaled errors are initialized in a similar manner. 
The MCMC iterations comprise the following steps.
We suppress the subscript $\bepsilon$ to keep the notation clean as in the main paper.

\begin{enumerate}[topsep=0pt,itemsep=-1ex,partopsep=2ex,parsep=2ex, leftmargin=0cm, rightmargin=0cm, wide=3ex]
\item {\bf Updating the parameters specifying $f_{\bX}$:}
For the MIW model the parameters specifying the density $f_{\bX}$ are updated using the following steps.
\vspace{-5ex}\\
\bse
(\bpi\vert \bzeta) &\sim& \Dir(\alpha/K+n_{1},\alpha/K+n_{2},\dots,\alpha/K+n_{K}),  \\
(C_{i}\vert  \bzeta) &\sim& \Mult(1, p_{i1},p_{i2},\dots,p_{iK}),\\
(\bmu_k\vert  \bzeta) &\sim& \MVN_{p}(\bmu_{k}^{(n)},\bSigma_{k}^{(n)}), \\
(\bSigma_{k} \vert \bzeta) &\sim& \IW_{p}\{n_{k}+\nu_{0},\textstyle\sum_{i:C_{i}=k}(\bX_{i}-\bmu_{k})(\bX_{i}-\bmu_{k})\trans+\bPsi_{0}\},
\ese
\vspace{-5ex}\\
where 
$n_{k} = \sum_{i}1(C_{i}=k)$, 
$p_{ik} \propto \pi_{k} \times \MVN_{p}(\bX_{i}\vert\bmu_{k},\bSigma_{k})$,
$\bSigma_{k}^{(n)} = (\bSigma_{0}^{-1}+n_{k}\bSigma_{k}^{-1})^{-1}$ 
and $\bmu_{k}^{(n)} =  \bSigma_{k}^{(n)} \left\{\bSigma_{k}^{-1}\textstyle\sum_{i:C_{i}=k}\bX_{i}+\bSigma_{0}^{-1}\bmu_{0}\right\}$. 
To update the parameters specifying the covariance matrices in the MLFA model, the sampler cycles through the following steps.
\vspace{-5ex}\\
\bse
(\blambda_{k,j}\vert \bzeta) &\sim& \MVN_{q}\{(\bD_{k,j}^{-1}+\sigma_{j}^{-2}\etam_{k}\trans\etam_{k})^{-1}\sigma_{j}^{-2}\etam_{k}\trans(\bX_{k}^{(j)}-\bmu_{k}^{(j)}),(\bD_{k,j}^{-1}+\sigma_{j}^{-2}\etam_{k}\trans\etam_{k})^{-1}\},  \\
(\etam_{i}\vert C_{i}=k,\bzeta) &\sim& \MVN_{q}\{(\Ind_{q}+\bLambda_{k}\trans\bOmega^{-1}\bLambda_{k})^{-1}\bLambda_{k}\trans\bOmega^{-1}(\bX_{i}-\bmu_{k}),(\Ind_{q}+\bLambda_{k}\trans\bOmega^{-1}\bLambda_{k})^{-1}\},  \\
(\sigma_{j}^{2}\vert \bzeta) &\sim& \IG  \left\{a_{\sigma}+n/2,b_{\sigma}+(1/2)\textstyle\sum_{i=1}^{n}(X_{ij}-\bmu_{C_{i},j}-\blambda_{C_i,j}\trans\etam_{i})^{2}\right\},   \\
(\phi_{k,jh} \vert \bzeta) &\sim& \Ga\{(\nu+1)/2,(\nu+\tau_{k,h}\lambda_{k,jh}^{2})/2\},  \\
(\delta_{k,h}\vert \bzeta) &\sim& \textstyle  \Ga\{a_{h}+p(q-h+1)/2,1+\sum_{\ell=1}^{q}\tau_{k,\ell}^{(h)} \sum_{j=1}^{p}\phi_{k,j\ell}\lambda_{k,j\ell}^{2}/2\},  
\ese
\vspace{-5ex}\\
where 
$D_{k,j}^{-1} = \diag(\phi_{k,j1}\tau_{k,1},\dots,\phi_{k,jq}\tau_{k,q})$, 
$\tau_{k,\ell}^{(h)} = \prod_{t=1,t\neq h}^{\ell}\delta_{k,t}$,
$\bX_{k}^{(j)} = (X_{i_{1} j},X_{i_{2} j},\dots,X_{i_{n_k} j})\trans$,
$\etam_{k}^{n_{k}\times q} = (\etam_{i_{1}},\etam_{i_{2}},\dots,\etam_{i_{n_k}})\trans$, $\{i_{1}, i_{2},\dots,i_{n_{k}}\} = \{i: C_{i}=k\}$.

\item {\bf Updating the parameters specifying $f_{\bepsilon}$:}
The unconstrained full conditionals of the parameters specifying $f_{\bepsilon}$ are very similar. 
For instance, for the MIW model they are given by
\vspace{-5ex}
\bse
(\bpi\vert \bzeta) &\sim& \Dir(\alpha/K+N_{1},\alpha/K+N_{2},\dots,\alpha/K+N_{K}),  \\
(C_{ij}\vert  \bzeta) &\sim& \Mult(1, p_{ij1},p_{ij2},\dots,p_{ijK}),\\
(\bmu_k\vert  \bzeta) &\sim& \MVN_{p}(\bmu_{k}^{(N)},\bSigma_{k}^{(N)}), \\
(\bSigma_{k} \vert \bzeta) &\sim& \IW_{p}\{N_{k}+\nu_{0},\textstyle\sum_{ij:C_{ij}=k}(\bepsilon_{ij}-\bmu_{k})(\bepsilon_{ij}-\bmu_{k})\trans+\bPsi_{0}\},
\ese
\vspace{-5ex}\\
where 
$N_{k} = \sum_{i,j}1(C_{ij}=k)$, 
$p_{ijk} \propto \pi_{k} \times \MVN_{p}(\bepsilon_{ij}\vert\bmu_{k},\bSigma_{k})$,
$\bSigma_{k}^{(N)} = (\bSigma_{0}^{-1}+N_{k}\bSigma_{k}^{-1})^{-1}$ 
and $\bmu_{k}^{(N)} =  \bSigma_{k}^{(N)} \left\{\bSigma_{k}^{-1}\textstyle\sum_{ij:C_{ij}=k}\bepsilon_{ij}+\bSigma_{0}^{-1}\bmu_{0}\right\}$. 
Samples from the constrained posterior $(\{\bmu_{k}\}_{k=1}^{K} \vert \sum_{k=1}^{K}\pi_{k}\bmu_{k}=0, \bzeta)$ 
are then obtained from the unconstrained full conditionals $(\bmu_{k} \vert \bzeta)$ given above 
using the simple additional steps described in Section \ref{sec: mvt density of heteroscedastic errors} of the main paper.
The steps to update the parameters specifying the covariance matrices in the MLFA model are similarly obtained and are excluded.

\item {\bf Updating the values of $\bX$:}
When the measurement errors are independent of $\bX$, the $\bX_{i}$ have closed form full conditionals given by
\vspace{-5ex}\\
\bse
(\bX_{i}\vert C_{\bX,i}=k,C_{\bepsilon,i1}=k_{1},\dots, C_{\bepsilon,im_{i}}=k_{m_{i}},\bzeta) &\sim& \MVN_{p}(\bmu_{\bX}^{(n)},\bSigma_{\bX}^{(n)}),
\ese
\vspace{-5ex}\\
where 
$\bSigma_{\bX}^{(n)} = (\bSigma_{\bX,k}^{-1} + \sum_{j=1}^{m_{i}}\bSigma_{\bepsilon,k_{j}}^{-1})^{-1}$ 
and $\bmu_{\bX}^{(n)} = \bSigma_{\bX}^{(n)} (\bSigma_{\bX,k}^{-1}\bmu_{\bX,k} + \sum_{j=1}^{m_{i}}\bSigma_{\bepsilon,k_{j}}^{-1}\bW_{ij})$. 
For conditionally heteroscedastic measurement errors, the full conditionals are given by
\vspace{-5ex}\\
\bse
&&\hspace{-1cm}(\bX_{i}\vert C_{\bX,i}=k,C_{\bepsilon,i1}=k_{1},\dots, C_{\bepsilon,im_{i}}=k_{m_{i}},\bzeta) \\
&&\propto \MVN_{p}(\bX_{i}\vert \bmu_{\bX,k},\bSigma_{\bX,k}) \times \textstyle\prod_{j=1}^{m_{i}} \MVN_{p}\{\bW_{ij} \vert \bX_{i}+\bS(\bX_{i}) \bmu_{\bepsilon,k_{j}},\bS(\bX_{i})\bSigma_{\bepsilon,k_{j}}\bS(\bX_{i})\},
\ese
\vspace{-5ex}\\
The full conditionals do not have closed forms. 
Metropolis-Hastings (MH) steps with multivariate truncated normal proposals are used within the Gibbs sampler.  

\item {\bf Updating the parameters specifying $s_{\ell}$: } 
When the measurement errors are conditionally heteroscedastic, 
we first estimate the variance functions $s_{\ell}^{2}(X_{i\ell})$ by fitting univariate submodels $W_{ij\ell} = X_{i\ell} + s_{\ell}(X_{i\ell})\epsilon_{ij\ell}$ for each $\ell$.
The details are provided in Section \ref{sec: mvt estimation of variance functions}.
The parameters characterizing other components of the full model are then sampled using the Gibbs sampler described above, keeping the estimates of the variance functions fixed. 

An alternative class of algorithms integrates out the mixture probabilities $\bpi$ and works with the resulting Polya urn scheme (Neal, 2000). 
We did not consider such algorithms as they render the labels $C_{i}$ a-priori dependent, requiring the prior conditionals $(C_{i}\vert \bC_{-i})$ to be recomputed each time any $C_{i}$ is updated. 
Importantly, we also need the sampled values of $\bpi$ to enforce the zero mean restriction $\sum_{k=1}^{K}\pi_{k}\bmu_{k}=0$ on the measurement errors. 

\end{enumerate}

\section{Estimation of the Variance Functions} \label{sec: mvt estimation of variance functions}

When the measurement errors are conditionally heteroscedastic, 
we need to update the parameters $\bxi_{\ell}$ that specify the variance functions $s_{\ell}^{2}(X_{i\ell})$.
These parameters do not have closed form full conditionals.
MCMC algorithms, where we tried to integrate MH steps for $\bxi_{\ell}$ with the sampler for the parameters specifying $f_{\bepsilon}$, 
were numerically unstable and failed to converge sufficiently quickly. 
We need to supply the values of the scaled errors $\epsilon_{ij\ell}$ to step 2 of the algorithm described in Section \ref{sec: mvt posterior computation}
and the instability stems from the operation $\bepsilon_{ij} = \bS(\bX_{i})^{-1}\bU_{ij}$ required to calculate the scaled residuals $\epsilon_{ij\ell}$, 
as we try to divide $U_{ij\ell}$ by the quantity $s_{\ell}(X_{i\ell})$, which may be very small for certain values of $X_{i\ell}$, 
for example, for values of $X_{i\ell}$ near zero for the EATS data application. See Figure \ref{fig: mvt EATS data results VFn}.

To solve the problem, we adopt a novel two-stage procedure. 
First, for each $k$, we estimate the functions $s_{\ell}^{2}(X_{i\ell})$ by fitting the univariate submodels $W_{ij\ell} = X_{i\ell} + s_{\ell}(X_{i\ell})\epsilon_{ij\ell}$.
The problem of numerical instability arising out of the operation to determine the values of the scaled errors remains in these univariate subproblems too. 
But the following lemma from Pelenis (2014), presented here for easy reference, provides us with an escape route by allowing us to avoid this operation in the first place.
\begin{Lem}\label{Lem: mvt lemma from Pelenis}
Let $\btheta_{1:K} = \{(\pi_{k},\mu_{k},\sigma_{k}^{2})\}_{k=1}^{K}$ be such that
\vspace{-5ex}\\
\be
\textstyle f_{1}(\epsilon\vert\btheta_{1:K}) = \sum_{k=1}^{K} \pi_{k}~\Normal(\epsilon\vert \mu_{k},\sigma_{k}^{2}),  ~~~\hbox{with}~~ \sum_{k=1}^{K} \pi_{k} = 1,  ~~~ \sum_{k=1}^{K} \pi_{k}\mu_{k} = 0. 
\label{eq: mvt density of scaled errors model 1}
\ee
\vspace{-5ex}\\
Then there exists a set of parameters $\btheta_{1:(K-1)}^{\star}=\{(\pi_{k}^{\star},p_{k,r}^{\star},\mu_{k,r}^{\star},\sigma_{k,r}^{\star 2})\}_{r=1,k=1}^{2,K-1}$ such that 
\vspace{-5ex}\\
\be
&&\textstyle \hspace{-0.85cm} f_{1}(\epsilon\vert\btheta_{1:K}) = f_{2}(\epsilon\vert \btheta_{1:(K-1)}^{\star})  =  \sum_{k=1}^{K-1} \pi_{k}^{\star}~\sum_{r=1}^{2} p_{k,r}^{\star}\Normal(\epsilon\vert \mu_{k,r}^{\star},\sigma_{k,r}^{\star 2}),   \label{eq: mvt density of scaled errors model 2}\\
&&\textstyle \sum_{k=1}^{K-1} \pi_{k}^{\star} = 1, ~~~ \sum_{r=1}^{2} p_{k,r}^{\star} = 1, ~~~ \sum_{r=1}^{2} p_{k,r}^{\star}\mu_{k,r}^{\star} = 0~ \forall k. \nonumber
\ee
\vspace{-10ex}\\
\end{Lem}
Lemma \ref{Lem: mvt lemma from Pelenis} implies that the univariate submodels for the density of the scaled errors given by (\ref{eq: mvt density of scaled errors model 1}) 
has a reparametrization (\ref{eq: mvt density of scaled errors model 2}) where each component is itself a two-component normal mixture with its mean restricted at zero. 
The reparametrization (\ref{eq: mvt density of scaled errors model 2}) thus replaces the zero mean restriction on (\ref{eq: mvt density of scaled errors model 1}) 
by similar restrictions on each of its components.
These restrictions also imply that each mixture component in (\ref{eq: mvt density of scaled errors model 2}) can be further reparametrized by only four free parameters.
One such parametrization could be in terms of $\wt\btheta_{k} = (\wt{p}_{k},\wt\mu_{k},\wt\sigma_{k,1}^{2},\wt\sigma_{k,2}^{2})$,
where $(p_{k,1}^{\star},\sigma_{k,1}^{\star2},\sigma_{k,2}^{\star 2}) = (\wt{p}_{k},\wt\sigma_{k,1}^{2},\wt\sigma_{k,2}^{2})$ 
and $\mu_{k,r}^{\star}=c_{k,r}\tmu_{k}$,
where $c_{k,1}=(1-\wt{p}_{k})/\{\wt{p}_{k}^2+(1-\wt{p}_{k})^{2}\}^{1/2}$ and $c_{k,2} = -\wt{p}_{k}/\{\wt{p}_{k}^2+(1-\wt{p}_{k})^{2}\}^{1/2}$.
Letting $p_{0}$ denote the prior assigned to $\wt\btheta_{k}$, the full conditional of 
$\wt\btheta_{k}$ in terms of the conditional likelihood $f_{U\vert X}$ is proportional to 
$P_{0} (\wt\btheta_{k}) \textstyle\prod_{ij: C_{\epsilon,ij\ell}=k}f_{U\vert X}(U_{ij\ell}\vert X_{i\ell}, \bxi_{\ell}, \wt\btheta_{k}, \bzeta)$. 
The problem of numerical instability can now be tackled by using MH steps 
to update not only the parameters $\bxi_{\ell}$ specifying the variance functions but also the parameters $\{\wt\btheta_{k}\}_{k}$ characterizing the density $f_{\epsilon}$ 
using the conditional likelihood $f_{U\vert X}$ (and not $f_{\epsilon}$ itself), thus escaping the need to separately determine the values of the scaled errors. 

\begin{figure}[!ht]
\begin{center}
\includegraphics[height=12cm, width=16cm, trim=0cm 0cm 0cm 0cm, clip=true]{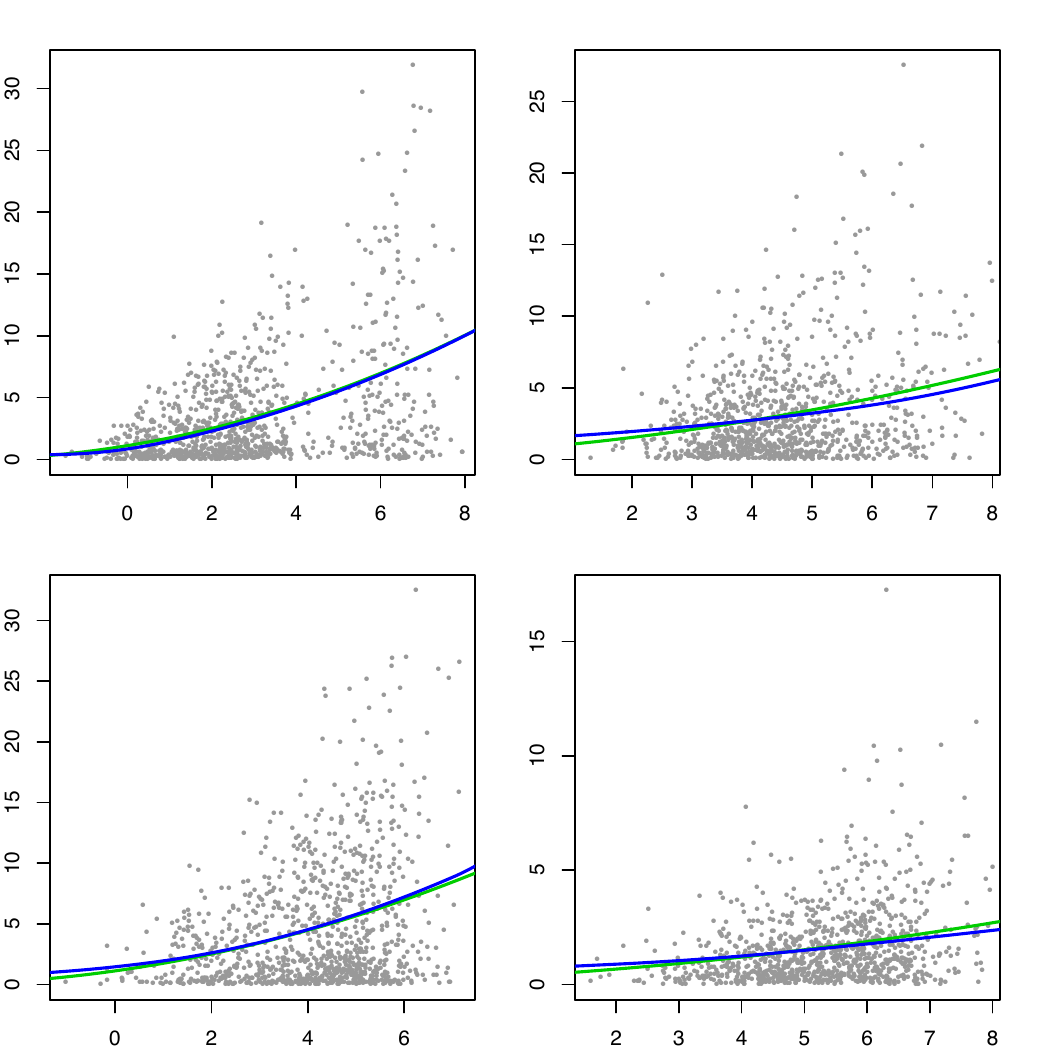}
\end{center}
\caption{\baselineskip=10pt Results for the variance functions $s^{2}(X)$ produced by the univariate density deconvolution method for each component of $\bX$ for the conditionally heteroscedastic error distribution $f_{\bepsilon}^{(2)}$ with sample size $n=1000$, $m_{i}=3$ replicates for each subject and component specific covariance matrices with autoregressive structure (AR). 
The results correspond to the data set that produced the median of the estimated integrated squared errors (ISE) out of a total of 100 simulated data sets for the MIW (mixtures with inverse Wishart priors) method.
For each component of $\bX$, the true variance function is $s^{2}(X) = (1+X/4)^{2}$.
See Section \ref{sec: mvt density of heteroscedastic errors} and Section \ref{sec: mvt estimation of variance functions} for additional details.
In each panel, the true (lighter shaded green lines) and the estimated (darker shaded blue lines) variance functions 
are superimposed over a plot of subject specific sample means vs subject specific sample variances.
The figure is in color in the electronic version of this article.
}
\label{fig: mvt simulation results VFn d4 n1000 m3 MIW X1 E1 AR}
\end{figure}

The priors and the hyper-parameters for the univariate submodels are chosen following the suggestions of Sarkar, et al. (2014) 
who used an infinite dimensional extension of this reparametrized finite dimensional submodel.  
The strategy of exploiting the properties of overfitted mixture models to determine the number of mixture components described in Section \ref{sec: mvt choice of hyper-parameters} 
can also be applied to the univariate subproblems. 
High precision estimates of the variance functions can be obtained using these reparametrized finite dimensional univariate deconvolution models.
See Figure \ref{fig: mvt simulation results VFn d4 n1000 m3 MLFA X1 E1 Ind} 
and also Figures \ref{fig: mvt simulation results VFn d4 n1000 m3 MIW X1 E1 AR} and \ref{fig: mvt simulation results VFn d4 n1000 m3 MIW X1 HT_E0 Ind} in the Supplementary Materials for illustrations.

A similar reparametrization exists for the multivariate problem too,
but the strategy would not be very effective in a multivariate set up as it would require updating the mean vectors and the covariance matrices involved in $f_{\bepsilon}$ through MH steps 
which are not efficient in simultaneous updating of large numbers of parameters. 
After estimating the parameters characterizing the variance functions from the univariate submodels, 
we therefore keep these estimates fixed and sample the other parameters using the Gibbs sampler described in Section \ref{sec: mvt posterior computation}. 
Additional details follow.

As discussed in Section \ref{sec: mvt density of heteroscedastic errors} of the main paper, 
the variance functions $s_{\ell}^{2}$'s can not be uniquely determined 
without additional identifiability restrictions on the variance of $\epsilon_{ij\ell}$. 
This, however, does not pose any problem to assess $\var(U_{ij\ell}\vert X_{i\ell})$ which can be estimated as
$\wh{v}_{\ell}(X_{i\ell}) = \sum_{m=1}^{M} v_{\ell}^{(m)}(X_{i\ell}) \var^{(m)}(\epsilon_{ij\ell})/M$, 
where $v_{\ell}^{(m)}(X_{i\ell})$ 
and $\var^{(m)}(\epsilon_{ij\ell})$ are estimates of $s_{\ell}^{2}(X_{i\ell})$ and $\var(\epsilon_{ij\ell})$ 
based on the $m\th$ sample drawn from the posterior of the $\ell\th$ univariate submodel in the first stage. 
The final estimate of $\bxi_{\ell}$ is then obtained as 
$\textstyle \wh\bxi_{\ell,opt} = \arg_{\bxi_{\ell}} \min \sum_{r=1}^{R_{\ell}}\left\{\wh{v}_{\ell}(X_{r\ell}^{\Delta}) - \bB_{q,J_{\ell},\ell}(X_{r\ell}^{\Delta}) \exp(\bxi_{\ell})\right\}^{2}$,
where $\{X_{r\ell}^{\Delta}\}_{r=1}^{R_{\ell}}$ is a set of grid points on the support $[A_{\ell},B_{\ell}]$ of the variance functions.  

In the second stage, we keep these estimates $\wh\bxi_{\ell,opt}$ fixed 
and sample the other parameters using the Gibbs sampler described in Section \ref{sec: mvt posterior computation}. 
At the $m\th$ MCMC iteration of the Gibbs sampler, the scaled errors to be used in step 2 of the algorithm are obtained as 
$\epsilon_{ij\ell}^{(m)}=(W_{ij\ell}-X_{i\ell}^{(m)})/\wh{s}_{\ell}(X_{i\ell}^{(m)})$, where $\wh{s}_{\ell}(X_{i\ell}^{(m)})=\{\bB_{q,J_{\ell},\ell}(X_{i\ell}^{(m)}) \exp(\wh{\bxi}_{\ell,opt})\}^{1/2}$ and $X_{i\ell}^{(m)}$ is sampled value of $X_{i\ell}$ at the $m\th$ iteration. 

Appropriate scale adjustments are made to make the estimate $\wh{f}_{\bepsilon}$ comparable to the true $f_{\bepsilon}$ in simulation experiments. 
Specifically, $\wh{f}_{\bepsilon} = \sum_{m=1}^{M}\pi_{k}^{(m)}\MVN(\bD\bmu_{k}^{(m)},\bD\bSigma_{k}^{(m)}\bD)/M$, 
where $\bD=\diag(\sigma_{true,1},\dots,\sigma_{true,p})$, 
$\sigma_{true,\ell}^{2}$ is the variance of $\epsilon_{ij\ell}$ under the true $f_{\bepsilon}$ used to generate them, 
and $\{\pi_{k}^{(m)},\bmu_{k}^{(m)},\Sigma_{k}^{(m)}\}_{k=1}^{K}$ are $m\th$ sampled values from the posterior of the parameters $\{\pi_{k},\bmu_{k},\Sigma_{k}\}_{k=1}^{K}$ specifying $f_{\bepsilon}$.

\section{The Two-Stage Sampler} \label{sec: mvt two-stage sampler}
\begin{figure}[!ht]
\begin{center}
\includegraphics[height=10cm, width=11cm, trim=0cm 0cm 0cm 0cm, clip=true]{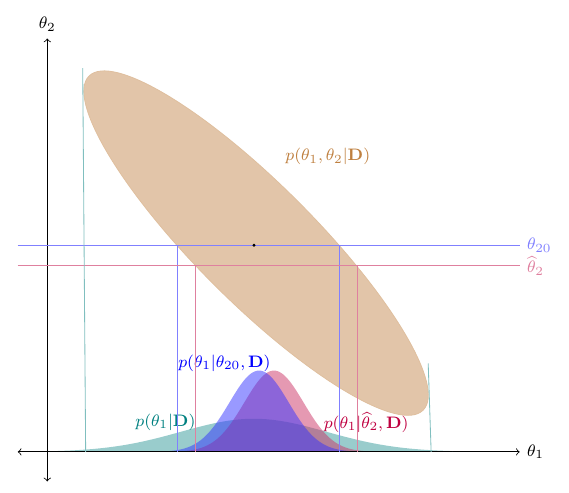}
\end{center}
\caption{\baselineskip=5pt Heuristics of the two-stage sampler. 
The brown elliptical region shows the joint posterior $p(\theta_{1},\theta_{2} \vert \bD)$ of two parameters $\theta_{1}$ and $\theta_{2}$ given data $\bD$. 
The light blue curve shows $p(\theta_{1} \vert \bD)$, the marginal posterior of $\theta_{1}$ given data $\bD$. 
The blue curve shows $p(\theta_{1} \vert \theta_{20}, \bD)$, the posterior of $\theta_{1}$, where $\theta_{20}$, the `true' value of $\theta_{2}$, is known. 
The red curve shows $p(\theta_{1} \vert \wh\theta_{2}, \bD)$, the pseudo-posterior of $\theta_{1}$ given $\wh\theta_{2}$, an estimate of $\theta_{2}$. 
$p(\theta_{1} \vert \wh\theta_{2}, \bD)$ will be close to $p(\theta_{1} \vert \theta_{20}, \bD)$ when $\wh\theta_{2}$ is close to $\theta_{20}$. 
}
\label{fig: mvt approximate sampler}
\end{figure}

Over the last two decades, MCMC techniques have remained at the forefront of Bayesian inference. 
The literature on the topic is already vast and is still rapidly expanding. 
While the research on exact MCMC methods is still highly active, 
owing to numerous practical challenges, 
approximate computation methods are becoming increasingly popular. 
For a recent review of traditional exact methods and more recent approximate tools, see Green, et al. (2015). 
The basic idea of the two-stage sampler described above, while being simple and intuitive, 
is a novel addition to the growing literature on the topic. 
We are studying its properties in greater detail in simpler settings in a separate manuscript. 
Figure \ref{fig: mvt approximate sampler} below provides some heuristics.

Consider the problem of drawing samples from the posterior $p(\theta_{1},\theta_{2} \vert \bD)$ of two parameters $\theta_{1}$ and $\theta_{2}$ given data $\bD$. 
The basic MCMC sampler iterates between sampling from 
(A) $p(\theta_{1} \vert \theta_{2}, \bD)$ and (B) $p(\theta_{2} \vert \theta_{1}, \bD)$. 
If, however, the `true' value of $\theta_{2}$ (in a frequentist sense), say $\theta_{20}$, is known, 
we only require step (A), which becomes $p(\theta_{1} \vert \theta_{20}, \bD)$. 
And if we substitute $\theta_{2}$ by a point estimate $\wh\theta_{2}$, step (A) becomes $p(\theta_{1} \vert \wh\theta_{2}, \bD)$.  
While an uncertainty assessment based on $p(\theta_{1} \vert \wh\theta_{2}, \bD)$ will be overly optimistic 
compared to that based on the actual marginal posterior $p(\theta_{1} \vert \bD)$, 
$p(\theta_{1} \vert \wh\theta_{2}, \bD)$ and $p(\theta_{1} \vert \theta_{20}, \bD)$ will be close when $\wh\theta_{2}$ is close to $\theta_{20}$, and 
samples drawn from $p(\theta_{1} \vert \wh\theta_{2}, \bD)$ may be used for approximate Bayesian inference on $\theta_{1}$.

The two-stage sampler can also be explained using the following heuristics. 
Under suitable regularity conditions and considering parametric models 
(observe that Bayesian nonparametric models are usually large parametric models), 
the posterior distribution $p(\theta_{1},\theta_{2} \vert \bD)$ can be approximated by a Gaussian distribution centered at the true value $\btheta_0 = (\theta_{10},  \theta_{20})$ 
and variance equal to the inverse of the Fisher information matrix $\bI(\btheta_0)$.  
The justification of this argument is usually tedious and follows from Bernstein von-Mises (BvM) theorems. 
Refer, for example, to Johnstone (2010), Bontemps (2011), Bickel and Kleijn (2012), Spokoiny (2013) and Castillo and Nickl (2014) for recent literature on BvM theorems 
in nonparametric Bayesian models and growing parametric Bayesian models. 
For the sake of convenience, let us assume such results are true for $p(\theta_{1},\theta_{2} \vert \bD)$.  Hence the marginal posterior distribution 
$p(\theta_{1} \vert \bD)$  is similar to a Gaussian distribution with mean $\theta_{10}$ and variance  
$[\bI(\theta_0)]^{-1}_{11}$, the $(1, 1)\th$ block of the inverse of $\bI(\theta_0)$.
Assuming $\wh\theta_{2}$ to be a consistent estimate of $\theta_{20}$, 
the conditional posterior distribution in step (A) can be approximated by $p(\theta_{1} \vert \theta_{20}, \bD)$ 
which in turn is similar to a Gaussian distribution centered at 
$\theta_{10}$ with precision matrix $\bI(\theta_{10} \vert \theta_{20})$, 
the conditional Fisher information matrix assuming $\theta_{20}$ to be known. 
In classical inference, it is well known that 
$[\bI(\theta_0)]^{-1}_{11} \geq [\bI(\theta_{10} \vert \theta_{20})]^{-1}$ in the sense that the difference is non-negative definite, 
since knowing $\theta_{20}$ results in a higher value of the `information'.  
While confidence intervals based on samples drawn by the two-stage algorithm will be optimistic, 
the draws will be centered around the true value $\theta_{10}$ 
and hence may be used for approximate `mean' inference on $\theta_{1}$.  
%

\pagebreak\newpage
\section{Comments on the Model for ${\bU \vert \bX}$} \label{sec: mvt comments on the model for U given X}

\vspace{-1ex}
As shown in Sarkar, et al. (2014), even in univariate deconvolution settings, due to the nonavailability of precise information about $X$, 
variations in higher order conditional moments of $(U\vert X)$ are extremely difficult to capture even in large data sets. 
Semiparametric approaches that focus separately on the first two moments, namely $E(U\vert X)=0$ and $\var(U\vert X)$, and the shape of $f_{U\vert X}$, 
are thus more efficient than possible fully nonparametric approaches even when the truth closely follows the setup of the nonparametric model. 
See their Section 4.3. 
This will certainly remain true in the significantly more difficult multivariate deconvolution problem. 
In building models for $f_{\bU\vert\bX}$, we may thus concentrate on the class of models that separates the problem of modeling $\cov(\bU\vert\bX)$ 
from that of modeling the shape and other properties of $f_{\bU\vert\bX}$. 
Recent advances in covariance regression models, 
where the covariance of the multivariate regression errors are allowed to vary flexibly with precisely measured and possibly multivariate predictors, 
provide us with clues about how this may be achieved. 
However, as we explain in the following section, there are major differences between conditionally varying multivariate regression errors 
and conditionally varying multivariate measurement errors. 
As an implication, covariance regression methods may not be exactly appropriate for modeling conditionally varying covariance matrices $\cov(\bU\vert\bX)$ in measurement error settings.

\vspace{-2ex}
\subsection{Regression Errors vs Measurement Errors}\label{sec: mvt regression vs measurement errors}

\begin{figure}[!ht]
\hspace*{1.5cm}\includegraphics[height=2cm, width=12cm, trim=1cm 1.25cm 1cm 1.25cm, clip=true]{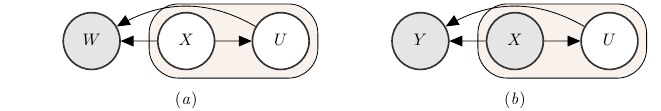}
\caption{
(a) Dependency structure in a univariate deconvolution model with latent variable of interest $X$, associated measurement errors $U$ and replicates $W$.
(b) Dependency structure in a univariate regression model with response $Y$, 
associated regression errors $U$ and a univariate observed predictor $X$. 
In both panels, the filled rectangular regions focus on the dependency structures between the conditionally varying errors $U$ and the conditioning variable $X$. 
The unfilled and the shaded nodes signify latent and observable variables, respectively. 
}
\label{fig: univriate graphical models supplementary materials}
\end{figure}

Consider the problem of flexible modeling of conditionally heteroscedastic regression errors where the response and the covariates are both univariate. 
Consider also the problem of modeling conditionally heteroscedastic measurement errors in a univariate deconvolution set up.
From a modeling perspective, Bayesian hierarchical framework allows us to treat these two problems on par 
by treating both the covariate in the regression problem and the variable of interest in the deconvolution problem simply as conditioning variables.  
Of course in the regression problem $X$ is precisely measured, whereas in the deconvolution problem $X$ would be latent, 
but in either case we are required to flexibly model the density of $(U\vert X)$ subject to $E(U\vert X) = 0$, where $U$, depending upon the context, denotes either regression or measurement errors. 
See Figure \ref{fig: univriate graphical models supplementary materials}.
Models for regression errors that allow their variance to vary with the values of the covariate (Pati and Dunson, 2013; Pelenis, 2014) can thus be tried as  
potential candidates for models for univariate conditionally heteroscedastic measurement errors.
Conversely, the models for conditionally heteroscedastic univariate measurement errors (Staudenmayer, et al. 2008;  Sarkar, et al. 2014) can also be employed to model univariate conditionally heteroscedastic regression errors.

\vskip 10pt

\begin{figure}[!ht]
\centering
\includegraphics[height=5.5cm, width=15cm, trim=2cm 1cm 1cm 1cm, clip=true]{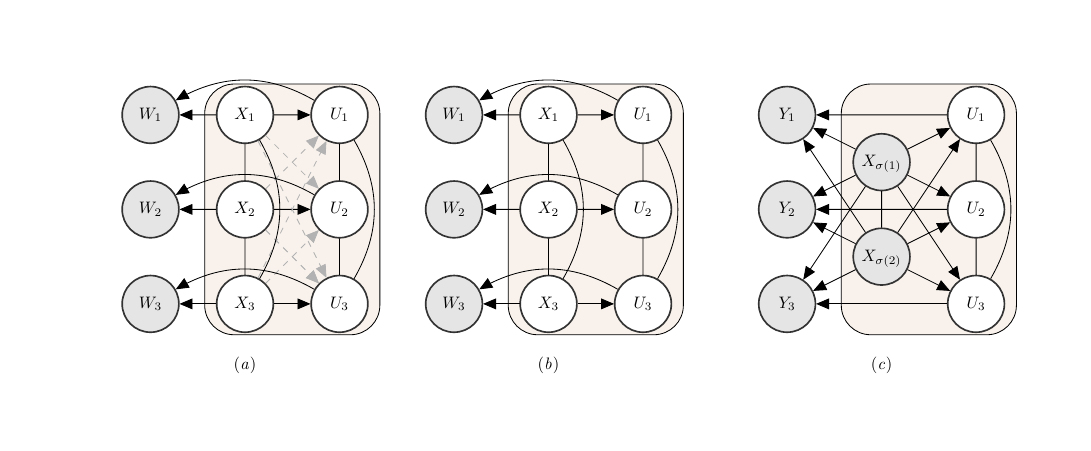}
\vskip -5pt
\caption{
(a) Dependency structure in a trivariate deconvolution model with latent variable of interest $\bX=(X_{1},X_{2},X_{3})\trans$, associated measurement errors $\bU=(U_{1},U_{2},U_{3})\trans$ and replicates $\bW=(W_{1},W_{2},W_{3})\trans$.
The solid black and the dashed gray edges signify strong and weak dependencies, respectively. 
(b) Dependence relationships in a trivariate deconvolution problem implied by the `separable' measurement error model $(\bU\vert \bX)=\bS(\bX)\bepsilon$ 
with $\bepsilon$ independent of $\bX$ and $\bS(\bX)=\diag\{s_{1}(X_{1}),s_{2}(X_{2}),s_{3}(X_{3})\}$. 
Unlike panel (a), possible weak relationships 
between $U_{\ell}$ and $\{X_{m}\}_{m\neq \ell}$ 
are ignored. 
(c) Dependency structure in a trivariate regression model with response $\bY=(Y_{1},Y_{2},Y_{3})$, 
associated regression errors $\bU=(U_{1},U_{2},U_{3})\trans$ and an observed bivariate predictor $\bX=(X_{1},X_{2})\trans$ 
where $\bX_{\sigma}=(X_{\sigma(1)},X_{\sigma(2)})\trans$ denotes arbitrary reordering of $\bX$.  
In both panels, the filled rectangular regions focus on the dependency structures between the conditionally varying errors $\bU$ and the conditioning variable $\bX$. 
The unfilled and the shaded nodes signify latent and observable variables, respectively. 
The directed and the undirected edges represent one-way and two-way relationships, respectively. 
}
\label{fig: graphical models supplementary materials}
\end{figure}

This is not quite true in a multivariate set up.
Interpreting the variables of interest $\bX$ broadly as conditioning variables, one can again loosely connect the problem of modeling conditionally heteroscedastic  multivariate measurement errors to the problem of covariance regression (Hoff and Niu, 2012; Fox and Dunson, 2016 etc.),
where the goal is to develop models that allow the covariance of multivariate regression errors to vary flexibly with precisely measured and possibly multivariate predictors.
In covariance regression problems, the dimension of the regression errors is typically unrelated to the dimension of the predictors. 
Different components of the regression errors are assumed to be equally influenced by different components of the predictors 
and hence independent reordering of the components of $\bX_{i}$ will not change the dependency structure. 
In multivariate deconvolution problems, in contrast, 
the $\ell\th$ component $U_{ij\ell}$ is the measurement error associated exclusively with $X_{i\ell}$. 
Here the dimension of $\bU_{ij}$ is the same as the dimension of $\bX_{i}$ 
and any reordering of the components of $\bX_{i}$ would require that 
the components of $\bU_{ij}$ and $\bW_{ij}$ be also reordered using the same relabeling scheme. 
See Figure \ref{fig: graphical models supplementary materials}. 
While different components of the measurement error vectors $\bU_{ij}$ may be correlated, 
this exclusive association between $U_{ij\ell}$ and $X_{i\ell}$ implies the plausibility that the dependence of $U_{ij\ell}$ on $\bX_{i}$ can be explained primarily through $X_{i\ell}$.
Figure \ref{fig: mvt EATS data results VFn}, for instance, suggests strong conditional heteroscedasticity patterns 
and it is plausible to assume that the conditional variability in $U_{ij\ell}$ can be explained primarily by $X_{i\ell}$ only.
%
%
The dependency structure of conditionally varying multivariate measurement errors are, therefore, different from that of conditionally varying multivariate regression errors. 
Additionally, the aforementioned covariance regression approaches all assume multivariate normality of the regression errors. 
As is well established in the literature, parametric distributional assumptions on the errors can be particularly restrictive in measurement error problems.

\vskip 0pt

These issues preclude direct application of existing covariance regression approaches to model conditionally heteroscedastic multivariate measurement errors. 
Models for conditionally varying multivariate measurement errors $(\bU \vert \bX)$ should highlight their unique features, 
accommodate distributional flexibility, enforce the mean zero restriction 
and, to be practically effective, should be computationally stable even in the absence of precise information on the conditioning variable $\bX$. 

While we reiterate that, for both modeling and computational reasons, 
the covariance regression methodology of Fox and Dunson (2016) is not be suitable for our purposes, 
they still provide clues about how the problems of flexible modeling $\cov(\bU\vert\bX)$ 
and that of modeling the shape of $f_{\bU\vert\bX}$ can be separated. 
The following section explains.

\vspace{-2ex}
\subsection{Latent Factor Models for Different Covariance Classes} \label{sec: latent factor models for different covariance classes} 
Lemma \ref{lem: cholesky} gives a slightly modified version of Lemma 2.1 of Fox and Dunson (2016). 

\begin{Lem} \label{lem: cholesky}
Any conditionally varying covariance matrix $\cov(\bU\vert\bX)= \bSigma(\bX)$ can be represented as $\bSigma(\bX)=\bLambda(\bX)\bLambda\trans(\bX)$ for some lower triangular matrix $\bLambda(\bX)=((\lambda_{\ell, m}(\bX)))$.
\end{Lem}
\begin{proof}
The proof follows from straightforward application of Cholesky factorization. 
\end{proof}

\vskip 5pt
Following Lemma \ref{lem: cholesky}, 
introducing a latent factor $\bepsilon$, we can write $(\bU\vert\bX,\bepsilon)=\bLambda(\bX)\bepsilon$, 
that is, $(U_{\ell}\vert\bX,\bepsilon)=\sum_{m=1}^{\ell}\lambda_{\ell,m}(\bX)\epsilon_{m}$, 
with $\bepsilon \perp \bX$ and $\cov(\bepsilon)=\bI_{p}$. 
Completely unrestricted covariance functions can thus be modeled via such latent variable framework by flexibly modeling $\bLambda(\bX)$. 
$E(\bU\vert\bX)=\bzero$ can be achieved by setting $E(\bepsilon)=\bzero$.  

The general nature of the latent factor formulation having been established, 
we formulate the subsequent results in terms of additional restrictions on such models. 
Following the discussion in Section \ref{sec: mvt regression vs measurement errors}, 
we now focus specifically on covariance functions $\cov(\bU\vert\bX)$ for measurement error problems, 
where $\bU$ and $\bX$ are of the same dimension, each component $U_{\ell}$ of $\bU$ being related to the corresponding component $X_{\ell}$ of the conditioning vector $\bX$. 
We consider first the situation when $(U_{\ell}\vert \bX, \bepsilon)$ depends exclusively on $X_{\ell}$ but not on $\{X_{m}\}_{m\neq\ell}$.

\begin{Lem} \label{lem: cholesky me} 
Let $(\bU\vert\bX,\bepsilon)=\bLambda(\bX)\bepsilon$, where $\bLambda(\bX)=((\lambda_{\ell, m}(\bX)))$ is lower-triangular, 
$\bepsilon \perp \bX$ and $\cov(\bepsilon)=\bI_{p}$. 
If $(U_{\ell} \vert \bX,\bepsilon) = (U_{\ell} \vert X_{\ell},\bepsilon)$ for all $\ell$, 
then $\lambda_{\ell, m}(\bX)=\lambda_{\ell, m}(X_{\ell})$ for all $\ell, m$.
\end{Lem}
\begin{proof}
The proof follows trivially by noting that $(U_{\ell}\vert\bX,\bepsilon)=\sum_{m=1}^{\ell} \lambda_{\ell, m}(\bX)\epsilon_{m} = (U_{\ell}\vert X_{\ell},\bepsilon)$, 
if and only if, for all $m\leq \ell$, $\lambda_{\ell, m}(\bX)$ is a function of $X_{\ell}$ only. 
\end{proof}

\vskip 5pt
As an immediate corollary of Lemma \ref{lem: cholesky me}, the conditional moments $m_{\ell}^{r}(\bX)=E (U_{\ell}^{r}\vert\bX)$ are functions of $X_{\ell}$ only 
and the conditional cross-moments $m_{\ell,m}^{r,s}(\bX)=E (U_{\ell}^{r}U_{m}^{s}\vert\bX)$ are functions of $X_{\ell}$ and $X_{m}$ only.
Modeling variations in the conditional cross-moments is a daunting task in multivariate settings, particularly in the absence of precise information on $\bX$. 
The next result allows the cross-moments $m_{\ell,m}^{r,s}(\bX)$ to vary with $X_{\ell}$ and $X_{m}$, 
but assumes the correlations $\corr(U_{\ell},U_{m} \vert \bX)$ to remain constant across $\bX$.  

\begin{Lem} \label{lem: cholesky diagonal} 
Let $(\bU\vert\bX,\bepsilon)=\bLambda(\bX)\bepsilon$, where $\bLambda(\bX)=((\lambda_{\ell, m}(\bX)))$ is lower-triangular, 
$\bepsilon \perp \bX$ and $\cov(\bepsilon)=\bI_{p}$. 
Also, let $(U_{\ell} \vert \bX,\bepsilon) = (U_{\ell} \vert X_{\ell},\bepsilon)$ for all $\ell$, 
and $\corr(U_{\ell},U_{m} \vert \bX)$ does not vary with $\bX$ for all $\ell\neq m$. 
Then, $\bLambda(\bX)=\bLambda_{1}(\bX)\bC$ for some diagonal matrix $\bLambda_{1}(\bX)=\diag\{\lambda_{1}(X_{1}),\dots,\lambda_{p}(X_{p})\}$ 
and some lower-triangular matrix $\bC$.
\end{Lem}
\begin{proof}
From Lemma \ref{lem: cholesky me}, we have 
$\lambda_{\ell, m}(\bX)=\lambda_{\ell, m}(X_{\ell})$ for all $\ell, m$, 
and $\corr(U_{\ell},U_{m}\vert\bX)$ varies with $X_{\ell}$ and $X_{m}$ only. 
Under the additional assumption of Lemma \ref{lem: cholesky diagonal}, 
we first prove that $ \lambda_{\ell,m}(X_{\ell}) = c_{\ell,m} \lambda_{\ell,\ell}(X_{\ell})$ for some constant $c_{\ell,m}$ for all $m < \ell$ and all $\ell=2,\dots,p$. 
Without loss of generality, we assume that $\corr(U_{\ell},U_{m}\vert \bX) = r_{\ell,m} \neq 0$ for all $\ell\neq m$. 
We have 
\vspace{-5ex}\\
\be
\corr(U_{1},U_{2}\vert \bX)=\frac{\lambda_{2,1}(X_{2})}{\{\lambda_{2,1}^{2}(X_{2})+\lambda_{2,2}^{2}(X_{2})\}^{1/2}} = r_{1,2}  ~ \Rightarrow ~
\lambda_{2,2}^{2}(X_{2}) = \frac{(1-r_{1,2}^{2})}{r_{1,2}^{2}} \lambda_{2,1}^{2}(X_{2}). ~ \label{eq: cholesky diagonal 1}
\ee
\vspace{-4ex}\\
So the proposition holds true for $\ell=2$. 
Next, assume that it holds for $\ell=2,\dots,h-1$ for some $h > 2$.  
Also, from (\ref{eq: cholesky diagonal 1}), $\var(U_{2}\vert\bX) = \textstyle\sum_{m=1}^{2} \lambda_{2,m}^{2}(X_{2}) = \lambda_{2,1}^{2}(X_{2}) / r_{1,2}^{2}$. 
This is, in fact, more generally true for all $\ell$. 
For instance, for $\ell=h$,
\vspace{-4ex}
\be
&& \corr(U_{1},U_{h}\vert \bX)=\frac{\lambda_{h,1}(X_{h})}{\{\sum_{m=1}^{h} \lambda_{h,m}^{2}(X_{h})\}^{1/2}} = r_{1,h}  ~~ \Rightarrow ~~
\sum_{m=2}^{h} \lambda_{h,m}^{2}(X_{h}) = \frac{(1-r_{1,h}^{2})}{r_{1,h}^{2}} \lambda_{h,1}^{2}(X_{h}) \nonumber\\
&& ~~ \Rightarrow ~~
\var(U_{h}\vert\bX) = \textstyle\sum_{m=1}^{h} \lambda_{h,m}^{2}(X_{h}) = \lambda_{h,1}^{2}(X_{h}) / r_{1,h}^{2}. \label{eq: cholesky diagonal 2}\\
&&\text{Then,} ~~
\corr(U_{2},U_{h}\vert \bX)=\frac{\lambda_{2,1}(X_{2})\lambda_{h,1}(X_{h})+\lambda_{2,2}(X_{2})\lambda_{h,2}(X_{h})}{\{\sum_{m=1}^{2} \lambda_{2,m}^{2}(X_{2})\}^{1/2} \{\sum_{m=1}^{h} \lambda_{h,m}^{2}(X_{h})\}^{1/2}} = r_{2,h}   \nonumber\\
&&~~ \Rightarrow~~ \frac{\lambda_{2,2}(X_{2})\{c_{2,1}\lambda_{h,1}(X_{h})+\lambda_{h,2}(X_{h})\}}{\abs{c_{2,1}\lambda_{2,2}(X_{2})} \abs{\lambda_{h,1}(X_{h})}} = \frac{r_{2,h}}{\abs{r_{1,2} r_{1,h}}}. \nonumber\\
&&~~ \Rightarrow~~ \lambda_{h,2}(X_{h}) = \wt{c}_{h,2} \lambda_{h,1}(X_{h})~\text{for some constant}~\wt{c}_{h,2}. \label{eq: cholesky diagonal 3}\\
&&\text{Next,} ~~
\corr(U_{3},U_{h}\vert \bX)=\frac{\sum_{m=1}^{3}\lambda_{3,m}(X_{3})\lambda_{h,m}(X_{h})}{\{\sum_{m=1}^{3} \lambda_{3,m}^{2}(X_{3})\}^{1/2} \{\sum_{m=1}^{h} \lambda_{h,m}^{2}(X_{h})\}^{1/2}} = r_{3,h}   \nonumber\\
&&~~ \Rightarrow~~ \frac{\lambda_{3,3}(X_{3})\{c_{3,1}\lambda_{h,1}(X_{h})+c_{3,2}\wt{c}_{h,2}\lambda_{h,1}(X_{h})+\lambda_{h,3}(X_{h})\}}{\abs{c_{3,1}\lambda_{3,3}(X_{3})} \abs{\lambda_{h,1}(X_{h})}} = \frac{r_{3,h}}{\abs{r_{1,3} r_{1,h}}}  \nonumber\\
&&~~ \Rightarrow~~ \lambda_{h,3}(X_{h}) = \wt{c}_{h,3} \lambda_{h,1}(X_{h})~\text{for some constant}~\wt{c}_{h,3}.  \label{eq: cholesky diagonal 4}\\
&&\text{Finally,}~~\corr(U_{h-1},U_{h}\vert \bX)=\frac{\sum_{m=1}^{h-1}\lambda_{h-1,m}(X_{h-1})\lambda_{h,m}(X_{h})}{\{\sum_{m=1}^{h-1} \lambda_{h-1,m}^{2}(X_{h-1})\}^{1/2} \{\sum_{m=1}^{h} \lambda_{h,m}^{2}(X_{h})\}^{1/2}} = r_{h-1,h}   \nonumber\\
&&~~ \Rightarrow~~ \frac{\lambda_{h-1,h-1}(X_{h-1})\{c_{h-1,1}\lambda_{h,1}(X_{h})+c_{h-1,2}\wt{c}_{h,2}\lambda_{h,1}(X_{h})+\dots+\lambda_{h,h}(X_{h})\}}{\abs{c_{h-1,1}\lambda_{h-1,1}(X_{h-1})} \abs{\lambda_{h,1}(X_{h})}} = \frac{r_{h-1,h}}{\abs{r_{1,h-1} r_{1,h}}}  \nonumber\\
&&~~ \Rightarrow~~ \lambda_{h,h-1}(X_{h}) = \wt{c}_{h,h-1} \lambda_{h,1}(X_{h})~\text{for some constant}~\wt{c}_{h,h-1}.  \label{eq: cholesky diagonal 5}
\ee
\vspace{-5ex}\\
Combining (\ref{eq: cholesky diagonal 3}), (\ref{eq: cholesky diagonal 4}), (\ref{eq: cholesky diagonal 5}) etc. with (\ref{eq: cholesky diagonal 2}), 
the proposition follows by principles of mathematical induction. 
This implies $\bLambda(\bX)=\bLambda_{1}(\bX)\bC$ where $\bLambda_{1}(\bX)=\diag\{\lambda_{1}(X_{1}),\dots,\lambda_{p}(X_{p})\}$ with $\lambda_{\ell}(X_{\ell})=\lambda_{\ell,\ell}(X_{\ell})$ for all $\ell$ and $\bC=((c_{\ell,m}))$ is a lower triangular matrix with $c_{\ell,\ell}=1$ for all $\ell$. 
\end{proof}

\vskip 5pt
Under the conditions of Lemma \ref{lem: cholesky diagonal}, we thus have $\cov(\bU\vert\bX) = \bSigma(\bX)= \bLambda_{1}(\bX) \bSigma_{1} \bLambda_{1}\trans(\bX)$ with $\bSigma_{1}=\bC\bC\trans$.
Introducing a latent factor $\bepsilon$, we can now write $(\bU\vert\bX,\bepsilon)=\bLambda_{1}(\bX)\bepsilon$ with $\bepsilon \perp \bX$ and $\cov(\bepsilon)=\bSigma_{1}$. 
Due to the diagonal nature of $\bLambda_{1}(\bX)$, each component $\epsilon_{\ell}$ of $\bepsilon$ is exclusively associated with the corresponding component $U_{\ell}$ of $\bU$ 
and may be treated as a scaled version of $U_{\ell}$.
Starting with a general latent factor model framework, 
with two additional restrictions that are particularly relevant in multivariate measurement error settings, 
we have now arrived at model (\ref{eq: mvt multiplicative structure}). 
The problems of modeling $\cov(\bU\vert\bX)$ and the shape of $f_{\bU\vert\bX}$ can now be achieved by separately modeling $\bLambda_{1}(\bX)$ and $f_{\bepsilon}$.
And $E(\bU\vert\bX)=\bzero$ can be achieved by enforcing $E(\bepsilon)=\bzero$.

\vspace{-2ex}
\subsection{Models for $\bU\vert \bX$ and $\cov(\bU\vert\bX)$}  \label{sec: mvt cov mat model}

In this section, we first revisit the models for conditionally varying measurement errors developed in Section \ref{sec: mvt density of errors} of the main paper.
A few plausible alternatives and generalizations, 
the implied covariance structures, their strengths, limitations and connections with the adopted model are also discussed. 

The model (\ref{eq: mvt multiplicative structure}) for conditionally varying measurement errors developed in Section \ref{sec: mvt density of errors} of the main paper assumes $(\bU_{ij}\vert\bX_{i})=\bS(\bX_{i})\bepsilon_{ij\ell}$ where $\bS(\bX_{i})=\diag\{s_{1}(X_{i1}),\dots,s_{p}(X_{ip})\}$ and $\bepsilon_{ij\ell}$ are distributed independently of $\bX$ with $E(\bepsilon_{ij})=\bzero$. 
This `separability' of $\bX_{i}$ and $\bepsilon_{ij}$ allows us to incorporate distributional flexibility and enforce the mean zero restriction using the techniques developed for independent errors in Section \ref{sec: mvt density of homoscedastic errors} in the main paper. 
The diagonal structure of $\bS$ highlights the exclusive associations between $U_{ij\ell}$ and $X_{i\ell}$ 
but ignores weak dependencies of $U_{ij\ell}$ on $\{X_{im}\}_{m\neq \ell}$. 
The general of shape of $f_{\bU\vert\bX}$ as well correlations between different components of $\bU_{ij}$ are inherited from $f_{\bepsilon}$. 
The associated dependency structure is summarized in Figure \ref{fig: graphical models supplementary materials}(b). 
The novel two-stage procedure described in Sections \ref{sec: mvt posterior computation} and \ref{sec: mvt estimation of variance functions} 
produces efficient and numerically stable posterior estimates. 

As discussed in Section \ref{sec: mvt multiplicative errors}, the model also arises naturally in multivariate multiplicative measurement error settings $\bW_{ij} = \bX_{i} \circ \wt\bU_{ij}$
where the errors $\wt\bU_{ij}$ are distributed independently of $\bX_{i}$ with $E(\wt\bU_{ij})=\bone$. 
The model can be reformulated as $\bW_{ij} = \bX_{i} + \bU_{ij}$, 
where $\bU_{ij}=\bS(\bX_{i})\bepsilon_{ij}$, $\bS(\bX_{i})=\diag\{X_{i1},\dots,X_{ip}\}$ and $\bepsilon_{ij}=(\wt\bU_{ij}-1)$ with $E(\bepsilon_{ij})=\bzero$. 
It thus conforms to the conditionally varying additive measurement error model (\ref{eq: mvt multiplicative structure}) described above.

These results and the ones provided in Section \ref{sec: latent factor models for different covariance classes} 
establish the fairly general nature of model (\ref{eq: mvt multiplicative structure}) and are also informative about cases outside its support. 
A few such cases that are particularly relevant to measurement error problems and form part of our research aspirations 
but are not pursued in detail in this article are briefly discussed below.

As informed by Lemma \ref{lem: cholesky me}, another class that implies $\var(U_{ij\ell}\vert\bX_{i})=s_{\ell}^{2}(X_{i\ell})$ 
and allows $\corr(U_{ij\ell},U_{ijm}\vert\bX_{i})$ to vary with $X_{i\ell}$ and $X_{im}$ 
is obtained by letting $\bU_{ij}=\bLambda(\bX_{i}) \bepsilon_{ij}$ with $\bLambda(\bX_{i})=((\lambda_{\ell,m}(X_{i\ell})))_{\ell=1,m=1}^{p,p}$. 
The model highlights the exclusive associations between $U_{ij\ell}$ and $X_{i\ell}$ 
- $\var(U_{ij\ell}\vert\bX_{i})$ depends on $X_{i\ell}$ and $\cov(U_{ij\ell},U_{ijm}\vert \bX_{i})$ depends on $X_{i\ell}$ and $X_{im}$. 
Modeling variations in conditional cross-moments is a daunting task in multivariate settings, 
more so in the absence of precise information about $\bX_{i}$. 
Towards a more parsimonious representation, the off-diagonal elements $\{\lambda_{\ell,m}(X_{i\ell})\}_{\ell \neq m}$ may be shrunk towards zero, 
resulting in a model that associates each $U_{ij\ell}$ with its own latent factor component $\epsilon_{ij\ell}$. 
That is, $\bLambda(\bX_{i})$ should be shrunk towards $\bLambda_{0}(\bX_{i})=\diag\{\lambda_{1,1}(X_{i1}),\dots,\lambda_{p,p}(X_{ip})\}$. 
This limiting case still allows $\var(U_{ij\ell}\vert \bX_{i})$ to vary flexibly with $X_{i\ell}$, and $\cov(U_{ij\ell},U_{ijm}\vert\bX)$ to vary with $X_{i\ell}$ and $X_{im}$, 
but assumes the correlations $\corr(U_{ij\ell},U_{ijm}\vert\bX_{i})$ to not vary with $\bX_{i}$.

Another flexible class of models for $(\bU_{ij}\vert\bX_{i})$ 
that conforms to the dependency structure depicted in Figure \ref{fig: graphical models supplementary materials}(a) 
is obtained by letting $\bU_{ij}=\bLambda(\bX_{i}) \bepsilon_{ij}$
with $\bLambda(\bX_{i})=((\lambda_{\ell,m}(X_{im})))_{\ell=1,m=1}^{p,p}$. 
The implied covariance structure is given by $\cov(\bU_{ij}\vert\bX_{i})=\bSigma(\bX_{i})=\bLambda(\bX_{i})\bSigma_{\bepsilon}\bLambda\trans(\bX_{i})$. 
Specifically, we have $(U_{ij\ell} \vert \bX_{i}) = \textstyle \sum_{m}\lambda_{\ell,m}(X_{im})\epsilon_{ijm}$ with
\vspace{-5ex}\\
\bse
&& \cov(U_{ij\ell_{1}},U_{ij\ell_{2}}\vert \bX_{i}) = \textstyle \sum_{m_{1},m_{2}} \lambda_{\ell_{1},m_{1}}(X_{im_{1}}) \lambda_{\ell_{2},m_{2}}(X_{im_{2}}) \sigma_{m_{1},m_{2}}\\
&& ~~~ = \textstyle \lambda_{\ell_{1},\ell_{1}}(X_{i\ell_{1}}) \lambda_{\ell_{2},\ell_{2}}(X_{i\ell_{2}}) \sigma_{\ell_{1},\ell_{2}} + \sum_{m_{1}\neq \ell_{1},m_{2} \neq \ell_{2}} \lambda_{\ell_{1},m_{1}}(X_{im_{1}}) \lambda_{\ell_{2},m_{2}}(X_{im_{2}}) \sigma_{m_{1},m_{2}} \\
&&\text{and}~~~\var(U_{ij\ell}\vert \bX_{i}) = \textstyle \lambda_{\ell,\ell}^{2}(X_{i\ell}) \sigma_{\ell,\ell} + \sum_{m_{1}\neq \ell,m_{2} \neq \ell} \lambda_{\ell,m_{1}}(X_{im_{1}}) \lambda_{\ell,m_{2}}(X_{im_{2}}) \sigma_{m_{1},m_{2}}.
 \ese 
\vspace{-5ex}\\
Ideally, to highlight the exclusive strong association between $U_{ij\ell}$ and $X_{i\ell}$, 
the diagonal elements of $\bLambda(\bX_{i})$, namely $\lambda_{\ell,\ell}(X_{i\ell})$, should dominate 
and the remaining off-diagonal elements $\{\lambda_{\ell,m}(X_{im})\}_{\ell \neq m}$ may be shrunk towards zero. 
That is, $\bLambda(\bX_{i})$ should be shrunk towards $\bLambda_{0}(\bX_{i})=\diag\{\lambda_{1,1}(X_{i1}),\dots,\lambda_{p,p}(X_{ip})\}$. 

Since measurement error problems are well known to be inherently computationally unstable, 
it is not clear whether any practical gain in efficiency can be achieved 
by modeling large number of off-diagonal functions in $\bLambda(\bX_{i})$ at the expense of significantly increased model complexity. 
Model (\ref{eq: mvt multiplicative structure}) considered in this article instead focuses on the special limiting cases with $\bS(\bX_{i})=\bLambda_{0}(\bX_{i})$. 

Another extension results from mixtures of multiplicative and independent additive errors. 
In univariate settings, such models were considered in Rocke a Durbin (2001) for studying gene expression levels measured by DNA slides. 
In multivariate settings, we have $\bU_{ij}=\bX_{i} \circ \bepsilon_{ij}^{(1)}+\bepsilon_{ij}^{(2)}$, 
where $\bepsilon_{ij}^{(k)}$, $k=1,2$ are distributed independently of $\bX_{i}$. 
With $\cov(\bepsilon_{ij}^{(k)})=\bSigma_{k}=((\sigma_{\ell,m}^{(k)}))_{m=1,\ell=1}^{p,p}$ for $k=1,2$, 
the implied covariance structure is given by $\cov(\bU_{ij}\vert\bX_{i})=\bS(\bX_{i})\bSigma_{1}\bS(\bX_{i})+\bSigma_{2}$, 
where $\bS(\bX_{i})=\diag\{X_{i1},\dots,X_{ip}\}$, as above. 
The model conforms to the dependency structure of Figure \ref{fig: graphical models supplementary materials}(b)
but can not be strictly written as model (\ref{eq: mvt multiplicative structure}). 
However, as can be seen from Figure \ref{fig: mvt EATS data results VFn}, 
in our motivating nutritional epidemiology application, 
smaller average consumptions naturally result in more precise 24 hour recalls, 
the variability approaching 0 as the true consumption approaches 0.  
Under the assumption of continuity, $\lim_{\bX\to\bzero}\bSigma(\bX) \to \bzero^{p\times p}$ implies $\bSigma_{2}=\bzero^{p\times p}$, 
resulting in model (\ref{eq: mvt multiplicative structure}). 

\subsection{Model Adequacy Checks}\label{sec: mvt model adequacy checks}
In Figure \ref{fig: mvt EATS data results VFn} in the main paper, we showed the plots of subject specific means $\overline{W}_{i\ell}$ of the replicates vs the corresponding subject-specific variances $S_{W,i\ell}^{2}$ for each of the four dietary components included in our analysis in Section \ref{sec: mvt data analysis}. 
These plots suggest very strong conditional heteroscedasticity patterns in the measurement errors.  
If we consider the plots of subject specific means $\overline{W}_{i\ell}$ vs subject specific variances $S_{W,im}^{2}$ for all possible pairs $(\ell,m)$, 
we will see similar monotone increasing patterns not just for the pairs with $\ell=m$, but in pairs with $\ell\neq m$ too. 
This can be explained by the high correlation between different components of $\bX_{i}$, see Figure \ref{fig: mvt EATS data results XS},  
and does not necessarily imply that the conditional variability in $U_{ij\ell}$ depends on other components of $\bX_{i}$, not just $X_{i\ell}$.  
As discussed in the previous subsections, 
since the $\ell\th$ component $U_{ij\ell}$ is the measurement error associated exclusively with $X_{i\ell}$, 
it is plausible to assume that the conditional variability of $U_{ij\ell}$ 
can be modeled mostly as a function of $X_{i\ell}$ only.

\begin{figure}[!ht]
\begin{center}
\includegraphics[height=16cm, width=16cm, trim=0cm 0cm 0cm 0cm, clip=true]{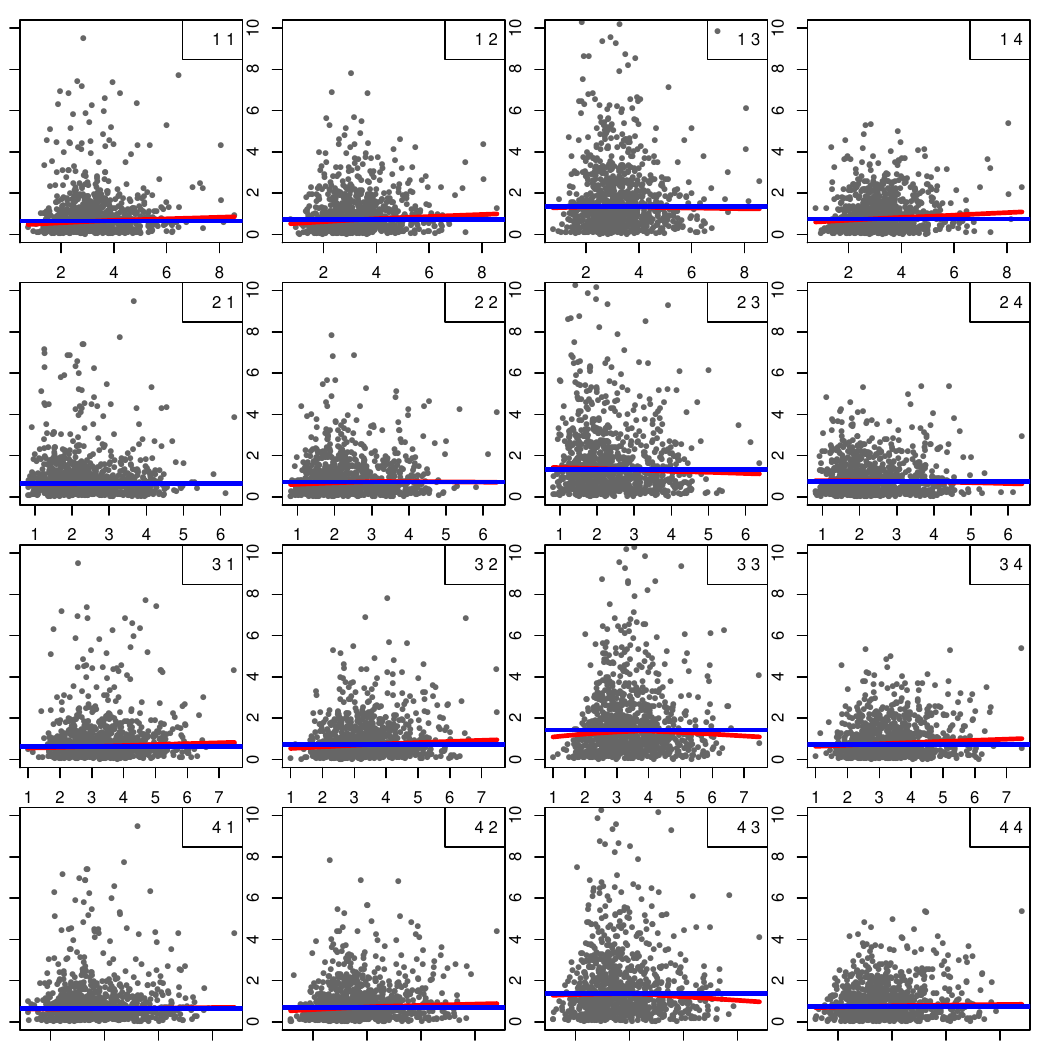}
\end{center}
\caption{\baselineskip=10pt Panel $(\ell,m)$ shows the plot of estimates $\wh{X}_{i\ell}$ of $X_{i\ell}$ vs subject specific variances $\wh{S}_{\epsilon,im}^{2}$ of scaled measurement error residuals $\wh\epsilon_{ijm}$, 
produced by univariate deconvolution methods. 
See Section \ref{sec: mvt model adequacy checks} of the Supplementary Materials for additional details. 
The darker horizontal lines in each panel represent the upper 10\% trimmed mean of the subject specific variances $\wh{S}_{\epsilon,i\ell}^{2}$.
The lighter solid lines in each panel represent nonparametric lowess fits.
}
\label{fig: mvt EATS data covariance model justification}
\end{figure}

We present here some diagnostic plots to further validate the practical adequacy of this structural assumption.
Figure \ref{fig: mvt EATS data covariance model justification} shows the plots of $\wh{X}_{i\ell}$ vs subject specific variances $\wh{S}_{\epsilon,im}^{2}$ of $\wh\epsilon_{ijm}$, 
where $\wh{X}_{i\ell}$ represent the posterior means of $X_{i\ell}$ values and $\wh\epsilon_{ijm} = (W_{ijm}-\wh{X}_{im})/\wh{s}_{m}(\wh{X}_{im})$ 
represent the corresponding scaled measurement error residuals 
produced by the univariate submodels for the EATS data set analyzed in Section \ref{sec: mvt data analysis} of the main paper. 
The figure indicates constant variance of the scaled measurement error residuals $\wh\epsilon_{ij\ell}$ over the entire range of $X_{im}$ values for all $(\ell,m)$ pairs. 
Nonparametric Eubank-Hart tests of no covariate effect (Eubank and Hart, 1992) applied to $(\wh{X}_{i\ell},\wh{S}_{\epsilon,im}^{2})$ for all $(\ell,m)$ pairs  
(treating $\wh{X}_{i\ell}$ as the covariate and $\wh{S}_{\epsilon,im}^{2}$ as the response) 
produced a minimum Benjamini-Hochberg adjusted p-value of $0.096$, suggesting that there is no residual heteroscedasticity left in $U_{ij\ell}$ 
after accounting for the variability in $U_{ij\ell}$ that can be sufficiently explained through $X_{i\ell}$ only. 
See Table \ref{tab: mvt Eubank-Hart p-values}.
It may thus be concluded that for the EATS data application model (\ref{eq: mvt multiplicative structure}) 
developed in Section \ref{sec: mvt density of heteroscedastic errors} of the main paper 
that implies $\var(U_{ij\ell}\vert \bX_{i}) = s_{\ell}^{2}(X_{i\ell})\var(\epsilon_{ij\ell})$ 
suffices to explain the conditional variability in the measurement errors.

Model (\ref{eq: mvt multiplicative structure}) also assumed that only the conditional variability of $\bU_{ij}$ depends on $\bX_{i}$, 
and derived other features of $\bU_{ij}$ like skewness, multimodality, heavy-tails etc. from the scaled errors $\bepsilon_{ij}$. 
As shown in Sarkar, et al. (2014), even in the much simpler univariate set up, in the absence of precise information on $X_{i\ell}$, 
variations in other features of $U_{ij\ell}$ for varying values of $X_{i\ell}$, if any, are extremely difficult to detect.   
More importantly, semiparametric methods that make the multiplicative structural assumption $(U_{ij\ell}\vert X_{i\ell}) = s_{\ell}(X_{i\ell})\epsilon_{ij\ell}$ 
are highly robust to departures from this assumption 
and significantly outperform possible nonparametric alternatives that allow all order moments of $U_{ij\ell}$ to vary flexibly with $X_{i\ell}$, not just the conditional variance,  
even in scenarios where the true data generating process closely conforms to these nonparametric alternatives.

\vskip 10pt
\begin{table}[!ht]
\centering
\begin{tabular}{|c|c|cccc|}
  \hline
 & Panel & p-values & BFN & BH & BY \\ 
  \hline
  1 & 1,1 & 0.991 & 1.000 & 0.991 & 1.000 \\ 
  2 & 1,2 & 0.764 & 1.000 & 0.873 & 1.000 \\ 
  3 & 1,3 & 0.251 & 1.000 & 0.446 & 1.000 \\ 
  4 & 1,4 & 0.129 & 1.000 & 0.446 & 1.000 \\ 
  5 & 2,1 & 0.598 & 1.000 & 0.736 & 1.000 \\ 
  6 & 2,2 & 0.266 & 1.000 & 0.446 & 1.000 \\ 
  7 & 2,3 & 0.037 & 0.592 & 0.197 & 0.667 \\ 
  8 & 2,4 & 0.990 & 1.000 & 0.991 & 1.000 \\ 
  9 & 3,1 & 0.224 & 1.000 & 0.446 & 1.000 \\ 
  10 & 3,2 & 0.012 & 0.192 & 0.096 & 0.325 \\ 
  11 & 3,3 & \bf{0.011} & \bf{0.176} & \bf{0.096} & \bf{0.325} \\ 
  12 & 3,4 & 0.497 & 1.000 & 0.692 & 1.000 \\ 
  13 & 4,1 & 0.519 & 1.000 & 0.692 & 1.000 \\ 
  14 & 4,2 & 0.163 & 1.000 & 0.446 & 1.000 \\ 
  15 & 4,3 & 0.279 & 1.000 & 0.446 & 1.000 \\ 
  16 & 4,4 & 0.244 & 1.000 & 0.446 & 1.000 \\ 
   \hline
\end{tabular}
\caption{\baselineskip=10pt 
The original and adjusted p-values (BFN=Bonferroni, BH=Benjamini-Hochberg, BY=Benjamini-Yekutli) returned by nonparametric Eubank-Hart tests of no covariate effect applied to $(\wh{X}_{i\ell},\wh{S}_{\epsilon,im}^{2})$ for all $(\ell,m)$ pairs  
treating $\wh{X}_{i\ell}$ as the covariate and $\wh{S}_{\epsilon,im}^{2}$ as the response. 
The minimum values corresponding to panel $(3,3)$ are highlighted. 
See Section \ref{sec: mvt model adequacy checks} and Figure \ref{fig: mvt EATS data covariance model justification} in the Supplementary Materials for additional details. 
}
\label{tab: mvt Eubank-Hart p-values}
\end{table}

\newpage
\thispagestyle{empty}

\newpage 
\section{Finite vs Infinite Mixture Models} \label{sec: mvt finite vs infinite mixture models}
In this article, we modeled the  $f_{\bX}$ and the density of the scaled measurement errors $f_{\bepsilon}$ 
using mixtures of fixed finite number of multivariate normal kernels.
Alternative approaches that escape the need to prespecify the number of mixture components include models with potentially infinite number of mixture components,
models induced by Dirichlet processes (Ferguson, 1973; Escobar and West, 1995) being perhaps the most popular among such techniques. 
Apart from flexibility, one major advantage of such techniques comes from the ability of associated MCMC machinery to perform model selection and model averaging implicitly and semiautomatically. 
Model averaging is achieved by allowing the number of mixture components to vary from one MCMC iteration to the other.
The number of mixture components that is visited the maximum number of times by the sampler then provides 
a maximum a-posteriori (MAP) estimate of the number of mixture components required to approximate the target density.  
However, in complicated multivariate set up like ours, MCMC algorithms for such infinite dimensional models become computationally highly intensive.
Mixtures based on fixed finite number of components, on the other hand, can greatly reduce computational complexity.
Recent studies of asymptotic properties of the posterior of overfitted mixture models (Rousseau and Mengersen, 2011) 
suggest that mixture models with sufficiently large number of components can perform automatic model selection and model averaging just like infinite dimensional models.
Additionally, as the proofs of the results in Section \ref{sec: mvt model flexibility} imply, 
the use of mixture models with fixed finite number of components does not necessarily imply a compromise on the issue of flexibility.
The approaches adopted in this article try to take the best from both worlds. 
Computational burden is reduced by keeping the number of mixture components fixed at some finite values.
At the same time, simultaneous semiautomatic model selection and model averaging is achieved by exploiting properties of overfitted mixture models.
We elaborate our arguments below,
pointing out the close connections and the subtle differences our adopted finite dimensional models have with the aforementioned infinite dimensional alternatives.

\subsection{Infinite Mixture Models as Limits of Finite Mixture Models}  \label{sec: mvt limits of finite mixture models}
Let $G_{K}=\sum_{k=1}^{K}\pi_{k}\delta_{\theta_{k}}$ with $(\pi_{1},\dots,\pi_{K})\sim \Dir(\alpha/K,\dots,\alpha/K)$ and $\theta_{k} \sim H$. 
Also, let $G_{\infty} \sim \DP(\alpha,H)$, a Dirichlet process with concentration parameter $\alpha$ and base measure $H$. 
Then, $G_{\infty}$ can be represented as $G_{\infty}=\sum_{k=1}^{\infty}\wt\pi_{k}\delta_{\theta_{k}}$ with $\wt\pi_{k} =V_{k}\prod_{\ell=1}^{k-1}(1-V_{\ell}), V_{\ell} \sim \Beta(1,\alpha)$ and $\theta_{k} \sim H$ (Sethuraman, 1994). 
As $K\to \infty$, $\int g(\theta) dG_{K}(\theta) \overset{d}{\to} \int g(\theta) dG_{\infty}(\theta)$ 
for any measurable function $g$ integrable with respect to $H$ (Ishwaran and Zarepour, 2000, 2002). 

The finite mixtures of multivariate normal kernels with symmetric Dirichlet priors 
that we used in this article to model both $f_{\bX}$ and the density of the scaled measurement errors $f_{\bepsilon}$
have close connections with infinite dimensional Dirichlet process based mixture models. 
Specifically, taking $g(\theta)=\MVN(\bmu,\bSigma)$ and appealing to the above result, we have $f_{\bX}=\sum_{k=1}^{K_{\bX}}\pi_{\bX,k} \MVN(\bmu_{\bX,k},\bSigma_{\bX,k}) \overset{d}{\to} \sum_{k=1}^{\infty}\wt\pi_{\bX,k}\MVN(\bmu_{\bX,k},\bSigma_{\bX,k})$ as $K_{\bX} \to \infty$.
Our proposed mechanism to enforce the mean zero restriction on $f_{\bepsilon}$ specifically requires a finite dimensional symmetric prior on the mixture probabilities
and therefore does not admit a straightforward infinite dimensional extension. 
But in the limit, as $K_{\bepsilon}\to \infty$, a reformulation of the model results in a complicated multivariate version of the infinite dimensional model of Sarkar, et al. (2014) 
(See Lemma \ref{Lem: mvt lemma from Pelenis} in Section \ref{sec: mvt estimation of variance functions}).

\subsection{Computational Complexity}  \label{sec: mvt computational complexity}

The implementation of complex infinite dimensional models, specially the complicated mean restricted model for the scaled errors, 
will be computationally intensive in a multivariate setting like ours. 
The computational simplicity of the finite dimensional methods proposed in this article make them particularly suitable for multivariate problems.

In this paragraph, we discuss additional mixing issues that render infinite dimensional models, 
particularly the ones with non or semiconjugate priors on the component specific parameters (like our MLFA model), unsuitable for multivariate applications. 
There are two main types of MCMC algorithms for fitting infinite dimensional mixture models - conditional methods and marginal methods. 
In the conditional scheme, the mixture probabilities are sampled. 
The mixture labels are then updated independently, conditional on the mixture probabilities.
The mixture probabilities in infinite dimensional mixture models can be stochastically ordered.
For instance, mixture probabilities in a  Dirichlet process mixture model satisfy $E(\wt\pi_{k})>E(\wt\pi_{k+1})$ and $\Pr(\wt\pi_{k}>\wt\pi_{k+1})>0.5$ for all $k\in\nN$.
This imposes weak identifiability on the mixture labels resulting in a complicated model space comprising many local modes of varying importance. 
Different permutations of the mixture labels are not equivalent and exploration of the entire model space becomes important for valid inference.
In high dimensional and large data settings it is difficult to achieve even by sophisticated MCMC algorithms with carefully designed label switching moves (Hastie, et al. 2013). 
The problem can be avoided with marginal methods (Neal, 2000) that integrate out the mixture probabilities and work with the resulting Polya urn scheme, 
rendering the mixture labels dependent but nonidentifiable. 
Unfortunately, such integration is possible only when conjugate priors are assigned to the component specific parameters.  
Typically for infinite dimensional models with non or semiconjugate priors on the component specific parameters, good mixing is thus difficult to achieve, 
particularly in complicated multivariate setup like ours.

Such issues also plague finite dimensional truncation based approximations to Dirichlet process mixture models 
where the mixture probabilities are constructed as $\wt\pi_{k} =V_{k}\prod_{\ell=1}^{k-1}(1-V_{\ell}), V_{\ell} \sim \Beta(1,\alpha), k=1,\dots,(K-1)$, and $V_{K}=1$ (Ishwaran and James, 2002) and the mixture components remain weakly identifiable. 

On the contrary, the issues of mixing and convergence become much less important for finite mixture models with symmetric priors 
$(\pi_{1},\dots,\pi_{K})\sim \Dir(\alpha/K,\dots,\alpha/K)$ on the mixture probabilities. 
With $K_{\bX}$ and $K_{\bepsilon}$ mixture components for the densities $f_{\bX}$ and $f_{\bepsilon}$, respectively, 
the posterior is still multimodal but comprises $K_{\bX}! \times K_{\bepsilon}!$ modal regions that are exact copies of each other. 
For inference on the overall density or any other functions of interest that are invariant to permutations of the mixture labels, 
it is only important that the MCMC sampler visits and explores at least one of the modal regions well 
and label switching (or the lack of it) does not present any problem (Geweke, 2007).

\subsection{Model Selection and Model Averaging}\label{sec: mvt model selection and model averaging}
\begin{figure}[!ht]
\begin{center}
\includegraphics[height=16cm, width=16cm, trim=0cm 0cm 0cm 0cm, clip=true]{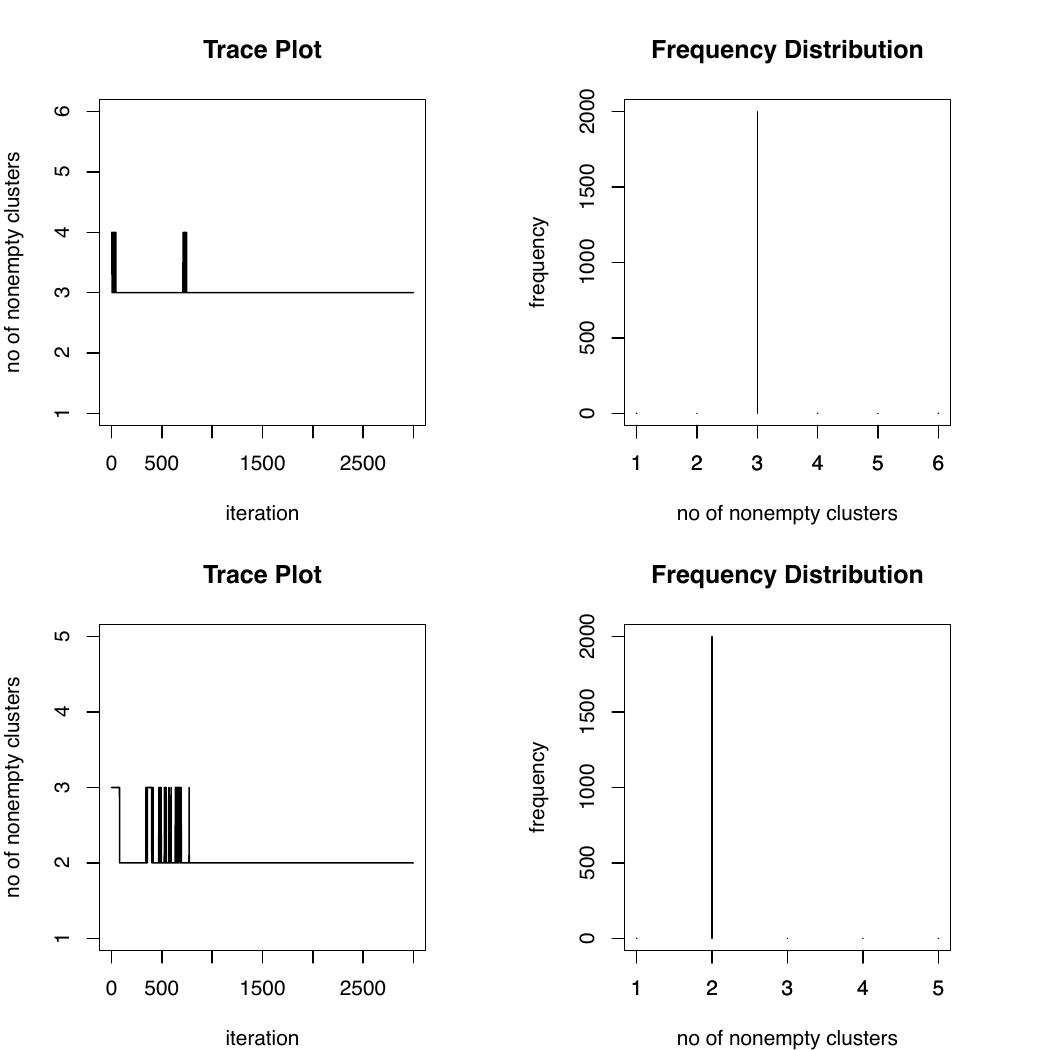}
\end{center}
\caption{\baselineskip=10pt 
Trace plots and frequency distributions of the number of nonempty clusters produced by the MLFA (mixtures of latent factor analyzers) method for the conditionally heteroscedastic error distribution $f_{\bepsilon}^{(2)}$ with sample size $n=1000$, $m_{i}=3$ replicates for each subject and identity matrix (I) for the component specific covariance matrices. 
See Section \ref{sec: mvt simulation studies} for additional details.
The results correspond to the simulation instance that produced the median of the estimated integrated squared errors (ISE) out of a total of 100 simulated data sets, 
when the number of mixture components for $f_{\bX}$ and $f_{\bepsilon}$ were kept fixed at $K_{\bX}=6$ and $K_{\bepsilon}=5$.
The upper panels are for the  $f_{\bX}$ and the lower panels are for the density of the scaled errors $f_{\bepsilon}$.
The true number of mixture components were $K_{\bX} = 3$ and $K_{\bepsilon} = 3$. 
As can be seen from Figure \ref{fig: mvt simulation results ES d4 n1000 m3 MLFA X1 E1 Ind},
a mixture model with $2$ nonempty clusters can approximate the true density of the scaled errors well.
}
\label{fig: mvt simulation results Trace Plots d4 n1000 m3 MLFA X1 E1 Ind}
\end{figure}

As mentioned at the beginning of Section \ref{sec: mvt finite vs infinite mixture models}, 
a major advantage of infinite dimensional mixture models is their ability to implicitly and semiautomatically perform model selection and model averaging.
Properties of overfitted mixture models can be exploited to achieve the same in finite dimensional models with sufficiently large number of components. 
Recently Rousseau and Mengersen (2011) studied the asymptotic behavior of the posterior for overfitted mixture models with Dirichlet prior $\Dir(\alpha_{1},\dots,\alpha_{K})$ on the mixture probabilities 
in a measurement error free set up and showed that 
the hyper parameter $(\alpha_{1},\dots,\alpha_{k})$ strongly influences the way the posterior handles overfitting.
In particular, when $\max_{k=1,\dots,K}\alpha_{k} < L/2$, where $L$ denotes the number of parameters specifying the component kernels, 
the posterior is asymptotically stable and concentrates in regions with empty redundant components. 
In this article, we chose symmetric Dirichlet priors $\Dir(\alpha/K,\dots,\alpha/K)$ on the mixture probabilities 
to model both the  $f_{\bX}$ and the density of the scaled measurement errors $f_{\bepsilon}$.
We set $\alpha_{\bX}=\alpha_{\bepsilon} =1$ so that the condition $\alpha/K < L/2$ is satisfied for both $f_{\bX}$ and $f_{\bepsilon}$.
In simulation experiments reported in Section \ref{sec: mvt simulation studies}, the behavior of the posterior was similar to that observed by Rousseau and Mengersen (2011) in measurement error free set up.
That is, when $K_{\bX}$ and $K_{\bepsilon}$ were assigned sufficiently large values, 
the MCMC chain quickly reached a stable stage where the redundant components became empty. 
See Figure \ref{fig: mvt simulation results Trace Plots d4 n1000 m3 MLFA X1 E1 Ind} in the main article 
and Figure \ref{fig: mvt simulation results Trace Plots d4 n1000 m3 MIW X1 E1 AR} and \ref{fig: mvt simulation results Trace Plots d4 n1000 m3 MLFA X1 E1 AR} in the Supplementary Materials 
for illustrations, where, with some abuse of nomenclature, the $k\th$ component is called empty if the associated mixture probability $\pi_{k} \leq 0.05$.  
Since such overfitted mixture models allow the number of nonempty mixture components to vary from one MCMC iteration to the next, 
model averaging is automatically achieved.
MAP estimates of the numbers of mixture components required to approximate the target densities are given by the numbers of components 
which are visited the maximum number of times by the MCMC sampler,
as in the case of infinite mixture models.

\begin{figure}[!ht]
\begin{center}
\includegraphics[height=14cm, width=16cm, trim=0cm 0cm 0cm 0cm, clip=true]{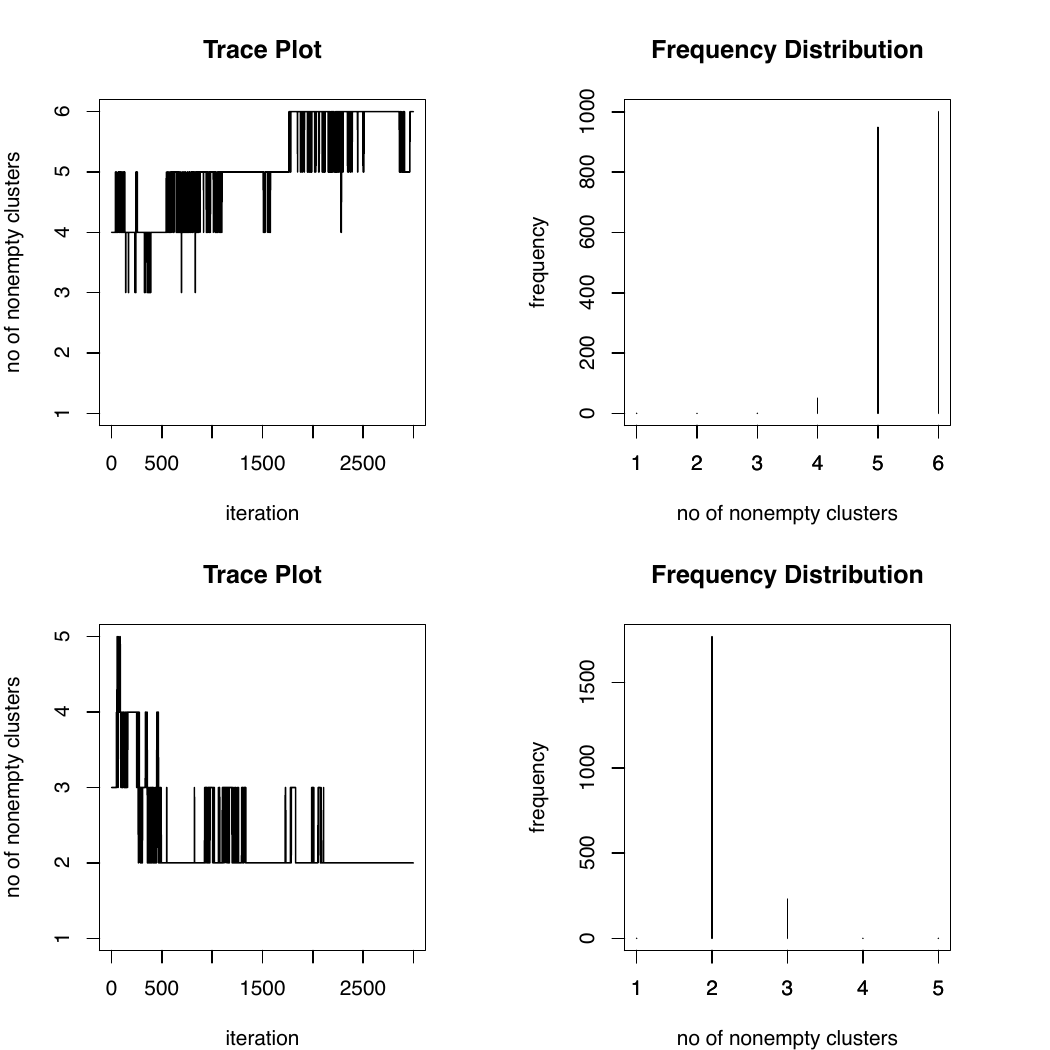}
\end{center}
\caption{\baselineskip=10pt 
Trace plots and frequency distributions of the number of nonempty clusters produced by the MIW (mixtures with inverse Wishart priors) method for the conditionally heteroscedastic error distribution $f_{\bepsilon}^{(2)}$ with sample size $n=1000$, $m_{i}=3$ replicates for each subject and identity matrix (I) for the component specific covariance matrices. 
See Section \ref{sec: mvt simulation studies} for additional details.
The results correspond to the simulation instance that produced the median of the estimated integrated squared errors (ISE) out of a total of 100 simulated data sets, 
when the number of mixture components for $f_{\bX}$ and $f_{\bepsilon}$ were kept fixed at $K_{\bX}=6$ and $K_{\bepsilon}=5$.
The upper panels are for the  $f_{\bX}$ and the lower panels are for the density of the scaled errors $f_{\bepsilon}$.
The true number of mixture components were $K_{\bX} = 3$ and $K_{\bepsilon} = 3$. 
As can be seen from Figure \ref{fig: mvt simulation results ES d4 n1000 m3 MIW X1 E1 Ind},
a mixture model with $2$ nonempty clusters can approximate the true density of the scaled errors well.
}
\label{fig: mvt simulation results Trace Plots d4 n1000 m3 MIW X1 E1 Ind}
\end{figure}

As discussed in the main paper, for the MIW method, when the measurement errors are conditionally heteroscedastic and the true covariance matrices are highly sparse, 
the strategy usually overestimates the number of non-empty mixture components required to approximate the target densities well. 
In these cases, the MIW method becomes highly numerically unstable and much larger sample sizes are required for the asymptotic results to hold. 
See Figure \ref{fig: mvt simulation results Trace Plots d4 n1000 m3 MIW X1 E1 Ind} in the main article for an illustration.
This may be regarded more as a limitation of the MIW method than a limitation of the adopted strategy to determine $K_{\bX}$ and $K_{\bepsilon}$. 
For the numerically more stable MLFA model, the asymptotic results are valid even for moderate sample sizes and such models are also more robust to overestimation of the number of nonempty clusters.

\subsection{Model Flexibility}\label{sec: mvt supp mat model flexibility}
The proofs of the support results presented in Section \ref{sec: mvt model flexibility} require that 
the number of mixture components of the corresponding mixture models be allowed to vary over the set of all positive integers.
However, as the technical details of the proofs reveal, 
the use of mixture models with fixed finite number of components does not necessarily imply a compromise on the issue of flexibility.
Indeed, a common recurring idea in the proofs of all these results, including those for the variance functions, is to show that 
any function coming from the target class can be approximated with any desired level of accuracy by the corresponding finite mixture models 
provided the models comprise sufficiently large number of mixture components and the function satisfies some fairly minimal regularly conditions. 
The requirement that the priors on the number of mixture components assign positive probability to all positive integers only helps us 
reach the final conclusions as immediate consequences.
For any given data set of finite size, the number of mixture components required to approximate a target density
will always be bounded above by the number of latent or observed variables generated by the target density.  
For most practical applications the required number would actually be much smaller than the number of variables generated by the target. 
Even if one applies mixture models that a-priori allow potentially infinitely many mixture components, 
the posterior will essentially concentrate on a finite set comprising moderately small positive integers.
This means that for all practical purposes, solutions based on finite mixture models with fixed but sufficiently large number of mixture components will essentially be as robust as 
solutions based on their infinite or varying dimensional counterparts 
while at the same time being significantly less burdensome from a computational viewpoint. 
The requirement that the priors on the number of mixture components assign positive mass on \emph{all} positive integers may thus be relegated to the requirement 
that the priors assign positive mass on sets of the form $\{1,\dots,K\}$, where $K$ is sufficiently large. 
Posterior computation for such models might be even much more intensive and complex requiring reversible jump moves. 
Since a mixture model with $K$ components is at least as flexible as a model with $(K-1)$ components, 
properties of overfitted mixture models discussed in Section \ref{sec: mvt model selection and model averaging} 
allow us to adopt a much simpler strategy.
We can simply keep the number of mixture components fixed at sufficiently large values for all MCMC iterations. 
Carefully chosen priors for the mixture probabilities then result in a posterior that concentrates in regions favoring empty redundant components, 
essentially eliminating the need to assign any priors on the number of mixture components.
We will still need some mechanism, preferably an automated and data adaptive one, to determine what values of $K$ would be sufficiently large.
This issue is discussed in the section on hyper-parameter choices in Section \ref{sec: mvt choice of hyper-parameters}.

The discussions of Section \ref{sec: mvt finite vs infinite mixture models} suggest that finite mixture models with 
sufficiently large number of mixture components and carefully chosen priors for the mixture probabilities
can essentially retain the major advantages of infinite dimensional alternatives including flexibility, automated model averaging and model selection
while at the same time being computationally much less burdensome, 
making them our preferred choice for complicated high dimensional problems.

\pagebreak\newpage
\section{Proofs of Theoretical Results of Section \ref{sec: mvt model flexibility}} \label{sec: mvt proofs of theoretical results}

\subsection{Proof of Lemma \ref{Lem: mvt KL support of the priors}}

Proof of part 1 of Lemma \ref{Lem: mvt KL support of the priors} follows mostly by modifications of the results of Norets and Pelenis (2012). 
We present here only the proof of part 2 that requires additional modifications along the lines of Pelenis (2014) 
to accommodate the mean zero restriction on the density of the measurement errors. 
The first step is to construct finite mixture models of the form 
\vspace{-5ex}\\
\bse
f_{m}(\bz\vert \btheta_{m}) =  \sum_{k=1}^{m+2}\pi_{m,k}~\MVN_{p}(\bz \vert \bmu_{m,k},\bSigma_{m,k}) ~~~ \text{with}~~~\sum_{k=1}^{m+2}\pi_{m,k} \bmu_{m,k} = \bzero \nonumber
\ese
\vspace{-5ex}\\
that can approximate any given density $f_{0}$ that has mean zero and satisfies Conditions \ref{cond: mvt regularity conditions on the density} with any desired level of accuracy.
The continuity of $f_{m}(\cdot\vert\btheta)$ implies that the KL distance between $f_{0}$ and $f_{m}$ remains small on sufficiently small open neighborhoods around $\btheta_{m}$. 
Both the MIW and the MLFA priors assign positive probability to open neighborhoods around $\btheta_{m}$. 
The conclusion of part 2 of Lemma \ref{Lem: mvt KL support of the priors} follows since the prior probability of having $(m+2)$ mixture components is also positive for all $m\in \nN$.

\begin{Lem} \label{Lem: mvt KL3}
For any $f_{0} \in \wt{\F}_{\bepsilon}$ and $\eta>0$, there exists $\btheta_{m}$ such that 
$d_{KL}\{f_{0}(\cdot),f_{m}(\cdot|\btheta_{m})\} < \eta$.
\end{Lem} 

\begin{proof}
Let $\{A_{m,k}\}_{k=1}^{m}$ be adjacent cubes with side length $h_m$, and $A_{m,0} = \rR^{p}-\cup_{k=1}^{m}A_{m,k}$ 
such that $h_{m}  \downarrow 0$ but $\cup_{k=1}^{m}A_{m,k} \uparrow \rR^{p}$ as $m \rightarrow \infty$.
So $\{A_{m,k}\}_{k=1}^{m}$ becomes finer but $\cup_{k=1}^{m}A_{m,k}$ covers more of $\rR^{p}$ as $m$ increases.
Additionally, let the partition be constructed in such a way that for all $m$ sufficiently large, if $\bepsilon \in A_{m,0}$, 
then $C_{r}(\bepsilon)\cap A_{m,0}$ contains a hypercube $C_{0}(\bepsilon)$ with side length $r/2$ and a vertex at $\bepsilon$;
and if $\bepsilon \notin A_{m,0}$, then $C_{r}(\bepsilon)\cap (\rR^{p}-A_{m,0})$ contains a hypercube $C_{1}(\bepsilon)$ with side length $r/2$ and a vertex at $\bepsilon$.
Consider the model 
\vspace{-5ex}\\
\bse
f_{m}(\bz) &=& f_{m}(\bz\vert \btheta_{m}) =  \sum_{k=1}^{m+2}\pi_{m,k}~\MVN_{p}(\bz \vert \bmu_{m,k},\bSigma_{m,k}).
\ese
\vspace{-5ex}\\
Set $\pi_{m,k} = \int_{A_{m,k}} f_{0}(\bz)d\bz$ for $k=1,2,\dots,m$ and $\pi_{m,k} = P_{f_0}(A_{m,0})/2 = \int_{A_{m,k}} f_{0}(\bz)d\bz/2$ for $k=(m+1),(m+2)$. 
Then $\sum_{k=1}^{m+2}\pi_{m,k} = \int_{\rR^p}f_{0}(\bz)d\bz = 1$.
Define $g(\bd)=\sum_{k=1}^{m}\pi_{m,k}(\bc_{m,k}+\bd) + \int_{A_{m,0}} \bz f_{0}(\bz) d\bz$, where $\bc_{m,k}$ is the center of $A_{m,k}$ for $k=1,2,\dots,m$. 
\vspace{-5ex}\\
\bse
g(h_{m}\bone_{p}/2) &=& \sum_{k=1}^{m}\pi_{m,k}(\bc_{m,k}+h_{m}\bone_{p}/2) + \int_{A_{m,0}}\bz f_{0}(\bz)d\bz \\
&=& \sum_{k=1}^{m} \int_{A_{m,k}} (\bc_{m,k}+h_{m}\bone_{p}/2) f_{0}(\bz)d\bz  + \int_{A_{m,0}}\bz f_{0}(\bz) d\bz \\
&\geq& \sum_{k=1}^{m} \int_{A_{m,k}} \bz f_{0}(\bz) d\bz  + \int_{A_{m,0}}\bz f_{0}(\bz) d\bz = \int_{\rR^p} \bz f_{0}(\bz)d\bz = \bzero.
\ese
\vspace{-5ex}\\
Similarly $g(-h_{m}\bone_{p}/2)\leq 0$. 
Since $g(\cdot)$ is continuous, there exists $\bd_{m}\in[-h_{m}/2,h_{m/2}]^{p}$ such that $g(\bd_{m})=\bzero$.
Set $\bmu_{m,k} = (\bc_{m,k}+\bd_{m})$ for $k=1,2,\dots,m$.
Also set $\bmu_{m,m+1} = 2 \int_{A_{m,0}} \bz f_{0}(\bz)d\bz/\int_{A_{m,0}} f_{0}(\bz)d\bz$ and $\bmu_{m,m+2} = \bzero$ 
when $\int_{A_{m,0}}f_{0}(\bz)d\bz>0$, and $\bmu_{m,0} = \bzero$ otherwise. 
Then $\sum_{k=1}^{m+2}\pi_{m,k}\bmu_{m,k} = g(\bd_{m}) = \bzero$.
Also set $\bSigma_{m,k} = \sigma_{m}^{2}\Ind_{p}$ for $k=1,2,\dots,m$ with $\sigma_{m}\to 0$, and $\Sigma_{m,m+1} = \Sigma_{m,m+2} = \sigma_{0}^{2}\Ind_p$.

Consider a sequence $\{\delta_{m}\}_{m=1}^{\infty}$ satisfying $\delta_{m}>6p^{1/2}h_{m}$ and $\delta_{m}\to 0$.
Fix $\bepsilon\in \rR^{p}$. 
Define $C_{\delta_{m}}(\bepsilon) = [\bepsilon-\delta_{m}\bone_{p}/2,\bepsilon+\delta_{m}\bone_{p}/2]$.
For $m$ sufficiently large $C_{\delta_{m}}(\bepsilon) \subseteq \cup_{k=1}^{m}A_{m,k}$, $C_{\delta_{m}}(\bepsilon) \cap A_{m,0} = \phi$ 
and the set $\{k: 1\leq k \leq m, A_{m,k}\subset C_{\delta_{m}}(\bepsilon)\}$ is non-empty.
For $k=1,\dots,m$, when $A_{m,k}\subset C_{\delta_{m}}(\bepsilon)$,
$\pi_{m,k} \geq \inf_{\bz\in C_{\delta_{m}}(\bepsilon)} f_{0}(\bz) h_{m}^{p}$.
Therefore,
\vspace{-5ex}\\
\bse
f_{m}(\bepsilon) &\geq& \sum_{\{k:1\leq k \leq m, A_{m,k}\subset C_{\delta_{m}}(\bepsilon)\}} ~ \pi_{m,k} ~ \MVN_{p}(\bepsilon \vert \bmu_{m,k},\sigma_{m}^{2}\Ind_{p}) \\
&\geq& \inf_{z\in C_{\delta_{m}}(\bepsilon)} f_{0}(\bz) \sum_{\{k:A_{m,k}\subset C_{\delta_{m}}(\bepsilon)\}} h_{m}^{p} ~ \MVN_{p}(\bepsilon\vert \bc_{m,k}+\bd_{m},\sigma_{m}^{2}\Ind_{p}) \\
&\geq& \inf_{z\in C_{\delta_{m}}(\bepsilon)} f_{0}(\bz) ~~ \left\{1-\frac{6p^{3/2}h_{m}\delta_{m}^{p-1}}{(2\pi)^{p/2}\sigma_{m}^{p}}-\frac{8p\sigma_{m}}{(2\pi)^{1/2}\delta_{m}}\right\},
\ese
where the last step follows from Lemma 1 and Lemma 2 of Norets and Pelenis (2012).
Let $h_{m},\delta_{m},\sigma_{m}$ further satisfy $h_{m}/\sigma_{m}^{p} \to 0, \sigma_{m}/\delta_{m} \to 0$.
Then for any $\eta>0$ there exists an $M_{1}$ large enough such that for all $m>M_{1}$
\vspace{-5ex}\\
\bse
f_{m}(\bepsilon) \geq \inf_{\bz\in C_{\delta_{m}}(\bepsilon)} f_{0}(\bz) \cdot (1-\eta).
\ese
\vspace{-5ex}\\
Without loss of generality, we may assume $f_{0}(\bepsilon)>0$. 
Since $f_{0}(\cdot)$ is continuous and $\delta_{m} \rightarrow 0$, there also exists an $M_{2}$ such that for all $m>M_{2}$ 
we have $\inf_{\bz\in C_{\delta_{m}}(\bepsilon)} f_{0}(\bz)>0$ and
\vspace{-5ex}\\
\bse
\frac{f_{0}(\bepsilon)}{\inf_{\bz\in C_{\delta_{m}}(\bepsilon)}f_{0}(\bz)} \leq (1+\eta).
\ese
\vspace{-5ex}\\
Therefore, for all $m>\max\{M_{1},M_{2}\}$, we have
\vspace{-5ex}\\
\bse
1 \leq \max \left\{1,\frac{f_{0}(\bepsilon)}{f_{m}(\bepsilon)} \right\} \leq \max\left\{1, \frac{f_{0}(\bepsilon)}{\inf_{z\in C_{\delta_{m}}(\bepsilon)}f_{0}(z) \cdot (1-\eta)}\right\} \leq \frac{(1+\eta)}{(1-\eta)}.
\ese
\vspace{-5ex}\\
Thus, $\log\max\{1,f_{0}(\bepsilon)/f_{m}(\bepsilon)\} \rightarrow 0$ as $m \rightarrow \infty$. 
Pointwise convergence is thus established.
Next, we will find an integrable upper bound for $\log\max\{1,f_{0}(\bepsilon)/f_{m}(\bepsilon)\}$.

For point wise convergence we can assume $\bepsilon\notin A_{m,0}$ for sufficiently large $m$.
But to find integrable upper bound, we have to consider both the cases $\bepsilon \in A_{m,0}$ and $\bepsilon \notin A_{m,0}$.
When $\bepsilon\in A_{m,0}$, we have $\hbox{P}_{f_0}(A_{m,0}) = \int_{A_{m,0}}f_0(\bz)d\bz 
\geq \int_{A_{m,0}\cap C_{r}(\bepsilon)}f_0(\bz)d\bz \geq \lambda\{A_{m,0}\cap C_{r}(\bepsilon)\} \inf_{\bz\in A_{m,0}\cap C_{r}(\bepsilon)}f_{0}(\bz)
\geq (r/2)^{p} \inf_{\bz\in C_{r}(\bepsilon)}f_{0}(\bz)$, since $\lambda\{A_{m,0}\cap C_{r}(\bepsilon)\} \geq \lambda\{C_{0}(\bepsilon)\} \geq (r/2)^{p}$.
Using part 4 of Conditions \ref{cond: mvt regularity conditions on the density} and Lemma 1 and Lemma 2 of Norets and Pelenis (2012) again, 
if $\bepsilon \notin A_{m,0}$, for $m$ sufficiently large 
\vspace{-5ex}\\
\bse
&&\hspace{-1cm} \sum_{\{k:A_{m,k}\subset C_{r}(\bepsilon)\}} h_{m}^{p}~\MVN_{p}(\bepsilon \vert \bmu_{m,k},\sigma_{m}^{2}\Ind_{p})
\geq \sum_{\{k:A_{m,k}\subset C_{1}(\bepsilon)\}} h_{m}^{p}~\MVN_{p}(\bepsilon \vert \bmu_{m,k},\sigma_{m}^{2}\Ind_{p})  \\
&& \geq \int_{C_{1}(\bepsilon)}~\MVN_{p}(\bz\vert\bepsilon,\sigma_{m}^{2}\Ind_{p})d\bz - \frac{3p^{3/2}(r/2)^{p-1}h_{m}}{(2\pi)^{p/2}\sigma_{m}^{p}}  \\
&& \geq \left\{\frac{1}{2^p}  -  \frac{8p\sigma_{m}}{2^{p}(2\pi)^{1/2}r}  -  \frac{3p^{3/2}h_{m}r^{p-1}}{2^{p-1}(2\pi)^{p/2}\sigma_{m}^{p}}\right\} \geq \frac{1}{2^{p+1}},
\ese
\vspace{-5ex}\\
This implies
\vspace{-5ex}\\
\bse
f_{m}(\bepsilon) &=& \sum_{k=1}^{m}P_{f_0}(A_{m,k})~\MVN_{p}(\bepsilon \vert \bmu_{m,k},\sigma_{m}^{2}\Ind_{p}) + \sum_{k=m+1}^{m+2}(1/2) P_{f_0}(A_{m,0})~\MVN_{p}(\bepsilon \vert \bmu_{m,k},\sigma_{0}^{2}\Ind_{p})  \\
&\geq& \sum_{k=1}^{m}P_{f_0}(A_{m,k})~\MVN_{p}(\bepsilon \vert \bmu_{m,k},\sigma_{m}^{2}\Ind_{p}) + (1/2) P_{f_0}(A_{m,0})~\MVN_{p}(\bepsilon \vert \bzero,\sigma_{0}^{2}\Ind_{p})  \\
&\geq& \{1-1(\bepsilon\in A_{m,0})\} ~ \inf_{\bz \in C_{r}(\bepsilon)} f_{0}(\bz)\sum_{\{k:A_{m,k}\subset C_{r}(\bepsilon)\}} \lambda(A_{m,k})~\MVN_{p}(\bepsilon \vert \bmu_{m,k},\sigma_{m}^{2}\Ind_{p}) \\
&& + ~ 1(\bepsilon\in A_{m,0}) (1/2) P_{f_0}(A_{m,0})~\MVN_{p}(\bepsilon \vert \bzero,\sigma_{0}^{2}\Ind_{p})    \\
&\geq& (1/2)\{1-1(\bepsilon\in A_{m,0})\} ~ \inf_{\bz \in C_{r}(\bepsilon)} f_{0}(\bz)  \\
&& + ~ 1(\bepsilon\in A_{m,0}) ~ (1/2) (r/2)^{p}~\MVN_{p}(\bepsilon \vert \bzero,\sigma_{0}^{2}\Ind_{p})~\inf_{\bz \in C_{r}(\bepsilon)} f_{0}(\bz)\\
&\geq& (1/2) (r/2)^{p}~\MVN_{p}(\bepsilon \vert \bzero,\sigma_{0}^{2}\Ind_{p})~\inf_{\bz \in C_{r}(\bepsilon)} f_{0}(\bz).
\ese
\vspace{-5ex}\\
The last step followed by choosing $\sigma_{0}^{2}$ large enough so that 
$(r/2)^{p}\sup_{\bepsilon\in \rR^p}\MVN_{p}(\bepsilon \vert \bzero,\sigma_{0}^{2}\Ind_{p}) < (r/2)^{p}~\sigma_{0}^{-p}<2^{-(p+1)}<1$.
Therefore,
\vspace{-5ex}\\
\bse
&&\hspace{-0.7cm} \log\max\left\{1,\frac{f_{0}(\bepsilon)}{f_{m}(\bepsilon)}\right\}  \leq \log\max\left\{1,\frac{f_{0}(\bepsilon)}{(1/2) (r/2)^{p}~\MVN_{p}(\bepsilon \vert \bzero,\sigma_{0}^{2}\Ind_{p})~\inf_{\bz \in C_{r}(\bepsilon)} f_{0}(\bz)}\right\}     \\
&&\hspace{-0.4cm}  \leq \log \left[\frac{1}{ (1/2) (r/2)^{p}~\MVN_{p}(\bepsilon \vert \bzero,\sigma_{0}^{2}\Ind_{p})}   \max\left\{ (1/2)  (r/2)^{p}~\MVN_{p}(\bepsilon \vert \bzero,\sigma_{0}^{2}\Ind_{p}), \frac{f_{0}(\bepsilon)}{\inf_{\bz \in C_{r}(\bepsilon)} f_{0}(\bz)}\right\}\right]  \\
&&\hspace{-0.4cm}  \leq - \log\left\{ (1/2) (r/2)^{p}~\MVN_{p}(\bepsilon \vert \bzero,\sigma_{0}^{2}\Ind_{p})\right\}  + \log \left\{\frac{f_{0}(\bepsilon)}{\inf_{\bz \in C_{r}(\bepsilon)} f_{0}(\bz)}\right\}.
\ese
\vspace{-5ex}\\
The first and the second terms are integrable by part 2 and part 3 of Conditions \ref{cond: mvt regularity conditions on the density}, respectively. 
Since $\int f_{0}(\bepsilon)\log\{f_{0\bepsilon}/f_{m}(\bepsilon)\} d\bepsilon \leq \int f_{0}(\bepsilon)\log \max \{1, f_{0\bepsilon}/f_{m}(\bepsilon)\} d\bepsilon$, 
the proof of Lemma \ref{Lem: mvt KL3} is completed applying dominated convergence theorem (DCT).
\end{proof}

Let $\eta>0$ be given. 
According to Lemma \ref{Lem: mvt KL3}, 
there exists $\btheta_{m}^{\star} = (\bpi_{1:(m+2)}^{\star},\bmu_{1:(m+2)}^{\star},\bSigma_{1:(m+2)}^{\star})$ 
with $\bSigma_{k}^{\star}=\sigma_{m}^{2\star}\Ind_{p}$ for $k=1,\dots,m$ and $\bSigma_{k}^{\star}=\sigma_{0}^{2\star}\Ind_{p}$ for $k=(m+1),(m+2)$ 
such that $d_{KL}\{f_{0}(\cdot),f_{m}(\cdot\vert\btheta_{m}^{\star})\}<\eta/2$.
We have, for any $\btheta_{m}$,
\vspace{-5ex}\\
\bse
&&\hspace{-1cm}\int f_{0}(\bepsilon) ~\log\left\{\frac{f_{0}(\bepsilon)}{f_{m}(\bepsilon\vert\btheta_{m})}\right\}d\bepsilon  
=  \int f_{0}(\bepsilon)~\log\left\{\frac{f_{0}(\bepsilon)}{f_{m}(\bepsilon\vert\btheta_{m}^{\star})}\right\}d\bepsilon  
+ \int f_{0}(\bepsilon)~\log\left\{\frac{f_{m}(\bepsilon\vert\btheta_{m}^{\star})}{f_{m}(\bepsilon\vert\btheta_{m})}\right\}d\bepsilon.
\ese
\vspace{-3ex}

Let the second term in the above expression be denoted by $g(\btheta_{m})$. 
The priors puts positive mass on arbitrarily small open neighborhoods around $\btheta_{m}^{\star}$.
The result will follow if there exists an open neighborhood $\N(\btheta_{m}^{\star})$ around $\btheta_{m}^{\star}$ such that 
$\sup_{\btheta_{m}\in \N(\btheta_{m}^{\star})}  g(\btheta_{m}) < \eta/2$. 
Since $g(\btheta_{m}^{\star}) = 0$, it suffices to show that the function $g(\btheta_{m})$ is continuous at $\btheta_{m}^{\star}$.
Now $g(\btheta)$ is continuous at $\btheta_{m}^{\star}$ if for every sequence $\{\btheta_{m,n}\}_{n=1}^{\infty}$ with $\btheta_{m,n}\to\btheta_{m}^{\star}$, 
we have $g(\btheta_{m,n}) \to g(\btheta_{m}^{\star})$.
For all $\bepsilon\in \rR^{p}$, we have $\log\{f_{m}(\bepsilon\vert \btheta_{m,n}^{\star})/f_{m}(\bepsilon\vert \btheta_{m})\} \to 0$ as $\btheta_{m,n} \to \btheta_{m}^{\star}$.
Continuity of $g(\btheta_{m})$ at $\btheta_{m}^{\star}$ will follow from DCT if we can show that $\abs{f_{m}(\bepsilon\vert\btheta_{m}^{\star})/f_{m}(\bepsilon\vert\btheta_{m,n})}$ has an integrable with respect to $f_{0}$ upper bound.

Since $\btheta_{m,n}\to \btheta_{m}^{\star}$, for any arbitrarily small open neighborhood $\N(\btheta_{m}^{\star})$ around $\btheta_{m}^{\star}$, 
we must have $\btheta_{m,n}\in \N(\btheta_{m}^{\star})$ for all $n$ sufficiently large. 
Let $\btheta_{m} = (\bpi_{1:(m+2)},\bmu_{1:(m+2)},\bSigma_{1:(m+2)}) \in \N(\btheta_{m}^{\star})$. 
Since the eigenvalues of a real symmetric matrix depend continuously on the matrix, 
we must have  
$(\lambda_{1}(\bSigma_{k}),\lambda_{p}(\bSigma_{k}))\subset (\underline\sigma_{m}^{2\star},\overline\sigma_{m}^{2\star})$ for $k=1,\dots,m$ and 
$(\lambda_{1}(\bSigma_{k}),\lambda_{p}(\bSigma_{k}))\subset (\underline\sigma_{0}^{2\star},\overline\sigma_{0}^{2\star})$ for $k=(m+1),(m+2)$,
where $\underline\sigma_{m}^{2\star} < \sigma_{m}^{2\star}< \overline\sigma_{m}^{2\star}$ and $\underline\sigma_{0}^{2\star} < \sigma_{0}^{2\star}< \overline\sigma_{0}^{2\star}$.
Let $\underline\sigma^{2\star} = \min\{\underline\sigma_{m}^{2\star},\underline\sigma_{0}^{2\star}\}$ and $\overline\sigma^{2\star} = \max\{\overline\sigma_{m}^{2\star},\overline\sigma_{0}^{2\star}\}$.
Then $(\lambda_{1}(\bSigma_{k}),\lambda_{p}(\bSigma_{k}))\subset (\underline\sigma^{2\star},\overline\sigma^{2\star})$ for $k=1,\dots,(m+2)$. 
Similarly, for some finite $\mu^{\star}$, we must have $\bmu_{m,k} \in (-\mu^{\star}\bone_{p},\mu^{\star}\bone_{p}) = \N_{\mu^{\star}}$ for $k=1,\dots,(m+2)$. 
For any real positive definite matrix $\bSigma$, we have $\bz\trans \bSigma^{-1} \bz \leq \lambda_{1}^{-1}(\bSigma) \norm{\bz}^{2}$. 
Therefore, for any $\bepsilon \in \rR^{p}$ and for all $k=1,\dots,(m+2)$, 
we must have $(\bepsilon - \bmu_{m,k})\trans \bSigma_{m,k}^{-1} (\bepsilon-\bmu_{m,k}) \leq \underline\sigma^{-2\star}\{1(\bepsilon\in \N_{\mu^{\star}})  2^{p}\mu^{\star p}+  1(\bepsilon\notin \N_{\mu^{\star}})\norm{\bepsilon + \sign(\bepsilon)\mu^{\star}}^{2}\}$, where $\sign(\bepsilon) = \{\sign(\epsilon_{1}),\dots,\sign(\epsilon_{p})\}\trans$. 
Therefore, for any $\btheta_{m}\in \N(\btheta_{m}^{\star})$, we have
\vspace{-5ex}\\
\bse
[1(\bepsilon\in \N_{\mu^{\star}}) \MVN_{p}(2\mu^{\star}\bone_{p}\vert\bzero,\underline\sigma^{2\star}\Ind_{p}) + 1(\bepsilon\notin \N_{\mu^{\star}}) \MVN_{p}\{\bepsilon + \sign(\bepsilon)\mu^{\star}\vert\bzero,\underline\sigma^{2\star}\Ind_{p}\}]/\overline\sigma^{\star} \\
\leq  f_{m}(\bepsilon\vert\btheta_{m})   \leq   1/\underline\sigma^{\star}.
\ese
\vspace{-5ex}\\
The upper bound is a constant and the logarithm of the lower bound is integrable 
since, by part 2 of Conditions \ref{cond: mvt regularity conditions on the density}, the second order moments of $\bepsilon$ exist.
An $f_{0}$ integrable upper bound for the function $\sup_{\btheta_{m}\in \N(\btheta_{m}^{\star})}\abs{f_{m}(\bepsilon\vert\btheta_{m})}$ thus exists.
Finally, DCT applies because 
\vspace{-5ex}\\
\bse
 \int f_{0}(\bepsilon)~\abs{\log\left\{\frac{f_{m}(\bepsilon\vert\btheta_{m}^{\star})}{f_{m}(\bepsilon\vert\btheta_{m,n})}\right\}}d\bepsilon 
&\leq& 
\sup_{\btheta_{m}\in \N(\btheta_{m}^{\star})} \int f_{0}(\bepsilon)~\abs{\log\left\{\frac{f_{m}(\bepsilon\vert\btheta_{m}^{\star})}{f_{m}(\bepsilon\vert\btheta_{m})}\right\}}d\bepsilon     \\
&\leq&  2 \sup_{\btheta_{m}\in \N(\btheta_{m}^{\star})} \int f_{0}(\bepsilon)~\abs{f_{m}(\bepsilon\vert\btheta_{m})}d\bepsilon.
\ese
\vspace{-5ex}\\ 
The conclusion of part 2 of Lemma \ref{Lem: mvt KL support of the priors} follows since 
the prior probability of having $(m+2)$ mixture components is positive for all $m\in \nN$.

\subsection{Proof of Lemma \ref{Lem: mvt sup norm support of priors on variance functions}}	\label{sec: mvt proof of sup norm support of priors on variance functions}
Given $q$, let $\Pi_{q}$ denote a prior on $\nN_{q} = \{q+1,q+2,\dots\}$ such that $\Pi_{q}(J) >0 ~ \forall J\in \nN_{q}$.
Let $||\cdot||_{2}$ denote the Euclidean norm.
Let $\rR^{+} = (0,\infty)$. Given $J\sim \Pi_{q}$, also let $\Pi_{\beta\vert J}$ be a prior on $\rR^{+J}$
such that $\Pi_{\beta\vert J}\{N_{\delta}(\bbeta_{0})\} >0 $ for any $\delta>0$ and any $\bbeta_{0}\in \rR^{J}$,
where $N_{\delta}(\bbeta_{0}) = \{\bbeta: \bbeta\in \rR^{+J}, ||\bbeta-\bbeta_{0}||_{2} < \delta\}$.
Define $\S_{q,J} = \{v_{s}: v_{s} = \bB_{q,J} \bbeta = \sum_{j=1}^{J}b_{q,j}\beta_{j} ~ \hbox{for some}~ \bbeta \in\rR^{+J}\}$.
Then $\Pi_{\bV} = \Pi_{q}\times \Pi_{\beta\vert J}$ is the induced prior on $\S_{q} = \cup_{J={q+1}}^{\infty}\S_{q,J}$.

Define $\psi(v_{0},h) = \sup_{X,X'\in[A,B], |X-X'|\leq h}|v_{0}(X)-v_{0}(X')|$.
Let $\lfloor\alpha\rfloor = \min\{n: n\in\nN, n\geq\alpha\}$.
For any $X$,
$(i)~ b_{q,j}(X) \geq 0 ~ \forall j$,
$(ii)~ \sum_{j=1}^{J}b_{q,j}(X) = 1$,
$(iii)~ b_{q,j}$ is positive only inside the interval $[t_{j},t_{j+q+1}]$,
and $(iv)$ for $j\in\{(q+1), (q+2), \dots, (q+K)\}$, for any $X\in (t_j,t_{j+1})$, only $(q+1)$ B-splines $b_{q,j-q}(X),b_{q,j-q+1}(X),\dots,b_{q,j}(X)$ are positive.
Using these local support properties of B-splines, the results on page 147 of de Boor (2000) can be modified to show that, for any $v_{0}\in\C_{+}[A,B]$,
\vspace{-5ex}\\
\bse
\inf_{v_{s}\in \S_{q,J}} ||v_{0}-v_{s}||_{\infty} \leq  \lfloor (q+1)/2\rfloor  ~ \psi(v_{0},\Delta_{\max}) \rightarrow 0 ~~\hbox{as}~\Delta_{\max} \rightarrow 0.
\ese
\vspace{-5ex}\\
Also, if $q\geq(\alpha-1)$, we can modify the results on page 149 of de Boor (2000) to show that,  for any $v\in \C_{+}^{\alpha}[A,B]$,
\vspace{-5ex}\\
\bse
\inf_{v_{s}\in \S_{q,J}} ||v_{0}-v_{s}||_{\infty} &\leq& c(q) c(q-1) \dots c(q-\alpha_0+1) ~ ||v_{0}^{(\alpha_0)}||_{\infty} ~ \Delta_{\max}^{\alpha_0},
\ese
\vspace{-5ex}\\
where $c(q) = \lfloor (q+1)/2\rfloor$.
For any two functions $g_1$ and $g_2$, $\sup| g_1g_2| \leq \sup |g_1|\sup|g_2|$.
Taking $g_1(X,X')= \{v_{0}^{(\alpha_0)}(X)-v_{0}^{(\alpha_0)}(X')\}/(X-X')^{(\alpha-\alpha_0)}$ and $g_2(X,X') = (X-X')^{(\alpha-\alpha_0)}$, we have
$||v_{0}^{(\alpha_0)}||_{\infty} \leq ||v_{0}||_{\alpha} (B-A)^{(\alpha-\alpha_0)}$.
Therefore,
\vspace{-5ex}\\
\bse
\inf_{v_{s}\in \S_{q,J}} ||v_{0}-v_{s}||_{\infty} &\leq& c(q,\alpha_{0}) ~  (B-A)^{(\alpha-\alpha_0)} ~ ||v_{0}||_{\alpha} ~ \Delta_{\max}^{\alpha_0}.
\ese
\vspace{-5ex}\\
Furthermore, when the knot points $\{t_{q+1+j}\}_{j=0}^{K}$ are equidistant
\vspace{-5ex}\\
\bse
\inf_{v_{s}\in \S_{q,J}} ||v_{0}-v_{s}||_{\infty} \leq c(q,\alpha_{0})  ||v_{0}^{(\alpha)}||_{\infty} \frac{(B-A)^{\alpha}}{K^{\alpha_0}}\leq  c(q,\alpha)||v_{0}||_{\alpha}K^{-\alpha}.
\ese
\vspace{-5ex}

Given any $v_{0}\in C_{+}[A,B] (\hbox{or}~C_{+}^{\alpha}[A,B])$ and $\delta>0$, find $J\in \nN_{q}$ and $\bbeta_{0}\in \rR^{+J}$ such that
$||v_{0}-\bB_{q,J}\bbeta_{0}||_{\infty} = \inf_{v_{s}\in \S_{q,J}} ||v_{0}-v_{s}||_{\infty}<\delta/2$.
Next consider a neighborhood $N_{\eta}(\bbeta_{0})$ such that for any $\bbeta\in N_{\eta}(\bbeta_{0})$, we have
$||\bB_{q,J}\bbeta-\bB_{q,J}\bbeta_{0}||_{\infty} < \delta/2$.
Then for any $\bbeta\in N_{\eta}(\bbeta_{0})$, we have
$||\bB_{q,J}\bbeta-v_{0}||_{\infty} \leq ||\bB_{q,J}\bbeta-\bB_{q,J}\bbeta_{0}||_{\infty} + ||\bB_{q,J}\bbeta_{0}-v_{0}||_{\infty} <\delta$.
Also $\Pi_{\bV}(||v-v_{0}||_{\infty} <\delta) \geq \Pi_{q}(J) ~ \Pi_{\beta\vert J}\{N_{\eta}(\bbeta_{0})\} > 0$.
Proof of Lemma \ref{Lem: mvt sup norm support of priors on variance functions} then follows as a special case
taking $\bbeta = \exp(\bxi)$ and taking $\Pi_{q}$ and $\Pi_{\beta\vert J}$ to be the priors on $J$ and $\bbeta$ induced by $P_{0}(K)$ and $P_{0}(\bxi|K,\sigma_{\xi}^{2})$, respectively.

\subsection{Proof of Lemma \ref{Lem: mvt KL support of the prior on the density of U|X}}
We first prove some additional lemmas to used in the proof of Lemma \ref{Lem: mvt KL support of the prior on the density of U|X}. 
\begin{Lem}
$\Pi_{\bV}(||v-v_{0}||_{\infty} <\delta)>0 ~ \forall \delta>0$ implies that
$\Pi_{\bV}(||g\circ {v}-g\circ {v}_{0}||_{\infty} <\delta)>0 ~ \forall \delta>0$ for every continuous function $g: \rR \rightarrow \rR$.
 \end{Lem}
\begin{proof}
Let ${v}:[A,B] \rightarrow [C_1,D_1]$ and ${v}_{0}:[A,B]\rightarrow [C_2,D_2]$.
Then $({v}-{v}_{0}):[A,B] \rightarrow [C_1-D_2,D_1-C_2]=[C,D]$, say.
Then $g:[C,D]\rightarrow \rR$ is a uniformly continuous function.
Therefore, given any $\delta>0$, there exists a $\eta>0$ such that $|g(Z_1)-g(Z_2)|<\delta$ whenever $|Z_1-Z_2|<\eta$.
Now let $||{v}-{v}_{0}||_{\infty}  = \sup_{X\in[A,B]} |{v}(X)-{v}_{0}(X)| <\eta$.
This implies, for all $X\in[A,B]$, $|{v}(X)-{v}_{0}(X)| <\eta$.
Therefore, for all $X\in[A,B]$, $|g\{{v}(X)\}-g\{{v}_{0}(X)\}|<\delta$, and hence $||g\circ{v}-g\circ{v}_{0}||_{\infty}\leq\delta$.
Hence the proof.
\end{proof}

\begin{Cor}  \label{Cor: mvt sup norm support of priors on sqrt of variance functions}
In particular, taking $g(Z)=Z^{1/2} ~ \forall Z>0$ and $g(Z) = 0$ otherwise, we have $\Pi_{\bV}( ||v^{1/2}-v_{0}^{1/2}||_{\infty} <\delta) = \Pi_{\bV}( ||s-s_{0}||_{\infty} <\delta)>0 ~ \forall \delta>0$ for all ${v}_{0}\in \C_{+}[A,B] (\hbox{or}~ \C_{+}^{\alpha}[A,B])$.
\end{Cor}

Let $P_{\bepsilon,K} \{(\bmu,\bSigma)\vert \bpi_{1:K},\bmu_{1:K},\bSigma_{1:K}\} = \sum_{k=1}^{K}\pi_{k}\delta_{(\bmu_{k},\bSigma_{k})}(\bmu,\bSigma)$, where $\delta_{\btheta}$ denotes a point mass at $\btheta$. 
We have, with the the hyper-parameters implicit, $P_{0}(\bpi_{1:K},\bmu_{1:K},\bSigma_{1:K}) = P_{0\pi}(\bpi_{1:K}) P_{0\mu}(\bmu_{1:K}\vert \bpi_{1:K}) P_{0\Sigma}(\bSigma_{1:K})$.
Denoting $P_{\bepsilon,K} \{(\bmu,\bSigma)\vert \bpi_{1:K},\bmu_{1:K},\bSigma_{1:K}\}$ simply by $P_{\bepsilon,K} (\bmu,\bSigma)$. 
Let $c$ be a generic for constants that are not of direct interest.
For any square matrix $\bA$ of order $p$, let $\lambda_{1}(\bA)\leq\dots\leq\lambda_{p}(\bA)$ denote the ordered eigenvalues of $\bA$.
The following lemma proves some properties of $P_{\bepsilon,K}$ and $f_{\bepsilon}$.

\begin{Lem} \label{Lem: mvt moments of fe}
1. $\int \norm{\bmu}_{2}^{2}dP_{\bepsilon,K}(\bmu,\bSigma) < \infty$ a.s.
~~~2. $\int \lambda_{1}^{-1}(\bSigma) dP_{\bepsilon,K}(\bmu,\bSigma) < \infty$ a.s. \\
3. $\int \abs{\bSigma}^{-1/2}dP_{\bepsilon,K}(\bmu,\bSigma) < \infty$ a.s. 
\end{Lem}
\begin{proof}
1. 
The prior $P_{0\mu}(\bmu_{1:K}\vert \bpi_{1:K})$ is of the form (\ref{eq: conditional mvt normal posterior of the mean vector}), that is, 
$P_{0\mu}(\bmu_{1:K}\vert  \bpi_{1:K})  = \MVN_{Kp}(\bzero,\bSigma^{0}-\bSigma_{1,R}^{0}\bSigma_{R,R}^{-1}\bSigma_{R,1}^{0})$, 
where $\bSigma^{0}$ is a $Kp\times Kp$ block-diagonal matrix independent of $\bpi_{1:K}$, all $k$ principal blocks of order $p\times p$ being $\bSigma_{0}$.
The matrix $\bSigma_{1,R}^{0}\bSigma_{R,R}^{-1}\bSigma_{R,1}^{0}$ depends on $\bpi_{1:K}$ and is nonnegative definite so that its diagonal elements are all nonnegative.
Let $\bSigma_{0}=((\sigma_{0,ij}))$ and $\bSigma_{1,R}^{0}\bSigma_{R,R}^{-1}\bSigma_{R,1}^{0} = ((\sigma_{R,ij}))$.
Then,  
$\int \norm{\bmu_{k}}_{2}^{2} dP_{0\mu}(\bmu_{1:K}\vert \bpi_{1:K}) = \left\{ \sum_{j=1}^{p}\sigma_{0,jj} - \sum_{j=(k-1)p+1}^{kp}\sigma_{R,jj} \right\} \leq \sum_{j=1}^{p}\sigma_{0,jj} = \trace(\bSigma_{0})$.
Therefore,
\vspace{-5ex}\\
\bse
&&\hspace{-1cm}
\int \int \norm{\bmu}_{2}^{2}dP_{\bepsilon,K}(\bmu,\bSigma) dP_{0}(\bpi_{1:K},\bmu_{1:K},\bSigma_{1:K})   
=  \sum_{k=1}^{K} \int \pi_{k} \norm{\bmu_{k}}_{2}^{2}   dP_{0\mu}(\bmu_{1:K}\vert \bpi_{1:K})   dP_{0\pi}(\bpi_{1:K})   \\
&& \leq \trace(\bSigma_{0}) <\infty.
\ese
\vspace{-5ex}

\noindent 2. 
We have $\int \int \lambda_{1}^{-1}(\bSigma) dP_{\bepsilon,K}(\bmu,\bSigma) dP_{0}(\bpi_{1:K},\bmu_{1:K},\bSigma_{1:K}) =  \int \lambda_{1}^{-1}(\bSigma) dP_{0\Sigma}(\bSigma)$.

When $\bSigma \sim \IW_{p}(\nu_{0},\bPsi_{0})$, we have $\bPsi_{0}^{-1/2}\bSigma^{-1}\bPsi_{0}^{-1/2} \sim \Wish_{p}(\nu_{0},\Ind)$ and $\trace(\bPsi_{0}^{-1}\bSigma^{-1}) = \trace(\bPsi_{0}^{-1/2}\bSigma^{-1}\bPsi_{0}^{-1/2}) \sim \chi^{2}_{p\nu_{0}}$.
Here $\Wish_{p}(\nu,\bPsi)$ denotes a Wishart distribution with degrees of freedom $\nu$ and mean $\nu\bPsi$. 
For any two positive semidefinite matrices $\bA$ and $\bB$, we have 
$\lambda_{1}(\bA)\trace(\bB) \leq \trace(\bA\bB) \leq \lambda_{p}(\bA) \trace(\bB)$.
Therefore,
$\lambda_{1}(\bPsi_{0}^{-1}) E \{\trace(\bSigma^{-1})\}  \leq E\{\trace(\bPsi_{0}^{-1}\bSigma^{-1})\} = p\nu_{0}$.
Hence, $\int \lambda_{1}^{-1}(\bSigma) dP_{0\Sigma}(\bSigma) = E \lambda_{p}(\bSigma^{-1}) \leq E\{\trace(\bSigma^{-1})\} < \infty$.

When $\bSigma = \bOmega + \bLambda \bLambda\trans$ with $\bOmega = \diag(\sigma_{1}^{2},\dots,\sigma_{p}^{2})$, we have 
 $\trace(\bSigma^{-1}) = \trace\{\bOmega^{-1}-\bOmega^{-1}\bGamma(\Ind_{p}+\bGamma\trans\bOmega^{-1}\bGamma)^{-1}\bGamma\trans\bOmega^{-1}\}
\leq \trace(\bOmega^{-1}) = \sum_{j=1}^{p}\sigma_{j}^{-2}$,
where $\bGamma$ is a $p\times p$ matrix satisfying $\bGamma\bGamma\trans = \bLambda\bLambda\trans$.
Thus, $\int \lambda_{1}^{-1}(\bSigma) dP_{0\Sigma}(\bSigma_{1:K}) = E \lambda_{p}(\bSigma^{-1}) \leq E\{\trace(\bSigma^{-1})\} \leq \sum_{j=1}^{p}E\sigma_{j}^{-2} < \infty$  
whenever $\sigma_{j}^{2}\sim\IG(a,b)$ with $a>1$.

\vspace{2ex}
\noindent 3. 
When $\bSigma \sim \IW_{p}(\nu_{0},\bPsi_{0})$, we have $\lambda_{1}^{p/2}(\bPsi_{0}^{-1}) E \{\trace(\bSigma^{-1})\}^{p/2}  \leq E\{\trace(\bPsi_{0}^{-1}\bSigma^{-1})\}^{p/2} < \infty$.
Hence, $\int \abs{\bSigma}^{-1/2} dP_{0\Sigma}(\bSigma) = \int \prod_{j=1}^{p}\lambda_{j}^{1/2}(\bSigma^{-1})dP_{0\Sigma}(\bSigma) 
\leq \int \lambda_{p}^{p/2}(\bSigma^{-1}) dP_{0\Sigma}(\bSigma) = E \lambda_{p}^{p/2}(\bSigma^{-1}) \leq E\{\trace(\bSigma^{-1})\}^{p/2} < \infty$.

For any two positive semidefinite matrix $\bA$ and $\bB$, we have $\abs{\bA+\bB}\geq \abs{\bA}$.
Therefore, when $\bSigma = \bOmega + \bLambda \bLambda\trans$, we have 
$\int \abs{\bSigma}^{-1/2} dP_{0\Sigma}(\bSigma_{1:K}) \leq \int \abs{\bOmega}^{-1/2}dP_{0\Sigma}(\bSigma_{1:K}) 
= \int \prod_{j=1}^{p}\sigma_{j}^{-1}dP_{0\Sigma}(\bSigma_{1:K}) = \prod_{j=1}^{p}E\sigma_{j}^{-1} < \infty$, whenever $\sigma_{j}^{2}\sim\IG(a,b)$ independently.
\end{proof}

\vspace{4ex}
The following lemma proves a property of $f_{\bepsilon}= \int \int f_{c\bepsilon}(\bepsilon\vert \bmu,\bSigma)dP_{\bepsilon,K}(\bmu,\bSigma)dP_{0}(K)$.
Here $P_{0}(K)$ denotes the prior on $K$, the number of mixture components.

\begin{Lem} \label{Lem: mvt limit of KL divergence of involving fe}
 Let $f_{0\bepsilon}\in\wt\F_{\bepsilon}$ and $f_{\bepsilon} \sim \Pi_{\bepsilon}$ and $\bD(\btau)=\diag(\tau_{1},\tau_{2},\dots,\tau_{p})$.
 Then
\vspace{-5ex}\\
\bse
\lim_{\btau\rightarrow \bone} \int  f_{0\bepsilon}(\bepsilon) ~ \log \left[ \frac{f_{\bepsilon}(\bepsilon)}{\abs{\bD(\btau)}^{-1}f_{\bepsilon}\{\bD(\btau) \bepsilon\}} \right]~ d\bepsilon = 0.
\ese
 \end{Lem}
\begin{proof}
We have $\abs{\bD(\btau)}^{-1} f_{c\bepsilon}\{\bD(\btau)\bepsilon\}  \to f_{c\bepsilon}(\bepsilon)$ as $\btau\to\bone$.
Since $\btau\rightarrow \bone$, without loss of generality, we may assume $\abs{\bD(\btau)}>1/2$.
Define $c = \int \abs{\bSigma}^{-1/2}dP_{\bepsilon,K}(\bmu,\bSigma)$. Then $c < \infty$.
Also $\int \abs{\bD(\btau)}^{-1} f_{c\bepsilon}\{\bD(\btau)\bepsilon\vert \btheta\} dP_{\bepsilon,K}(\bmu,\bSigma) \leq \int 2(2\pi)^{-p/2}\abs{\bSigma}^{-1/2} dP_{\bepsilon,K}(\bmu,\bSigma) < 2c < \infty$.
Applying DCT, $\abs{\bD(\btau)}^{-1} f_{\bepsilon}\{\bD(\btau) \bepsilon\} \rightarrow f_{\bepsilon}(\bepsilon)$ as $\btau \rightarrow \bone$.
Therefore, for any $\bepsilon\in \rR$,
\vspace{-5ex}\\
\bse
\log \left[\frac{f_{\bepsilon}(\bepsilon)}{\abs{\bD(\btau)}^{-1}f_{\bepsilon}\{\bD(\btau) \bepsilon\}}\right] \rightarrow 0 ~~~\hbox{as}~ \btau \rightarrow \bone.
\ese
\vspace{-5ex}\\
To find an integrable with respect to $f_{0\bepsilon}$ upper bound for 
$\log \left[\abs{\bD(\btau)}f_{\bepsilon}(\bepsilon)/f_{\bepsilon}\{\bD(\btau) \bepsilon\}\right]$, we use Lemma \ref{Lem: mvt moments of fe}.
To do so, we can ignore the prior $P_{0}(K)$ 
since the upper bounds obtained in Lemma \ref{Lem: mvt moments of fe} do not depend on the specific choice of $K$. 
We have, using part 3 of Lemma \ref{Lem: mvt moments of fe},
\vspace{-5ex}\\
\bse
 \int \abs{\bSigma}^{-1/2}\exp\left[-\frac{1}{2}\{\bD(\btau) \bepsilon-\bmu\}\trans\bSigma^{-1}\{\bD(\btau)\bepsilon-\bmu\}\right]  dP_{\bepsilon,K}(\bmu,\bSigma)  \\
 \leq  \int \abs{\bSigma}^{-1/2} dP_{\bepsilon,K}(\bmu,\bSigma) \leq c. 
\ese
\vspace{-5ex}\\
Since $\btau \rightarrow \bone$, without loss of generality we may also assume $\tau_{k}<2$ for all $k$.
Therefore, 
\vspace{-5ex}\\
\bse
&& \hspace{-0.5cm} |\log ~ f_{\bepsilon}\{\bD(\btau) \bepsilon\}|     \\
	&& \leq   \log (2\pi)^{p/2} + \left| \log \int \abs{\bSigma}^{-1/2}\exp\left[-\frac{1}{2}\{\bD(\btau) \bepsilon-\bmu\}\trans\bSigma^{-1}\{\bD(\btau)\bepsilon-\bmu\}\right]  dP_{\bepsilon,K}(\bmu,\bSigma) \right|  \\
&&\leq \log (2\pi)^{p/2} + |\log~c|  \\
	&&~~~~~~~~~~~  -   \log \int c^{-1}\abs{\bSigma}^{-1/2}\exp\left[-\frac{1}{2}\{\bD(\btau) \bepsilon-\bmu\}\trans\bSigma^{-1}\{\bD(\btau)\bepsilon-\bmu\}\right]  dP_{\bepsilon,K}(\bmu,\bSigma)  \\
&&\leq \log \{c(2\pi)^{p/2}\} + |\log~c|   \\
	&&~~~~~~~~~~~ + \frac{1}{2}  \int \log \abs{\bSigma} dP_{\bepsilon,K}(\bmu,\bSigma) +  \frac{1}{2} \int \{\bD(\btau) \bepsilon-\bmu\}\trans\bSigma^{-1}\{\bD(\btau)\bepsilon-\bmu\}  dP_{\bepsilon,K}(\bmu,\bSigma)  \\
&&\leq \log \{c(2\pi)^{p/2}\} + |\log~c|   \\
	&&~~~~~~~~~~~ + \frac{1}{2}  \int \log \abs{\bSigma} dP_{\bepsilon,K}(\bmu,\bSigma)  + \frac{1}{2} \int \norm{\bD(\btau) \bepsilon-\bmu}_{2}^{2}\lambda_{1}^{-1}(\bSigma)dP_{\bepsilon,K}(\bmu,\bSigma)  \\
&&\leq \log \{c(2\pi)^{p/2}\} + |\log~c|   \\
	&&~~~~~~~~~~~ + \frac{1}{2}  \int \log \abs{\bSigma} dP_{\bepsilon,K}(\bmu,\bSigma)  + \int \{\norm{\bD(\btau) \bepsilon}_{2}^{2}+\norm{\bmu}_{2}^{2}\}\lambda_{1}^{-1}(\bSigma)dP_{\bepsilon,K}(\bmu,\bSigma) \\
&&\leq \log \{c(2\pi)^{p/2}\} + |\log~c|   + \frac{1}{2}  \int \log \abs{\bSigma} dP_{\bepsilon,K}(\bmu,\bSigma)  \\
	&&~~~~~~~~~~~+ \norm{2\bepsilon}_{2}^{2} \int \lambda_{1}^{-1}(\bSigma)dP_{\bepsilon,K}(\bmu,\bSigma) + \int \norm{\bmu}_{2}^{2}dP_{\bepsilon,K}(\bmu,\bSigma) \int \lambda_{1}^{-1}(\bSigma)dP_{\bepsilon,K}(\bmu,\bSigma),
\ese
\vspace{-5ex}\\
where the third step followed from application of Jensen's inequality on $g(Z) = -\log~Z$.
The regularity assumptions on $f_{0\bepsilon}$ and Lemma \ref{Lem: mvt moments of fe} imply that the RHS above is $f_{0\bepsilon}$ integrable.
The conclusion of Lemma \ref{Lem: mvt limit of KL divergence of involving fe} follows from an application of DCT again.
\end{proof}

\vspace{.5cm}
To prove Lemma \ref{Lem: mvt KL support of the prior on the density of U|X}, let $f_{\bU\vert \bS}$ denote the density of $\bU = \bS(\bX) \bepsilon$, 
where $\bS = \diag(s_{1},\dots,s_{p})$.
Then $f_{\bU\vert \bX} = f_{\bU\vert \bS(\bX)}$.
We have $f_{\bU\vert \bS}(\bU) = \abs{\bS}^{-1} f_{\bepsilon}(\bS^{-1}\bU)$.
This implies
\vspace{-5ex}\\
\bse
\int f_{0\bU\vert \bS_{0}}(\bU) \log\frac{f_{0\bU\vert \bS_{0}}(\bU)}{f_{\bU\vert \bS}(\bU)} d\bU = \int f_{0\bU\vert \bS_{0}}(\bU) \log\frac{f_{0\bU\vert \bS_{0}}(\bU)}{f_{\bU\vert \bS_{0}}(\bU)} d\bU  +  \int f_{0\bU\vert \bS_{0}}(\bU) \log\frac{f_{\bU\vert \bS_{0}}(\bU)}{f_{\bU\vert \bS}(\bU)} d\bU \\
= \int f_{0\bepsilon}(\bepsilon) \log\frac{f_{0\bepsilon}(\bepsilon)}{f_{\bepsilon}(\bepsilon)} d\bepsilon  +  \int f_{0\bepsilon}(\bepsilon) \log \frac{f_{\bepsilon}(\bepsilon)}{\abs{\bS}^{-1}\abs{\bS_{0}}f_{\bepsilon}(\bS^{-1}\bS_{0}\bepsilon)} d\bepsilon.
\ese
\vspace{-5ex}\\
Let $\delta>0$ be given.
By part 2 of Lemma \ref{Lem: mvt KL support of the priors}, $\Pi_{\bepsilon}\{f_{\bepsilon}: d_{KL}(f_{0\bepsilon},f_{\bepsilon})<\delta/2\}>0$.
Let $\bs = (s_{1},\dots,s_{p})\trans$ and $\bs_{0} = (s_{01},\dots,s_{0p})\trans$.
By Lemma \ref{Lem: mvt limit of KL divergence of involving fe}, there exists $\eta>0$ such that $\norm{\bs_{0}-\bs}_{\infty}<\eta$ implies
$\int f_{0\bepsilon}(\bepsilon) ~ \log [f_{\bepsilon}(\bepsilon)/\{\abs{\bS}^{-1}\abs{\bS_{0}} f_{\bepsilon}(\bS^{-1}\bS_{0}\bepsilon)\}] ~ d\bepsilon<\delta/2$ for every $f_{\bepsilon}\sim \Pi_{\bepsilon}$.
Using a straightforward multivariate extension of Corollary \ref{Cor: mvt sup norm support of priors on sqrt of variance functions}, 
we have $\Pi_{\bV}(||\bs_{0}-\bs||_{\infty}<\eta) > 0$.
Combining these results,
$\Pi_{\bU\vert \bV}\{\sup_{\bX\in\X}d_{KL}(f_{0\bU\vert \bX},f_{\bU\vert \bX})<\delta\} \geq \Pi_{\bepsilon}\{d_{KL}(f_{0\bepsilon},f_{\bepsilon})<\delta/2\} ~ \Pi_{\bV}(||{\bs}_{0}-\bs||_{\infty}<\eta) > 0$.
Hence the proof of part 2 of Lemma \ref{Lem: mvt KL support of the prior on the density of U|X}.

Part 1 of Lemma \ref{Lem: mvt KL support of the prior on the density of U|X} follows trivially from part 2 of Lemma \ref{Lem: mvt KL support of the prior on the density of U|X} 
since $||\bs_{0}-\bs||_{\infty}<\eta$ implies $\norm{\bs_{0}(\bX)-\bs(\bX)}_{\infty}<\eta$ for any $\bX\in \X$.

To prove part 3 of Lemma \ref{Lem: mvt KL support of the prior on the density of U|X}, note that
\vspace{-5ex}\\
\bse
&& \hspace{-1cm} d_{KL}(f_{0,\bX,\bU},f_{\bX,\bU}) = \int_{\X\times \rR^{p}} f_{0,\bU\vert \bX}(\bU\vert \bX) f_{0,\bX}(\bX) ~\log\frac{f_{0,\bU\vert \bX}(\bU\vert \bX)f_{0,\bX}(\bX)}{f_{\bU\vert \bX}(\bU\vert \bX)f_\bX(\bX)} d\bX d\bU \\
&=& \int_{\X} f_{0,\bX}(\bX) \int_{\rR^{p}}f_{0,\bU\vert \bX}(\bU\vert \bX) ~\log\frac{f_{0,\bU\vert \bX}(\bU\vert \bX)}{f_{\bU\vert \bX}(\bU\vert \bX)} d\bU d\bX + \int_{\X} f_{0,\bX}(\bX) ~\log\frac{f_{0,\bX}(\bX)}{f_{\bX}(\bX)} d\bX \\
&\leq& \sup_{\bX\in\X} d_{KL}\{f_{0,\bU\vert \bX}(\bU\vert \bX),f_{\bU\vert \bX}(\bU\vert \bX)\} + d_{KL}(f_{0\bX},f_{\bX}).
\ese
\vspace{-5ex}\\
Part 3 of Lemma \ref{Lem: mvt KL support of the prior on the density of U|X} now follows from part 2 of Lemma \ref{Lem: mvt KL support of the prior on the density of U|X} 
and part 1 of Lemma \ref{Lem: mvt KL support of the priors}.

\subsection{Proof of Theorem \ref{Thm: mvt L1 support of induced prior on density of W}}
Let $d_{H}(f_{0},f) = [\int \{f_{0}^{1/2}(\bZ) - f^{1/2}(\bZ)\}^{2}d\bZ]^{1/2}$ denote the Hellinger distance between any two densities $f_{0}$ and $f$.
From Chapter 1 of Ghosh and Ramamoorthi (2010), we have
\vspace{-5ex}\\
\be
d_{H}^{2}(f_{0},f) \leq ||f_{0}-f||_{1} \leq 2~d_{KL}^{1/2}(f_{0},f). \label{eq: mvt inequalities among distances 1}
\ee
\vspace{-5ex}\\
Using (\ref{eq: mvt inequalities among distances 1}), we have,
\vspace{-5ex}\\
\bse
&~&\hspace{-1cm} ||f_{0\bW}-f_{\bW}||_{1} = \int |f_{0\bW}(\bW)-f_{\bW}(\bW)|d\bW \\
&=& \int \left|\int f_{0\bX}(\bX)f_{0\bW\vert \bX}(\bW)d\bX - \int f_{\bX}(\bX)f_{\bW\vert \bX}(\bW)d\bX\right| d\bW \\
&\leq& \int \left|\int f_{0\bX}(\bX)f_{0\bW\vert \bX}(\bW)d\bX - \int f_{\bX}(\bX)f_{0\bW\vert \bX}(\bW)d\bX\right|d\bW \\
	&& + \int\left|\int f_{\bX}(\bX)f_{0\bW\vert \bX}(\bW)d\bX - \int f_{\bX}(\bX)f_{\bW\vert \bX}(\bW)d\bX\right|d\bW \\
&\leq& \int \int |f_{0\bX}(\bX)-f_{\bX}(\bX)| f_{0\bW\vert \bX}(\bW)d\bX d\bW \\
 	&& + \int\int f_{\bX}(\bX)|f_{0\bW\vert \bX}(\bW) - f_{\bW\vert \bX}(\bW)| d\bX d\bW \\
&=& \int |f_{0\bX}(\bX)-f_{\bX}(\bX)| d\bX + \int f_{\bX}(\bX)  \int  |f_{0\bW\vert \bX}(\bW) - f_{\bW\vert \bX}(\bW)| d\bW d\bX \\
&=& \int |f_{0\bX}(\bX)-f_{\bX}(\bX)| d\bX + \int f_{\bX}(\bX)  \int  |f_{0\bU\vert \bX}(\bW-\bX) - f_{\bU\vert \bX}(\bW-\bX)| d\bW d\bX \\
&\leq& ||f_{0\bX} - f_{\bX}||_{1} + \sup_{\bX\in\X} ||f_{0\bU\vert \bX} - f_{\bU\vert \bX}||_{1} \\
&\leq& 2~d_{KL}^{1/2}(f_{0\bX},f_{\bX}) + 2 \sup_{\bX\in\X} d_{KL}^{1/2}(f_{0\bU\vert \bX},f_{\bU\vert \bX}).
\ese
\vspace{-5ex}\\
The proof of Theorem \ref{Thm: mvt L1 support of induced prior on density of W} follows by
combining part 1 of Lemma \ref{Lem: mvt KL support of the priors} and part 2 of Lemma \ref{Lem: mvt KL support of the prior on the density of U|X}.


\section{Additional Figures} \label{sec: mvt additional figures}
We first present, in Subsection \ref{sec: mvt additional figures Ind}, some additional figures summarizing the results of the simulation experiments for diagonal covariance matrices discussed in Section \ref{sec: mvt simulation studies} of the main paper.  
Then in Subsection \ref{sec: mvt additional figures Ind}, we present figures that summarize the results of simulation experiments for covariance matrices with AR structure. 
Finally in Subsection \ref{sec: mvt additional figures for  EATS data set}, we present some additional figures summarizing the results of the EATS data set analyzed in Section \ref{sec: mvt data analysis} of the main paper.


\newpage
\subsection{Additional Figures Summarizing the Results of the Simulation Experiments for Diagonal Covariance Structure} \label{sec: mvt additional figures Ind}

\newpage
\thispagestyle{empty}

\subsection{Additional Figures Summarizing the Results of the Simulation Experiments for AR Covariance Structure} \label{sec: mvt additional figures AR}

\newpage
\thispagestyle{empty}

\begin{figure}[!ht]
\begin{center}
\includegraphics[height=16cm, width=16cm, trim=0cm 0cm 0cm 0cm, clip=true]{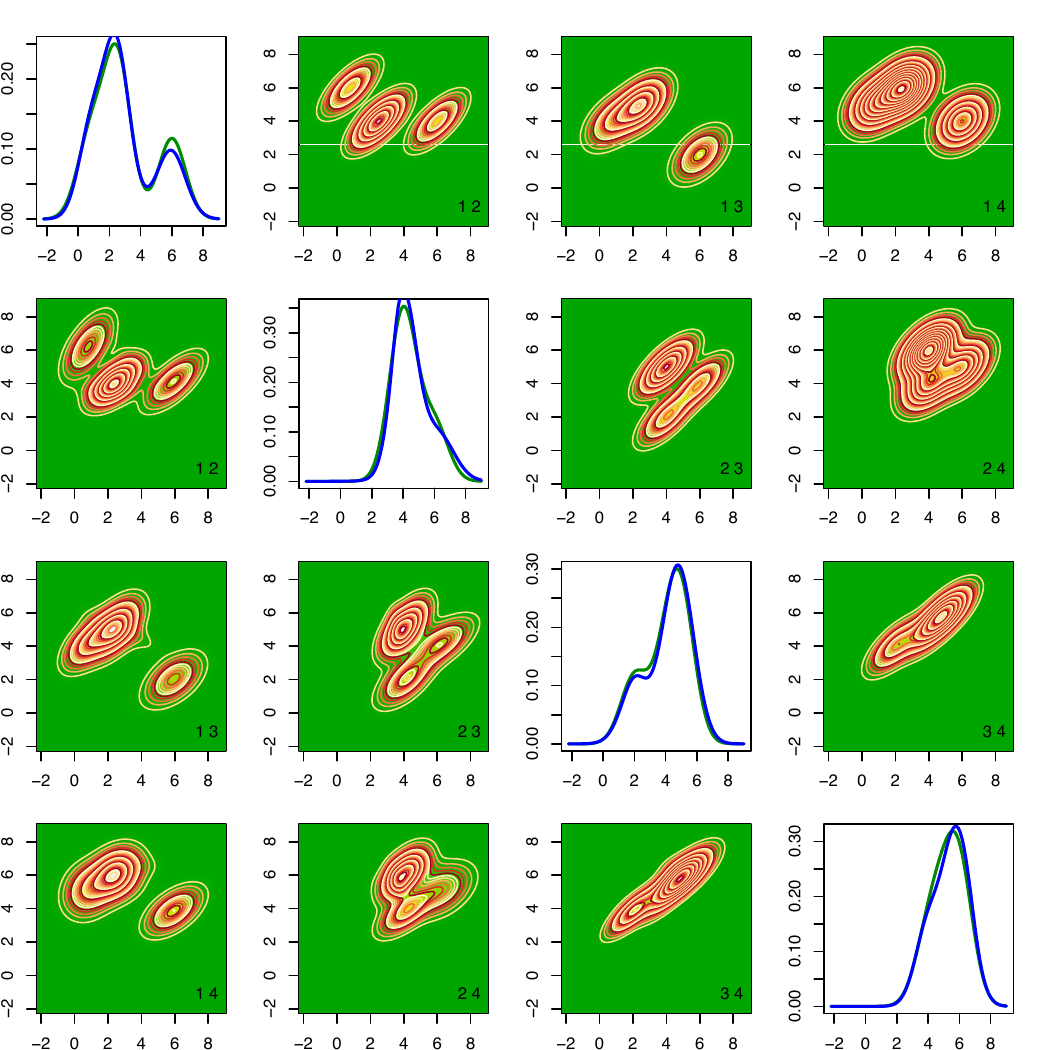}
\end{center}
\caption{\baselineskip=10pt Results for the  $f_{\bX}$ produced by the MIW (mixtures with inverse Wishart priors) method for the conditionally heteroscedastic error distribution $f_{\bepsilon}^{(2)}$ with sample size $n=1000$, $m_{i}=3$ replicates for each subject and component specific covariance matrices with autoregressive structure (AR). 
The results correspond to the data set that produced the median of the estimated integrated squared errors (ISE) out of a total of 100 simulated data sets.
See Section \ref{sec: mvt simulation studies} for additional details.
The upper triangular panels show the contour plots of the true two dimensional marginal densities.
The lower triangular diagonally opposite panels show the corresponding estimates.
The numbers $i,j$ at the bottom right corners of the off-diagonal panels show that the marginal densities $f_{X_{i},X_{j}}$ are plotted in those panels. 
The diagonal panels show the true (lighter shaded green lines) and the estimated (darker shaded blue lines) one dimensional marginals.
The figure is in color in the electronic version of this article.
}
\label{fig: mvt simulation results XS d4 n1000 m3 MIW X1 E1 AR}
\end{figure}

\newpage
\thispagestyle{empty}

\begin{figure}[!ht]
\begin{center}
\includegraphics[height=16cm, width=16cm, trim=0cm 0cm 0cm 0cm, clip=true]{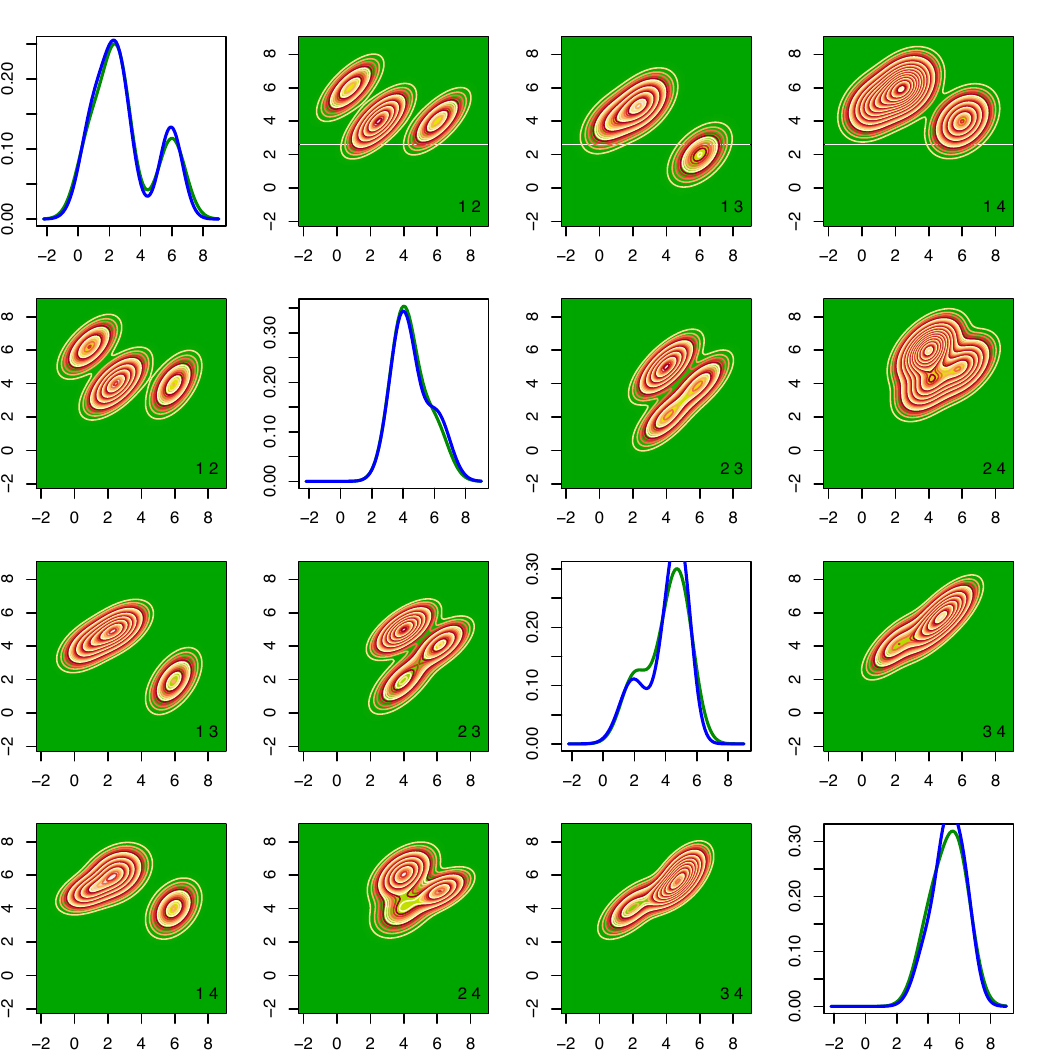}
\end{center}
\caption{\baselineskip=10pt Results for the  $f_{\bX}$ produced by the MLFA (mixtures of latent factor analyzers) method for the conditionally heteroscedastic error distribution $f_{\bepsilon}^{(2)}$ with sample size $n=1000$, $m_{i}=3$ replicates for each subject and component specific covariance matrices with autoregressive structure (AR). The results correspond to the data set that produced the median of the estimated integrated squared errors (ISE) out of a total of 100 simulated data sets.
See Section \ref{sec: mvt simulation studies} for additional details.
The upper triangular panels show the contour plots of the true two dimensional marginal densities.
The lower triangular diagonally opposite panels show the corresponding estimates.
The numbers $i,j$ at the bottom right corners of the off-diagonal panels show that the marginal densities $f_{X_{i},X_{j}}$ are plotted in those panels. 
The diagonal panels show the true (lighter shaded green lines) and the estimated (darker shaded blue lines) one dimensional marginals.
The figure is in color in the electronic version of this article.
}
\label{fig: mvt simulation results XS d4 n1000 m3 MLFA X1 E1 AR}
\end{figure}

\newpage
\thispagestyle{empty}

\begin{figure}[!ht]
\begin{center}
\includegraphics[height=16cm, width=16cm, trim=0cm 0cm 0cm 0cm, clip=true]{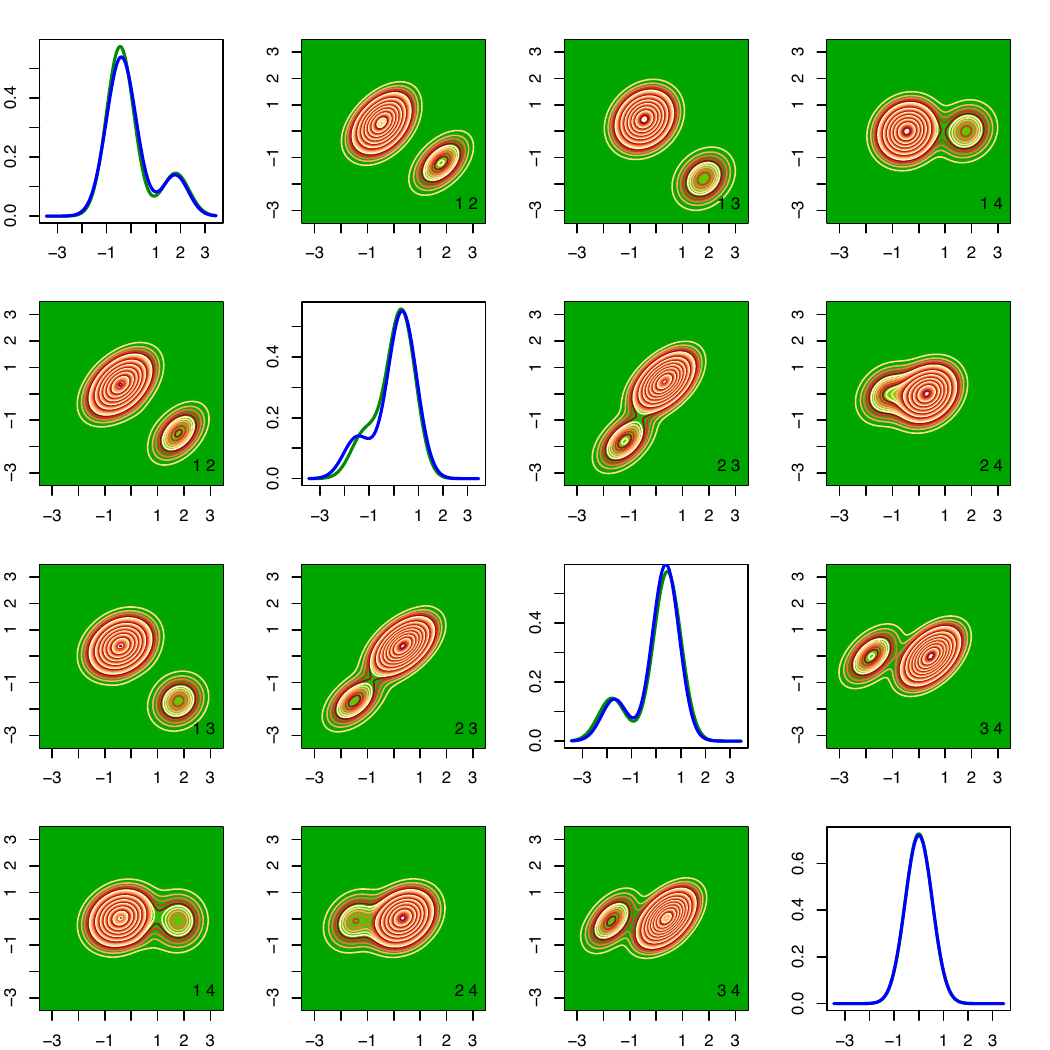}
\end{center}
\caption{\baselineskip=10pt Results for the density of the scaled errors $f_{\bepsilon}$ produced by the MIW (mixtures with inverse Wishart priors) method for the conditionally heteroscedastic error distribution $f_{\bepsilon}^{(2)}$ with sample size $n=1000$, $m_{i}=3$ replicates for each subject and component specific covariance matrices with autoregressive structure (AR). The results correspond to the data set that produced the median of the estimated integrated squared errors (ISE) out of a total of 100 simulated data sets.
See Section \ref{sec: mvt simulation studies} for additional details.
The upper triangular panels show the contour plots of the true two dimensional marginal densities.
The lower triangular diagonally opposite panels show the corresponding estimates.
The numbers $i,j$ at the bottom right corners of the off-diagonal panels show that the marginal densities $f_{\epsilon_{i},\epsilon_{j}}$ are plotted in those panels. 
The diagonal panels show the true (lighter shaded green lines) and the estimated (darker shaded blue lines) one dimensional marginals.
The figure is in color in the electronic version of this article.
}
\label{fig: mvt simulation results ES d4 n1000 m3 MIW X1 E1 AR}
\end{figure}

\newpage
\thispagestyle{empty}

\begin{figure}[!ht]
\begin{center}
\includegraphics[height=16cm, width=16cm, trim=0cm 0cm 0cm 0cm, clip=true]{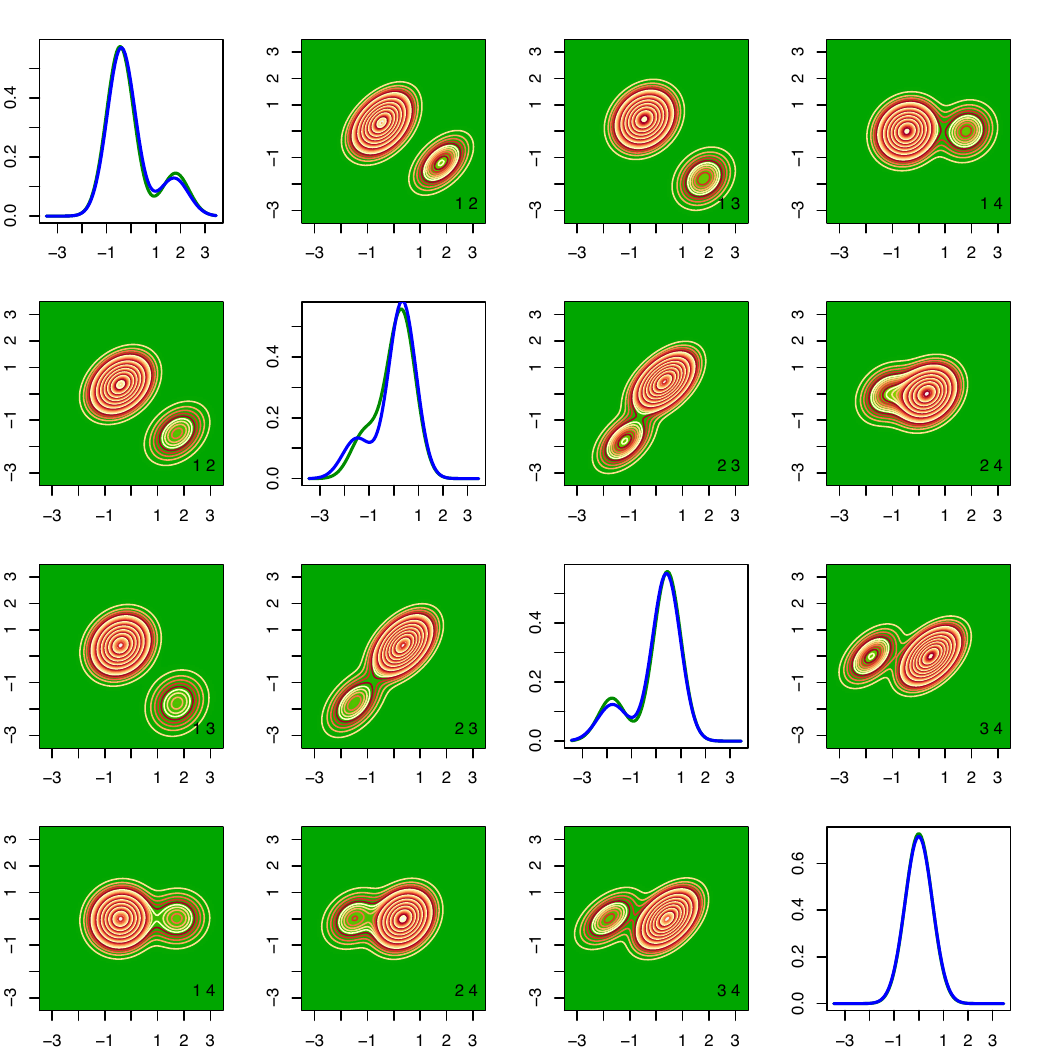}
\end{center}
\caption{\baselineskip=10pt Results for the density of the scaled errors $f_{\bepsilon}$ produced by the MLFA (mixtures of latent factor analyzers) method for the conditionally heteroscedastic error distribution $f_{\bepsilon}^{(2)}$ with sample size $n=1000$, $m_{i}=3$ replicates for each subject and component specific covariance matrices with autoregressive structure (AR). 
The results correspond to the data set that produced the median of the estimated integrated squared errors (ISE) out of a total of 100 simulated data sets.
See Section \ref{sec: mvt simulation studies} for additional details.
The upper triangular panels show the contour plots of the true two dimensional marginal densities.
The lower triangular diagonally opposite panels show the corresponding estimates.
The numbers $i,j$ at the bottom right corners of the off-diagonal panels show that the marginal densities $f_{\epsilon_{i},\epsilon_{j}}$ are plotted in those panels. 
The diagonal panels show the true (lighter shaded green lines) and the estimated (darker shaded blue lines) one dimensional marginals.
The figure is in color in the electronic version of this article.
}
\label{fig: mvt simulation results ES d4 n1000 m3 MLFA X1 E1 AR}
\end{figure}

\newpage
\thispagestyle{empty}

\begin{figure}[!ht]
\begin{center}
\includegraphics[height=16cm, width=16cm, trim=0cm 0cm 0cm 0cm, clip=true]{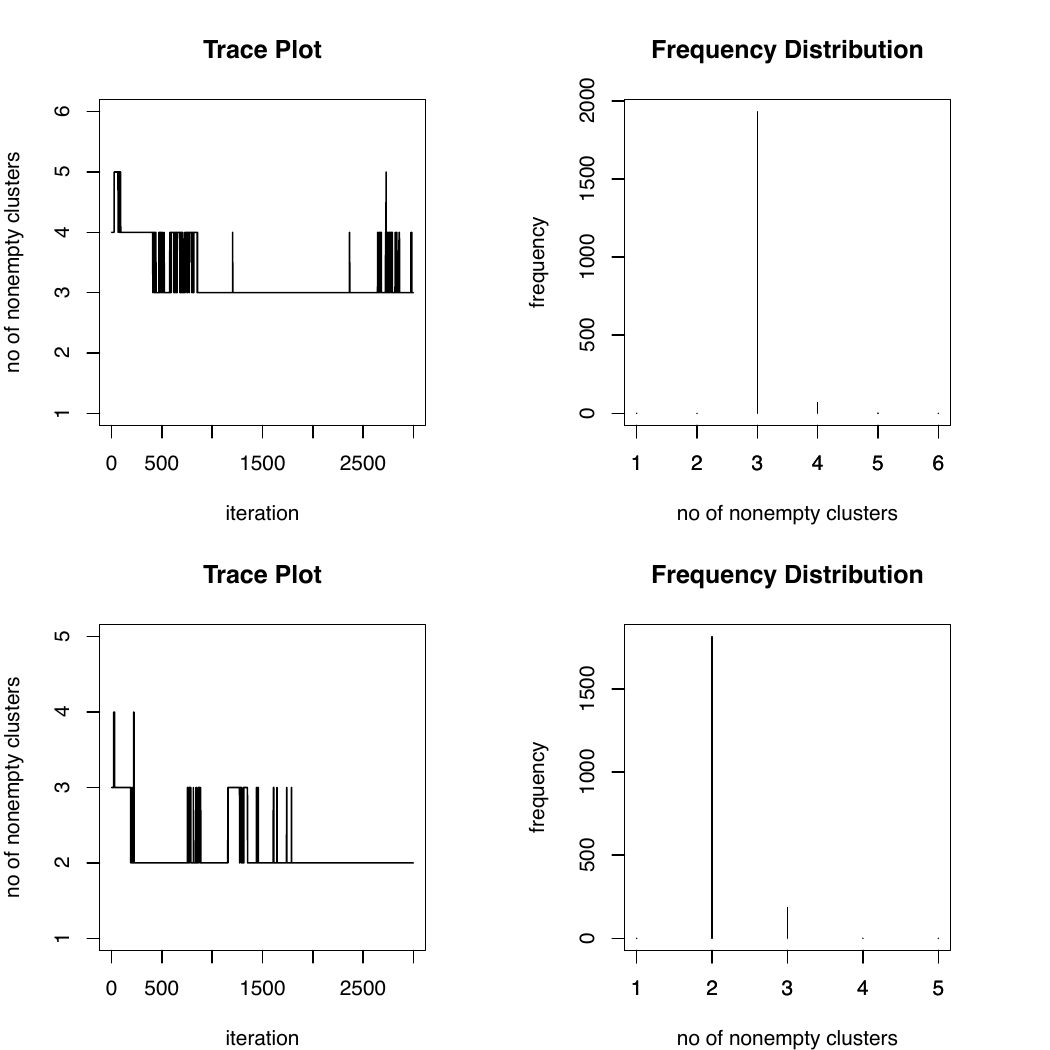}
\end{center}
\caption{\baselineskip=10pt 
Trace plots and frequency distributions of the number of nonempty clusters produced by the MIW (mixtures with inverse Wishart priors) method for the conditionally heteroscedastic error distribution $f_{\bepsilon}^{(2)}$ with sample size $n=1000$, $m_{i}=3$ replicates for each subject and component specific covariance matrices with autoregressive structure (AR). 
See Section \ref{sec: mvt simulation studies} for additional details.
The results correspond to the simulation instance that produced the median of the estimated integrated squared errors (ISE) out of a total of 100 simulated data sets, 
when the number of mixture components for both $f_{\bX}$ and $f_{\bepsilon}$ were kept fixed at $K_{\bX}=6$ and $K_{\bepsilon}=5$.
The upper panels are for the  $f_{\bX}$ and the lower panels are for the density of the scaled errors $f_{\bepsilon}$.
The true number of mixture components were $K_{\bX} = 3$ and $K_{\bepsilon} = 3$. 
As can be seen from Figure \ref{fig: mvt simulation results ES d4 n1000 m3 MIW X1 E1 AR},
a mixture model with $2$ nonempty clusters can approximate the true density of the scaled errors well.
}
\label{fig: mvt simulation results Trace Plots d4 n1000 m3 MIW X1 E1 AR}
\end{figure}

\newpage
\thispagestyle{empty}

\begin{figure}[!ht]
\begin{center}
\includegraphics[height=16cm, width=16cm, trim=0cm 0cm 0cm 0cm, clip=true]{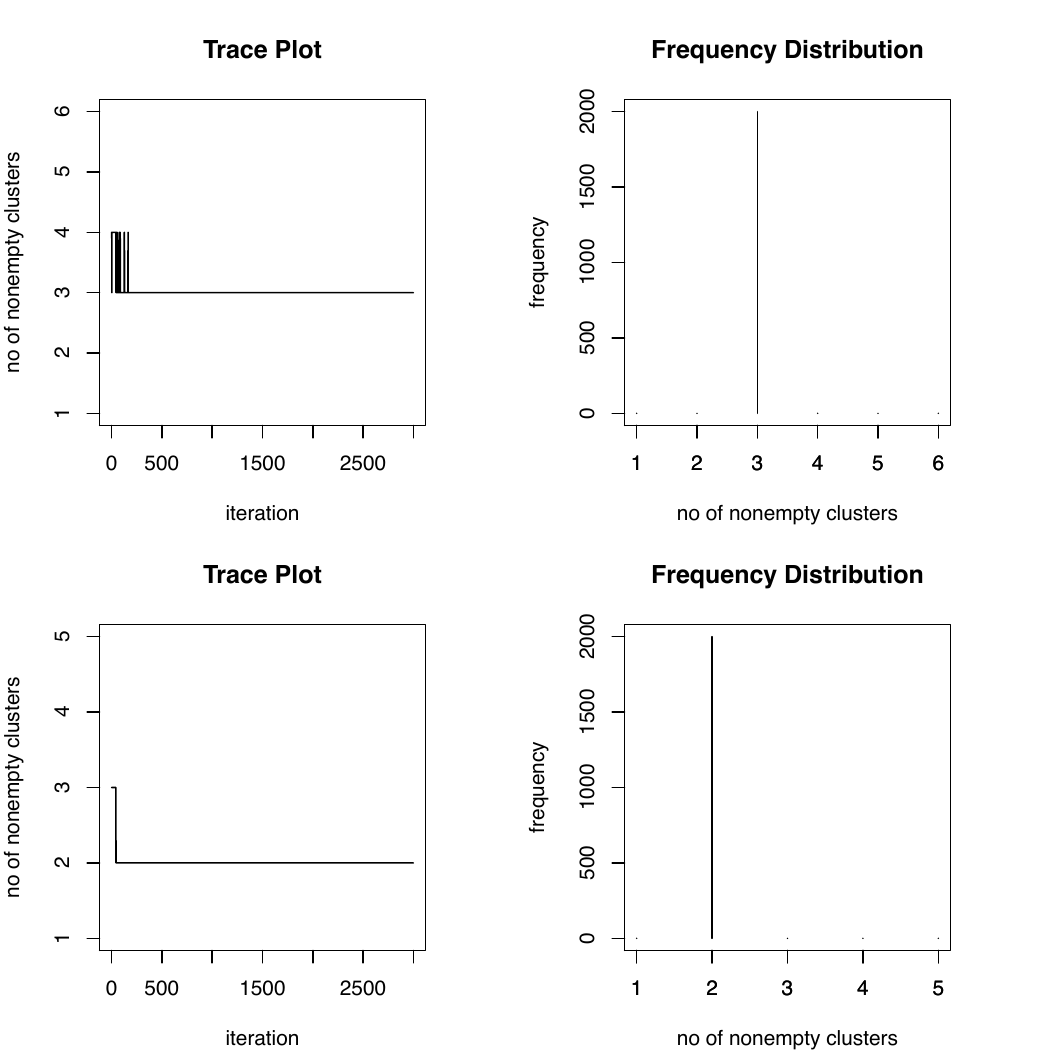}
\end{center}
\caption{\baselineskip=10pt 
Trace plots and frequency distributions of the number of nonempty clusters produced by the MLFA (mixtures of latent factor analyzers) method for the conditionally heteroscedastic error distribution $f_{\bepsilon}^{(2)}$ with sample size $n=1000$, $m_{i}=3$ replicates for each subject and component specific covariance matrices with autoregressive structure (AR).  
See Section \ref{sec: mvt simulation studies} for additional details.
The results correspond to the simulation instance that produced the median of the estimated integrated squared errors (ISE) out of a total of 100 simulated data sets, 
when the number of mixture components for $f_{\bX}$ and $f_{\bepsilon}$ were kept fixed at $K_{\bX}=6$ and $K_{\bepsilon}=5$.
The upper panels are for the  $f_{\bX}$ and the lower panels are for the density of the scaled errors $f_{\bepsilon}$.
The true number of mixture components were $K_{\bX} = 3$ and $K_{\bepsilon} = 3$. 
As can be seen from Figure \ref{fig: mvt simulation results ES d4 n1000 m3 MLFA X1 E1 AR},
a mixture model with $2$ nonempty clusters can approximate the true density of the scaled errors well.
}
\label{fig: mvt simulation results Trace Plots d4 n1000 m3 MLFA X1 E1 AR}
\end{figure}

\clearpage\newpage
\subsection{Additional Figures Summarizing the Results for the EATS Data Set Analyzed in Section \ref{sec: mvt data analysis} of the Main Paper} \label{sec: mvt additional figures for  EATS data set}

\begin{figure}[!ht]
\begin{center}
\includegraphics[height=16cm, width=16cm, trim=0cm 0cm 0cm 0cm, clip=true]{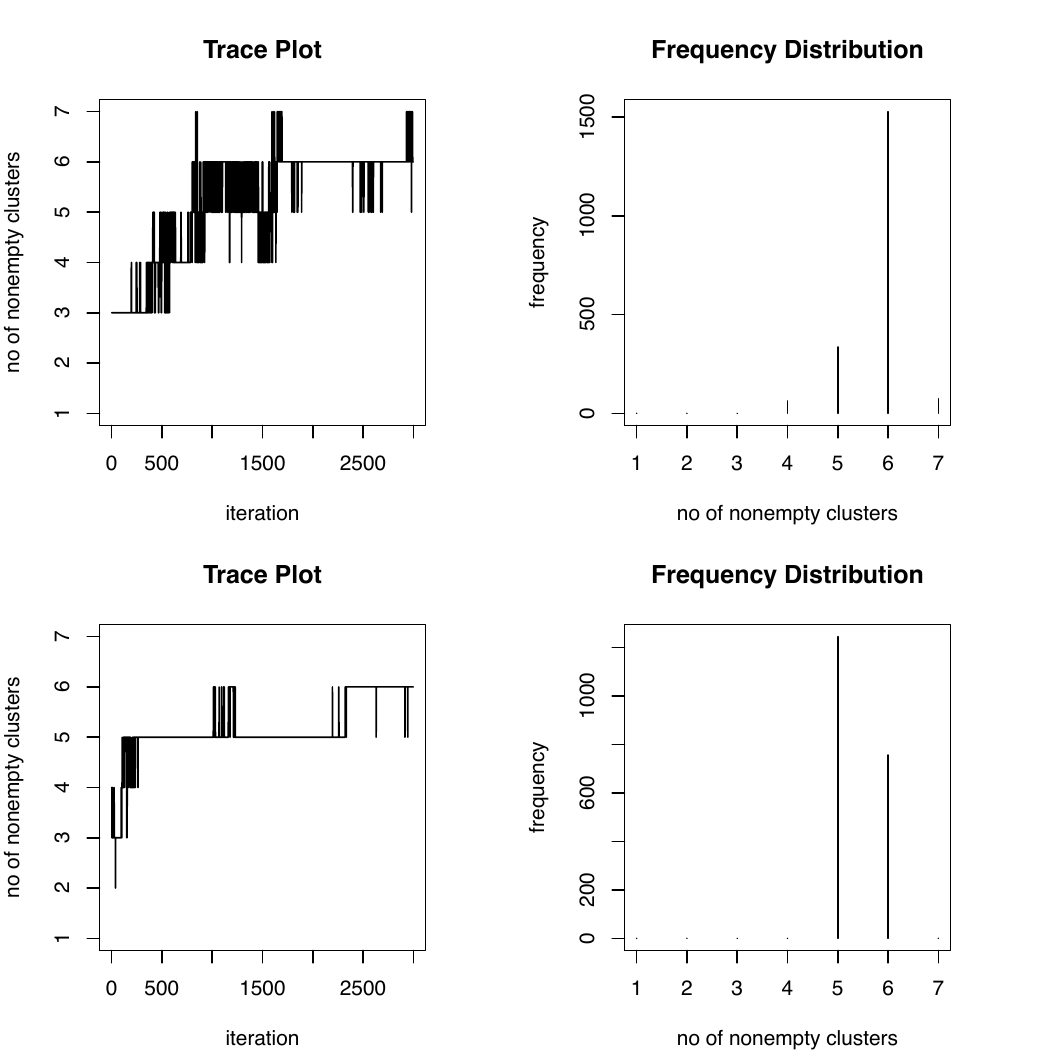}
\end{center}
\caption{\baselineskip=10pt 
Trace plots and frequency distributions of the number of nonempty clusters produced by the MIW (mixtures with inverse Wishart priors) method for the EATS data example. 
See Section \ref{sec: mvt data analysis} for additional details.
The number of mixture components for both $f_{\bX}$ and $f_{\bepsilon}$ were kept fixed at $K_{\bX}=K_{\bepsilon}=7$.
The upper panels are for the  $f_{\bX}$ and the lower panels are for the density of the scaled errors $f_{\bepsilon}$.
}
\label{fig: mvt EATS  data results Trace Plots MIW}
\end{figure}

\newpage
\thispagestyle{empty}

\begin{figure}[!ht]
\begin{center}
\includegraphics[height=16cm, width=16cm, trim=0cm 0cm 0cm 0cm, clip=true]{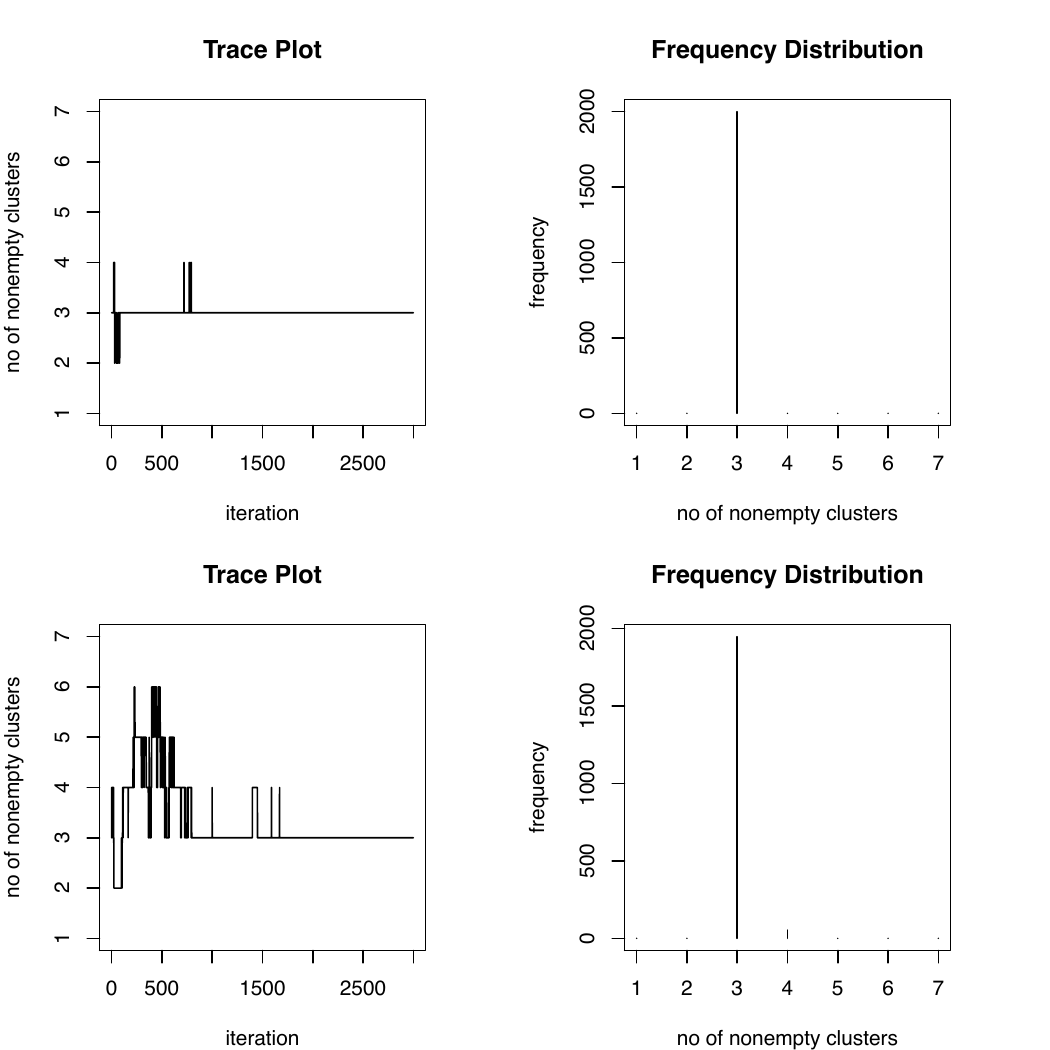}
\end{center}
\caption{\baselineskip=10pt 
Trace plots and frequency distributions of the number of nonempty clusters produced by the MLFA (mixtures of latent factor analyzers) method for the EATS data example. 
See Section \ref{sec: mvt data analysis} for additional details.
The number of mixture components for both $f_{\bX}$ and $f_{\bepsilon}$ were kept fixed at $K_{\bX}=K_{\bepsilon}=7$.
The upper panels are for the  $f_{\bX}$ and the lower panels are for the density of the scaled errors $f_{\bepsilon}$.
}
\label{fig: mvt EATS  data results Trace Plots MLFA}
\end{figure}


\clearpage\newpage
\section{Additional Simulation Experiments} \label{sec: mvt additional simulation studies}
This section presents the results of additional simulation experiments for multivariate t and multivariate Laplace distributed measurement errors. 
Cases when $f_{\bX}$ is multivariate t or mixture of multivariate t are also considered. 
For easy reference, brief descriptions of these distributions are provided below.   

\vspace{-2.5ex}
\subsection{Multivariate t Distribution}
A random variable $Z$ following a Student's t-distribution with degrees of freedom $\nu$, mean $\mu$ and variance $\nu b/(\nu-2)$ can be represented as $Z=\mu+\nu^{1/2}b^{1/2}X/Y^{1/2}$, 
where $Y$ and $X$ are independent, $Y$ follows a chi-square distribution with $\nu$ degrees of freedom, denoted by $Y\sim \chi^{2}_{\nu}$, and $X$ follows a standard normal distribution.
A natural extension to multivariate set up is given by 
$\bZ = \bmu + \nu^{1/2} \bSigma^{1/2}\bX/Y^{1/2}$, 
where $Y\sim \chi^{2}_{\nu}$ and $\bX\sim \MVN_{p}(\bzero,\bI)$ independently. 
The random vector $\bZ$ is then said to follow a multivariate t-distribution (Kotz and Nadarajah, 2004) with degrees of freedom $\nu$, mean $\bmu$ and covariance $\nu\bSigma/(\nu-2)$, denoted by $\MVT_{p}(\nu,\bmu,\bSigma)$.
The above characterization can be used to sample from a $\MVT_{p}(\nu,\bmu,\bSigma)$ density.
The density of $\bZ$ is given by
\vspace{-5ex}\\
\bse
f_{\bZ}(\bz) = \frac{\Gamma{\{(\nu+p)/2\}}}{\Gamma{(\nu/2)}(\nu\pi)^{p/2}\abs{\bSigma}^{1/2}} \cdot {\{1+(\bz-\bmu)\trans\bSigma^{-1}(\bz-\bmu)/\nu\}^{-(\nu+p)/2}}. 
\ese
\vspace{-4ex}\\
The characteristic function is given by 
\vspace{-5ex}\\
\bse
\phi(\bt) = \exp(i\bt\trans\bmu) \cdot \frac{|| \nu^{1/2}\bSigma^{1/2}\bt ||^{\nu/2}}{2^{\nu/2-1}\Gamma(\nu/2)}\cdot H_{\nu/2}(|| \nu^{1/2}\bSigma^{1/2}\bt ||),~~~\bt\in \rR^{p},
\ese
\vspace{-5ex}\\
where $H_{\alpha}$ denotes a McDonald's function of order $\alpha(>1/2)$ and admits the integral representation 
\vspace{-7ex}\\
\bse
H_{\alpha}(t)  =  (2/t)^{\alpha}  \cdot  \frac{\Gamma(\alpha+1/2)}{\pi^{1/2}} \int_{0}^{\infty}(1+u^{2})cos(tu)du,~~~t>0.
\ese
\vspace{-5ex}\\
When $\bSigma=\bI$, the identity matrix, the components $Z_{i}$ and $Z_{j}$ are uncorrelated, but not statistically independent.
With $\bmu=(\mu_{1}\dots,\mu_{p})\trans$ and $\bSigma = ((\sigma_{ij}))$, 
the $i\th$ random variable $Z_{i}$ marginally follows a univariate Student's t-distribution with degrees of freedom $\nu$, mean $\mu_{i}$ and variance $\nu\sigma_{ii}/(\nu-2)$.

\vspace{-2.5ex}
\subsection{Multivariate Laplace Distribution}
A random variable $Z$ following a Laplace distribution with mean $\mu$ and variance $b$ has the density 
\vspace{-5ex}\\
\bse
f_{Z}(z) = (2b)^{-1/2} \exp(-2^{1/2}b^{-1/2}\abs{z-\mu}).
\ese
\vspace{-5ex}\\
$Z$ can be represented as $Z=\mu+Y^{1/2}b^{1/2}X$, 
where $Y$ and $X$ are independent and follow standard exponential and standard normal distributions, respectively.
A natural extension to multivariate set up is given by 
$\bZ = \bmu + Y^{1/2} \bSigma^{1/2}\bX$, 
where $Y$ follows a standard exponential density and $\bX\sim \MVN_{p}(\bzero,\bI)$ independently of $Y$. 
The random vector $\bZ$ is then said to follow a multivariate Laplace distribution (Eltoft, et al. 2006) with mean $\bmu$ and covariance $\bSigma$, denoted by $\MVL_{p}(\bmu,\bSigma)$.
The above characterization can be used to sample from a $\MVL_{p}(\bmu,\bSigma)$ density.
The density of $\bZ$ is then given by
\vspace{-5ex}\\
\bse
f_{\bZ}(\bz) = \frac{2}{(2\pi)^{p/2}\abs{\bSigma}^{1/2}} \cdot \frac{K_{p/2-1}\{2^{1/2}h^{1/2}(\bz)\}}{\{h(\bz)/2\}^{p/4-1/2}},
\ese
\vspace{-5ex}\\
where $h(\bz) = (\bz-\bmu)\trans\bSigma^{-1}(\bz-\bmu)$ and $K_{m}$ denotes modified Bessel functions of the second kind of order $m$.
Using asymptotic formula for the Bessel functions, namely $K_{m}(z) = \{\pi/(2z)\}^{1/2}\exp(-z)$ as $\abs{z}\to \infty$, we have 
\vspace{-5ex}\\
\bse
f_{\bZ}(\bz) 
\approx \frac{\pi^{1/2}}{(2\pi)^{p/2}\abs{\bSigma}^{1/2}} \cdot \frac{2^{(p-1)/4}}{h^{(p-1)/4}(\bz)} \cdot \exp\{-2^{1/2}h^{1/2}(\bz)\}.
\ese
\vspace{-5ex}\\
The characteristic function is given by $\phi(\bt) = \exp(i\bt\trans\bmu)(1+\bt\trans\bSigma\bt/2)^{-1}$ for $\bt\in \rR^{p}$.
For $p>1$, the density has a singularity at $\bmu$.
When $\bSigma=\bI$, the identity matrix, the components $Z_{i}$ and $Z_{j}$ are uncorrelated, but not statistically independent.
With $\bmu=(\mu_{1}\dots,\mu_{p})\trans$ and $\bSigma = ((\sigma_{ij}))$, 
the $i\th$ random variable $Z_{i}$ marginally follows a univariate Laplace distribution with mean $\mu_{i}$ and variance $\sigma_{ii}$.

\subsection{Summary of Results}
\begin{figure}[!ht]
\begin{center}
\includegraphics[width=16cm, trim=0cm 0cm 0cm 0cm, clip=true]{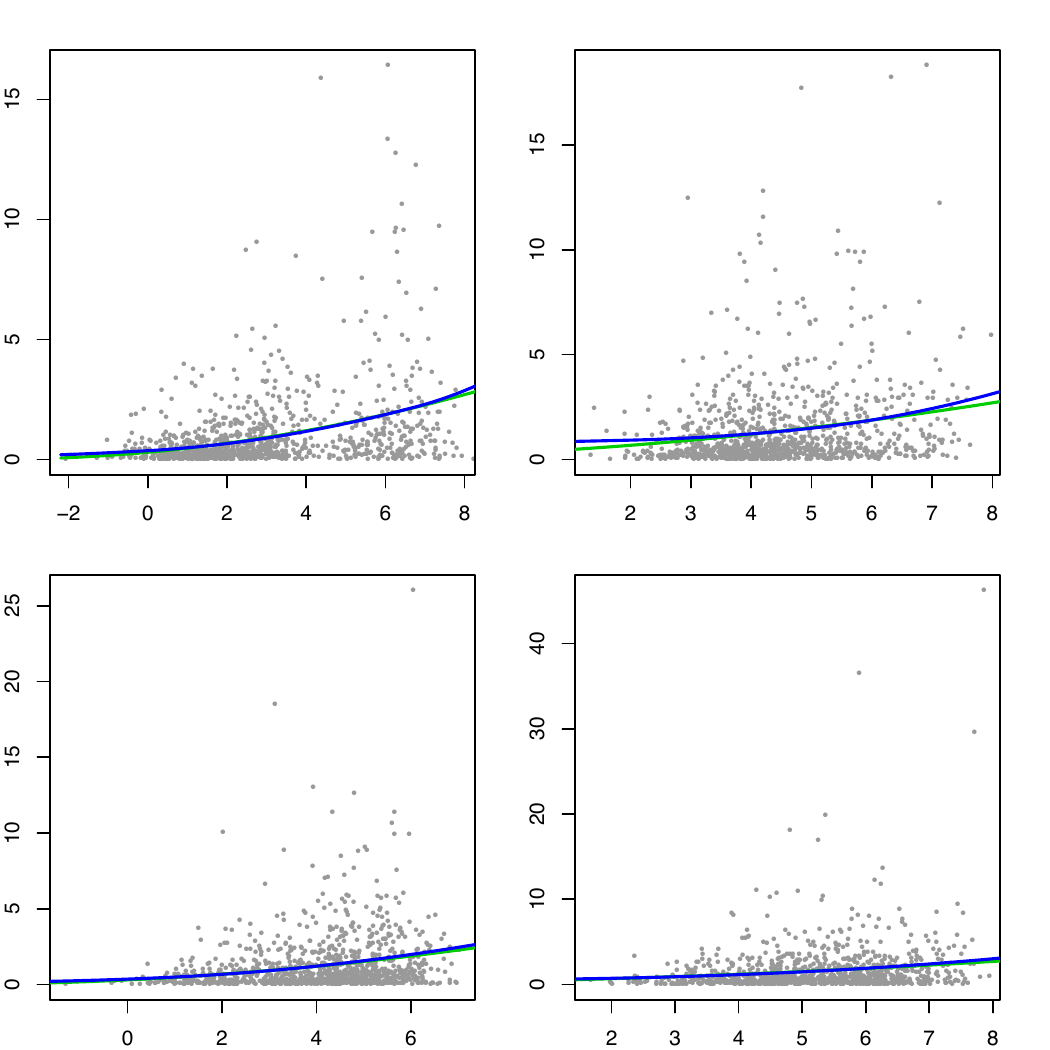}
\end{center}
\caption{\baselineskip=10pt Results for the variance functions $s^{2}(X)$ 
produced by the univariate density deconvolution method for each component of $\bX$ for conditionally heteroscedastic multivariate Laplace ($f_{\bepsilon}^{(4)}$) distributed measurement errors with sample size $n=1000$, $m_{i}=3$ replicates for each subject and identity matrix (I) for the component specific covariance matrices. 
The results correspond to the data set that produced the median of the estimated integrated squared errors (ISE) out of a total of 100 simulated data sets for the MIW (mixtures with inverse Wishart priors) method.
For each component of $\bX$, the true variance function is $s^{2}(X) = (1+X/4)^{2}$.
See Section \ref{sec: mvt density of heteroscedastic errors} and Section \ref{sec: mvt estimation of variance functions} for additional details.
In each panel, the true (lighter shaded green lines) and the estimated (darker shaded blue lines) variance functions 
are superimposed over a plot of subject specific sample means vs subject specific sample variances.
The figure is in color in the electronic version of this article.
}
\label{fig: mvt simulation results VFn d4 n1000 m3 MIW X1 HT_E0 Ind}
\end{figure}

\begin{figure}[!ht]
\begin{center}
\includegraphics[width=16cm, trim=0cm 0cm 0cm 0cm, clip=true]{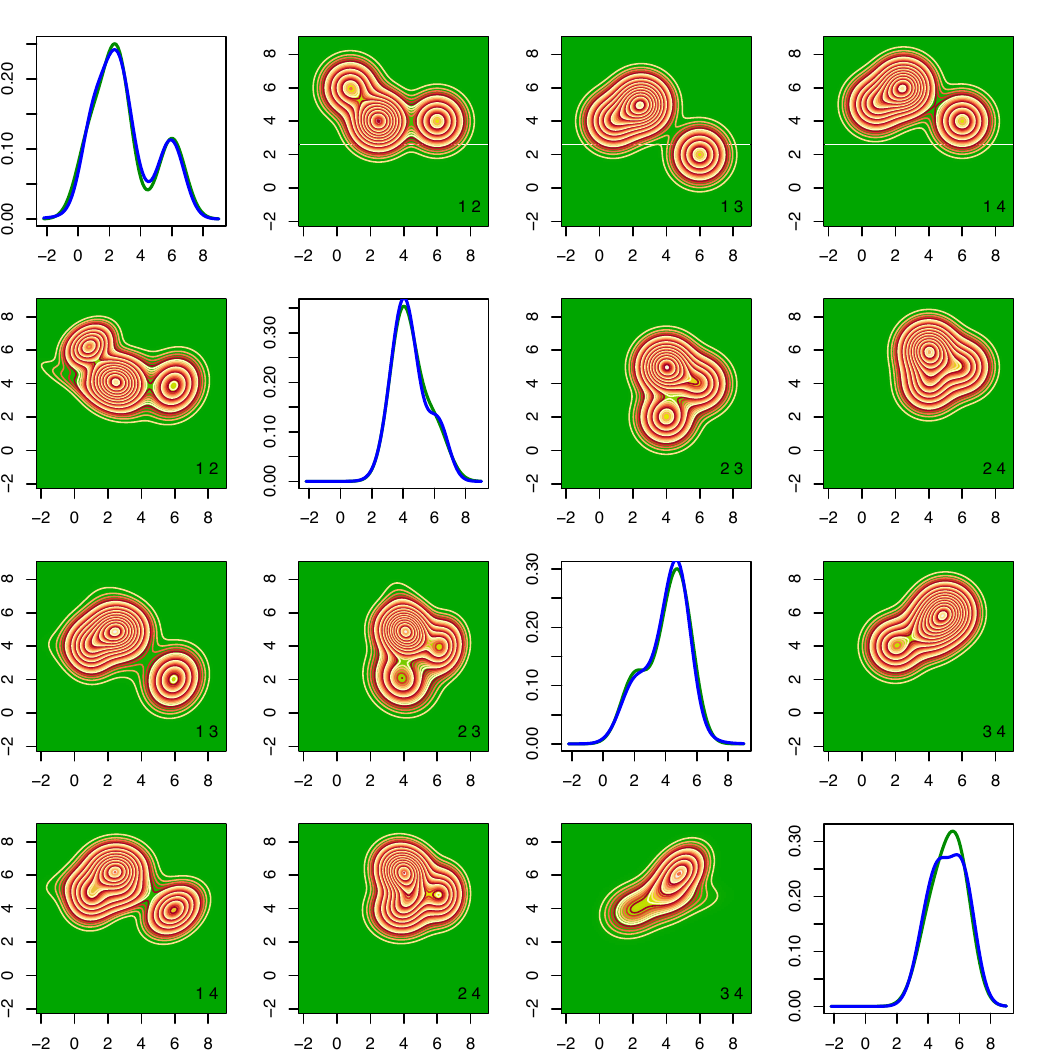}
\end{center}
\caption{\baselineskip=10pt Results for the  $f_{\bX}$ 
produced by the MIW (mixtures with inverse Wishart priors) method for conditionally heteroscedastic multivariate Laplace ($f_{\bepsilon}^{(4)}$) distributed measurement errors with sample size $n=1000$, $m_{i}=3$ replicates for each subject and identity matrix (I) for the component specific covariance matrices. 
The results correspond to the data set that produced the median of the estimated integrated squared errors (ISE) out of a total of 100 simulated data sets.
See Section \ref{sec: mvt simulation studies} and Section \ref{sec: mvt additional simulation studies} for additional details.
The upper triangular panels show the contour plots of the true two dimensional marginal densities.
The lower triangular diagonally opposite panels show the corresponding estimates.
The numbers $i,j$ at the bottom right corners of the off-diagonal panels show that the marginal densities $f_{X_{i},X_{j}}$ are plotted in those panels. 
The diagonal panels show the true (lighter shaded green lines) and the estimated (darker shaded blue lines) one dimensional marginals.
The figure is in color in the electronic version of this article.
}
\label{fig: mvt simulation results XS d4 n1000 m3 MIW X1 HT_E0 Ind}
\end{figure}

\begin{figure}[!ht]
\begin{center}
\includegraphics[width=16cm, trim=0cm 0cm 0cm 0cm, clip=true]{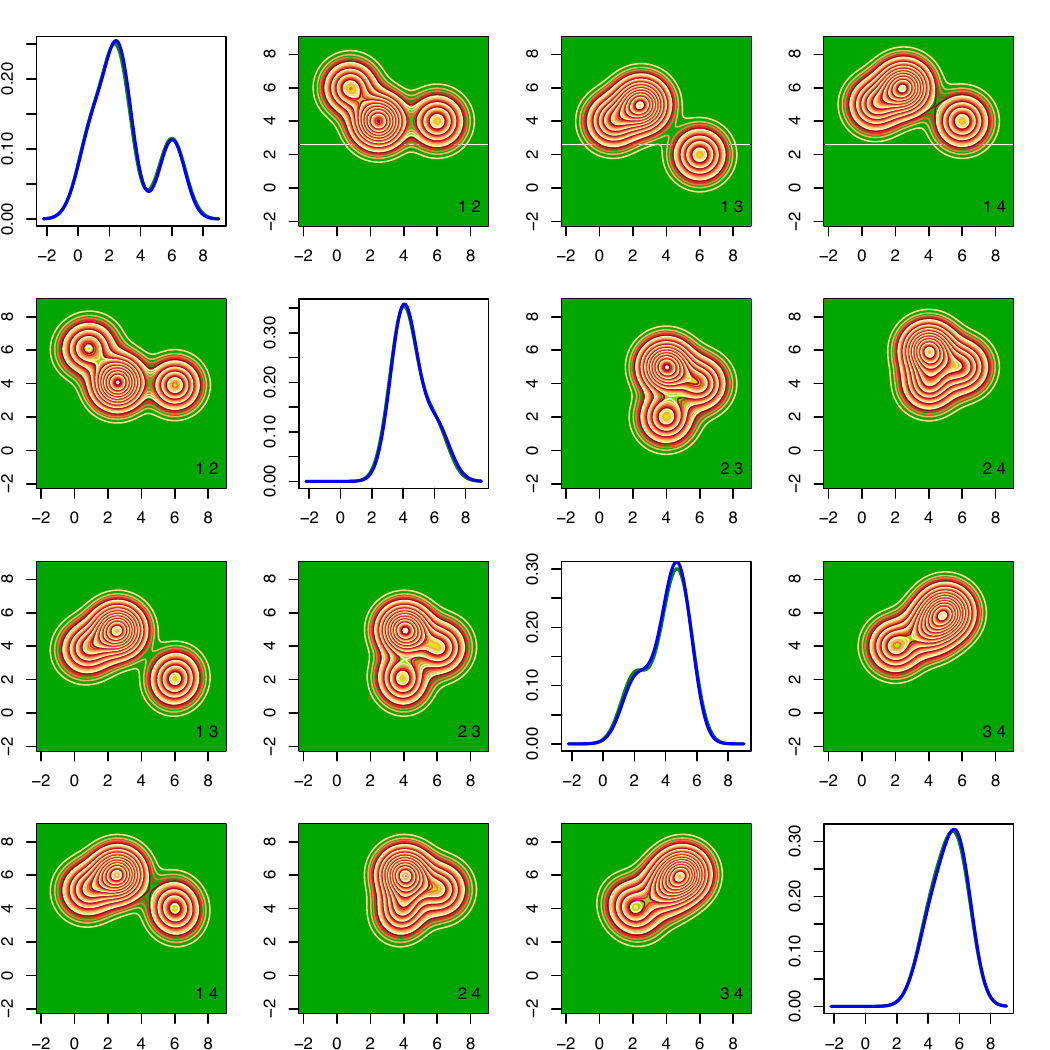}
\end{center}
\caption{\baselineskip=10pt Results for the  $f_{\bX}$ 
produced by the $\text{MLFA}_{2}$ (mixtures of latent factor analyzers) method for conditionally heteroscedastic multivariate Laplace ($f_{\bepsilon}^{(4)}$) distributed measurement errors with sample size $n=1000$, $m_{i}=3$ replicates for each subject and identity matrix (I) for the component specific covariance matrices. 
The results correspond to the data set that produced the median of the estimated integrated squared errors (ISE) out of a total of 100 simulated data sets.
See Section \ref{sec: mvt simulation studies} and Section \ref{sec: mvt additional simulation studies} for additional details.
The upper triangular panels show the contour plots of the true two dimensional marginal densities.
The lower triangular diagonally opposite panels show the corresponding estimates.
The numbers $i,j$ at the bottom right corners of the off-diagonal panels show that the marginal densities $f_{X_{i},X_{j}}$ are plotted in those panels. 
The diagonal panels show the true (lighter shaded green lines) and the estimated (darker shaded blue lines) one dimensional marginals.
The figure is in color in the electronic version of this article.
}
\label{fig: mvt simulation results XS d4 n1000 m3 MLFA X1 HT_E0 Ind}
\end{figure}

\begin{figure}[!ht]
\begin{center}
\includegraphics[width=16cm, trim=0cm 0cm 0cm 0cm, clip=true]{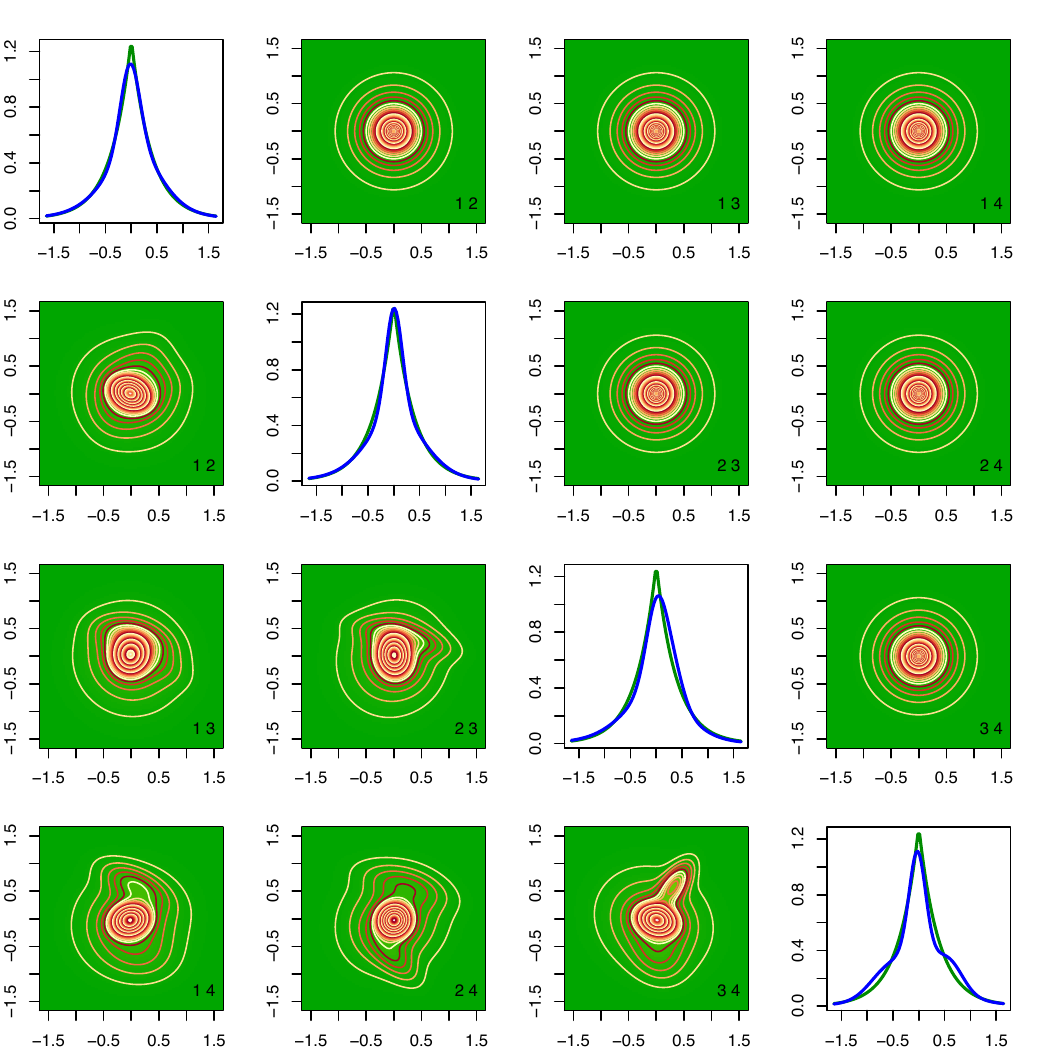}
\end{center}
\caption{\baselineskip=10pt Results for the density of the scaled errors $f_{\bepsilon}$ 
produced by the MIW (mixtures with inverse Wishart priors) method for conditionally heteroscedastic multivariate Laplace ($f_{\bepsilon}^{(4)}$) distributed measurement errors with sample size $n=1000$, $m_{i}=3$ replicates for each subject and identity matrix (I) for the component specific covariance matrices. 
The results correspond to the data set that produced the median of the estimated integrated squared errors (ISE) out of a total of 100 simulated data sets.
See Section \ref{sec: mvt simulation studies} and Section \ref{sec: mvt additional simulation studies} for additional details.
The upper triangular panels show the contour plots of the true two dimensional marginal densities.
The lower triangular diagonally opposite panels show the corresponding estimates.
The numbers $i,j$ at the bottom right corners of the off-diagonal panels show that the marginal densities $f_{\epsilon_{i},\epsilon_{j}}$ are plotted in those panels. 
The diagonal panels show the true (lighter shaded green lines) and the estimated (darker shaded blue lines) one dimensional marginals.
The figure is in color in the electronic version of this article.
}
\label{fig: mvt simulation results ES d4 n1000 m3 MIW X1 HT_E0 Ind}
\end{figure}

\begin{figure}[!ht]
\begin{center}
\includegraphics[width=16cm, trim=0cm 0cm 0cm 0cm, clip=true]{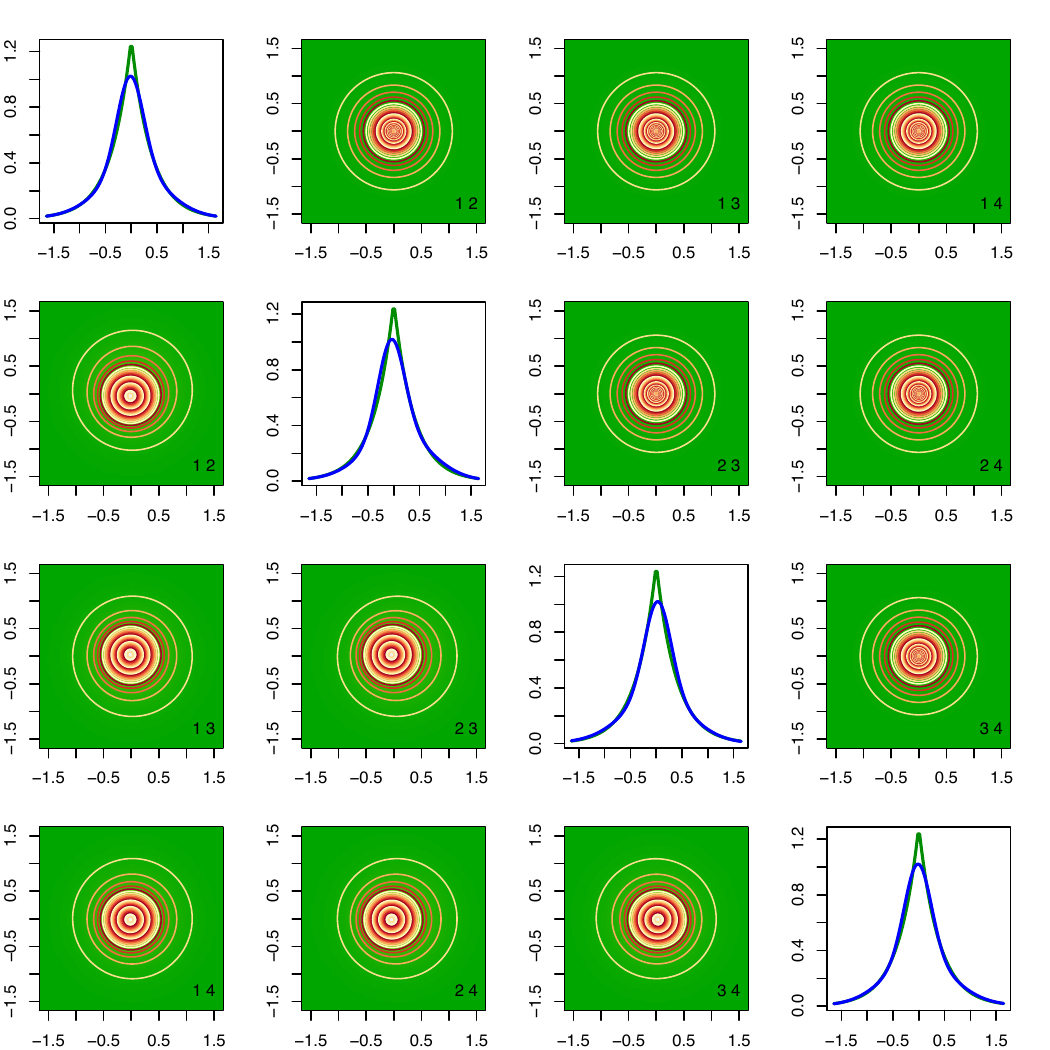}
\end{center}
\caption{\baselineskip=10pt Results for the density of the scaled errors $f_{\bepsilon}$ 
produced by the $\text{MLFA}_{2}$ (mixtures of latent factor analyzers) method for conditionally heteroscedastic multivariate Laplace ($f_{\bepsilon}^{(4)}$) distributed measurement errors with sample size $n=1000$, $m_{i}=3$ replicates for each subject and identity matrix (I) for the component specific covariance matrices. 
The results correspond to the data set that produced the median of the estimated integrated squared errors (ISE) out of a total of 100 simulated data sets.
See Section \ref{sec: mvt simulation studies} and Section \ref{sec: mvt additional simulation studies} for additional details.
The upper triangular panels show the contour plots of the true two dimensional marginal densities.
The lower triangular diagonally opposite panels show the corresponding estimates.
The numbers $i,j$ at the bottom right corners of the off-diagonal panels show that the marginal densities $f_{\epsilon_{i},\epsilon_{j}}$ are plotted in those panels. 
The diagonal panels show the true (lighter shaded green lines) and the estimated (darker shaded blue lines) one dimensional marginals.
The figure is in color in the electronic version of this article.
}
\label{fig: mvt simulation results ES d4 n1000 m3 MLFA X1 HT_E0 Ind}
\end{figure}

\begin{figure}[!ht]
\begin{center}
\includegraphics[height=16cm, width=16cm, trim=0cm 0cm 0cm 0cm, clip=true]{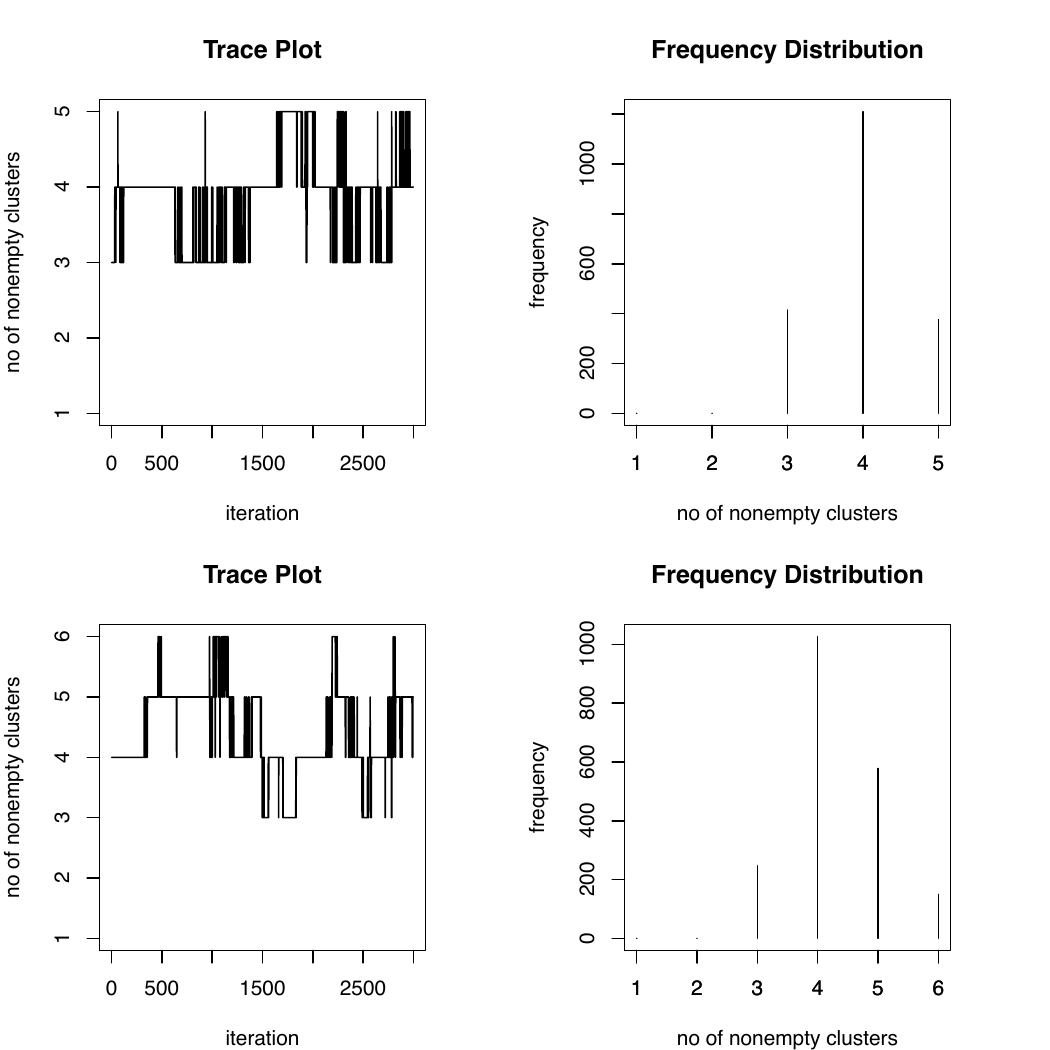}
\end{center}
\caption{\baselineskip=10pt 
Trace plots and frequency distributions of the number of nonempty clusters produced by the MIW (mixtures with inverse Wishart priors) method for conditionally heteroscedastic multivariate Laplace ($f_{\bepsilon}^{(4)}$) distributed measurement errors with sample size $n=1000$, $m_{i}=3$ replicates for each subject and identity matrix (I) for the component specific covariance matrices. 
See Section \ref{sec: mvt simulation studies} and Section \ref{sec: mvt additional simulation studies} for additional details.
The upper panels are for the  $f_{\bX}$ and the lower panels are for the density of the scaled errors $f_{\bepsilon}$.
The results correspond to the simulation instance that produced the median of the estimated integrated squared errors (ISE) out of a total of 100 simulated data sets, 
when the number of mixture components for both $f_{\bX}$ and $f_{\bepsilon}$ were kept fixed at $K_{\bX}=5$ and $K_{\bepsilon}=6$, respectively.
}
\label{fig: mvt simulation results Trace Plots d4 n1000 m3 MIW X1 HT_E0 Ind}
\end{figure}

\begin{figure}[!ht]
\begin{center}
\includegraphics[height=16cm, width=16cm, trim=0cm 0cm 0cm 0cm, clip=true]{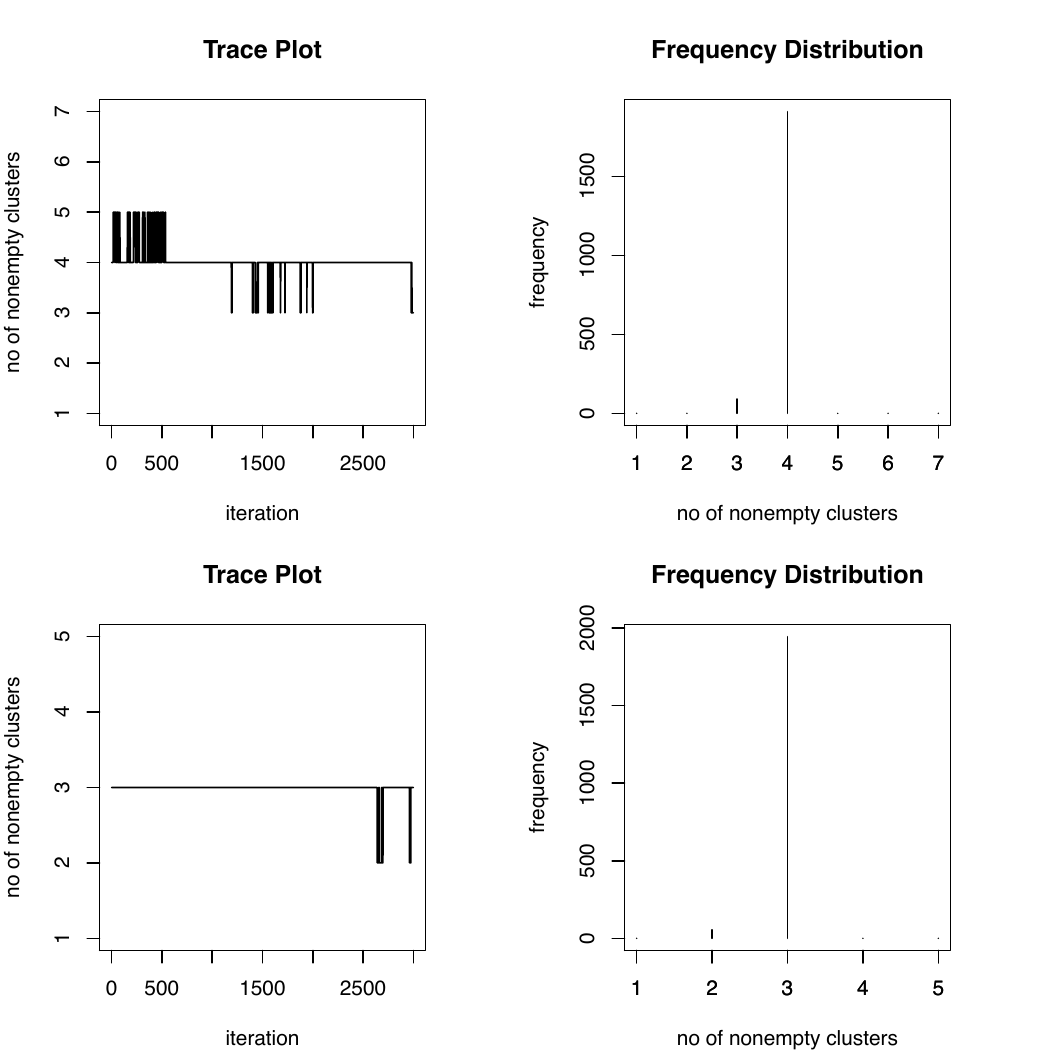}
\end{center}
\caption{\baselineskip=10pt 
Trace plots and frequency distributions of the number of nonempty clusters produced by the $\text{MLFA}_{2}$ (mixtures of latent factor analyzers) method for conditionally heteroscedastic multivariate Laplace ($f_{\bepsilon}^{(4)}$) distributed measurement errors with sample size $n=1000$, $m_{i}=3$ replicates for each subject and identity matrix (I) for the component specific covariance matrices.  
See Section \ref{sec: mvt simulation studies} and Section \ref{sec: mvt additional simulation studies} for additional details.
The upper panels are for the  $f_{\bX}$ and the lower panels are for the density of the scaled errors $f_{\bepsilon}$.
The results correspond to the simulation instance that produced the median of the estimated integrated squared errors (ISE) out of a total of 100 simulated data sets, 
when the number of mixture components for $f_{\bX}$ and $f_{\bepsilon}$ were kept fixed at $K_{\bX}=7$ and $K_{\bepsilon}=5$, respectively.
}
\label{fig: mvt simulation results Trace Plots d4 n1000 m3 MLFA X1 HT_E0 Ind}
\end{figure}

The results of the simulation experiments the measurement errors are distributed according to $f_{\bepsilon}^{(3)}=\MVT_{4}(6,\bzero,\bSigma)$ 
and $f_{\bepsilon}^{(4)}=\MVL_{4}(\bzero,\bSigma)$ probability laws independently of $\bX$ are presented in Table \ref{tab: mvt MISEs homoscedastic t and Laplace}.
The results for conditionally heteroscedastic measurement errors are presented in Table \ref{tab: mvt MISEs heteroscedastic t and Laplace}.  
In both cases, $\bX$ is distributed according to the mixture of multivariate normals described in Section \ref{sec: mvt simulation studies} of the main paper. 
As in the main paper, in each case four different choices for the covariance matrix $\bSigma$ were considered. 
The general patterns of the estimated MISEs are similar to that observed in Table \ref{tab: mvt MISEs heteroscedastic} of the main paper  
where the true measurement error distributions were finite mixtures of multivariate normal kernels.  
While in theory the MLFA model described in the main paper can approximate distributions like the multivariate Laplace that puts significant mass around the origin, 
in practice, since it assumes $\bOmega_{k}=\bOmega=\diag\{\sigma_{1}^{2},\dots,\sigma_{p}^{2}\}$ for all $k$, 
it often smooths out the spikes at the origin. 
A mild variation, referred to as the $\text{MLFA}_{2}$ model, that instead assumes $\bOmega_{k}=\sigma_{k}^{2}\Ind_{p}$ 
and results in slight improvement in the MISE performance is also included in Table \ref{tab: mvt MISEs homoscedastic t and Laplace} and Table \ref{tab: mvt MISEs heteroscedastic t and Laplace}. 
For the simulation experiments and the real data analysis presented in the main text, the two versions of the MLFA model perform very similarly and the latter version was not included. 
Results for conditionally heteroscedastic multivariate Laplace errors with diagonal covariance structure are summarized in Figures \ref{fig: mvt simulation results VFn d4 n1000 m3 MIW X1 HT_E0 Ind}-\ref{fig: mvt simulation results Trace Plots d4 n1000 m3 MLFA X1 HT_E0 Ind} 
with observations similar to those discussed in Section \ref{sec: mvt simulation studies} of the main paper.

\newpage
\thispagestyle{empty}
\newgeometry{left=2cm,right=2.5cm,top=2.5cm,bottom=0.1cm}

\begin{table}[!ht]\footnotesize
\begin{center}
\begin{tabular}{|c|c|c|c c c c|}
\hline
\multirow{2}{80pt}{True Error Distribution} & \multirow{2}{50pt}{Covariance Structure} & \multirow{2}{*}{Sample Size}	& \multicolumn{4}{|c|}{MISE $\times 10^4$} \\ \cline{4-7}
				&												&		& $\text{MLFA}_{2}$ 		& MLFA		& MIW		& Naive	  \\  \hline\hline
\multirow{8}{80pt}{(c)  Multivariate t}	& \multirow{2}{50pt}{\centering I}	
																& 500	& \bf{1.06}		& 1.38		& 3.98 		& 12.32		\\
				&												& 1000	& \bf{0.53}		& 0.65	 	& 1.54		& 9.91		\\ \cline{2-7}
				& \multirow{2}{50pt}{\centering LF}  						& 500	& \bf{6.62}		& 8.26	 	& 7.57		& 47.22		\\
				&												& 1000	& 4.73		& 5.78 		& \bf{3.65} 	& 45.70		\\  \cline{2-7}
				& \multirow{2}{50pt}{\centering AR}  						& 500	& 12.69		& 13.56		& \bf{6.14}		& 40.76		\\
				&												& 1000	& 11.36		& 9.16		& \bf{3.45}		 & 39.59		\\ \cline{2-7}
				& \multirow{2}{50pt}{\centering EXP}  					& 500	& 7.84		& 8.42		& \bf{5.00}		& 26.85		\\
				&												& 1000	& 6.26		& 6.64		& \bf{2.38}		& 26.04		\\ \hline\hline
\multirow{8}{80pt}{(d) Multivariate Laplace}	& \multirow{2}{50pt}{\centering I}	
																& 500	& \bf{1.08}		& 1.32		& 3.08		& 8.22		\\
				&												& 1000	& \bf{0.50}		& 0.63		& 1.20		& 6.25		\\ \cline{2-7}
				& \multirow{2}{50pt}{\centering LF}  						& 500	& \bf{4.41}		& 5.57		& 5.66		& 32.31		\\
				&												& 1000	& \bf{2.38}		& 3.53		& 2.84		& 31.10		\\  \cline{2-7} 
				& \multirow{2}{50pt}{\centering AR}  						& 500	& 8.38		& 8.72		& \bf{5.14}		& 27.30		\\
				&												& 1000	& 6.08		& 6.19		& \bf{2.56}		& 26.19		\\ \cline{2-7}
				& \multirow{2}{50pt}{\centering EXP}  					& 500	& 5.24		& 5.67		& \bf{4.14}		& 17.57		\\
				&												& 1000	& 3.58		& 4.17		& \bf{1.98}		& 16.86		\\ \hline

\end{tabular}
\caption{\baselineskip=10pt 
Mean integrated squared error (MISE) performance
of MLFA (mixtures of latent factor analyzers) and MIW (mixtures with inverse Wishart priors) density deconvolution models 
for {\bf homoscedastic} errors compared with a naive method that ignores measurement errors for different measurement error distributions.
See Section \ref{sec: mvt density deconvolution models} and Section \ref{sec: mvt additional simulation studies} for additional details. 
The minimum value in each row is highlighted.
}
\label{tab: mvt MISEs homoscedastic t and Laplace}
\end{center}
\end{table}

\begin{table}[!ht]\footnotesize
\begin{center}
\begin{tabular}{|c|c|c|c c c c|}
\hline
\multirow{2}{80pt}{True Error Distribution} & \multirow{2}{50pt}{Covariance Structure} & \multirow{2}{*}{Sample Size}	& \multicolumn{4}{|c|}{MISE $\times 10^4$} \\ \cline{4-7}
				&												&		& $\text{MLFA}_{2}$		& MLFA 	& MIW		& Naive	  \\  \hline\hline
\multirow{8}{80pt}{(c)  Multivariate t}	& \multirow{2}{50pt}{\centering I}	
																& 500	& \bf{2.78}		& 3.25	& 24.48 	& 19.10		\\
				&												& 1000	& \bf{1.39} 	& 1.53	& 13.40	& 17.75		\\ \cline{2-7}
				& \multirow{2}{50pt}{\centering LF}  						& 500	& \bf{12.65} 	& 14.72	& 52.77	& 69.64		\\
				&												& 1000	& \bf{6.71} 	& 8.43	& 25.66 	& 66.49		\\  \cline{2-7}
				& \multirow{2}{50pt}{\centering AR}  						& 500	& \bf{20.54}	& 23.2	& 43.22	& 64.07		\\
				&												& 1000	& \bf{13.53}	& 18.41	& 21.42	& 59.81		\\ \cline{2-7}
				& \multirow{2}{50pt}{\centering EXP}  					& 500	& \bf{11.56}	& 14.12	& 37.68	& 43.57		\\
				&												& 1000	& \bf{8.19}		& 11.97	&18.22	& 41.66		\\ \hline\hline
\multirow{8}{80pt}{(d) Multivariate Laplace}	& \multirow{2}{50pt}{\centering I}	
																& 500	& \bf{1.81}		& 2.32	& 9.60	& 10.31			\\
				&												& 1000	& \bf{0.97}		& 1.20	& 4.20	& 8.86			\\ \cline{2-7}
				& \multirow{2}{50pt}{\centering LF}  						& 500	& \bf{7.33}		& 10.30	& 17.52	& 41.89			\\
				&												& 1000	& \bf{3.99}		& 5.28	& 7.65	& 40.93			\\  \cline{2-7} 
				& \multirow{2}{50pt}{\centering AR}  						& 500	& \bf{9.79}		& 14.13	& 15.64	& 35.50			\\
				&												& 1000	& \bf{5.54}		& 9.32	& 6.59	& 34.91			\\ \cline{2-7}
				& \multirow{2}{50pt}{\centering EXP}  					& 500	& \bf{7.26}		& 9.90	& 13.93	& 23.71			\\
				&												& 1000	& \bf{3.90}		& 5.12	& 5.19	& 22.78			\\ \hline

\end{tabular}
\caption{\baselineskip=10pt 
Mean integrated squared error (MISE) performance
of MLFA (mixtures of latent factor analyzers) and MIW (mixtures with inverse Wishart priors) density deconvolution models 
for {\bf conditionally heteroscedastic} errors compared with a naive method that ignores measurement errors for different measurement error distributions.
See Section \ref{sec: mvt density deconvolution models} and Section \ref{sec: mvt additional simulation studies} for additional details. 
The minimum value in each row is highlighted.
}
\label{tab: mvt MISEs heteroscedastic t and Laplace}
\end{center}
\end{table}
\restoregeometry

\pagebreak\newpage
\thispagestyle{empty}
\baselineskip=17pt

We also extend the simulation experiments to scenarios when $\bX$ is distributed according to
(B) $f_{\bX}^{(3)} = \MVT_{4}(6,\bmu_{\bX},\bSigma_{\bX}), \bmu_{\bX}=(2,2,2,2)\trans$, 
(C) $f_{\bX}^{(4)} = \sum_{k=1}^{2}\pi_{\bX,k} \MVT_{4}(6,\bmu_{\bX,k},\bSigma_{\bX})$, $\bpi_{\bX}=(0.75,0.25)\trans, \bmu_{\bX,1}=(2,4,2,2)\trans, \bmu_{\bX,2}=(4,2,4,2)\trans$.
In each case, four different choices for $\bSigma_{\bX}$ are considered as in Section \ref{sec: mvt simulation studies} of the main paper. 
We focus on the case when the measurement errors are conditionally heteroscedastic. 
Results are presented in Tables \ref{tab: mvt MISEs heteroscedastic mvt} and \ref{tab: mvt MISEs heteroscedastic mixture mvt}.

\begin{table}[!ht]\footnotesize
\begin{center}
\begin{tabular}{|c|c|c|c|c c c|}
\hline
\multirow{2}{85pt}{True Distribution of Interest $f_{\bX}$} & \multirow{2}{75pt}{True Error Distribution $f_{\bepsilon}$} & \multirow{2}{50pt}{Covariance Structure} & \multirow{2}{*}{Sample Size}	& \multicolumn{3}{|c|}{MISE $\times 10^4$} \\ \cline{5-7}
				&&&												& $\text{MLFA}_{2}$		& MIW		& Naive	  \\  \hline\hline
\multirow{32}{85pt}{(B) Multivariate t}							& \multirow{8}{75pt}{(a) Multivariate Normal}	& \multirow{2}{50pt}{\centering I}	
																& 500	& \bf{4.35}		& 20.36		& 18.17	\\
				& &												& 1000	& \bf{2.36}		& 13.14		& 12.65	\\ \cline{3-7}
				& & \multirow{2}{50pt}{\centering LF}						& 500	& \bf{21.31}	& 78.22		& 75.42	\\
				& &												& 1000	& \bf{15.57}	& 52.73		& 67.77	\\  \cline{3-7} 
				& & \multirow{2}{50pt}{\centering AR}  					& 500	& \bf{33.18}	& 59.77		& 63.33	\\
				& &												& 1000	& \bf{29.29}	& 51.11		& 53.40	\\  \cline{3-7}
				& & \multirow{2}{50pt}{\centering EXP}  					& 500	& \bf{19.58}	& 40.72		& 44.83 	\\
				& &												& 1000	& \bf{17.78}	& 32.01		& 37.58	\\ \cline{2-7}
				& \multirow{8}{75pt}{(b) Mixture of Multivariate Normals}		& \multirow{2}{50pt}{\centering I}	
																& 500	& \bf{5.16}		& 27.21 		& 38.03	\\
				& &												& 1000	& \bf{2.87}		& 18.17		& 35.99	\\ \cline{3-7}
				& & \multirow{2}{50pt}{\centering LF}  					& 500	& \bf{27.89}	& 73.75 		& 159.29	\\
				& &												& 1000	& \bf{19.27}	& 53.66		& 161.77	\\  \cline{3-7}
				& & \multirow{2}{50pt}{\centering AR}  					& 500	& \bf{38.41}	& 81.77		& 159.34	\\
				& &												& 1000	& \bf{34.22}	& 55.25		& 156.05	\\  \cline{3-7}
				& & \multirow{2}{50pt}{\centering EXP}  					& 500	& \bf{21.95}	& 45.76		& 100.33	\\
				& &												& 1000	& \bf{18.14}	& 37.72		& 99.09	\\ \cline{2-7}
				& \multirow{8}{80pt}{(c)  Multivariate t}					& \multirow{2}{50pt}{\centering I}	
																& 500	& \bf{4.16}		& 27.73		& 23.42	\\
				& &												& 1000	& \bf{2.34} 	& 19.87		& 20.36	\\ \cline{3-7}
				& & \multirow{2}{50pt}{\centering LF}  					& 500	& \bf{22.83} 	& 91.04		& 90.39	\\
				& &												& 1000	& \bf{14.03} 	& 85.33  		& 89.31 	\\  \cline{3-7}
				& & \multirow{2}{50pt}{\centering AR}  					& 500	& \bf{40.60}	& 76.40 		& 86.87	\\
				& &												& 1000	& \bf{36.93}	& 70.76		& 75.19	\\ \cline{3-7}
				& & \multirow{2}{50pt}{\centering EXP}  					& 500	& \bf{26.36}	& 55.65 		& 61.25	\\
				& &												& 1000	& \bf{18.51}	& 40.46		& 49.52	\\ \cline{2-7}
				& \multirow{8}{80pt}{(d) Multivariate Laplace}				& \multirow{2}{50pt}{\centering I}	
																& 500	& \bf{3.93}		& 16.48		& 16.14 	\\
				& &												& 1000	& \bf{1.81}		& 6.85		& 14.02	\\ \cline{3-7}
				& & \multirow{2}{50pt}{\centering LF}  					& 500	& \bf{16.36}	& 47.19		& 70.22	\\
				& &												& 1000	& \bf{12.13}	& 27.64		& 59.48	\\  \cline{3-7} 
				& & \multirow{2}{50pt}{\centering AR}  					& 500	& \bf{29.46}	& 42.44		& 63.79	\\
				& &												& 1000	& \bf{18.81}	& 21.19		& 47.92	\\ \cline{3-7}
				& & \multirow{2}{50pt}{\centering EXP}  					& 500	& \bf{19.00}	& 34.74		& 39.64	\\
				& &												& 1000	& \bf{13.30}	& 16.24		& 32.76	\\ \hline
\end{tabular}
\caption{\baselineskip=11pt 
Mean integrated squared error (MISE) performance
of MLFA (mixtures of latent factor analyzers) and MIW (mixtures with inverse Wishart priors) density deconvolution models 
for {\bf conditionally heteroscedastic} errors compared with a naive method that ignores measurement errors for different measurement error distributions.
See Section \ref{sec: mvt density deconvolution models} and Section \ref{sec: mvt additional simulation studies} for additional details. 
The minimum value in each row is highlighted.
}
\label{tab: mvt MISEs heteroscedastic mvt}
\end{center}
\end{table}

\newpage
\thispagestyle{empty}

\begin{table}[!ht]\footnotesize
\begin{center}
\begin{tabular}{|c|c|c|c|c c c|}
\hline
\multirow{2}{85pt}{True Distribution of Interest $f_{\bX}$} & \multirow{2}{75pt}{True Error Distribution $f_{\bepsilon}$} & \multirow{2}{50pt}{Covariance Structure} & \multirow{2}{*}{Sample Size}	& \multicolumn{3}{|c|}{MISE $\times 10^4$} \\ \cline{5-7}
				&&&												& $\text{MLFA}_{2}$		& MIW		& Naive	  \\  \hline\hline
\multirow{32}{85pt}{(C) Mixture of Multivariate t}							& \multirow{8}{75pt}{(a) Multivariate Normal}	& \multirow{2}{50pt}{\centering I}	
																& 500	& \bf{4.84}		& 13.68		& 12.43 	\\
				& &												& 1000	& \bf{2.82}		& 7.41		& 10.15 	\\ \cline{3-7}
				& & \multirow{2}{50pt}{\centering LF}						& 500	& \bf{21.62}	& 30.01			& 47.95 	\\
				& &												& 1000	& \bf{13.40}	& 19.72		& 44.97 	\\  \cline{3-7} 
				& & \multirow{2}{50pt}{\centering AR}  					& 500	& \bf{22.56}	& 29.35		& 43.99	 \\
				& &												& 1000	& \bf{19.80}	& 25.59		& 39.63	 \\  \cline{3-7}
				& & \multirow{2}{50pt}{\centering EXP}  					& 500	& \bf{18.36}	& 27.27		& 28.00	 \\
				& &												& 1000	& \bf{13.41}	& 17.73		& 25.14 	\\ \cline{2-7}
				& \multirow{8}{75pt}{(b) Mixture of Multivariate Normals}		& \multirow{2}{50pt}{\centering I}	
																& 500	& \bf{5.39}		& 14.64		& 22.90	\\
				& &												& 1000	& \bf{2.80}		& 10.77		& 21.55	\\ \cline{3-7}
				& & \multirow{2}{50pt}{\centering LF}  					& 500	& \bf{24.48}	& 32.87		& 98.00	\\
				& &												& 1000	& \bf{15.62}	& 20.52		& 98.79	\\  \cline{3-7}
				& & \multirow{2}{50pt}{\centering AR}  					& 500	& \bf{26.73}	& 31.09		& 90.78	\\
				& &												& 1000	& \bf{23.44}	& 29.06		& 91.24	\\  \cline{3-7}
				& & \multirow{2}{50pt}{\centering EXP}  					& 500	& \bf{19.56}	& 25.39		& 58.83	\\
				& &												& 1000	& \bf{13.90}	& 18.29		& 59.93	\\ \cline{2-7}
				& \multirow{8}{80pt}{(c)  Multivariate t}					& \multirow{2}{50pt}{\centering I}	
																& 500	& \bf{4.91}		& 18.09 	& 16.30		\\
				& &												& 1000	& \bf{2.89} 	& 11.59	& 14.00		\\ \cline{3-7}
				& & \multirow{2}{50pt}{\centering LF}  					& 500	& \bf{23.50} 	& 33.79 	& 60.18		\\
				& &												& 1000	& \bf{15.85} 	& 25.83	& 58.20 		\\  \cline{3-7}
				& & \multirow{2}{50pt}{\centering AR}  					& 500	& \bf{26.98}	& 33.78	& 54.07		\\
				& &												& 1000	& \bf{22.04}	& 29.77	& 51.64		\\ \cline{3-7}
				& & \multirow{2}{50pt}{\centering EXP}  					& 500	& \bf{18.62}	& 24.00	& 36.26		\\
				& &												& 1000	& \bf{12.64}	& 18.57	& 33.61		\\ \cline{2-7}
				& \multirow{8}{80pt}{(d) Multivariate Laplace}				& \multirow{2}{50pt}{\centering I}	
																& 500	& \bf{4.76}		& 9.34	& 15.96 		\\
				& &												& 1000	& \bf{2.33}		& 5.04	& 13.96		\\ \cline{3-7}
				& & \multirow{2}{50pt}{\centering LF}  					& 500	& \bf{16.59}	& 22.54	& 65.33		\\
				& &												& 1000	& \bf{11.69}	& 13.41	& 59.25		\\  \cline{3-7} 
				& & \multirow{2}{50pt}{\centering AR}  					& 500	& \bf{24.73}	& 26.21	& 58.87		\\
				& &												& 1000	& \bf{15.71}	& 17.48	& 47.62		\\ \cline{3-7}
				& & \multirow{2}{50pt}{\centering EXP}  					& 500	& \bf{14.26}	& 19.12	& 34.53		\\
				& &												& 1000	& \bf{10.96}	& 13.25	& 32.47		\\ \hline
\end{tabular}
\caption{\baselineskip=11pt 
Mean integrated squared error (MISE) performance
of MLFA (mixtures of latent factor analyzers) and MIW (mixtures with inverse Wishart priors) density deconvolution models 
for {\bf conditionally heteroscedastic} errors compared with a naive method that ignores measurement errors for different measurement error distributions.
See Section \ref{sec: mvt density deconvolution models} and Section \ref{sec: mvt additional simulation studies} for additional details. 
The minimum value in each row is highlighted.
}
\label{tab: mvt MISEs heteroscedastic mixture mvt}
\end{center}
\end{table}


\newpage
\thispagestyle{empty}

\newpage
\thispagestyle{empty}

\newpage
\thispagestyle{empty}

\newpage
\thispagestyle{empty}

\newpage
\thispagestyle{empty}

\newpage
\thispagestyle{empty}

\newpage
\thispagestyle{empty}

\newpage
\section{Potential Impact on Nutritional Epidemiology}\label{sec: mvt potential impact}

The joint distribution of long-term average intakes of different dietary components allows nutritionists 
to study the dietary habits of the population of interest in fine detail. 
The plots of pairwise marginal distributions presented in Figure \ref{fig: mvt EATS data results XS}, for instance, 
provide detailed information on the joint consumption patterns of different pairs of dietary components. 
While such graphical summaries of the joint distributions may not be available for more than two components, 
numerical summaries of the joint distribution can provide answers to important questions 
such as what proportion of the population consume certain dietary components above, between or below certain amounts etc. 
The last question is particularly important as it relates to the proportion of the population that are deficient in certain dietary components. 
Focusing again on a two-dimensional case for illustration, namely Fiber and Potassium, 
Figure \ref{fig: mvt fiber and potassium cdf} below shows their marginal and joint cumulative distribution function (CDF) on a set of grid points 
from which such proportions can be readily obtained. 
Dietary components are often reported in different measurement units. 
The figures presented in Section \ref{sec: mvt data analysis} are based on a linear scale transformation $W_{ij\ell}=20 \times \{W_{ij\ell,obs}-W_{ij\ell,obs,min}\} / \{W_{ij\ell,obs, max} - W_{ij\ell,obs,min}\}$ 
so that the $W_{ij\ell}$ for different components are unitless and fall between 0 and 20 units. 
Figure \ref{fig: mvt fiber and potassium cdf} report the marginal and the joint CDF of fiber and potassium on a set of grid points in their original measurement units. 
We can readily see that, considered jointly, 
approximately $59\%$ of adult Americans consume less than 20.55 grams of fiber and 3338.55 milligrams of potassium, 
whereas the corresponding marginal values are $71.2\%$ and $67.6\%$, respectively.

The focus of the nutritional epidemiology examples considered in this article were 
on the estimation of joint consumption patterns of a set of regularly consumed dietary components  
whose reported intakes were all continuously measured.
In contrast, for dietary components that are consumed episodically, the reported intakes equal zero on non-consumption days, and are positive on consumption days. 
The methodology developed in this article paves the way to more sophisticated deconvolution methods that can accommodate such zero inflated data. 
We are pursuing this problem as the subject of a separate study, with promising preliminary results.
This will be a crucial step forward towards providing a highly flexible statistical framework for estimating the distribution of the U.S. Department of Agriculture's Healthy Eating Index (HEI, \href{www.cnpp.usda.gov/HealthyEatingIndex.htm}{www.cnpp.usda.gov/HealthyEatingIndex.htm}). 
HEI is a measure of diet quality that involves six episodically and seven regularly consumed dietary components and is used to assess compliance with the U.S. Dietary Guidelines for Americans (\href{www.health.gov/dietaryguidelines}{www.health.gov/dietaryguidelines}) and monitor changes in dietary patterns.  
Efficient estimation of the distribution of HEI will allow nutritionists to answer public health questions that have important policy implications. 
We expect successful implementation of our methods  
to eventually replace the currently popular NCI method 
\href{www.riskfactor.cancer.gov/diet/usualintakes/method.html}{(www.riskfactor.cancer.gov/diet/usualintakes/method.html)} for estimation of HEI.

\pagebreak\newpage
\thispagestyle{empty}

\newgeometry{left=2cm,right=2.5cm,top=1cm,bottom=1cm}
\begin{figure}[ht]
\begin{center}
\includegraphics[height=6cm, width=18cm, trim=0cm 0cm 1.5cm 1cm, clip=true]{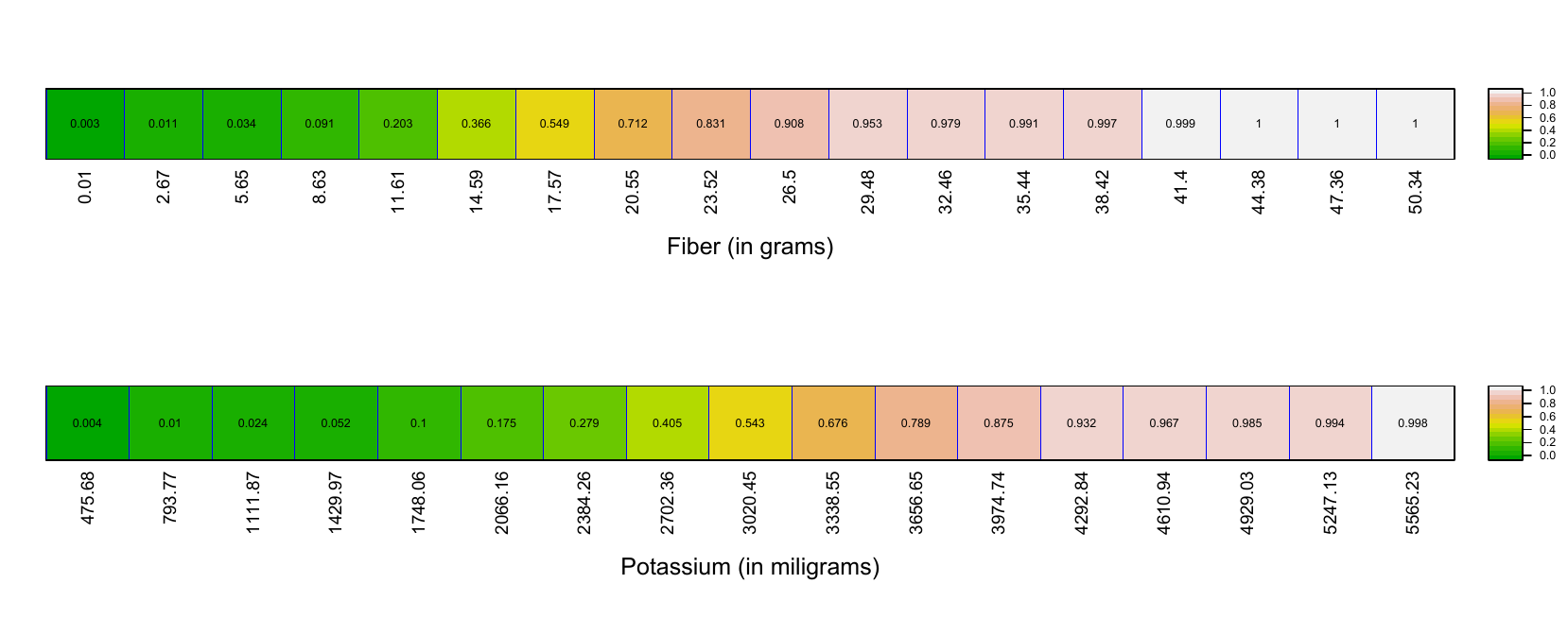}
\includegraphics[height=14cm, width=18cm, trim=0cm 0.25cm 0cm 0.5cm, clip=true]{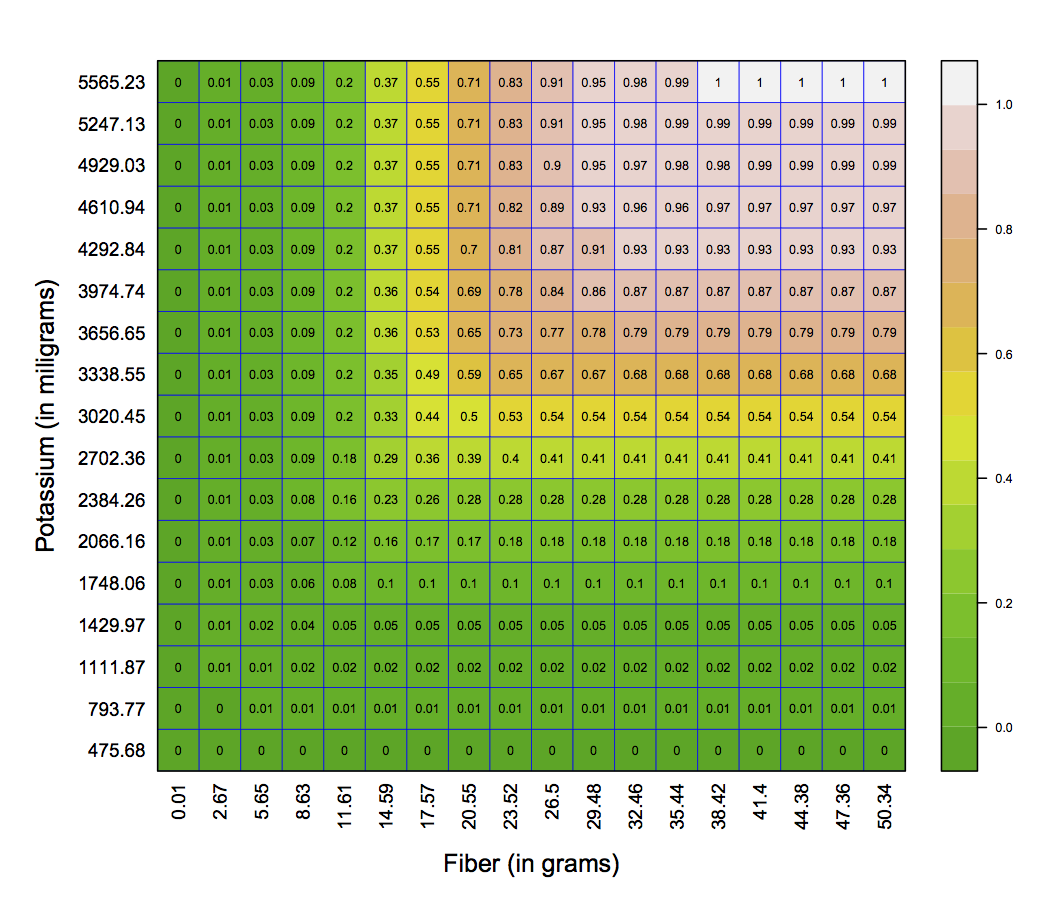}
\end{center}
\caption{\baselineskip=10pt Results for Fiber and Potassium in their commonly used measurement units. 
The top two panels show their marginal cumulative distribution functions. 
The bottom panel shows their joint cumulative distribution function for a set of grid points. 
The figure is in color in the electronic version of this article.
}
\label{fig: mvt fiber and potassium cdf}
\end{figure}
\restoregeometry

\newpage
\baselineskip=17pt
\section*{Additional References}
\vspace{-2ex}
\refmark
Bickel, P. J. and Kleijn, B. J. K. (2012). The semiparametric Bernstein-von Mises theorem.
\ANNALS, 40, 206-237.
\refmark
Bontemps, D. (2011). Bernstein-von Mises theorems for Gaussian regression with increasing number of regressors.
\ANNALS, 39, 2557-2584.
\refmark
Carroll, R. J.,  Chen X. and Hu, Y. (2010). Identification and estimation of nonlinear models using two samples with nonclassical measurement errors.
\JNS, 22, 379-399.
\refmark
Castillo, I. and Nickl, R. (2014). On the Bernstein-von Mises phenomenon for nonparametric Bayes procedures. 
\ANNALS, 42, 1941-1969.
\refmark
de Boor, C. (2000). \emph{A Practical Guide to Splines}.
New York: Springer.
\refmark
 d'Haultfoeuille, X. (2011). On the completeness condition in nonparametric instrumental problems. 
\ECTH, 27, 460-471.
\refmark
Eltoft, T., Kim, T. and Lee, T. W. (2006). On the multivariate Laplace distribution.
\IEEESPL, 13, 300-303. 
\refmark
Escobar, M. D. and West, M. (1995). Bayesian density estimation and inference using mixtures.
\JASA, 90, 577-588.
\refmark
Eubank, R. L. and Hart, J. D. (1992). Testing goodness-of-fit in regression via order selection criteria. 
\ANNALS, 20, 1412-1425. 
\refmark
Ferguson, T. F. (1973). A Bayesian analysis of some nonparametric problems.
\ANNALS, 1, 209-230.
%
%
\refmark
Fraley, C. and Raftery, A. E. (2007). Model-based methods of classification: using the mclust software in chemometrics.
\JSS, 18, 1-13.
%
%
%
\refmark
Ghosh, J. K. and Ramamoorthi, R. V. (2010). \emph{Bayesian Nonparametrics}. 
New York: Springer.
\refmark
Goldberg, R. R . (1961). \emph{Fourier transforms}. Volume 32. London: Cambridge.
\refmark
Green, J. P., Latuszynski, K. Pereyra, M. and Roberts, C. P. (2015). Bayesian computation: summary of the current state, and samples backwards and forwards.
\SaC, 25, 835-862.
\refmark
Hastie, D. I., Liverani, S. and Richrdson, S. (2015). Sampling from Dirichlet process mixture models with unknown concentration parameter: mixing issues in large data implementations.
\SaC, 25, 1023-1037.
%
%
\refmark
Ishwaran, H. and James, L. F. (2002). Approximate Dirichlet process computing in finite normal mixtures: smoothing and prior information. 
\JCGS, 11, 508-532.
\refmark
Ishwaran, H. and Zarepour, M. (2000). Markov chain Monte Carlo in approximate Dirichlet and beta two-parameter process hierarchical models. 
\BIOK, 87, 371-390.
\refmark
Ishwaran, H. and Zarepour, M. (2002). Exact and approximate sum-representations for the Dirichlet process. 
\CANADAJS, 30, 269-283.
\refmark
Johnstone, I. M. (2010). High dimensional Bernstein-von Mises: simple examples. 
{\it Institute of Mathematical Statistics Collections}, 6, 87-98.
\refmark
Kotz, S. and Nadarajah, S. (2004). \emph{Multivariate t Distributions and Their Applications}. 
Cambridge: Cambridge University Press. 
%
%
\refmark
Neal, R. M. (2000). Markov chain sampling methods for Dirichlet process mixture models.
\JCGS, 9, 249-265.
%
%
\refmark
Norets, A. and Pelenis, J. (2012). Bayesian modeling of joint and conditional distributions.
\JECM, 168, 332-346. 
%
%
\refmark
Pati, D. and Dunson, D. (2013). Bayesian nonparametric regression with varying residual density.
\ANNALSISM, 66, 1-13.
\refmark
Pelenis, J. (2014). Bayesian Regression with Heteroscedastic Error Density and Parametric Mean Function.
\JECM, 178, 624-638.
\refmark
Rocke, D. and Durbin, B. (2001). {A model for measurement error for gene expression arrays}.
\emph{Journal of Computational Biology}, 8, 557-569.
\refmark
Rousseau, J. and Mengersen, K. (2011). Asymptotic behavior of the posterior distribution in overfitted mixture models
\JRSSB, 73, 689-710.
%
%
\refmark
Sethuraman, J. (1994). A constructive definition of Dirichlet priors. 
\SSNC, 4, 639-650.
%
%
\refmark
Spokoiny, V. (2013). Bernstein-von Mises theorem for growing parameter dimension. 
{\it arXiv preprint arXiv:1302.3430}.

\end{document}